\documentclass[11pt]{article}

\usepackage{cite}
 \usepackage{subfigure}
\usepackage{multirow}
\usepackage{helvet}
\usepackage{amsmath}
\usepackage{amssymb}
\usepackage{setspace}
\usepackage{graphicx}
\usepackage{empheq}
\usepackage{soul}
\usepackage{tabularx}
\usepackage{array}
\usepackage{float}
\usepackage{braket}
\usepackage[utf8]{inputenc}
\usepackage{url}
\usepackage{hyperref}
\usepackage{slashed}
\usepackage[font=footnotesize,labelfont=bf]{caption}
\usepackage{color}

\def\be{\begin{equation}}
\def\ee{\end{equation}}

\usepackage[a4paper,top=3cm,bottom=3cm,left=2.5cm,right=2.5cm,bindingoffset=0mm]{geometry}

\begin{document}
\hspace*{\fill} DESY-24-101

\begin{center}

{\Large \bf A Stress Test of Global PDF Fits:  \\ \vspace*{0.15cm} Closure Testing the MSHT PDFs and a First Direct  \\ \vspace*{0.15cm} Comparison to the  Neural Net Approach}

\vspace*{1cm}
L. A. Harland-Lang$^{a}$, T. Cridge$^b$ 
and R.S. Thorne$^a$\\                                               
\vspace*{0.5cm}                                                    
$^a$ Department of Physics and Astronomy, University College London, London, WC1E 6BT, UK \\   
$^b$ Deutsches Elektronen-Synchrotron DESY, Notkestr. 85, 22607 Hamburg, Germany 

\begin{abstract}
\noindent We present a first global closure test of the fixed parameterisation (MSHT) approach to PDF fitting. We find that the default MSHT20 parameterisation can reproduce the features of the input set in such a closure test to well within the textbook uncertainties. This provides strong evidence that parameterisation inflexibility in the MSHT20 fit is not a significant issue in the data region. We also present the first completely like--for--like comparison between two global PDF fits, namely MSHT and NNPDF, where the only difference is guaranteed to be due to the fitting methodology. To achieve this, we present a fit to the NNPDF4.0 data and theory inputs, but with the MSHT fixed parameterisation. We find that this gives a moderately, but noticeably, better fit quality than the central NNPDF4.0 fits, both with perturbative and fitted charm, and that this difference persists at the level of the PDFs and benchmark cross sections. The NNPDF4.0 uncertainties are found to be broadly in line with the MSHT results if a textbook $T^2=1$ tolerance is applied, but to be significantly smaller if a tolerance typical of the MSHT20 fit is applied. This points to an inherent inconsistency between these approaches.  We discuss the need for an enlarged tolerance criterion in global PDF fits in detail, and demonstrate the impact of data/theory inconsistencies in the closure test setting; namely, these do not lead to any increase in the $T^2=1$ PDF uncertainty. We also investigate the impact of restricting the PDF parameterisation to have fewer free parameters than the default MSHT20 case, and find this can be significant at the level of both closure tests and the full fit. 

\end{abstract}

\end{center}
 
\begin{spacing}{1.2}
\clearpage
\tableofcontents
\clearpage
\end{spacing}

\section{Introduction}

With the wealth of data being taken at the Large Hadron Collider (LHC), and the advanced theoretical calculations being performed to match this, the LHC physics programme now lies firmly in the high precision regime. A key ingredient, and indeed bottleneck, in almost all precision LHC analyses derives from our knowledge of the structure of the proton, as encoded in the parton distribution functions (PDFs). Indeed, for many recent experimental analyses, most notably Electroweak precision studies, PDFs represent the dominant, or one of the most dominant, sources of uncertainty~\cite{ATLAS:2018gqq,LHCb:2021bjt,ATLAS:2023lhg,ATLAS:2024erm,CMS:2024aps}. More broadly they play a significant role in Higgs physics~\cite{Cepeda:2019klc,Amoroso:2022eow,Jones:2023uzh,Andersen:2024czj} and searches for physics beyond the Standard Model, see e.g.~\cite{Carrazza:2019sec,Gao:2022srd,Amoroso:2022eow}.

Motivated by this, there has been much progress in recent years towards precision determinations of the PDFs in global fits from the CT, MSHT and NNPDF groups~\cite{Hou:2019efy,Bailey:2020ooq,NNPDF:2021njg} (see also~\cite{H1:2015ubc,Alekhin:2017kpj,ATLAS:2021vod} for other determinations). These extract the PDFs by fitting to a wide range of data, from low energy fixed--target experiments, to HERA and the Tevatron and LHC hadron colliders. Within this, high precision data from the LHC is now in particular playing an increasingly significant role in such determinations. The theoretical predictions entering such fits are 
almost without exception now at the level of next--to--next--leading--order (NNLO) in the QCD coupling, while electroweak and QED corrections are also accounted for, in the latter case currently via dedicated fits~\cite{Xie:2021equ,NNPDF:2024djq,Cridge:2021pxm,Cridge:2023ryv}. More recently, these analyses have been extended to approximate ${\rm N}^3$LO (a${\rm N}^3$LO) order by the MSHT and then the NNPDF groups~\cite{McGowan:2022nag,NNPDF:2024nan}. These combine the significant amount of known information about the ${\rm N}^3$LO results while  including approximations for the unknown parts, with corresponding theoretical uncertainties associated with these and included in the PDF fit. Given the amount of known ${\rm N}^3$LO information available, these allow for an increased level of accuracy in comparison to previous NNLO PDF determinations. A variety of follow-up studies on different aspects of the a${\rm N}^3$LO PDFs have also been performed \cite{Jing:2023isu,Cridge:2023ozx,Cridge:2024exf,Cooper-Sarkar:2024crx} showing reduced data tensions, a preference in the fits for this level of theoretical precision and consistency amongst the two groups for the PDF evolution.

The progress in each of the individual global PDF fits, as described above, is therefore without doubt  significant. However, there remain important  questions when the comparison of the different fits is considered that so far remain unresolved.  In particular, there are key differences in the fitting methodologies that can in principle have a major impact on the resulting PDFs and their uncertainties. 

Two particularly relevant, and connected, differences relate to the parameterisation of the PDFs and the definition of their corresponding uncertainties. In the NNPDF fits a neural network (NN) is used to  parameterise the PDFs at the input scale, with in NNPDF4.0~\cite{NNPDF:2021njg} there being 763 free parameters in the NN, albeit such that these are effectively constrained via the usual process of training and validation a well as other prior constraints. In the CT and MSHT fits, on the other hand, a fixed polynomial basis is used to parameterise the PDFs. For MSHT20~\cite{Bailey:2020ooq} a basis of Chebyshev polynomials is used, with 52 free parameters in total, while in CT18 the baseline set is parameterised in terms of Bernstein polynomials, and has rather fewer free parameters.

In terms of the PDF uncertainties, rather different approaches are also taken by all three groups. In the CT and MSHT fits, the PDF uncertainties are defined directly in terms of an expansion of the Hessian matrix of second derivatives of the $\chi^2$ with respect to the PDF parameters around the  minimum of the fit to the global dataset. These uncertainties are in particular not defined using the textbook $\Delta \chi^2=T=1$ criterion, which would be applicable to the ideal scenario of complete statistical compatibility between the multiple datasets entering the fit, a completely faithful evaluation of the experimental uncertainties within each dataset, and theoretical calculations that match these exactly. There is a great deal of evidence that the first two situations do not hold in a PDF fit (see e.g.~\cite{Collins:2001es,Pumplin:2009sc,Kovarik:2019xvh}).
 For example, at the simplest level of the global fit quality, even in the most recent approximate ${\rm N}^3$LO fits~\cite{McGowan:2022nag,NNPDF:2024nan}, which come from both fixed parameterisation and neural network approaches, the $\chi^2$ per number of data points is $\sim 7$ standard deviations away from unity. Moreover, it is of course well known that the fixed--order theoretical predictions are not exact; although as described above there has been recent progress in evaluating the uncertainty due to this~\cite{McGowan:2022nag,NNPDF:2024dpb,NNPDF:2024nan}, the above  $\chi^2$ already accounts for this.

Given this, there is  much support for using  an enlarged `tolerance' $\Delta \chi^2=T^2$  in such global PDF fits~\cite{Collins:2001es,Pumplin:2001ct,Martin:2009iq,Pumplin:2009sc,Watt:2012tq,Kovarik:2019xvh}. The motivation for this has been arrived at from various perspectives, but is in general based upon observation of the global and dataset fit qualities in  global PDF fits, with often significant departures observed from the behaviour one might expect if the more stringent $T^2=1$ parameter fitting criterion were to apply. Moreover, when fits to subsets of the global datasets are performed with a tolerance of $T^2=1$, these are often found to be statistically in tension~\cite{Watt:2012tq,Kovarik:2019xvh,Hou:2019efy}.
In both the CT and MSHT fits, an enlarged tolerance is therefore taken. In the MSHT20 fit, a `dynamic' tolerance criterion is applied to evaluate the most appropriate enlarged tolerance to apply, with $T^2\sim 10$ on average. For CT18 a different approach is taken, which arrives at a somewhat larger value (see~\cite{Hou:2019efy} for details) and includes the effect of varying the parameterisation basis from their baseline (which is more restricted than in MSHT20).

In the NNPDF fit, on the other hand, no explicit  tolerance is applied. Indeed, here the method of error propagation itself is not directly amenable to the methods applied by the CT and MSHT collaborations for including such an enlarged tolerance. The NNPDF approach relies upon fitting a set of pseudodata replicas derived from fluctuating the global dataset by its corresponding uncertainties, with the ensemble of resulting fits providing the final PDF set and its uncertainties. This Monte Carlo (MC) replica error generation can equally be applied using a fixed parameterisation as in CT/MSHT, and indeed it is has long been established~\cite{Watt:2012tq} in such a case that the MC replica error generation is closely equivalent to the Hessian one, but only if $T^2=1$ is taken. In the NNPDF approach, on the other hand, given a more flexible NN parameterisation is used and a training/validation split of the data performed to avoid overfitting, this direct connection is less transparent. Indeed, the fit results and most notably the size of the PDF uncertainties are observed to change quite sensitively with changes to the underlying NN methodology~\cite{NNPDF:2021njg}.

Given the above discussion, we may reasonably expect the different approaches to global PDF fitting to lead to different results for the PDFs and their uncertainties, even if the data and theory underlying the fit are the same. Indeed this was  observed in~\cite{PDF4LHCWorkingGroup:2022cjn,Cridge:2021qjj}, for a set of benchmark NNLO fits, where each collaboration performed their default fit but to a common reduced dataset, with common cuts, and with the theory settings (quark masses, perturbative charm etc) also unified as much as possible. In this way any residual differences could be largely assigned to those of methodology alone. While the results for the PDFs were broadly compatible, differences in the PDF uncertainties were evident, with the NNPDF3.1 uncertainty in particular being on average markedly lower. Given the updated methodology of the NNPDF4.0 fit~\cite{NNPDF:2021njg} results in a further reduction in PDF uncertainties for the same dataset, this difference is certain to be seen even more clearly in any updated benchmarking. Indeed, such a difference at the level of the PDF uncertainties is observed in the comparison between NNPDF4.0 and MSHT20 and CT18, even if here the underlying datasets and the treatment of them is also different. 

These differences between the PDF sets, due to methodology alone, are therefore certainly significant and an understanding of them is arguably as relevant to the LHC precision programme as the continued important progress being made within each fit discussed above. It is the aim of this paper to begin to address these questions directly.

Focusing on the MSHT fit, we are left with three possible explanations of the differences observed in the benchmark fits~\cite{PDF4LHCWorkingGroup:2022cjn}, which may in principle all be true to some extent. Namely, the NNPDF  uncertainties may be too aggressive (i.e. too small), the MSHT uncertainties may be too large (i.e. too conservative) or the MSHT fit may be less accurate, due to the less flexibly underlying fixed parameterisation, and hence an enlarged PDF uncertainty is required. The latter explanation in particular relates to the question of the enlarged tolerance, and to what extent this is required by parameterisation inflexibility, rather than any inherent feature of the data/theory comparison itself. Any  contribution from these inherent features, in particular, should also in general be accounted for in a NN fit.  

To address these questions, in this paper we  perform for the first time a global closure test of the MSHT20 fixed parameterisation approach. Namely, we generate pseudodata, which is by construction self--consistent, corresponding to a global PDF dataset and perform a fit using the MSHT20 parameterisation. Such closure tests have long been  used by the NNPDF collaboration in order to assess the faithfulness of the NN approach in these conditions~\cite{NNPDF:2014otw}, but have thus far not been applied to global PDF fits with fixed parameterisations. We will in particular find that the default MSHT20 parameterisation can reproduce the features of the input set in such a closure test to well within the $T^2=1$ uncertainties (which is the appropriate definition for a, by construction self--consistent, closure test). This provides strong evidence that parameterisation inflexibility in the MSHT20 fit is not a significant issue in the data region, and that it should not be a major contribution in any enlarged tolerance. 

The above closure tests are performed by making use of the publicly available NNPDF code~\cite{nnpdfweb}. While this is set up by default to allow the user to perform PDF fits within the NNPDF framework, it is readily amenable to instead using the MSHT20 fixed parameterisation for the PDFs at the input scale. We implement this modification in the current paper, which  allows us to perform a full global PDF fit to precisely the same data and theory settings that enter the NNPDF4.0 NNLO fit,  with the only difference being due to the methodology of how the PDFs are parameterised and the uncertainties defined.

We perform this fit with both fitted and perturbative charm, and then  compare the resulting fit qualities, PDFs and their uncertainties that come from fitting the NNPDF4.0 data/theory with the MSHT parameterisation with the result of the public NNPDF4.0 fit itself. Crucially, we find that the fit quality in the MSHT case is somewhat {\it better} than that of  NNPDF4.0, for both treatments of the charm PDF. More significantly, the resulting PDFs and the predicted benchmark cross sections, are often not compatible within the nominal NNPDF uncertainties. 

Given the  in principle increased flexibility of the NN fit, this result may appear somewhat counter-intuitive, and can in part be explained by additional prior constraints on the PDFs imposed by NNPDF, most notably positivity of the low $x$ gluon in the perturbative charm fit. However, even accounting for this, non--negligible differences remain.  We therefore study in detail the resulting PDFs, and find that indeed the MSHT fit have moderately more flexibility associated with them, which therefore results in a better fit to the data being achieved. On the other hand, from a close examination of the breakdown in the fit quality between datasets, the form of the underlying PDFs, and the impact of restricting or extending the number of free parameters in the fit we find no particular evidence of overfitting (which is controlled against  in the NNPDF fits, given the more flexible NN architecture).

Putting the reasons for this difference aside, the fact that the MSHT fit quality is better than the central NNPDF4.0 case is again strong evidence that parameterisation inflexibility is not playing a major in any enlarged tolerance that is required, or more precisely an increased role in comparison to the NNPDF fit. Further to this, we also compare the corresponding PDF uncertainties and indeed find that with respect to the quark flavour decomposition the NNPDF4.0 uncertainties are rather closely in line with the MSHT $T^2=1$ uncertainties, while the gluon and quark singlet uncertainties are somewhat larger, but still significantly lower than the $T^2=10$ uncertainties, which are representative of the enlarged tolerance applied in e.g. MSHT20. This demonstrates in a completely unambiguous way what was already suggested by the comparison of~\cite{PDF4LHCWorkingGroup:2022cjn}, but highlights that it is the tolerance that is playing the major role here; if a textbook $T^2=1$ uncertainty were instead used in the MSHT analysis, then the PDF uncertainties would be much more closely in line with the NNPDF ones. Given we have provided strong evidence that parameterisation inflexibility is not a major factor in this for the MSHT fit, this points to an inherent inconsistency between the approaches that requires resolution. In particular, while MSHT20 uncertainties are quoted with a tolerance of $T^2 \sim 10$, the NNPDF4.0 uncertainties correspond more closely to a tolerance of $T^2 \sim 1$, while there is no evidence that the differing methodologies should require such a difference.

Indeed, as described above, there are very good reasons for including an enlarged tolerance in global PDF fits. In the current study,  we discuss in detail the role of departures from textbook statistics in a global PDF fit, and the tensions between datasets that are a feature of this\footnote{As mentioned above, these can be due to the incomplete nature of the theoretical predictions in the fit, an incomplete evaluation of the experimental uncertainties and their correlations in particular datasets, other deficiencies in the data, or to all of these effects. Our analysis is agnostic with respect to this question.}. We explicitly demonstrate the impact of including such tensions in a global closure test and show that these have essentially no impact on the resulting PDF uncertainties. Hence, these are found to be unrepresentative, as they do not account for the increased spread in the PDF error due to these tensions. This is completely in line with first principles considerations, and  demonstrates the need for a modification of the textbook uncertainty definition in either the Hessian or MC replica approach, in the presence of such tensions. This has recently also been discussed in~\cite{Yan:2024yir}. We also compare the result of the public MSHT20 fit with the MSHT fits presented here. In other words, these compare the difference due to the change in dataset and theory setting alone (keeping the fitting methodology fixed) in the resulting PDFs. We find broad consistency at the level of the PDFs and benchmark cross sections if an enlarged $T^2=10$ definition is used, but crucially not if  the $T^2=1$ definition is used, when evident significant tensions appear. This provides further support for the need for an enlarged error definition, as provided by the tolerance. 

Finally, we also address the question of parameterisation flexibility from the point of view of restricting the number of free parameters to be less than the nominal 52 in the MSHT20 case. This number, and indeed the basis of Chebyshev polynomials that is used, was motivated by the original study of~\cite{Martin:2012da} and in particular the observation that this choice allows a fit with sub--percent level precision, it being expected that this would be required by the increasingly high precision LHC data now entering the fit. By restricting the PDF parameterisation to instead have 28 or 40 free parameters, we confirm that these more restricted parameterisations are insufficient to match the input of the global closure test within the $T^2=1$ (and even in some cases $T^2=10$) uncertainties. Hence, in such cases parameterisation inflexibility would clearly play a more significant role in requiring an enlarged tolerance, but this can be largely avoided by simply taking a suitably flexible parameterisation. Similarly large differences are also seen in the fit to the real data when these more restricted parameterisations are used. These results are directly relevant to e.g. the ABMP16, ATLASpdf21,  CT18 and HERAPDF2.0 sets~\cite{H1:2015ubc,Alekhin:2017kpj,Hou:2019efy,ATLAS:2021vod}), where the baseline parameterisation has fewer free parameters than in MSHT20. Conversely, we have investigated  the impact of increasing the number of free parameters moderately, and find no particular evidence that overfitting is occurring. {\it A fortiori}, the fact that the more restricted parameterisations cannot faithfully describe the input of a global closure indicates that overfitting will not be expected to occur in these cases.

The outline of this paper is as follows. In Section~\ref{sec:gen} we outline the general approach and the MSHT PDF parameterisation used, as well as the method of uncertainty propagation. In Section~\ref{sec:closure} we present a set of closure tests: in Section~\ref{subsec:PDF} we briefly present the result of closure tests to directly PDF--level pseudodata; in Section~\ref{sec:closureun} we present the result of an `unfluctuated' global closure test, namely without the pseudodata being fluctuated by their corresponding errors; in Section~\ref{subsec:fluc} we present the result of a `fluctuated' closure test, with these fluctuations included. In Section~\ref{sec:tol} we consider the role of the tolerance, and of dataset tensions in it: in Section~\ref{subsec:tolgen} we first provide some general discussion of the impact of tensions on PDF uncertainties within a toy model, while in Section~\ref{subsec:tolclos} we present a set of closure tests with dataset tensions included to demonstrate this impact in a full PDF fit. In Section~\ref{sec:fullfit} we present the results of a full global PDF fit within the NNPDF4.0 data/theory settings, but applying the MSHT parameterisation: in Section~\ref{sec:pcharm} we consider the perturbative charm case, and in Section~\ref{subsec:fcharm} the fitted charm case. In Section~\ref{sec:overfit} we examine the role of parameterisation flexibility, focussing on restricting the number of free parameters in both the closure test and full PDF fit. In Section~\ref{sec:cross} we present some LHC benchmark cross section predictions that come from the MSHT fits, and compare them to the NNPDF4.0 results. In Section~\ref{sec:availability} we describe the public availability of PDF sets related to the current study. Finally, in Section~\ref{sec:conc} we conclude, and in the appendices we present a more detailed study of the treatment of the PDF uncertainty eigenvectors in the Hessian approach, consider the role of PDF integrability in the MSHT fit, and present further fit quality and PDF comparisons.

\section{General Approach}\label{sec:gen}

This study will make use of the publicly available \texttt{NNPDF} fitting code~\cite{NNPDF:2021uiq}, described in detail here: 
\\
\\
\href{https://docs.nnpdf.science}{\texttt{https://docs.nnpdf.science}}
\\
\\
This is set up to allow the user to perform PDF fits within the NNPDF framework, i.e. via a neural net (NN) parameterisation of the PDFs. However, the code is modular, and it is in particular possible to evaluate the fit quality within the NNPDF framework (i.e. with their theory and choice of dataset, or any subset of it) but for an arbitrary PDF set at the input scale $Q_0$. In more detail, the NNPDF code is set up to make use of their \texttt{EKO}~\cite{Candido:2022tld} and \texttt{FastKernel} (FK) methodologies (see~\cite{NNPDF:2014otw} and references therein), namely via interpolation grids that provide  the theoretical prediction in  a PDF independent way, but taking as an input the PDFs at input scale $Q_0$. That is, the effect of DGLAP evolution is also accounted for in the FK grids. 

While the NNPDF public code is set up to apply the FK grids to input PDFs that come from a NN parameterisation, they can equally well be applied to those parameterised in an arbitrary fixed polynomial basis to produce the corresponding theory predictions in such a case. Combined with the publicly available treatment of the matching input datasets for these predictions, it is then possible to evaluate a data/theory comparison, and a fit quality $\chi^2$, for exactly the same data and theory settings as in the NNPDF fits (or some variation of them), but with the only difference being in the parameterisation of the underlying PDFs at input scale $Q_0$. We can then in the usual way use this data/theory comparison and $\chi^2$ evaluation as the basis for a PDF fit, where the $\chi^2$ is minimised and errors evaluated according to a given procedure.

This is the approach taken in the current study. To be specific, the parameterisation of the PDFs used in the MSHT20~\cite{Bailey:2020ooq,Thorne:2019mpt} global PDF fit is used. Here the PDF basis 
is given in terms of 
\be
u_V,\quad d_V, \quad S, \quad s_+, \quad s_-, \quad \overline{d}/\overline{u},\quad g\;,
\ee
where
\be
q_V = q-\overline{q},\quad s_\pm = s\pm \overline{s}, \quad S = 2\left(\overline{u} + \overline{d}\right) + s_+\;.
\ee
Each PDF is then given in terms of Chebyshev polynomials, following the original study of~\cite{Martin:2012da}.  That is, with the exception of the gluon and $s_-$, the PDFs are parameterised in the general form
\be\label{eq:cheb6_gen}
x f(x,Q_0) = A x^{\delta} (1-x)^{\eta}\left( 1+ \sum_{i=1}^6 a_{i}T_i(y(x))\right)\;,
\ee
where $T_i$ is the $i$th Chebyshev polynomial, and $y=1-2\sqrt{x}$. As described in more detail in~\cite{Martin:2012da}, the motivation for using this basis is that it naturally avoids large cancellations between subsequent terms in the expansion with $n$, increasing the stability of the result. For the $\overline{d}/\overline{u}$ we fix $\delta=0$ so that this tends to a constant value (in principle, left free in the fit), at low $x$. For the gluon, a somewhat modified form is taken to allow greater flexibility at low $x$, such that
\be\label{eq:gluonpar}
x g(x,Q_0) =A_g x^{\delta_g} (1-x)^{\eta_g}\left( 1+ \sum_{i=1}^4 a_{g,i}T_i(y(x))\right)+A_{g_-} x^{\delta} (1-x)^{\eta_{g_-}}x^{\delta_{g_-}}\;.
\ee
Historically, when only a small number of terms was present in the polynomial in the gluon PDF two powers were required to provide flexibility at very low $x$ values, but when a large number of polynomials, some of high order, are used this is no longer necessary, and indeed, may arguably give too much flexibility at low $x$.
Therefore, we will where relevant also present results where the parameterisation of the form \eqref{eq:cheb6_gen} is instead used, in particular in Section~\ref{sec:overfit}. Jumping ahead, this gives a very similar fit quality and resulting PDFs, but with a somewhat smaller PDF uncertainty at low $x$. 
For the relatively poorly constrained $s_-$ we take
\be
x s_-(x,Q_0) =A_{s_-} x^{\delta_{s_-}}  (1-x)^{\eta_{s_-}}\left(1-\frac{x}{x_0}\right)\;,
\ee
although there is no in principle issue with instead using a parameterisation of the form \eqref{eq:cheb6_gen}, with e.g. a smaller basis of Chebyshev polynomials and one coefficient $a_i$ suitably fixed to satisfy the zero strangeness sum rule.

After imposing sum rule constraints, the default MSHT20 parameterisation has 52 free parameters. With the basis and PDF parameterisation above, we can then evaluate the fit quality for precisely the same data and theory settings as in the NNPDF fit. We emphasise again that as well as the data entering the fit, that the theory settings, i.e. the heavy flavour scheme and quark masses, form of the DGLAP evolution from $Q_0$ that is implicit in the FK grids, value of the strong coupling and the precise perturbative cross section calculation for each process, are completely identical to those used in the corresponding NNPDF fit. Therefore, while the MSHT20 parameterisation is used, this will  otherwise differ from the result of the MSHT20 fit. In this way, the comparison is completely like--for--like, with the only difference being due to the PDF parameterisation. We will also use this framework as the basis for the closure tests that we will first consider in the following sections, before moving to the result of a full global fit.

Finally, for the uncertainty evaluation we will as usual follow the Hessian approach described in e.g.~\cite{Bailey:2020ooq}. Namely, the deviation of the fit quality from the best fit value is expanded around the  minimum of the global fit quality as
\be\label{eq:chiglob}
\Delta \chi^2_{\rm global} = H_{ij} (p_i-p_i^0)(p_j-p_j^0)\;,
\ee
where $p_i$ is the $i$th input parameter, with value $p_i^0$ at the global minimum, and the Hessian matrix has components
\be
H_{ij} = \frac{1}{2}\frac{\partial^2 \chi^2_{\rm global}}{\partial p_i \partial p_j}\bigg|_{\rm min}\;.
\ee
The PDF uncertainties are defined by rotating to a basis that diagonalises the Hessian matrix and then requiring that the deviations in fit quality along these orthogonal directions match the required  `tolerance' $T=\sqrt{\Delta \chi^2_{\rm global}}$ for the corresponding confidence interval. Namely, we have
\be\label{eq:hesseigenvalue}
H_{ij} v_{j k} = \lambda_k v_{ik}\;,
\ee
where $\lambda_k$ is the $k$th eigenvalue and $v_{ik}$ is the $i$th component of the $k$th eigenvector of $H$. The PDF parameter displacements are then given in terms of the rescaled eigenvectors 
\be\label{eq:eij}
e_{ij} \equiv \frac{v_{ik}}{\sqrt{\lambda_k}}\;,
\ee
so that a given eigenvector set $S_k^\pm$ is produced with parameters given by
\be\label{eq:ti}
a_i(S_k^\pm) = a_i^0 \pm t_i e_{ik}\;.
\ee
The rescaling of the eigenvectors according to \eqref{eq:eij} implies that in the Gaussian approximation (i.e. where \eqref{eq:chiglob} is exact) taking $t_i=1$ would correspond to a deviation of $\Delta \chi^2_{\rm global}=T=1$, though to be precise (and as is essential for higher eigenvectors where deviation from Gaussian behaviour becomes more pronounced) $t$ is chosen in order to match the desired value of $T$. In the textbook case we would have $T=1$ at 68\% confidence, but as discussed in the introduction, and further in Section~\ref{sec:tol}, there is much evidence that this is not sufficient for global PDF fits, and an enlarged tolerance needs to be taken.

In the MSHT20 study the value of $T$ is chosen according to the `dynamic tolerance' procedure (see~\cite{Bailey:2020ooq} for a detailed explanation) for which $T$ takes a different value for each eigenvector direction. However, it is has been noted in~\cite{Jing:2023isu} (and elsewhere) that this gives a result that is on average quite close to taking a fixed value of $T^2=10$, as we will verify later on. We will therefore, where appropriate, show either the result of the fixed $T^2=10$ or the dynamic tolerance criteria when we wish to demonstrate the size of the PDF uncertainties in the MSHT approach.

We  note that, as discussed in~\cite{Martin:2002aw,Martin:2009iq}, when it comes to considering variations of the PDF parameters around the global $\chi^2$ minimum in the Hessian formalism presented above, it is common for there to be  a certain amount of  redundancy between some of these PDF parameters, such that small changes in the values of some parameters can be largely compensated for by changes in other parameters. As a result of this high degree of correlation, the behaviour of certain PDF eigenvectors about the $\chi^2$ minimum can become highly non--quadratic. A practical solution to this issue is presented in~\cite{Martin:2002aw,Martin:2009iq} and is applied in the MSHT20 fits. Namely, certain PDF parameters are fixed at their best fit values when evaluating the Hessian, if these exhibit a significant degree of correlation with other PDF parameters. In this way, a set of more quadratic eigenvectors is arrived at, and a more stable application of the Hessian approach becomes possible.

This is discussed in detail in Appendix~\ref{app:PDFparfixing}, where it is emphasised that this redundancy in certain PDF parameters about the minimum, and the fixing of some of these in the MSHT20 approach at the uncertainty evaluation stage, does not imply that there is any reduced parametric freedom in the overall fit. That is, the full set of PDF parameters are allowed to be free in the minimisation stage. In the current paper we in fact take a slightly different approach, and now allow all PDF parameters to be free in the eigenvector evaluation, without fixing any PDF parameters. However, as discussed in detail in Appendix~\ref{app:PDFparfixing} this has a very minor impact on the PDF uncertainties, as the contribution to this from the non--quadratic eigenvectors with lower eigenvalues is very small.

\section{Closure Tests}\label{sec:closure}

\subsection{PDF Level Closure Test}\label{subsec:PDF}

We begin by briefly repeating, and extending, the closure test performed in~\cite{Martin:2012da}. That is, we do not yet consider the cross section observables that enter a PDF fit, but instead simply generate pseudodata corresponding directly to the values of the PDF themselves at different $x$ points. We in particular take the NNPDF4.0 NNLO perturbative charm central replica as our input, and generate 500 pseudodata points spread evenly in $\ln x$ between $10^{-5}$ and $0.99$, scattered by a 1\% error on the size of the input PDF. However, the precise results do not depend sensitively on this choice, e.g. on the size of the error, whether scattering is including or not, and the number of datapoints. We note that in what follows we will usually omit the `NNLO' label for brevity, but this is always implied.

The parameterisation is as described above, and corresponds to that used in the MSHT20 fit\footnote{In fact, this is with the minor exception that we allow the low $x$ power, $\delta_{s_-}$, of the strangeness asymmetry, $s_-$, to be free, whereas in~\cite{Bailey:2020ooq} it was fixed.}. We generate pseudodata for the 7 PDF combinations that enter the MSHT20 parameterisation directly, namely $u_V$, $d_V$, $S=2(\overline{u} +\overline{d}) + s + \overline{s}$, $s_+=s+\overline{s}$, $\overline{d}/\overline{u}$, $s_-=s-\overline{s}$, $g$ although generating these in other combinations will lead to similar results. We then perform a fit to these pseudodata, imposing the number and momentum sum rules in the usual way.

\begin{figure}
\begin{center}
\includegraphics[scale=0.65]{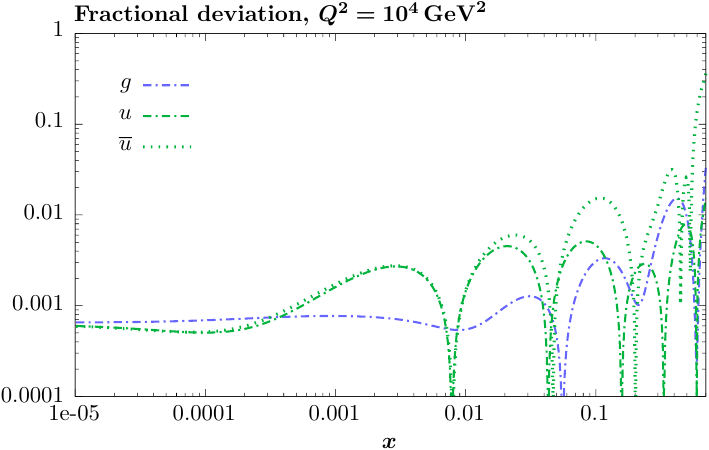}
\includegraphics[scale=0.65]{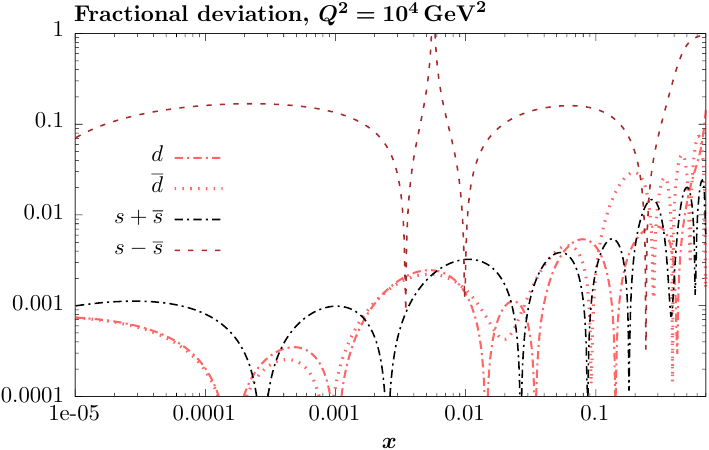}
\caption{\sf Fractional deviations of the fit to direct PDF pseudodata relative to the NNPDF4.0 (p. charm) input.}
\label{fig:pdfcl}
\end{center}
\end{figure}

The fractional deviations from the input NNPDF set are shown in Fig.~\ref{fig:pdfcl} for various PDF combinations. We can see in all cases other than $s_-$ that these lie between 1 per mille and 1\% across most of the $x$ range, with the exception of the highest $x$ values, where this can be somewhat larger. Thus, consistently with the results in~\cite{Martin:2012da}, a sub--percent level description of a set of PDFs generated with an in principle more flexible underlying set,  is achieved in many regions with the default MSHT20 parameterisation. The one exception to this is the $s_-$, for which we have only 3 free parameters. In this case, we can see that only $\sim 10\%$ level precision is achieved, away from the artificial peak at $x\sim 0.005$ where the PDF passes through zero. As we will see below, however, when it comes to a global closure test this is improved rather at larger scales and in the region where $s_-$ is most significantly different from zero. Nonetheless, this indicates that as expected a more flexible parameterisation will be needed when greater precision is required in the description of this particular distribution.

\subsection{Unfluctuated Global Closure Test}\label{sec:closureun}

We now move on to considering a closure test at the level of the observables that enter a global PDF fit. We in particular generate pseudodata corresponding to the NNPDF4.0 global dataset, and with the same input of the NNPDF4.0 (perturbative charm) central replica as above. In this section, we will  consider a closure test without shifting the pseudodata by their underlying experimental uncertainties. In other words, if the PDF parameterisation in the fit is flexible enough, then a perfect fit (with $\chi^2=0$) is in principle achievable. However, as the NNPDF4.0 set is  due to an in principle more flexible underlying parameterisation than the fixed MSHT20 parameterisation, we will expect some departure from a perfect fit, with the $\chi^2$ being somewhat larger. 

We note that this is similar to the `level 0' closure test first used by the NNPDF collaboration in~\cite{NNPDF:2014otw}, where this also refers to the cases of fitting unfluctuated pseudodata. However it is not identical, as we will only consider the result of a single fit to the unfluctuated pseudodata, whereas in the original level 0 closure test multiple fits are considered, with different random initialisation points; we will comment on this point further at the end of the section. We will also evaluate the corresponding PDF error that comes from propagating through the experimental uncertainties on the unfluctuated data in the fit. This in particular is something that is only explored in the context of fluctuated data as part of the NNPDF `level 2' closure test. Given there is no exact one--to--one correspondence between the labelling of the closure tests used by NNPDF and those considered here, we will not use this terminology but will instead refer to the current closure test as an `unfluctuated' closure test, for preciseness.

Performing the above closure test, we find a fit quality of $\chi^2=2.4$ for the 4627 points in the fit\footnote{Note this is higher than the quoted 4618 points quoted in~\cite{NNPDF:2021njg} for the fitted charm case, due to the more stringent $Q^2$ cut that is imposed for $F_2^c$ then.} i.e. $\sim 0.0005$ per point. While indeed non--zero, this is still encouragingly low.  Indeed, in~\cite{DelDebbio:2021whr} a $\chi^2/{\rm N}_{\rm pts}$ of 0.002 (0.012) in the NNPDF 4.0 (3.1) methodology is quoted for the L0 closure tests, i.e. $O(10)$ ($O(50-60)$) in total. So this level of agreement is rather similar to that found by NNPDF with their default setup in this study, although the precise value in that case depends on the specific parameters of the text, e.g. if a larger training length of the NN were taken a lower value could be achieved.

\begin{figure}
\begin{center}
\includegraphics[scale=0.6]{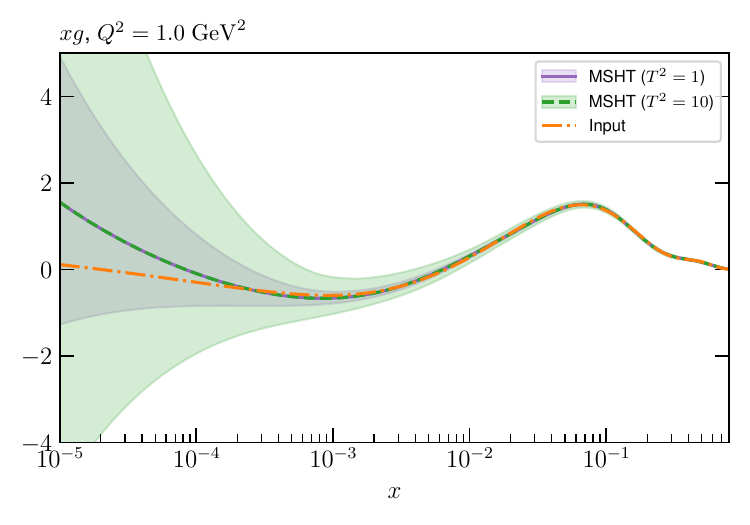}
\includegraphics[scale=0.6]{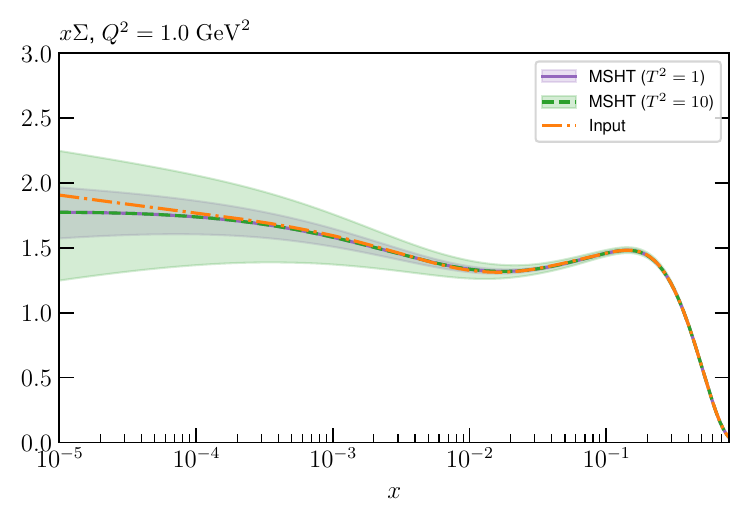}
\caption{\sf The input ($Q^2 = 1 \,{\rm GeV}^2$) gluon and quark singlet PDFs that result from an unfluctuated closure test fit to the NNPDF4.0 dataset,  using the MSHT20 parameterisation. The PDF uncertainties calculated with a $T^2=1$ ($T^2=10$) fixed tolerance are shown in purple (green) and  the NNPDF4.0 (p. charm) input is given by the dashed red line. The $T^2=10$ is shown for the sake of comparison, whereas the $T^2=1$ case is the appropriate definition for such a self--consistent closure test.}
\label{fig:glcl_abs}
\end{center}
\end{figure}

While this gives some indication that the MSHT20 parameterisation is performing  well for this specific closure test, a clearer picture is only found by looking at results at the level of the PDFs. In Fig.~\ref{fig:glcl_abs} we start by showing two representative PDFs at input scale $Q_0 = 1$ GeV, with both the textbook $T^2=1$ and an enlarged $T^2=10$ tolerance shown. As discussed in Sections~\ref{sec:gen} and~\ref{sec:tol} there is much evidence that an enlarged tolerance is required to provide an appropriate error estimate when performing a genuine PDF fit, with $T^2=10$ being on average rather similar to that found in the MSHT20 fit. A significant cause of this, and arguably the dominant one, is due to inconsistencies between the data and theory as well as between different datasets in the fit which render the application of the textbook $T^2=1$ criterion overly aggressive. Another possible reason for this, within the fixed parameterisation approach, is that the parameterisation itself may not be sufficiently flexible, such that the enlarged tolerance effectively accounts for part of the uncertainty due to the fixed choice of parameterisation, as in e.g.~\cite{Hou:2019efy}. In the context of this closure test the former cause, i.e. that of data/theory and dataset inconsistencies, will clearly not be present, as we have perfect consistency by construction. However, by showing the $T^2=10$ error, and in particular examining the level of any disagreement between the input set and the output fit relative to this, we can therefore evaluate the extent to which parameterisation inflexibility may play a role, at least in the context of the closure test. 

Clearly the fit results match the  input set in the case of Fig.~\ref{fig:glcl_abs} rather well, with some deviation observed at very low $x$, but always safely within the $T^2=1$ uncertainty band, and well within the $T^2=10$ case. The PDFs in this region are however not directly constrained by data, and it is only at higher $x$ and/or scale, and by plotting PDF ratios that a more precise and representative comparison can be made. 

\begin{figure}
\begin{center}
\includegraphics[scale=0.6]{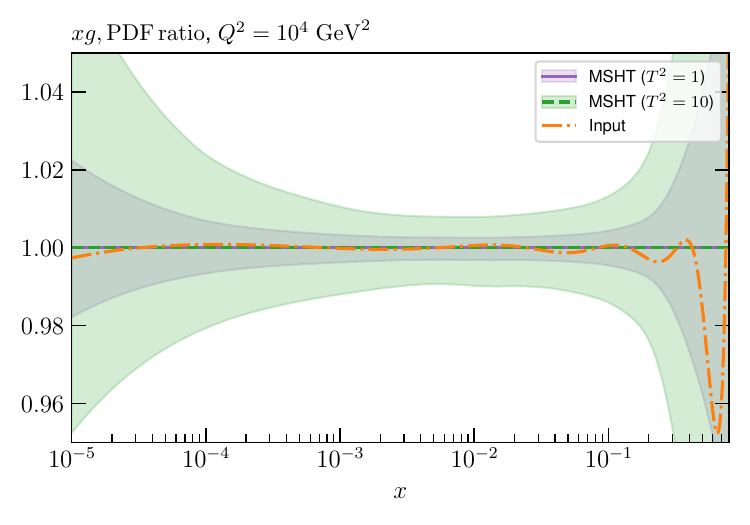}
\includegraphics[scale=0.6]{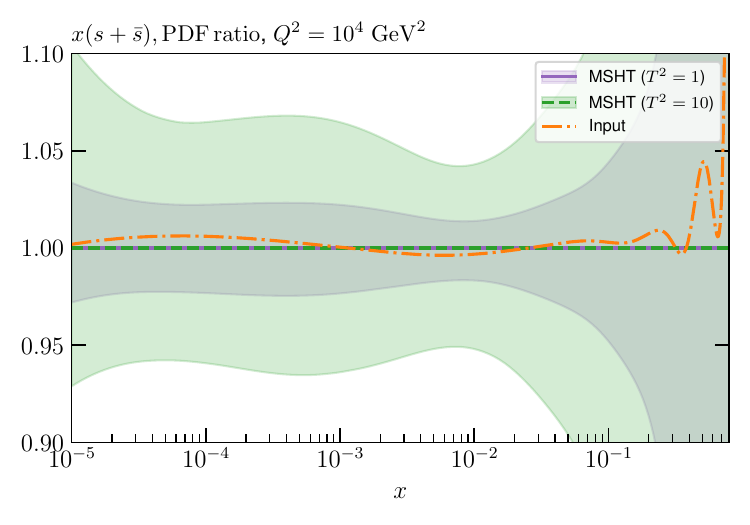}
\includegraphics[scale=0.6]{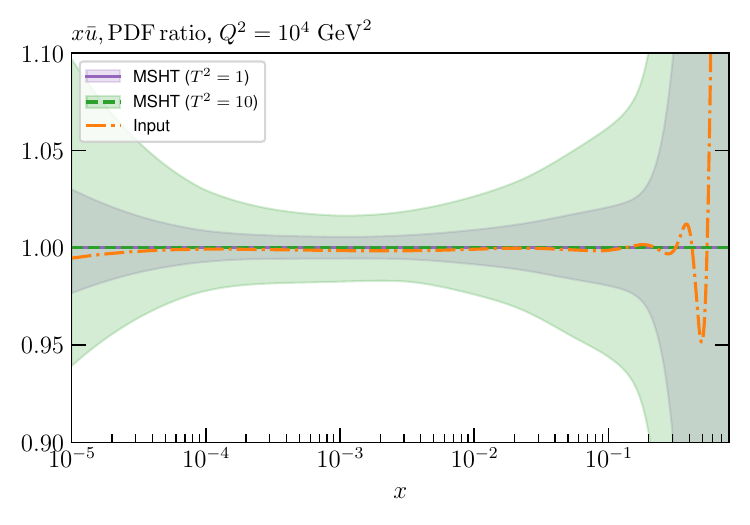}
\includegraphics[scale=0.6]{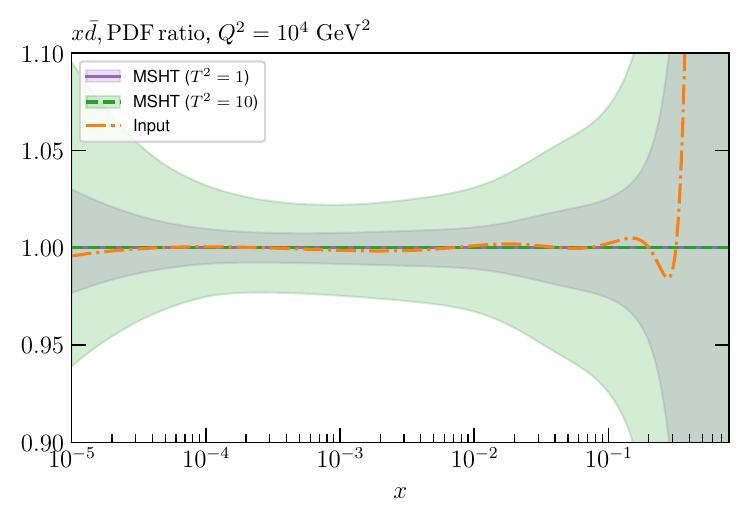}
\includegraphics[scale=0.6]{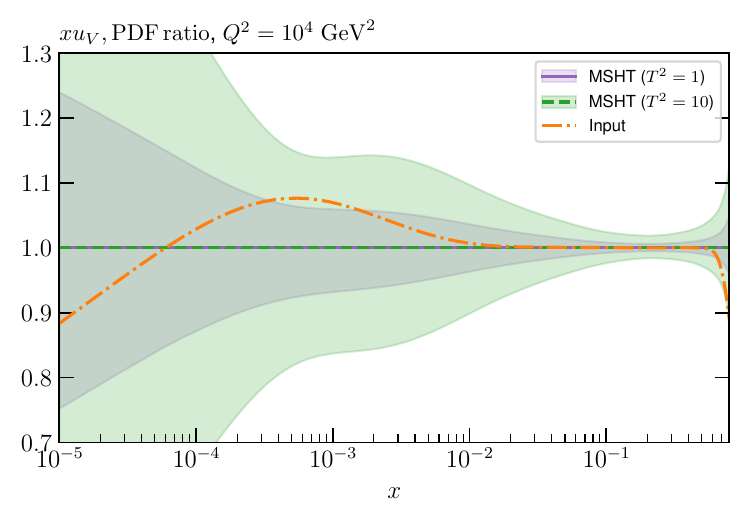}
\includegraphics[scale=0.6]{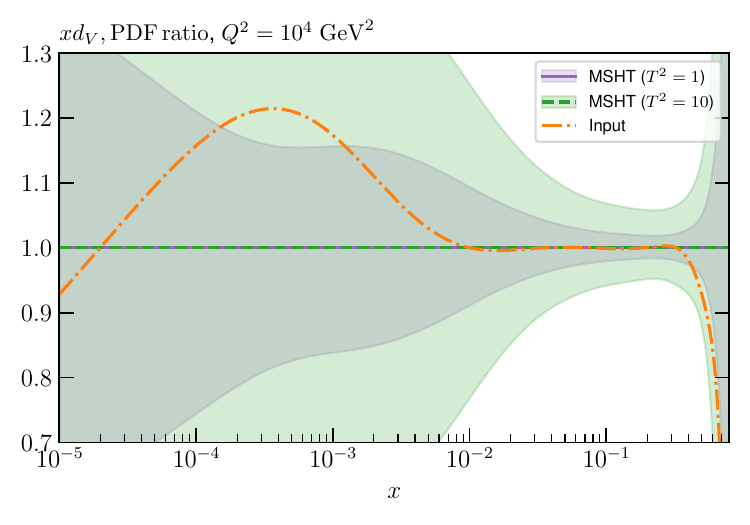}
\includegraphics[scale=0.6]{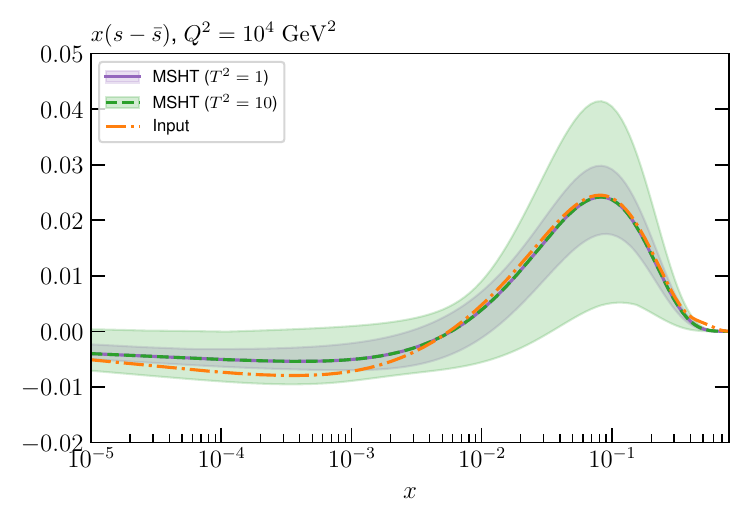}
\caption{\sf A selection of PDFs at $Q^2=10^4 \, {\rm GeV^2}$  that result from an unfluctuated closure test fit to the NNPDF4.0 dataset,  using the MSHT20 parameterisation. The PDF uncertainties calculated with a $T^2=1$ ($T^2=10$) fixed tolerance are shown in purple (green) and  the NNPDF4.0 (p. charm) input is given by the dashed orange line. Results are shown as a ratio to the MSHT fits (the central value for the tolerance choices is by definition the same), with the exception of the $s_-$, where the absolute PDFs are plotted. The $T^2=10$ is shown for the sake of comparison, whereas the $T^2=1$ case is the appropriate definition for such a self--consistent closure test.}
\label{fig:glcl_rat}
\end{center}
\end{figure}

These are shown in Fig.~\ref{fig:glcl_rat},  at $Q^2=10^4 \, {\rm GeV^2}$ and with both the textbook $T^2=1$ and an enlarged $T^2=10$ tolerance again shown. The agreement between the input and fit result is in general very good in the data region. At low to intermediate $x$ the deviation is in general at the per mille level, being largest (a few per mille) for the quark flavour decomposition, e.g. in the case of the strangeness. The deviation is generally at the $\sim 10\%$ level of the overall $T^2=1$ uncertainty, but can approach $\sim 50\%$ in some $x$ regions, but generally at rather higher $x$, i.e. towards the extrapolation region, which we will discuss further below. For the $s-\overline{s}$ the deviation is somewhat larger, but again most notably at rather low $x$ where current constraints are limited; some increased flexibility may be preferable for future fits, when these increase.

\begin{figure}
\begin{center}
\includegraphics[scale=0.6]{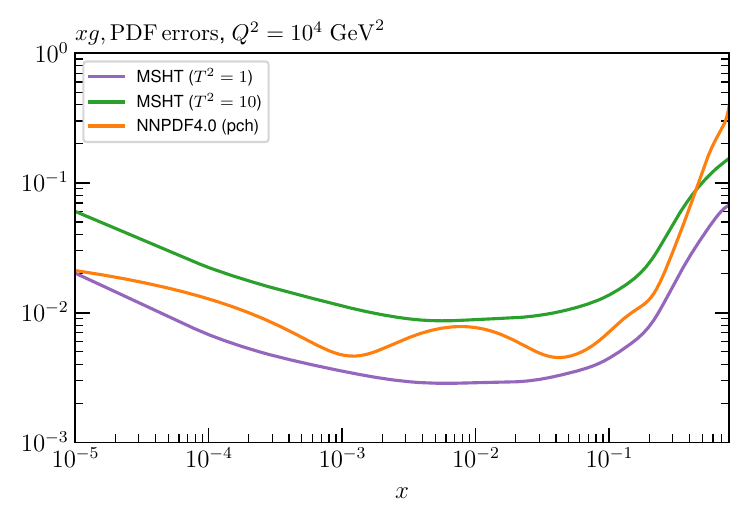}
\includegraphics[scale=0.6]{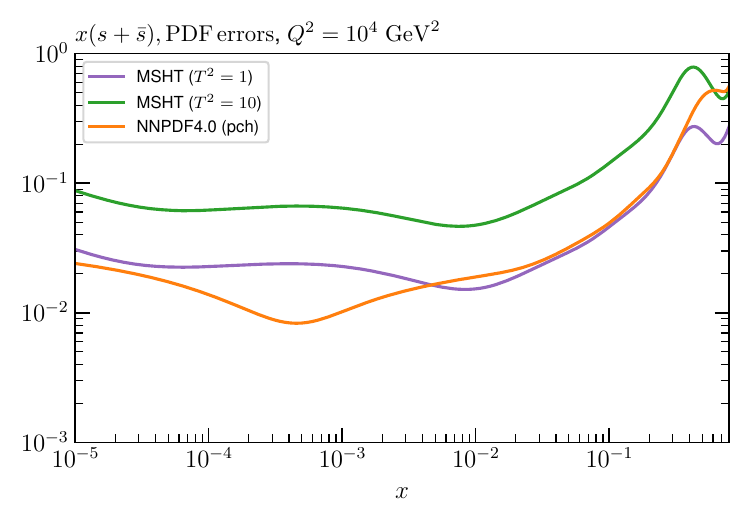}
\includegraphics[scale=0.6]{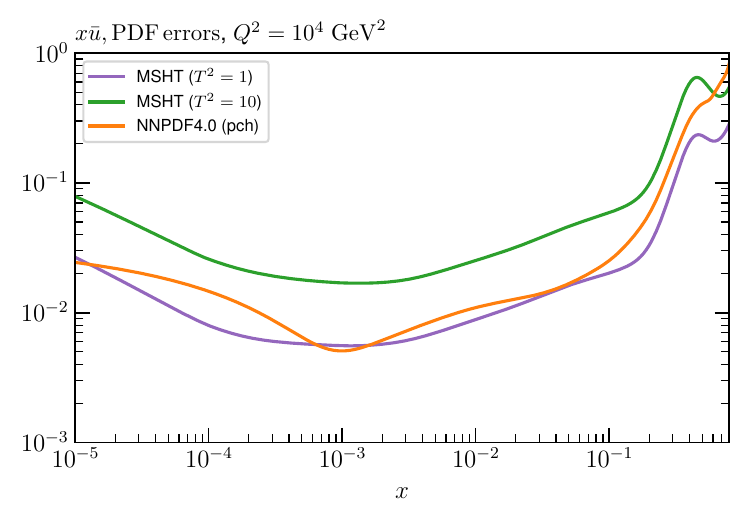}
\includegraphics[scale=0.6]{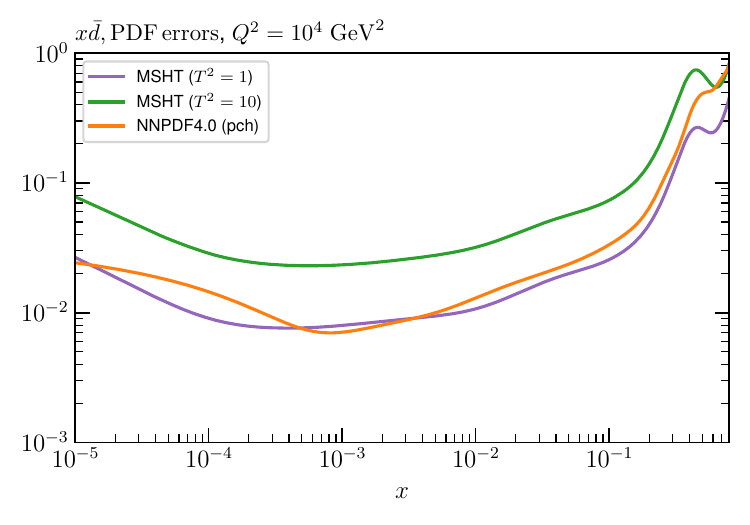}
\includegraphics[scale=0.6]{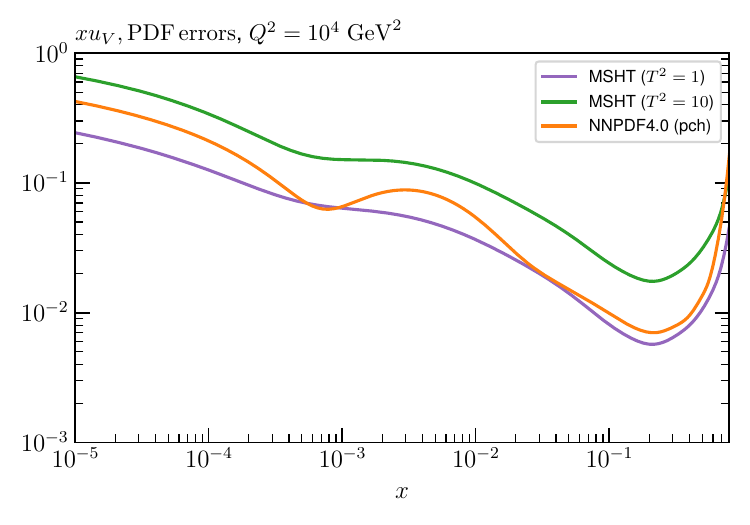}
\includegraphics[scale=0.6]{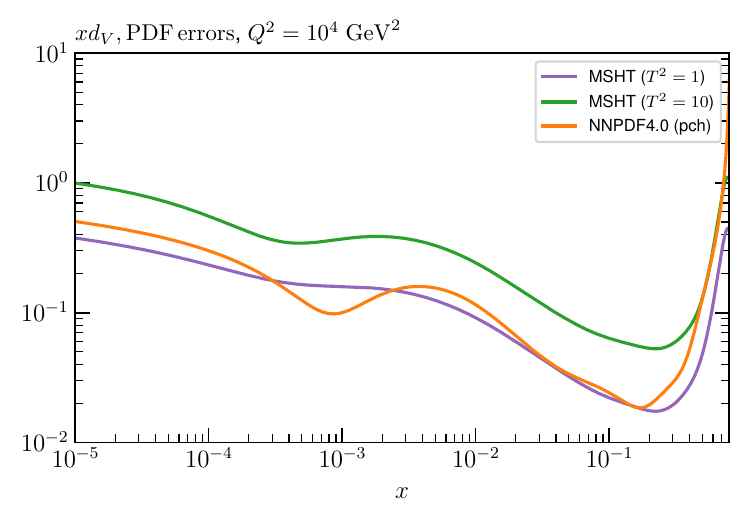}
\includegraphics[scale=0.6]{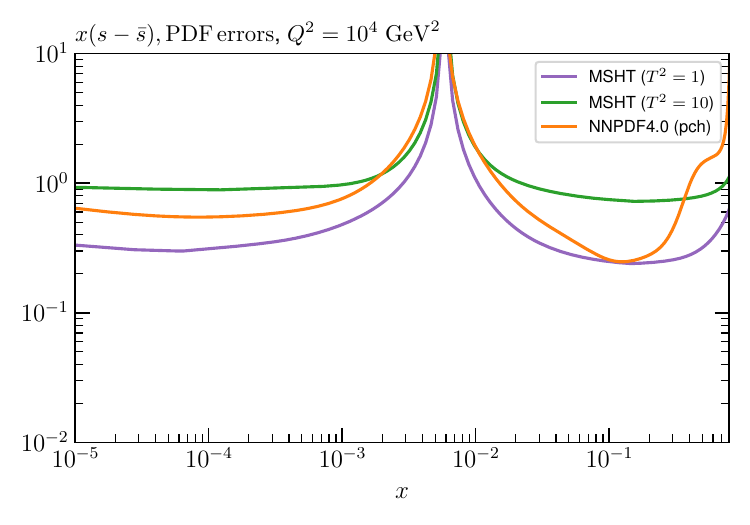}
\includegraphics[scale=0.6]{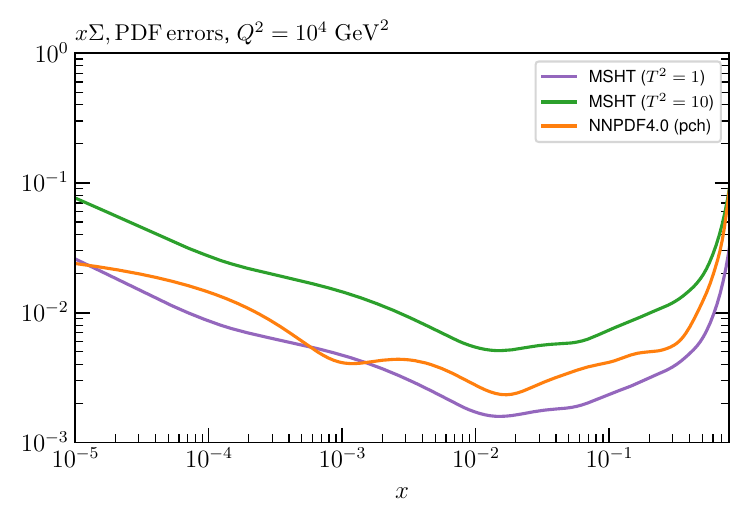}
\caption{\sf PDF uncertainties at $Q^2=10^4 \, {\rm GeV^2}$  that result from an unfluctuated closure test fit to the NNPDF4.0 dataset,  using the MSHT20 parameterisation. The PDF uncertainties calculated with a $T^2=1$ ($T^2=10$) fixed tolerance are shown as well as the NNPDF4.0 (p. charm) input for comparison, although the latter plays no role in the closure test.}
\label{fig:glcl_errs}
\end{center}
\end{figure}

We also consider the valence $u_V$ and $d_V$ distributions in Fig.~\ref{fig:glcl_rat}. In the $x\sim 0.01 - 0.3$ region where the valence distributions are largest, we can see that the level of deviation is very small. However, at low $x$ and for the $d_V$ at high $x$ this is no longer the case. It has been shown in~\cite{Bailey:2020ooq} that the valence quarks in similar regions are affected quite significantly by the extension of the MMHT14~\cite{Harland-Lang:2014zoa} parameterisation, such that 6 rather than 4 Chebyshev polynomials are used, and the $\overline{d}/\overline{u}$ combination, rather than $\overline{d}-\overline{u}$, is parameterised. More precisely, the deviations here occur at rather lower $x$ than  the changes observed in~\cite{Bailey:2020ooq}  due to the extended parameterisation. These are therefore occurring in a region of $x$ where direct constraints on (and indeed the phenomenological impact of) the PDFs are rather limited, but clearly in these regions parameterisation inflexibility is playing an increasing role, as we would expect.

Considering the $T^2=10$ uncertainties, also shown, these are as expected larger (by $\sim 3$) than the $T^2=1$ case. The absolute difference between the $T^2=1$ and $T^2=10$ uncertainty bands is almost universally significantly larger than any deviation between the output PDFs and the input set, while the ratio of the deviation to the $T^2=10$ uncertainty is almost universally at the level of a few percent. This therefore provides good evidence that in the significant majority of cases the size of the $T^2=10$ PDF uncertainties is not required by any parameterisation inflexibility in the MSHT20 fit. The exceptions are again the $s_-$, where we can see that the enlarged tolerance uncertainty now encompasses the deviation between the output PDF and the input away from the peak region, and similarly for the $u_V$, $d_V$ at low and very high $x$, as well as other distributions at high enough $x$.

We can also make a first comparison between the actual size of the $T^2=1$ and $T^2=10$ PDF uncertainties in the unfluctatued closure test and those in the NNPDF4.0 (perturbative charm) global fit. The size of the PDF uncertainties in the latter cases are determined by a fit to the real data, rather than in a closure test, and hence this is clearly not an entirely like--for--like comparison. Nonetheless, this comparison can serve as a first indication of the size of the PDF uncertainties in the NNPDF and MSHT fixed parameterisation approaches. These are shown in Fig.~\ref{fig:glcl_errs}, and we can see that at intermediate to higher $x$, for the quark flavour decomposition the NNPDF uncertainty is very similar in size to the $T^2=1$ case. The exception to this is most notably the gluon, and (related to this) the quark singlet, with some difference also seen in the $u_V$ and for various distributions at high $x \gtrsim 0.3-0.4$, where the NNPDF uncertainty is rather larger, and somewhere between the $T^2=1$ and $T^2=10$ cases. For the $s_-$ we can also see that the NNPDF uncertainty is larger than the $T^2=1$ case away from the $x\sim 0.1$ peak region, as we would expect. This is therefore already an indication that in many cases the NNPDF uncertainty may be roughly equivalent to taking a $T^2=1$ tolerance in the data region; we will verify this later in the context of a full fit. For the most significant exception of the gluon, we note that the PDF  uncertainty itself  is on average rather smaller in the data region than for other partons. In such as case, the uncertainty may be driven more by the particular NNPDF methodology.

\begin{figure}
\begin{center}
\includegraphics[scale=0.6]{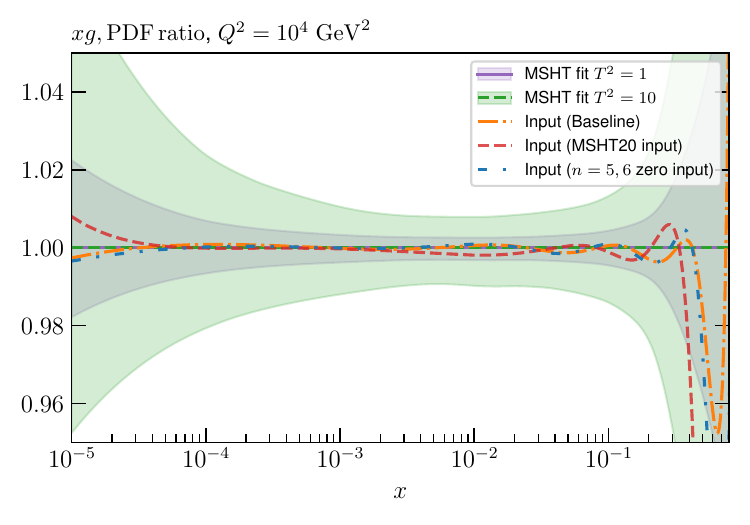}
\includegraphics[scale=0.6]{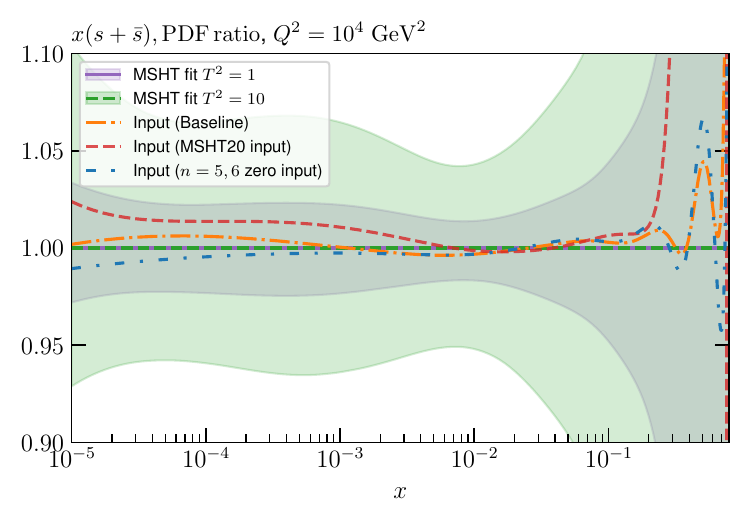}
\includegraphics[scale=0.6]{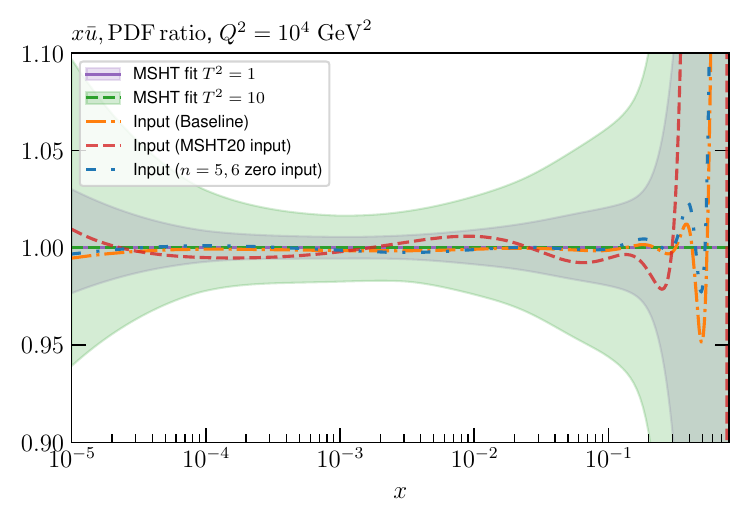}
\includegraphics[scale=0.6]{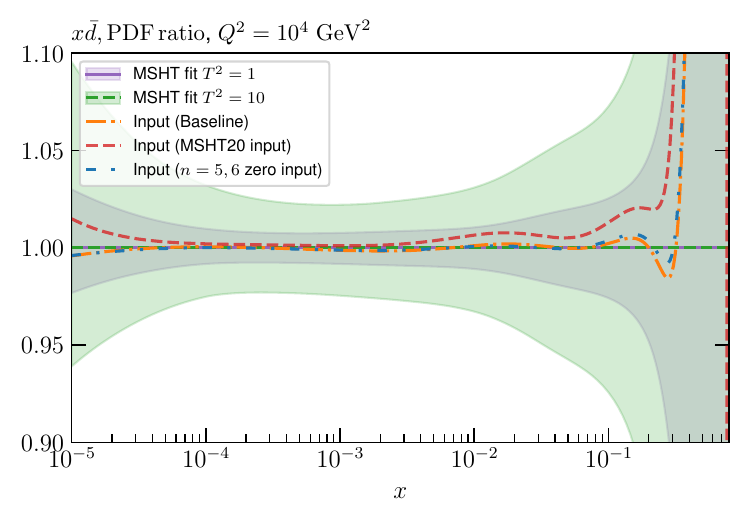}
\includegraphics[scale=0.6]{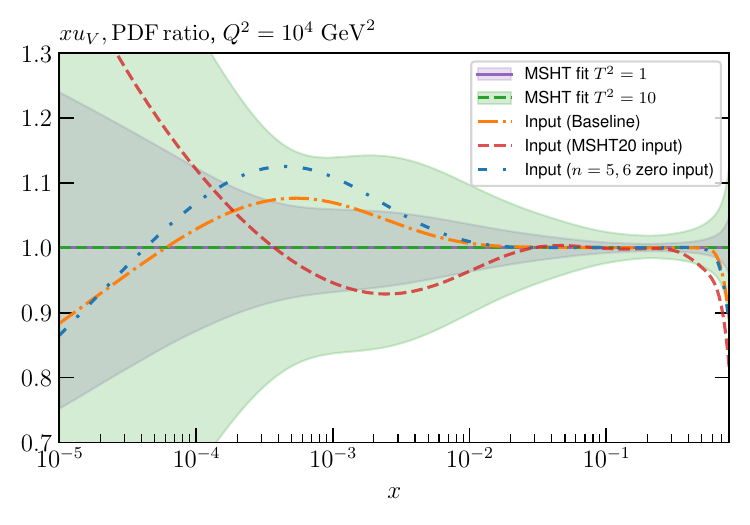}
\includegraphics[scale=0.6]{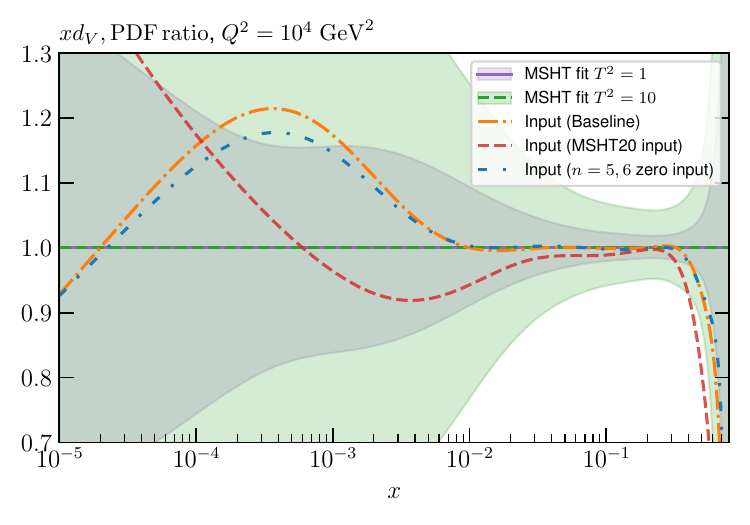}
\includegraphics[scale=0.6]{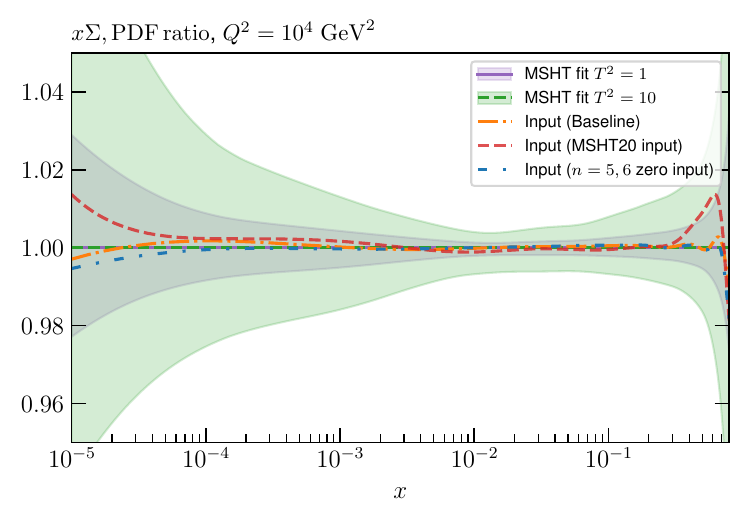}
\caption{\sf As in Fig.~\ref{fig:glcl_rat} but with an additional selection of closure test fit results shown, where the fits are started with different input PDF parameters, as indicated in the caption. Namely, the previous baseline is shown, as well as the result of taking this result and setting all $n=5,6$ Chebyshev coefficients to zero (but allowing them to be non--zero in the subsequent fit), and of starting from the same input as the MSHT20 NNLO set~\cite{Bailey:2020ooq}. In all cases, the lines indicate the ratio of the NNPDF4.0 (perturbative charm) input set to the fit, while the uncertainties bands correspond to the baseline fit; these are rather similar for the other fits. Note we show a somewhat different selection of PDFs to Fig.~\ref{fig:glcl_rat} for clearer illustration.}
\label{fig:glcl_minvar}
\end{center}
\end{figure}

We end this section with a few comments and caveats, and some further analysis based on these. First, we comment on a difference between the level 0 closure tests performed by NNPDF and the unfluctuated closure tests performed here. Namely, the fact that we show the result of a single fit to the unfluctuated pseudodata, whereas in the NNPDF level 0 closure test multiple fits are considered, with different random initialisation points. This is sensible in the context of a NNPDF fit, where the significantly increased flexibility of the NN architecture allows for multiple solutions with similar fit qualities. However,  the role of such multiple minima will be greatly reduced within the context of a single fixed parameterisation basis, and indeed in principle there will be a unique global minimum, given the input NNPDF set derives from a more flexible underlying PDF parameterisation than the MSHT20 fixed basis we use. In practice, for a given fit starting point it may be that the minimisation stops at some distance away from this global minimum, e.g. at a saddle point or a local distinct minima. More generally, by adopting a different input fixed parameterisation the location of the minimum will also change, and one can consider the variation that this entails, although we do not attempt a study of this question here.

To explore this point further, in Fig.~\ref{fig:glcl_minvar} we show a similar comparison to Fig.~\ref{fig:glcl_rat} but with an additional selection of closure test fit results, where the fits are started with different input PDF parameters. We note that the baseline fit takes as its input the PDF--level fit discussed in the previous section. In addition to this, we then first consider the case where as an input the baseline fit result is taken, but with all $n=5,6$ Chebyshev coefficients then set to zero (and allowing them to be non--zero in the subsequent fit). While all other free PDF parameters are as in the baseline best fit, this still results in a very different set of PDFs, with an extremely poor fit quality of $\sim 34$ per point, i.e. $\sim 160$ thousand in total. Nonetheless, on refitting we can see from  Fig.~\ref{fig:glcl_minvar} that we arrive back at a set of PDFs that lie very close to the baseline fit; the $\chi^2$ is $\sim 2.7$, i.e. very close to the baseline value of 2.4, although these precise numbers will depend mildly on the choice of stopping criterion etc in the minimisation algorithm. 

On the other hand, the potential issue of additional saddle points and/or local minima is in fact illustrated if we instead take the MSHT20 NNLO set~\cite{Bailey:2020ooq} as in input. The input fit quality is $\sim 650$, i.e. relatively poor, but not dramatically so, as we may expect from the fact the MSHT20 and NNPDF4.0 fits share many similar features in terms of experimental and theoretical inputs. After refitting, the final fit quality is now $\sim 20$, that is still relatively low but still rather higher than the baseline value of 2.4. The impact of this is evident in Fig.~\ref{fig:glcl_minvar}: while there remains a very close matching between the input and fit in the singlet sector, some deviations exist in terms of the quark flavour decomposition, albeit such that these remain generally well below the $T^2=1$ uncertainty band. We note that this minimum is genuinely distinct from the baseline case, in the sense that if we e.g. consider a line in PDF parameter space that connects the two minima, then the fit quality rapidly deteriorates as we move in either direction along this. As is apparent from the PDF comparisons, it is largely a feature of quark flavour decomposition and hence presumably of the extent to which the baseline dataset can allow a somewhat different decomposition with a similar (if slightly worse) fit quality. In addition, if a subset of the baseline input PDFs are set to the MSHT20 case then the fit tends to converge on the lower baseline minimum. Hence the presence of the distinct minimum is a feature of the differing balance in the quark flavour decomposition that exists in the complete MSHT20 input.

This question of possible saddle points and/or local minima is therefore certainly an important issue that has to be confronted in such fixed parameterisation fits. In the context of the MSHT20 fits (and earlier versions), care is always taken to perform multiple fits from different starting points to moderate the impact of this effect. Nonetheless, it remains encouraging that even for the case of the higher local minimum discussed above, the resulting PDFs generally lie well within the $T^2=1$ uncertainty, and certainly the $T^2=10$ one. Therefore, even if the enlarged error definition may to some extent account for this effect, there is no evidence that is close to the dominant contribution to it.

In the context of the current closure test, we have  in addition briefly explored to possibility of taking a randomly fluctuated set of input PDF parameters, by fluctuating these around the baseline best fit, subject to the usual constraints on low and high $x$ powers, and that the calculated gluon momentum fraction that results lies within a reasonable range, and is crucially not negative, as can often occur for unphysical combinations of PDF parameters that may result. In such a case, for 10 fits we find that 4 converge to $\chi^2$ values below 20, with 3 lying very close to the original best fit. The remaining fits show distinct issues with convergence, indicating that the input sets may be rather too extreme, even within the constraints discussed above. Nonetheless, the key point is that within such a context a global minimum close to the baseline is achieved for some of the fits, and indeed most of those which converge at all.

\begin{figure}
\begin{center}
\includegraphics[scale=0.6]{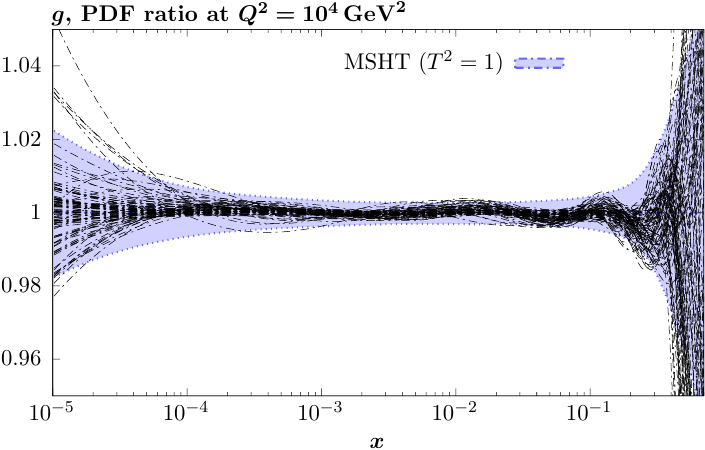}
\includegraphics[scale=0.6]{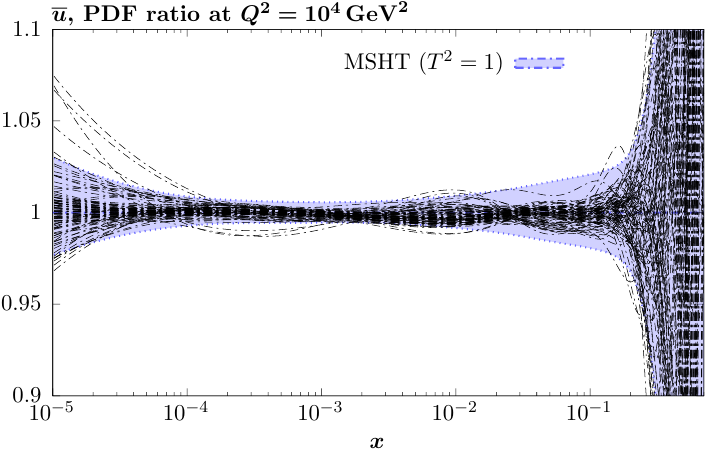}
\includegraphics[scale=0.6]{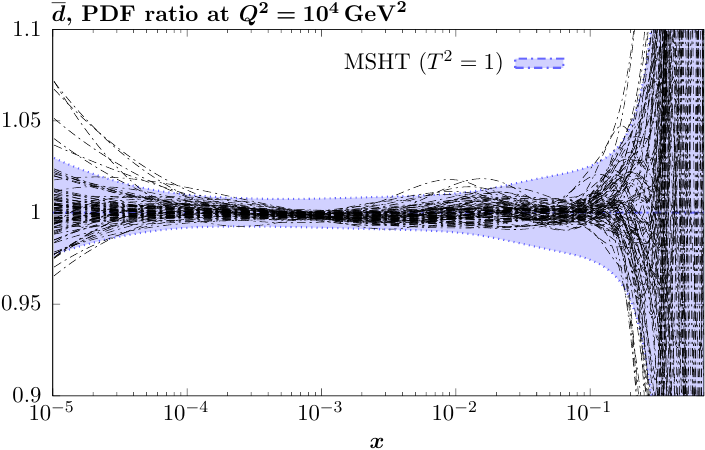}
\includegraphics[scale=0.6]{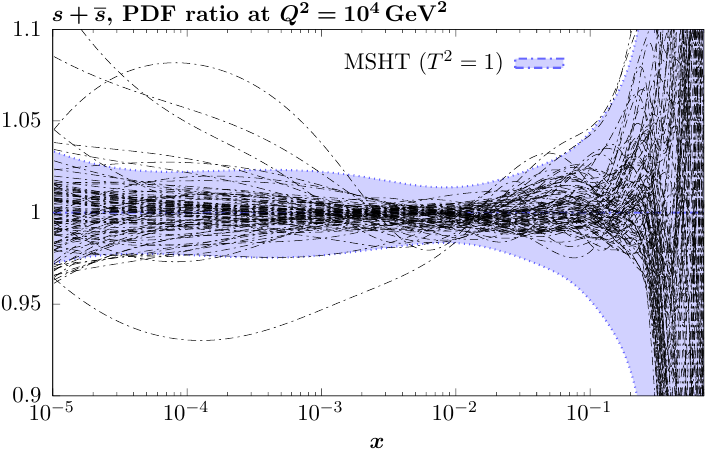}
\caption{\sf PDF ratios at $Q^2=10^4 \, {\rm GeV^2}$ of the result of  unfluctuated closure tests, using the MSHT20 parameterisation, to the corresponding NNPDF4.0 (perturbative charm) input. Each black curve corresponds to such a fit using a given replica from the NNPDF input set as input, such that there are 100 shown in total. For guidance, the Hessian PDF uncertainties calculated with a $T^2=1$  fixed tolerance that results from the fit to the central input set is shown, as in e.g. Fig.~\ref{fig:glcl_rat}. For clarity a limited, but representative, selection of 25 replicas is shown.}
\label{fig:glcl_nnreps_closures}
\end{center}
\end{figure}

\begin{figure}
\begin{center}
\includegraphics[scale=0.6]{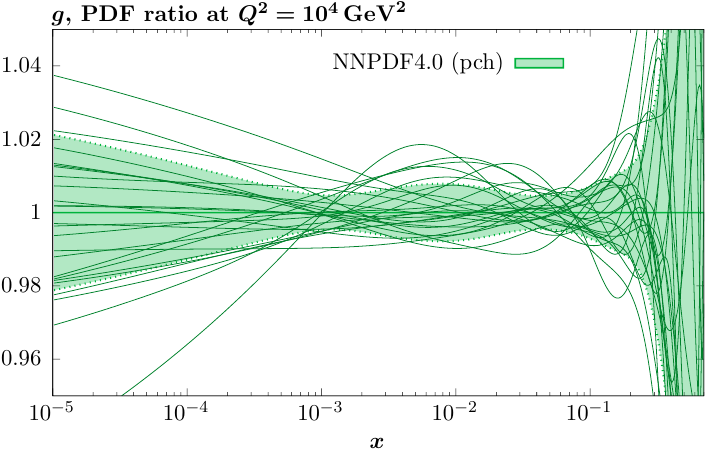}
\includegraphics[scale=0.6]{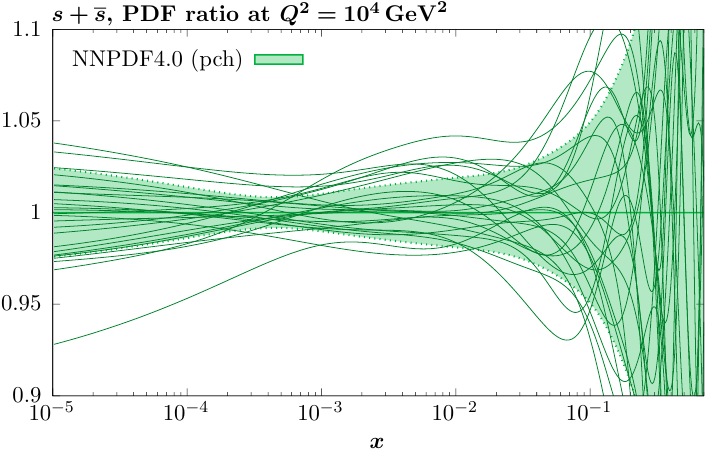}
\caption{\sf The ratio of the input replicas to the central NNPDF set (i.e. the replica average), with the MC replica uncertainty given for guidance, for the gluon and strangeness PDFs.}
\label{fig:glcl_nnreps_closures_abs}
\end{center}
\end{figure}

\begin{figure}
\begin{center}
\includegraphics[scale=0.6]{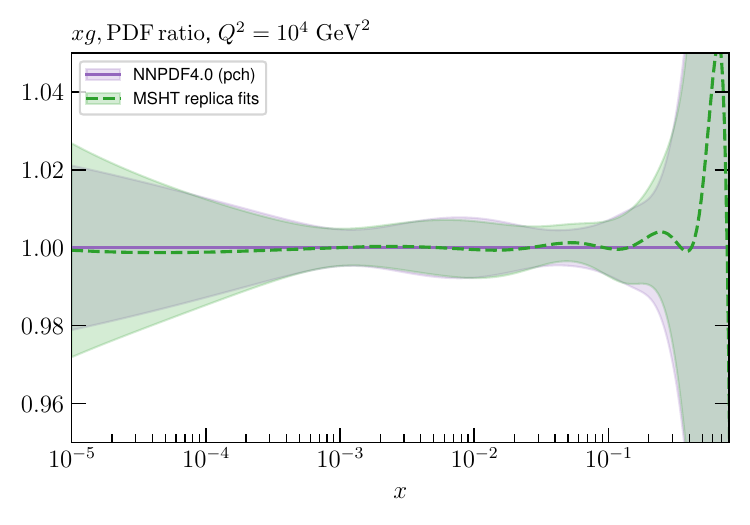}
\includegraphics[scale=0.6]{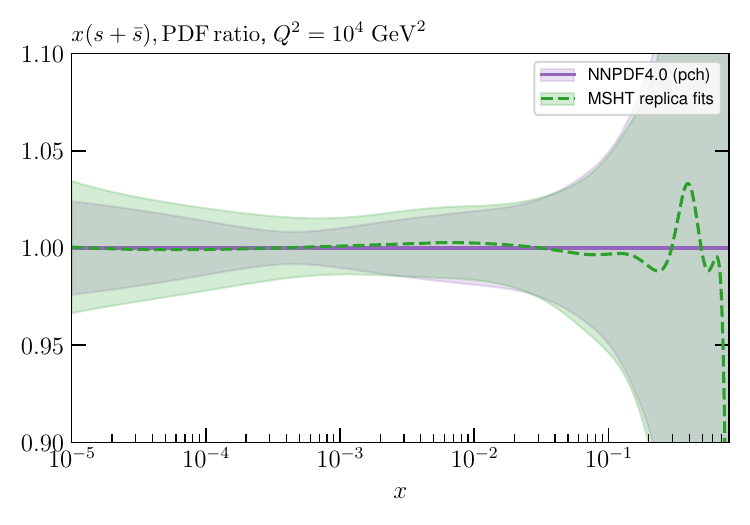}
\includegraphics[scale=0.6]{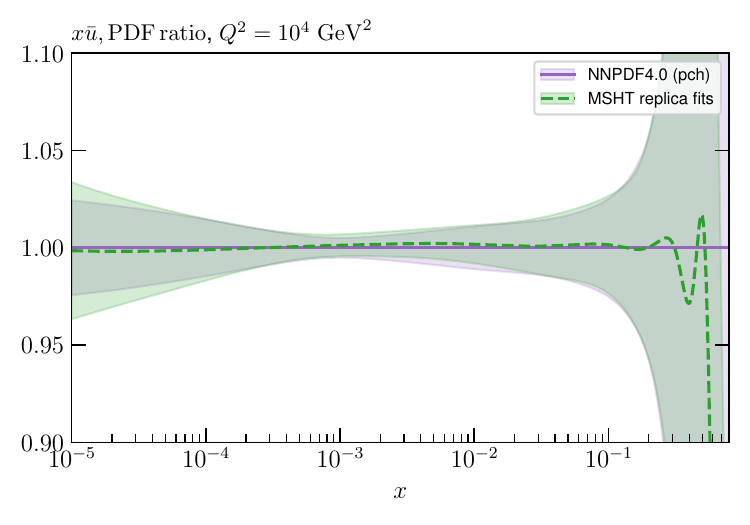}
\includegraphics[scale=0.6]{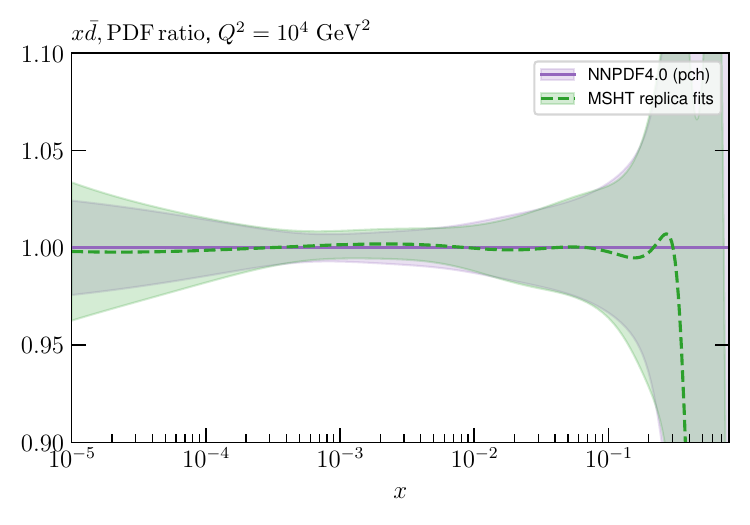}
\includegraphics[scale=0.6]{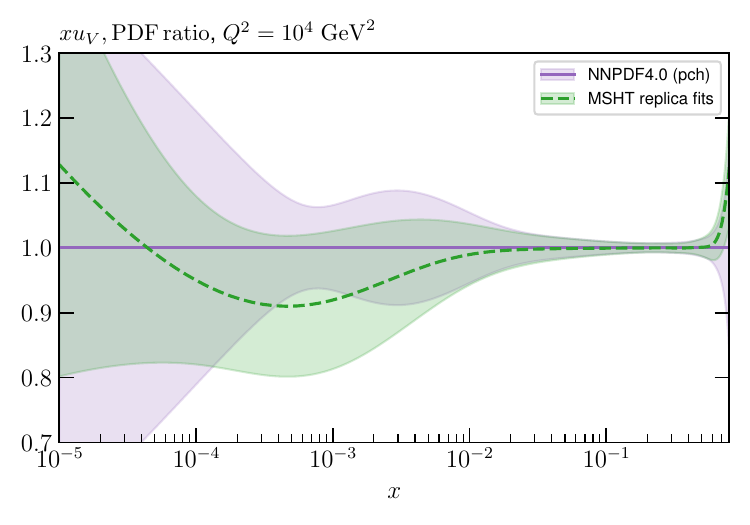}
\includegraphics[scale=0.6]{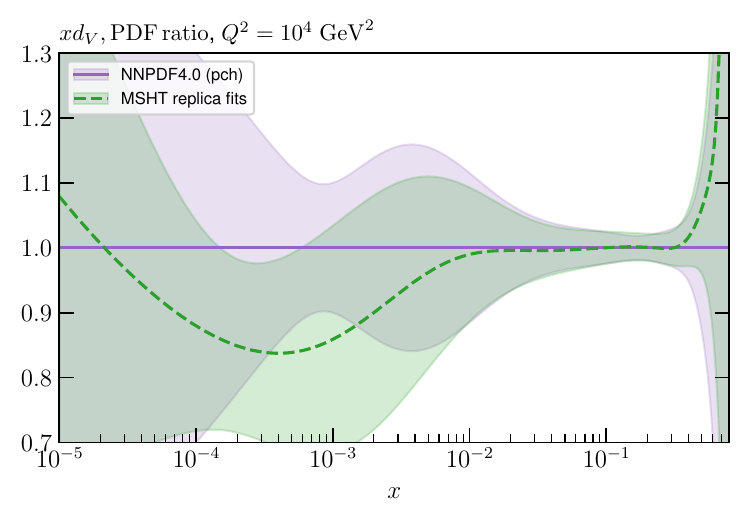}
\caption{\sf PDF ratios at $Q^2=10^4 \, {\rm GeV^2}$ of the result of  unfluctuated closure tests, using the MSHT20 parameterisation, to the corresponding NNPDF4.0 (pch) input.  The results of the same 100 fits shown in Fig.~\ref{fig:glcl_nnreps_closures} but now with the average and MC uncertainty on the replica ensemble shown both for the set of closure test fits (`MSHT replica fits') and the NNPDF4.0 replica inputs.}
\label{fig:glcl_nnreps_closures_ensemble}
\end{center}
\end{figure}

We next discuss the fact that we have for concreteness so far only considered a closure test to a single input PDF set. While this corresponds in principle to a more flexible underlying set, such that the good level of agreement observed between the input set and the fits is indeed non--trivial,  clearly if a different input where used (or a non--central replica of this set) our results will change at some level. Indeed, as we will show explicitly in Section~\ref{sec:overfit}, the PDFs that result from a fit with the MSHT parameterisation to the real data show moderately more variation then the central NNPDF4.0 set.  With this in mind, we  also consider the results of the same unfluctuated closure fit to that above, but where the input sets are now individual replicas of the NNPDF4.0 perturbative charm set. There are 100 of these in total, such that the central set corresponds to their average. Given this, we can expect the variability of the individual replicas to be greater and hence for the agreement between the fit and input set to be somewhat less good than in the central set case. However, we note that the flexibility of the MSHT20 set is chosen in order for the central best fit PDF to match the level of precision required by a global PDF fit to data of the sort contained within the NNPDF4.0 dataset. Given this,  the most appropriate comparison remains that for which the central NNPDF set is used as in input, and not of any individual replica, where the variability in the set is generally greater. 

For these fits, the best fit $\chi^2$ will vary from input to input, but is on average somewhat higher than in the central case, with $\langle \chi^2 \rangle \sim 15$, i.e. $\sim 0.003$ per point, though this is still of the same order of the value quoted in~\cite{DelDebbio:2021whr}. In Fig.~\ref{fig:glcl_nnreps_closures}  we show the ratio of the MSHT unfluctuated closure test fits to the 100 NNPDF replica inputs for a selection of PDFs, with the $T^2=1$ uncertainty from the fit to the central set given for guidance; for any individual fit the uncertainty will be similar.  We can see that on average the agreement between the fit and input remains very good, and well within this uncertainty in the data region. However, clearly the spread is larger and for a handful of sets the agreement can lie outside the quoted uncertainty band in some regions. In Fig.~\ref{fig:glcl_nnreps_closures_abs} we show for illustration the ratio of the input replicas to the central NNPDF set for the gluon and strangeness PDFs, along with the corresponding uncertainty band, purely for guidance. We can see that for the degree of variability in these replicas is indeed as expected often quite a bit larger than for the central set. Given this, it can be expected that the MSHT20 parameterisation  does not always achieve as good a matching to the input set as to the central set, even if in general it remains good.

Indeed, another interesting way to demonstrate the average matching (or lack of it) between these fits and the input replica sets is to simply compare the original NNPDF4.0 PDF set to that of the set of MSHT closure test fits, with the central (average) value and the uncertainty calculated from the corresponding MC ensembles. This is shown in Fig.~\ref{fig:glcl_nnreps_closures_ensemble} and we can see that the matching is rather good, albeit with some inflation in the spread of MSHT replicas in particular at low $x$, as a result of the noise present due to the mismatch between the individual fits and the input replicas that can occur, and tends to be larger in such less well constrained regions. 

Finally, while we have shown that the MSHT parameterisation is sufficiently flexible to perform well in a closure test (both unfluctuated, and as we will see fluctuated), these assume by construction complete consistency between data and theory and between the datasets in the fit. It is certainly possible that parameterisation flexibility may play more of a role in the more complicated, but more realistic context of a global fit and/or closure tests without such consistency built in. We will  consider this point further, which is inevitably tied up with the question of the PDF uncertainty treatment, in Section~\ref{sec:tol}. However, we note here that an issue that is  related to this is that in the real global fit it can be  necessary to fix certain poorly constrained parameters, such that the low $x$ power of the strangeness to be the same as that of the sea, in order to avoid unexpected behaviour in the quark flavour decomposition at low $x$. Indeed, such unphysical behaviour is often observed to be driven by an attempt to compensate for the non--ideal nature of the underlying fit. This can e.g. prefer for there to be some violation of the momentum sum rule from the PDFs in the data region, which can be achieved by rather unphysical behaviour at low $x$ outside of this region. At the level of closure tests, on the other hand (where such behaviour is not present by construction, due to the behaviour of the input PDF set), we find this is not the case.

\subsection{Fluctuated Global Closure Test}\label{subsec:fluc}

\subsubsection{PDF--level comparisons}

We now consider the case where the pseudodata are fluctuated by their  uncertainties due to the experimental covariance matrix. This is clearly closer to the case of a real PDF fit, as in principle a perfect fit to the data is no longer achievable. Indeed we expect, and find, a $\chi^2/N_{\rm pts} \sim 1$ for theory predictions corresponding to the original PDF input that has generated the underlying pseudodata, or more precisely within $1 \pm 2/\sqrt{N_{\rm pts}}$ at 68\% confidence. On the other hand, we are still assuming that the data errors are completely faithful, i.e. that there are no tensions between datasets beyond those due to statistical fluctuations, and in particular the theory used to produce the pseudodata (prior to fluctuations) exactly matches that used to evaluate the fit quality.

In the previous section we note that we evaluated the PDF uncertainty due to the unfluctuated closure test purely as a means to measure the significance of any deviation between the fit and input PDF set. In other words, it assessed the faithfulness of the MSHT fixed parameterisation approach in fitting the central value of the input set. The aim of performing a fluctuated closure test, on the other hand, is to validate the faithfulness of the PDF uncertainties that come out of the fit. This is directly analogous to the `level 2' closure tests performed by the NNPDF collaboration~\cite{NNPDF:2014otw}. In particular, if the uncertainties are faithful, then we expect the input PDF set to lie within the 1$\sigma$ uncertainty band of the fitted PDFs with close to 68\% confidence. As we will see, this is indeed the case.

To show this we begin by generating 100 pseudodata sets due to the same NNPDF4.0 (p. charm) input set as before, in each case with the pseudodata fluctuated with a different random number seed. Fitting the ensemble of these we can therefore generate a set of 100 fluctuated fits. The corresponding PDF uncertainty on the original unfluctuated fit can then be evaluated according to the MC  error propagation procedure, rather than the Hessian one considered in the previous section. The general expectation here is that the MC  uncertainty should match the Hessian uncertainty with $T^2=1$ in regions and for PDFs where the Gaussian approximation for the $\chi^2$ minimum in the PDF parameters is good, while for regions where this is less good, i.e. in the unconstrained extrapolation regions, the matching may be less close.

\begin{figure}
\begin{center}
\includegraphics[scale=0.6]{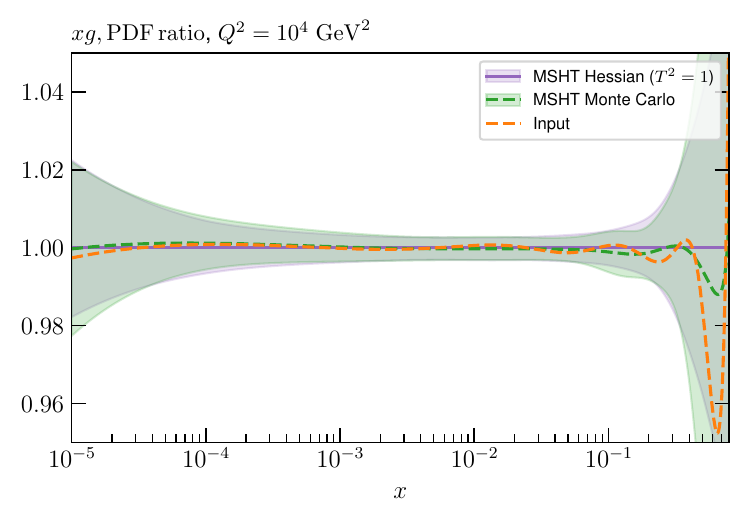}
\includegraphics[scale=0.6]{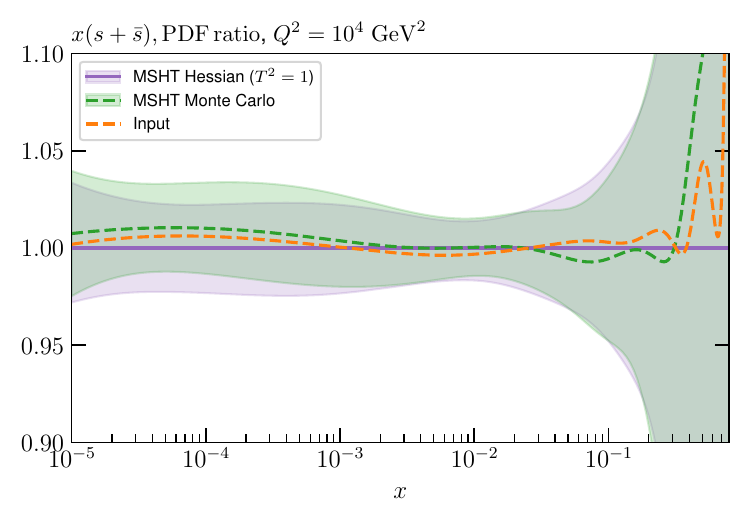}
\includegraphics[scale=0.6]{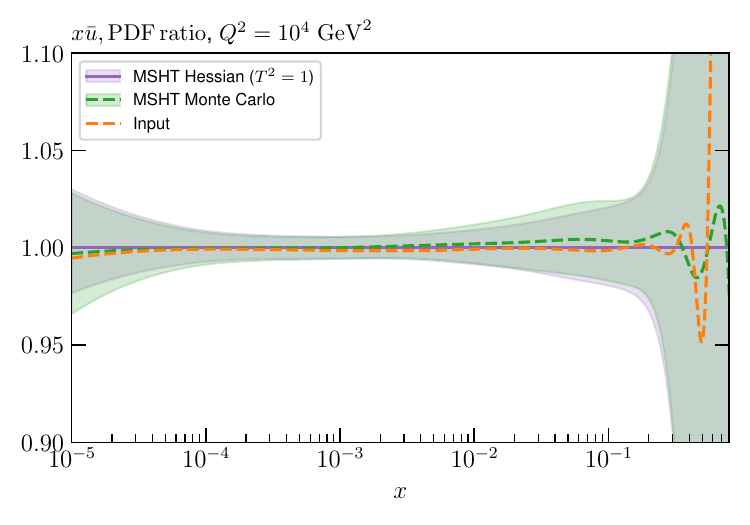}
\includegraphics[scale=0.6]{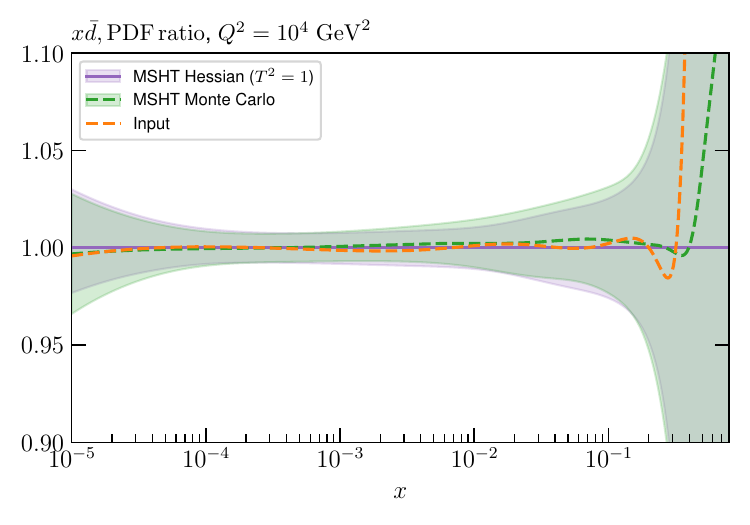}
\includegraphics[scale=0.6]{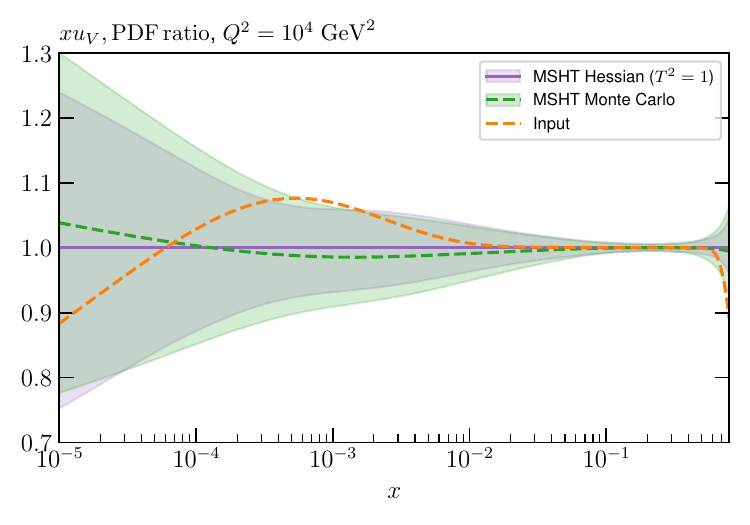}
\includegraphics[scale=0.6]{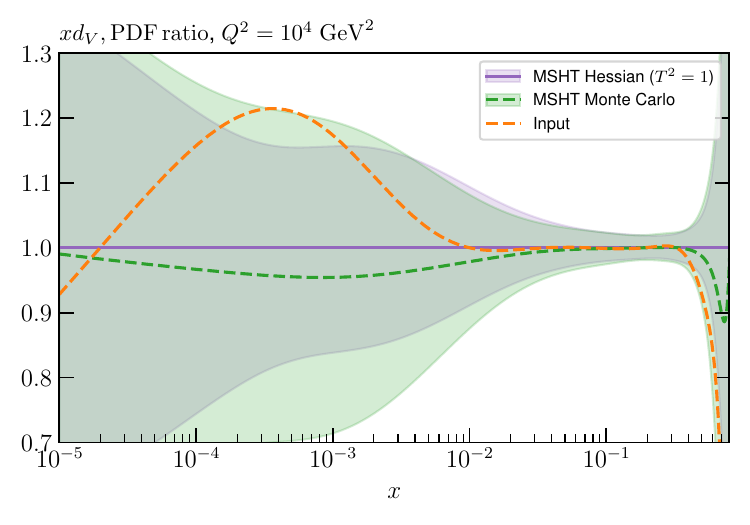}
\caption{\sf A selection of PDFs at $Q^2=10^4 \, {\rm GeV^2}$  that result from a fluctuated closure test fit to the NNPDF4.0 dataset,  using the MSHT20 parameterisation. The Hessian PDF uncertainties calculated with a $T^2=1$  fixed tolerance that result from the unfluctuated closure test are shown in purple (and are as in Fig.~\ref{fig:glcl_rat}), while the result of MC error generation are shown in green. The NNPDF4.0 (p. charm) input is given by the dashed red line. Results are shown as a ratio to the MSHT ($T^2=1$) fit.}
\label{fig:glcl_mcrep}
\end{center}
\end{figure}

The results of this comparison are shown in Fig.~\ref{fig:glcl_mcrep} and we can indeed see that in the most constrained regions of $x$ the matching between the MC  and Hessian result is rather encouraging. The MC  results are calculated using the textbook result for the $1\sigma$ standard deviation of the ensemble of fits, though we have confirmed that the results are very similar if instead the 68\% C.L. is calculated directly. For clarity we recall that the central value of the MC  result is derived from the average of the  fits to the fluctuated pseudodata, while the Hessian result derives from a fit to the unfluctuated pseudodata. Hence there is no strict requirement for the central values to agree, due either to finite sample effects or a breakdown in the Gaussian approximation. Indeed differences are observed, most dominantly in the less constrained high $x$ region, and for the valence distributions at low $x$, where again constraints are limited. 

\begin{figure}
\begin{center}
\includegraphics[scale=0.6]{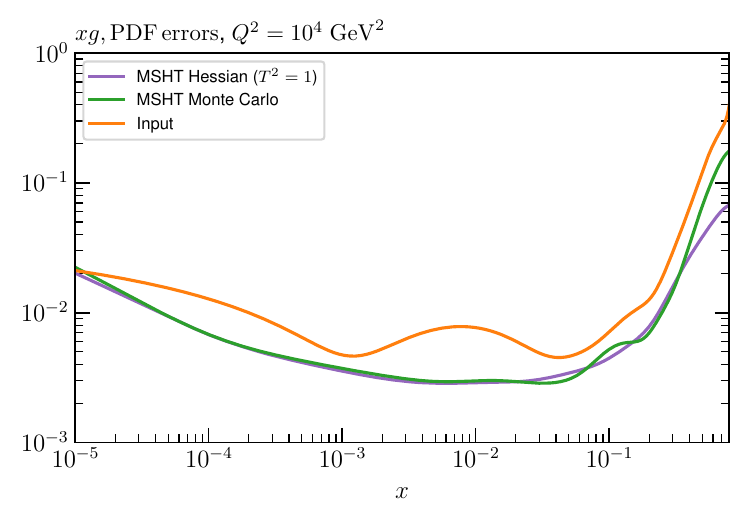}
\includegraphics[scale=0.6]{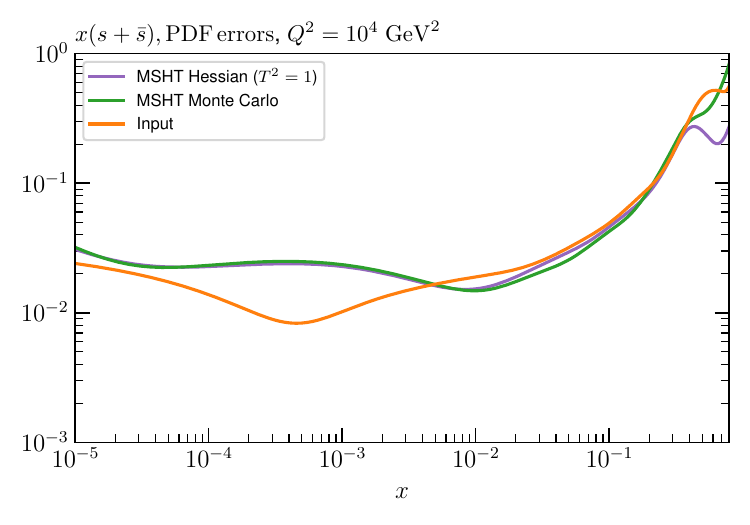}
\includegraphics[scale=0.6]{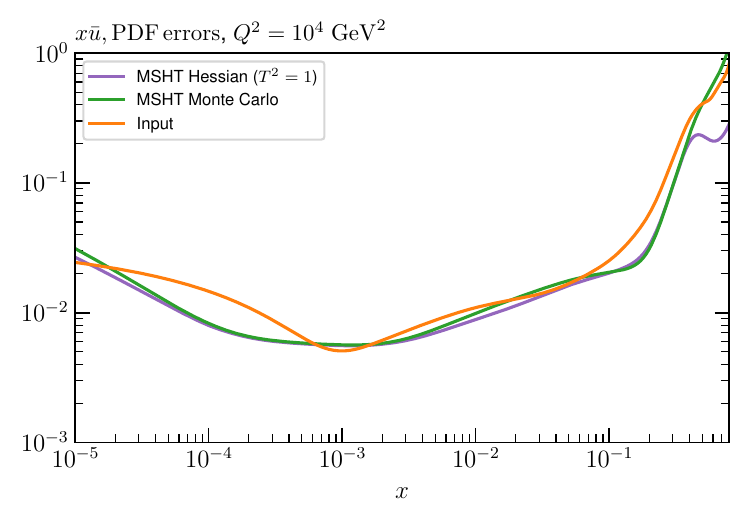}
\includegraphics[scale=0.6]{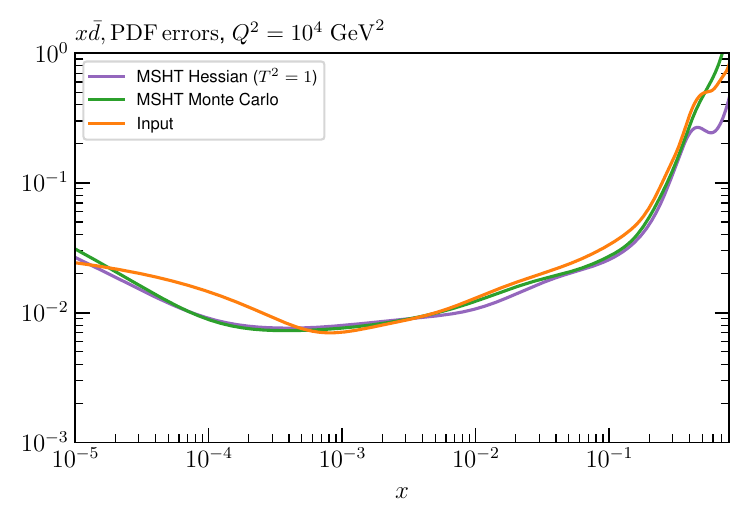}
\includegraphics[scale=0.6]{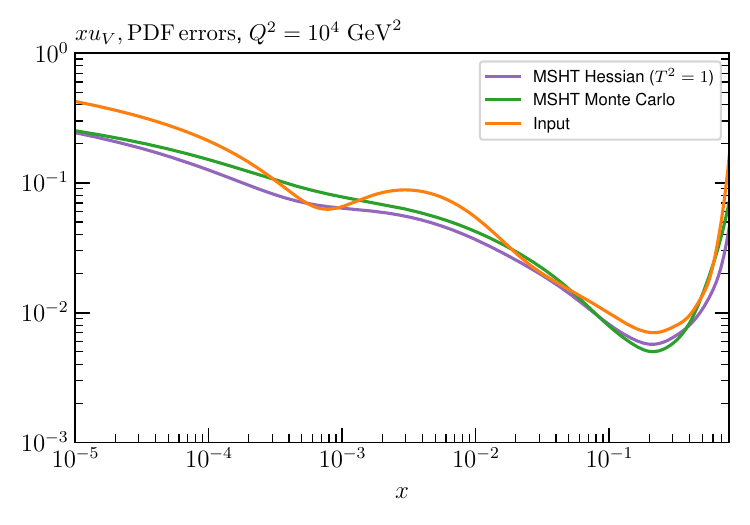}
\includegraphics[scale=0.6]{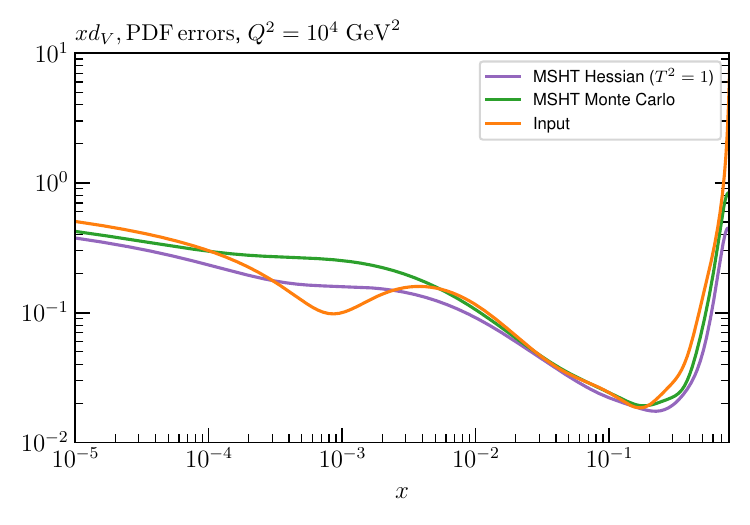}
\caption{\sf PDF uncertainties at $Q^2=10^4 \, {\rm GeV^2}$  that result from a fluctuated closure test fit to the NNPDF4.0 dataset,  using the MSHT20 parameterisation. The Hessian PDF uncertainties calculated with a $T^2=1$  fixed tolerance are shown in purple (and are as in Fig.~\ref{fig:glcl_errs}), while the result of MC error generation are shown in green. The NNPDF4.0 (p. charm) input is also shown for comparison, although this plays no role in the closure test.}
\label{fig:glcl_err_mcrep}
\end{center}
\end{figure}

\begin{figure}
\begin{center}
\includegraphics[scale=0.6]{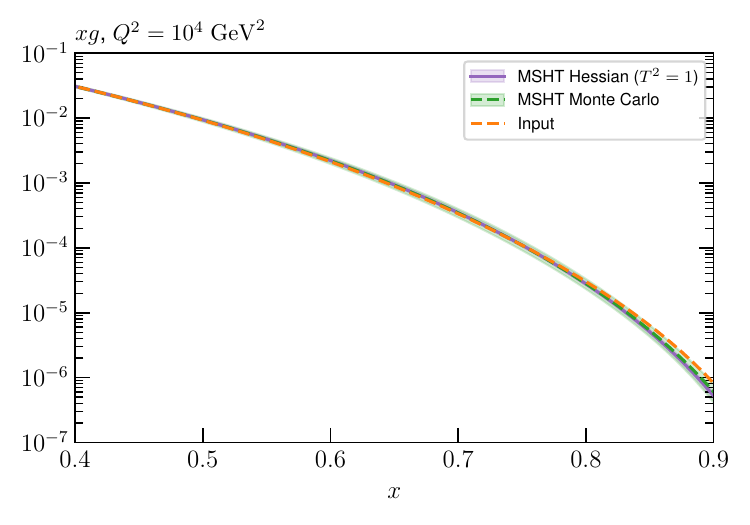}
\includegraphics[scale=0.6]{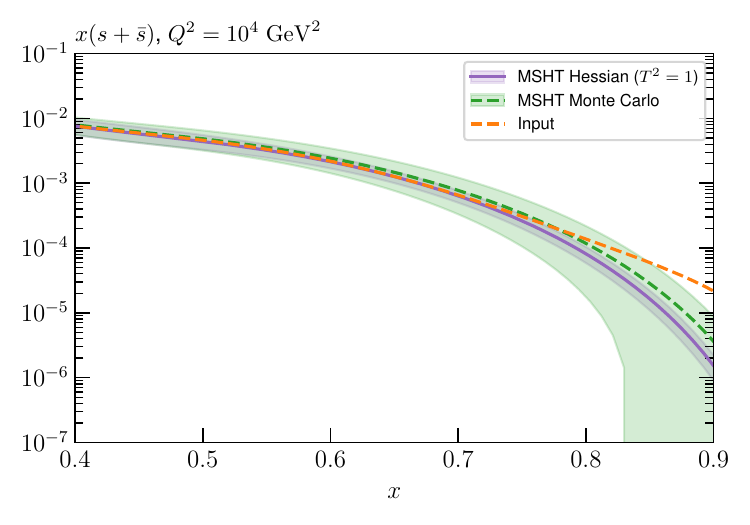}
\includegraphics[scale=0.6]{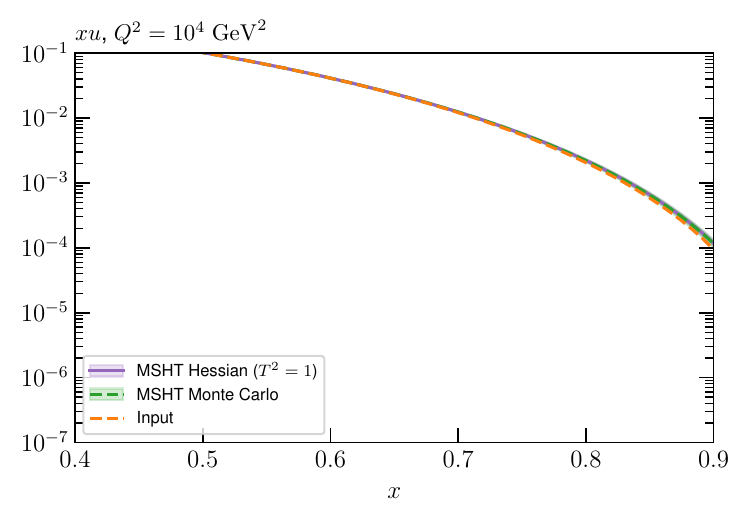}
\includegraphics[scale=0.6]{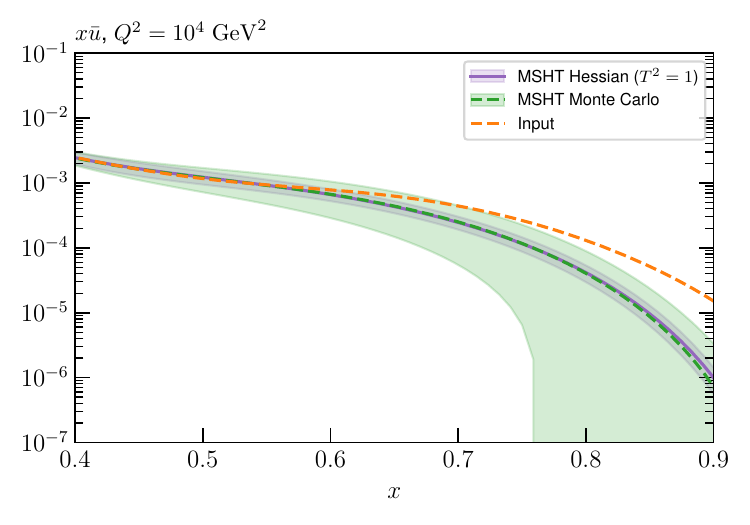}
\includegraphics[scale=0.6]{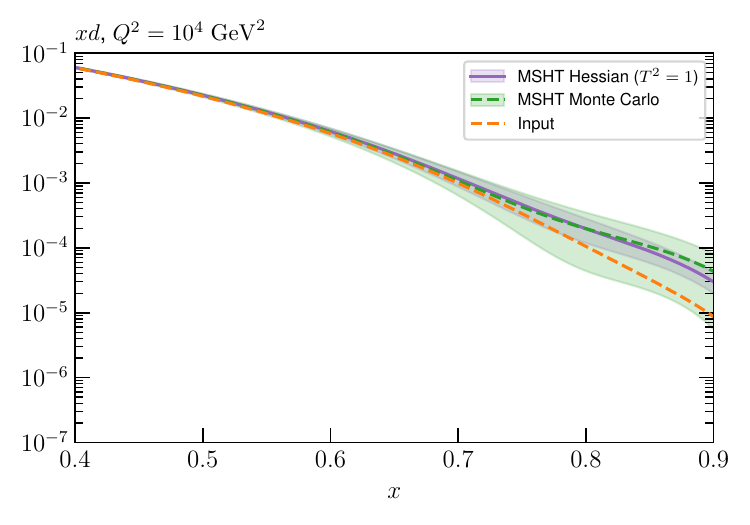}
\includegraphics[scale=0.6]{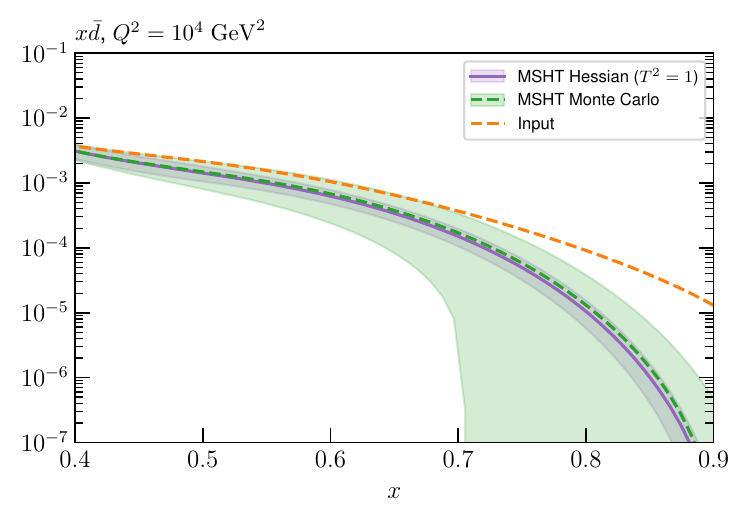}
\caption{\sf A selection of PDFs at $Q^2=10^4 \, {\rm GeV^2}$  that result from an unfluctuated closure test fit to the NNPDF4.0 dataset,  using the MSHT20 parameterisation, now focussing on the high $x$ region. The Hessian PDF uncertainties calculated with a $T^2=1$  fixed tolerance are shown in purple (and are as in Fig.~\ref{fig:glcl_rat}), while the result of the MC error generation are shown in green. The NNPDF4.0 (p. charm) input is given by the dashed red line.}
\label{fig:glcl_mcrep_highx}
\end{center}
\end{figure}

In Fig.~\ref{fig:glcl_err_mcrep} we show the corresponding PDF uncertainties, along with the NNPDF4.0 (p. charm) input PDF uncertainty for comparison, although as discussed in the previous section this plays no role in the closure test and is instead derived from the NNPDF fit to the real data. Again for most PDFs and in most regions of $x$, notably in the most directly constrained regions, the agreement between the Hessian ($T^2=1$) and MC  uncertainties is very good. 

As described above, however, there are regions where the agreement between the Hessian and MC  results is less good, in particular at high $x$, where direct constraints start to run out. We can see from Fig.~\ref{fig:glcl_err_mcrep} that in this region the MC  uncertainty tends to be larger, and interestingly tends to match the NNPDF4.0 input set rather more closely, though we emphasise again that this is a comparison of the NNPDF uncertainty that comes from a fit to the real data to the MSHT fixed parameterisation to pseudodata, so is not like for like. Nonetheless, this may indicate that the MC  error propagation, irrespective of whether a suitably flexible fixed polynomial basis or a neural network is used to parameterise the PDFs, may be a significant and perhaps even the primary factor in the fact that the NNPDF uncertainty in the high $x$ extrapolation region tends to be more conservative than a Hessian one, as in the case of e.g. MSHT20.

To examine this further we focus on the high $x$ region in Fig.~\ref{fig:glcl_mcrep_highx}, where the absolute PDFs are shown, and a number of features are clear. We can see that 
while for the  gluon and light quarks the input set is consistent with the Hessian output of the closure test within the  $T^2=1$ uncertainties  out to rather high $x$, for the less well determined antiquarks this is less true. In particular, clear deviations are evident at $x \gtrsim 0.4-0.5$ for the $\overline{u}$ and $\overline{d}$, which is precisely where direct data constraints on these PDFs run out, and hence is largely in the extrapolation region. This highlights one of the potential shortcomings of the fixed parameterisation approach, at least with a Hessian uncertainty evaluation, namely that in such regions the corresponding PDF uncertainties may not be conservative enough (although we note that for the $T^2=10$ uncertainties the deviations are at somewhat, though not dramatically, higher $x$). This issue is not one that the MSHT20 dynamic tolerance  is designed to address directly, given this is driven by a hypothesis testing criterion that is by construction applied to the data region, and hence this may indicate a future avenue of improvement.

On the other hand, we can indeed see again that the MC  uncertainties are significantly larger than the Hessian ones in this region. This may be due to the particularly non--linear dependence of the PDFs on the PDF parameters and indeed non--linear effects in the data fluctuations themselves (given the data errors can be comparable to the size of the data) at high $x$, in this region where the PDFs are very small. 

We note that in~\cite{Costantini:2024wby} a recent critical discussion of the MC  method is presented, where some indication that it can prove unreliable in certain scenarios, notably when non--linearity is present, is demonstrated. In particular, differences between the MC  approach and a Bayesian method are seen in a range of toy PDF fits. Here, the non--linearity is restricted by construction to be due purely to the quadratic dependence of hadronic observables on the PDFs, rather than in the parameterisation itself. This may explain why in such a case, the difference is instead focused at low $x$. In a more realistic scenario, such an approach may shed light on the differences observed here.

\begin{figure}
\begin{center}
\includegraphics[scale=0.6]{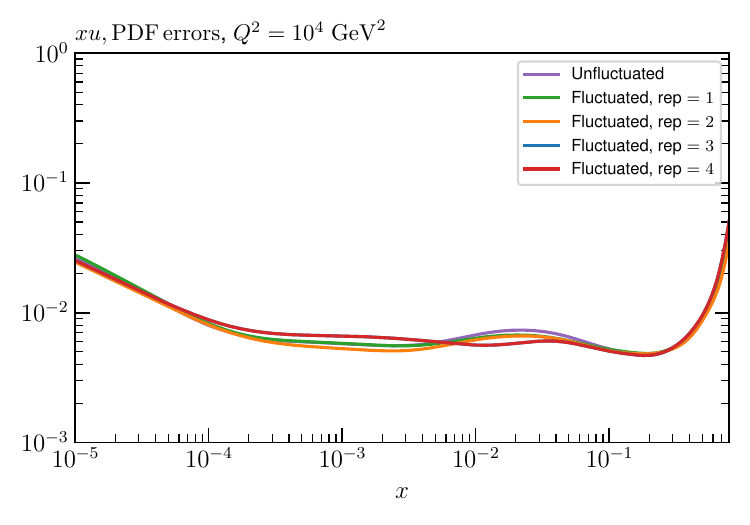}
\includegraphics[scale=0.6]{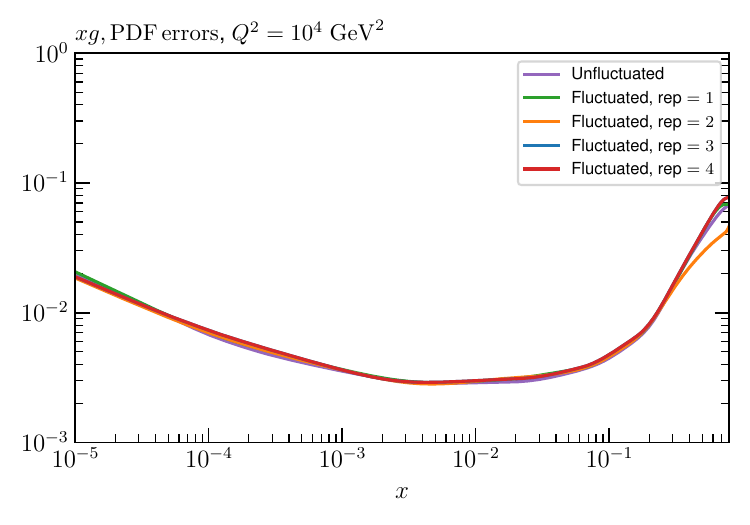}
\caption{\sf PDF uncertainties at $Q^2=10^4 \, {\rm GeV^2}$  that result from a fluctuated closure test fit to the NNPDF4.0 dataset,  using the MSHT20 parameterisation, for the up quark and gluon PDFs. The result of the unfluctuated closure test, as shown in Fig.~\ref{fig:glcl_errs}, is given as well as the Hessian uncertainty on the fit to four of the fluctuated pseudodata fits that enter the fluctuated closure test. All uncertainties are for $T^2=1$.}
\label{fig:glcl_rep_errorcomp}
\end{center}
\end{figure}

We next turn to the overall faithfulness of the PDF uncertainties in the context of the level 2 closure test discussed above, namely whether the input set is consistent with each of the 100 fluctuated fits within their corresponding (Hessian) 1$\sigma$ uncertainty with 68\% confidence. To demonstrate this, we in principle would have to perform a PDF eigenvector scan for each of the 100 fits above in order to evaluate their corresponding uncertainty. However, this is not in fact necessary, as for any given such fit, if we evaluate its Hessian uncertainty, we find the size of this is very stable across the fits, and indeed matches very closely with the Hessian uncertainty from the fit to the unfluctuated pseudodata shown in Fig.~\ref{fig:glcl_err_mcrep}. This is demonstrated explicitly in Fig.~\ref{fig:glcl_rep_errorcomp} for a small selection of fluctuated fits, where it is clear that the relative Hessian uncertainty is very stable, and similar to the result of the unfluctuated fit. We note a similar observation, namely of the stability of the uncertainty between different fluctuated fits, is made in~\cite{NNPDF:2014otw}, see the last paragraph of Section 4.3. The above observations are then in fact sufficient to demonstrate that the closure test has been passed. In particular: the central value of any of the 100 fluctuated fits will by definition lie within the (green) MC uncertainty band in Fig.~\ref{fig:glcl_mcrep}, which is very closely centred on the input set in the data region, with 68\% probability; this uncertainty band very closely matches to the Hessian uncertainty from the unfluctuated fit and hence  the relative Hessian uncertainty on any individual fluctuated best fit; therefore, if a given central value of one of the 100 fits lies within the 68\% MC uncertainty band it will also be consistent with the input within its 1$\sigma$ Hessian uncertainty. Putting this all together, each of these 100 fits will be consistent within its 1$\sigma$ Hessian uncertainty with the input set with 68\% frequency, as required.

As mentioned above, the above closure test is directly analogous to the `level 2' closure tests performed by the NNPDF collaboration. There are procedural distinctions in how these are implemented, but we note that these simply reflect the fact that different underlying fitting methodologies, and corresponding PDF uncertainty definitions, are being tested. In particular,  both the current and the NNPDF closure tests start with a particular set of fluctuated pseudodata and perform a standard fit with the PDF fitting framework to this. In the NNPDF case, this requires fluctuating these pseudodata a second (multiple) set of times, to generate a replica ensemble that is used to define the PDF uncertainties. In the MSHT  methodology, on the other hand, we start with the same pseudodata that has been fluctuated once, but as the PDF uncertainties are defined via the Hessian methodology, these are not fluctuated a second time. However, the fundamental test is the same.

\subsubsection{Statistical Estimators}

To demonstrate the above results quantitatively, we next consider a range of statistical estimators described in~\cite{DelDebbio:2021whr,NNPDF:2021njg}. We start with the PDF--level estimator defined in~\cite{NNPDF:2014otw}:
\be\label{eq:xipdf}
\xi_{n \sigma}^{\rm (pdf)} = \frac{1}{n_{\rm flav}} \frac{1}{n_{x}} \frac{1}{n_{\rm fit}} \sum_{i=1}^{n_{\rm flav}} \sum_{j=1}^{n_{x}} \sum_{l=1}^{n_{\rm fit}}I_{[-n \sigma^{i(l)}(x_j),n \sigma^{i(l)}(x_j)]}\left(q^{i(l)}(x_j)- q^{i(l)}_{\rm input}(x_j)\right)\;.
\ee
Here, $q^{i(l)}(x_j)$ is the resulting PDF of flavour $i$ for the $l$th fluctuated closure test at a point $x_j$, while $q^{i(l)}_{\rm input}(x_j)$ is the corresponding (NNPDF4.0 perturbative charm) input set. The standard deviation PDF uncertainty on the fit PDF is denoted by $\sigma^{i(l)}(x_j)$, while $I_A(x)$ is only non--zero, and equal to unity, if its argument (i.e. the difference between the fit PDF and the input) lies within the relevant $\pm n \sigma$ interval. In other words, this gives the fraction of the fit PDFs, summed over different flavours and at some specified $x$ points that are consistent with the input set within their $n\sigma$ uncertainty.

We will consider the $n=1$ case here, for which we will expect $\xi_{1\sigma}^{\rm PDF}\approx 0.68$ if the uncertainties are faithful. However, there are various caveats to this~\cite{NNPDF:2021njg}, notably that it assumes a Gaussian distribution in the fluctuated fit results at a given $x$ value, which will be less true away from the data region, and it does not account for the correlation between neighbouring $x$ points. Nonetheless, it is a useful indicator. With this in mind, we evaluate $\xi_{1\sigma}^{\rm PDF}$ for 20 $x$ points spaced logarithmically between $10^{-5}$ and $0.3$, while we have $n_{\rm fit}=100$, as described above. In order to avoid performing an eigenvector scan for all 100 of these fits, we use in our quoted results that the $1\sigma$ uncertainty on any given fluctuated fit is constant and  identical to that on the unfluctuated fit, which as described above is observed to hold very closely. However, we have confirmed for a smaller (but sufficient) number of fits, $n_{\rm fit}=25$, that evaluating these uncertainties exactly gives a very similar result, within the larger statistical uncertainty on $\xi$ that comes from this smaller $n_{\rm fit}$. The results are shown in Table~\ref{tab:xi} for $Q^2=10^4$ ${\rm GeV}^2$, and we can see that indeed  $\xi_{1\sigma}^{\rm PDF}\approx 0.68$ is satisfied rather well. There is  some very slight tendency for the overall estimator to lie below the textbook $1\sigma$ value, but given the imperfect nature of this estimator, as described above, we judge this to be more than sufficient to demonstrate a successful closure test. We note that if the evaluation is instead made at e.g. the input scale $Q_0$, and/or with a larger $x$ range, then the value of $\xi_{1\sigma}^{\rm PDF}$ is relatively stable.

\addtolength{\tabcolsep}{-0.3em}
\begin{table}
\begin{center}
  \scriptsize
  \centering
   \renewcommand{\arraystretch}{1.4}
\begin{tabular}{Xrcccccccc}\hline 
&$s$& $u$& $d$& $g$& $\overline{d}$& $\overline{u}$& $\overline{s}$&{\bf Total}
\\ \hline
$\xi_{1\sigma}^{\rm PDF}$&$0.67\pm 0.02$&$0.64\pm 0.02$&$0.66\pm 0.02$&$0.64\pm 0.02$&$0.65\pm 0.02$&$0.69\pm 0.02$&$0.65\pm 0.02$&$\mathbf{0.656\pm 0.006}$\\
\hline
\end{tabular}
\end{center}
\caption{\sf The $\xi_{1\sigma}^{\rm PDF}$ estimator at $Q^2=10^4$ ${\rm GeV}^2$, described in the text, for different parton flavours and for the sum over all flavours. The statistical uncertainty due to the finite number of points in $x$, flavour and fit space is given.}
\label{tab:xi}
\end{table}
\addtolength{\tabcolsep}{0.3em}

We next consider the equivalent estimator defined in data space
\be\label{eq:xidata}
\xi_{n \sigma}^{\rm (data)} = \frac{1}{N_{\rm data}}\frac{1}{n_{\rm fit}} \sum_{i=1}^{N_{\rm data}}  \sum_{l=1}^{n_{\rm fit}}I_{[-n \sigma^{i(l)},n \sigma^{i(l)}]}\left(g_i^{(l)} - f_i\right)\;.
\ee
Here $f_i$ is the theoretical prediction for the $i$th datapoint using the (NNPDF4.0 perturbative charm) input set, without fluctuation, i.e. exactly corresponding to the `true' value. The $g_i^{(l)}$ are the predictions from the $l$th fluctuated closure test fit and $\sigma^{i(l)}$ is the PDF uncertainty on this prediction, calculated using the PDF eigenvectors from this fit in the usual way. To be precise, we apply the appropriate asymmetric error definition, with
\begin{align}\nonumber
 \sigma_{+}^{i(l)}&=\sqrt{\sum_{k=1}^{N_{\rm eig}} \left\{  {\rm max} \, \left [ g_i^{(l)}(S_k^+)-  g_i^{(l)}(S_0),\,g_i^{(l)}(S_k^-)-  g_i^{(l)}(S_0),\,0 \right] \right\}^2}\;,\\ \label{eq:pdferrasym}
 \sigma_{-}^{i(l)}&=\sqrt{\sum_{k=1}^{N_{\rm eig}} \left\{  {\rm max} \, \left [   g_i^{(l)}(S_0)-g_i^{(l)}(S_k^+),\, g_i^{(l)}(S_0)-g_i^{(l)}(S_k^-) ,\,0 \right] \right\}^2}\;,
\end{align}
where $S_{k}^\pm $ corresponds to the $k$th $\pm$ eigenvector and $S_0$ the central PDF set. We can then either symmetrise the above positive and negative uncertainties to give $\sigma^{i(l)}$ or alternatively modify \eqref{eq:xidata} to count the fraction of the data points for which $f_i$ lies above or below $g_i^{(l)}$ within the corresponding $\pm$ uncertainty. These give very similar results, however we note that if the uncertainty is instead symmetrised at the eigenvector level, i.e.
\be\label{eq:pdferrsym}
\sigma^{i(l)}_{\rm sym}=\frac{1}{2}\sqrt{\sum_{k=1}^{N_{\rm eig}} \left[ g_i^{(l)}(S_k^+)- g_i^{(l)}(S_k^-) \right]^2}\;,
\ee
then as we will see the results differ somewhat. For recent discussion of the Hessian PDF uncertainty definition in the presence of asymmetric uncertainties see~\cite{Zhan:2024tic}.

This estimator therefore gives the fraction of the predicted datapoints that are consistent with the `true' input values within the quoted PDF uncertainty. To be consistent with the discussion in~\cite{NNPDF:2021njg} we present this estimator in the basis that diagonalises the experimental covariance matrix, such that the sum over the number of datapoints becomes a sum over the eigenvectors of the experimental covariance matrix. 

We present results for this estimator evaluated on the same full dataset as that which enters into the fit. This in contrast to~\cite{NNPDF:2021njg}, where the closure test fits are performed to a subset of the NNPDF4.0 dataset, namely a NNPDF3.1--like dataset. While in principle this may provide a more general test of the fitting procedure, we note in practice that as discussed in~\cite{NNPDF:2021njg} the kinematic coverage of the testing dataset in that  study is similar to the dataset entering the fit, and indeed this is necessary in order to ensure Gaussianity. That is, the testing data should be sensitive to the PDFs in a region where they are well constrained by the fitted data, and hence by construction the estimator is less reliable in any regions where this is not the case, e.g. in regions of extrapolation from the fitted data. Therefore, given the  test and fitted data are generated from the same underlying true input PDF set and  the relevant quantity in~\cite{NNPDF:2021njg} is the value of the statistical estimator due to the test data in a region that is constrained by the fitted data, it is arguable how much can be gained from performing this division. Moreover, this necessarily implies reducing the size of the dataset entering the closure test fit, which is undesirable from the point of view of producing a direct comparison with the results above. Nonetheless, an extension in this direction may be of use in the future.

\begin{table}
\begin{center}
  \scriptsize
  \centering
   \renewcommand{\arraystretch}{1.4}
\begin{tabular}{Xrccccc}\hline 
&DIS& DY& Top& Jets & {\bf Total} 
\\ \hline
$\xi_{1\sigma}^{\rm Data}$&$0.64\pm 0.01$&$0.68\pm 0.01$&$0.67\pm 0.02$&$0.64\pm 0.02$&$\mathbf{0.65\pm 0.01}$\\
\hline
${\rm erf}(1/\sqrt{2}R_{bv})$&$0.66\pm 0.01$&$0.72\pm 0.01$&$0.70\pm 0.03$&$0.62\pm 0.02$&$\mathbf{0.69\pm 0.005}$\\
\hline
\end{tabular}
\end{center}
\caption{\sf The $\xi_{1\sigma}^{\rm Data}$ estimators for different dataset types and for the total global dataset, in the basis that diagonalises the experimental covariance matrix. The statistical uncertainty due to the finite number of datapoints and fits is given. Results with and without the PDF uncertainties symmetrised are shown, along with the corresponding values of ${\rm erf}(R_{vb}/\sqrt{2})$.}
\label{tab:xi_data}
\end{table}

The corresponding values of the $1\sigma$ estimator, along with the statistical uncertainty due to the finite number of fits (now taken with $n_{\rm fit}=40$ for improved precision\footnote{As the PDF uncertainties are evaluated using the eigenvectors for each fit, the full set of 100 fits used in the PDF--level estimator is not useable here.}) for the total dataset and some representative divisions of it, are shown in Table~\ref{tab:xi_data}. We can see that the total value is slightly lower than the textbook $\sim 0.68$, indicating that the uncertainties may be slightly smaller than is required, but rather close to it in a manner that can give us some confidence in the overall faithfulness of the PDF uncertainties according to this estimator. This is particular true given that the value of this estimator is $\sim 1$ if we e.g. instead take PDF uncertainties evaluated according to a $T^2=10$ fixed tolerance, that is the test would fail dramatically in such a case, with the uncertainties being significantly too conservative. Considering the data subsets, we can see that there is some variation between these, but overall rather good consistency with expectations is observed.

\begin{figure}
\begin{center}
\includegraphics[scale=0.6]{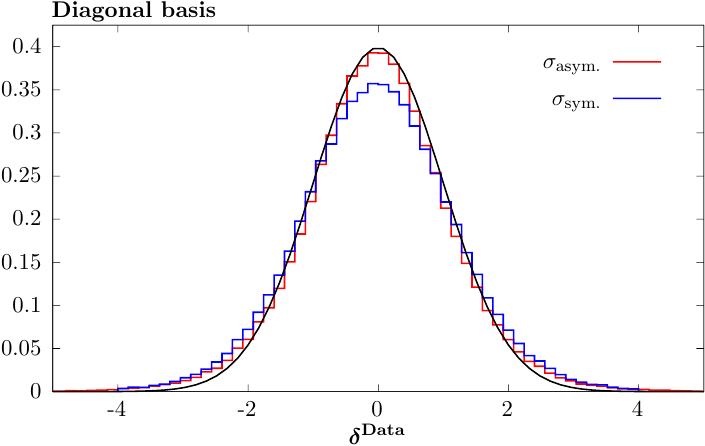}
\caption{\sf The normalized distribution of relative differences $\delta_{\rm data}$ in data  space in the basis that diagonalises the experimental covariance matrix. $\sigma_{\rm asym}$ corresponds to the appropriate asymmetric PDF uncertainty definition, while for $\sigma_{\rm sym}$ these are symmetrised at the eigenvector level. The latter case is shown for the sake of comparison. The black curve corresponds to a normal distribution with unit variance}
\label{fig:dels_diag_asym_sym_comp}
\end{center}
\end{figure}

Looking in more detail, in Fig.~\ref{fig:dels_diag_asym_sym_comp} we plot the values of
\be
\delta_I^{(l)} = \frac{g_i^{(l)} - f_i}{\sigma^{i(l)}}\;,
\ee
where the quantities on the right hand side are as defined in \eqref{eq:xidata} and we again evaluate these in the basis that diagonalises the experimental covariance matrix. If the PDF uncertainties are faithful, then these should be normally distributed with unit variance, and we can see that this is indeed very close to being true. There is a very mild tendency for the plotted values to have a larger tail than the normal distribution, consistent with the value of $\xi_{n \sigma}^{\rm (data)}$ being slightly lower than $0.68$, but this is clearly a small effect. To be precise, $\sigma^{i(l)}$ in the above expression is calculated by symmetrising the result of the asymmetric error definition \eqref{eq:pdferrasym}. However, if the symmetric definition \eqref{eq:pdferrsym} is instead used the distribution deviates more, as is is shown in the figure for comparison, indicating that this definition is indeed not to be preferred.

\begin{table}
\begin{center}
  \scriptsize
  \centering
   \renewcommand{\arraystretch}{1.4}
\begin{tabular}{Xrccccccc}\hline 
&DIS& DY& Top& Jets & {\bf Total} 
\\ \hline
$R_{bv}$&$1.04\pm 0.02$&$0.92\pm 0.02$&$0.97\pm 0.06$&$1.15\pm 0.05$&$\mathbf{0.98\pm 0.01}$\\
\hline
\end{tabular}
\end{center}
\caption{\sf The bias--to--variance ratios for different dataset types and for the total global dataset. The statistical uncertainty due to the finite number of datapoints and fits is given. Results with and without the PDF uncertainties symmetrised are shown, as described in the text.}
\label{tab:Rbv}
\end{table}

Also shown in Table~\ref{tab:xi_data} is the value of
\be\label{eq:erfrbv}
{\rm erf}\left(\frac{1}{\sqrt{2}R_{bv}}\right) \approx \xi_{1\sigma}^{\rm Data}\;,
\ee
where this approximate equality is expected to hold for successful closure tests\footnote{We note that in~\cite{DelDebbio:2021whr,NNPDF:2021njg} the $R_{bv}$ in \eqref{eq:erfrbv} is in the numerator of the erf function, however  this has been confirmed by the authors as a typo~\cite{DelDebbio:private}.}. Here, $R_{bv}$ is the bias--to--variance ratio defined in~\cite{DelDebbio:2021whr}, which is given by
\be\label{eq:rbvdef}
R_{bv} = \sqrt{\frac{{\bf E}_{\rm fits}[{\rm bias}]}{{\bf E}_{\rm fits}[{\rm variance}]}}\;,
\ee
where 
\begin{align}\label{eq:biasdef}
{\bf E}_{\rm fits}[{\rm bias}] &= \frac{1}{N_{\rm data}} \frac{1}{n_{\rm fit}} \sum_{l=1}^{n_{\rm fit}}(g^{(l)}- f)_i C_{ij}^{-1} (g^{(l)}- f)_j \;,\\ \label{eq:variancedef}
{\bf E}_{\rm fits}[{\rm variance}] &= \frac{1}{N_{\rm data}} \frac{1}{n_{\rm fit}} \sum_{l=1}^{n_{\rm fit}} \Sigma^{\rm var}\left[g^{(l)}_i ,g^{(l)}_j\right] C_{ij}^{-1}\;,
\end{align}
are defined in analogy to the quantities defined in~\cite{DelDebbio:2021whr} but for the Hessian uncertainty definition. Here the variance $\Sigma^{\rm var}$ is given in the case of a symmetric PDF uncertainty definition by
\be\label{eq:sigmavar}
\Sigma^{\rm var}\left[g^{(l)}_i ,g^{(l)}_j\right]= \frac{1}{4} \sum_{k=1}^{N_{\rm eig}}(g^{(l)}_i(S_k^+)- g_i^{(l)}(S_k^-)) (g^{(l)}_j(S_k^+)- g_j^{(l)}(S_k^-))\;.
\ee
However, given as noted above this is observed to be a less faithful uncertainty definition, we instead present results with the asymmetric treatment. This is possible by evaluating \eqref{eq:variancedef} after diagonalising the covariance matrix, in which case we simply have
\be
\Sigma^{\rm var}\left[g^{(l)}_i ,g^{(l)}_j\right]  \rightarrow \left(\sigma^{i(l)}\right)^2\;,
\ee
defined in the basis that diagonalises the experimental covariance matrix, and this can be evaluated using the asymmetric error definition. We note that if the symmetric error definition were used this would give the same results as applying  \eqref{eq:sigmavar} in the data basis, given \eqref{eq:variancedef} is basis independent. Results are shown in  Table~\ref{tab:Rbv}, again for the total dataset and for some subsets of it. For faithful PDF uncertainties we expect $R_{bv} \approx 1$ and we can indeed see that this holds rather closely. The value of ${\rm erf}(1/\sqrt{2}R_{bv})$ corresponding to these is given in Table~\ref{tab:xi_data} and we can see that the agreement is reasonable, though not perfect.

\addtolength{\tabcolsep}{-0.4em}
\begin{table}
\begin{center}
  \scriptsize
  \centering
   \renewcommand{\arraystretch}{1.4}
\begin{tabular}{Xrcccccc}\hline 
&DIS NC& DIS CC& DY& Top& Jets & {\bf Total} 
\\ \hline
${\bf E}_{\rm fits}[\Delta_{\chi^2}^{(l)}] $&$-0.0059\pm 0.0006$&$-0.0054\pm 0.0009$&$-0.0184\pm 0.0013$&$-0.0136\pm 0.047$&$-0.0065\pm 0.0014$&$\mathbf{-0.0084\pm 0.0003}$\\
\hline
\end{tabular}
\end{center}
\caption{\sf The value of ${\bf E}_{\rm fits}[\Delta_{\chi^2}^{(l)}] $ for different dataset types and for the total global dataset. The statistical uncertainty due to the finite number of datapoints and fits is given.}
\label{tab:delchi2}
\end{table}

Finally, in Table~\ref{tab:delchi2} we show the expectation over the same $n_{\rm fit}=40$ fits of 
\be
{\bf E}_{\rm fits}[\Delta_{\chi^2}^{(l)}] = \frac{1}{n_{\rm fits}} \frac{1}{N_{\rm data}} \sum_{l=1}^{n_{\rm fit}} \left( \chi^2(g^{(l)}, z^{(l)})- \chi^2(f, z^{(l)})\right)\;.
\ee
as defined in~\cite{NNPDF:2021njg}, where the $\chi^2$ values are evaluated according to the fluctuated pseudodata $z$ with theory inputs given by the closure test fit value $g$ or the true values $f$.  As discussed in this reference, this provides a measure of the amount of under-- or over--fitting. In particular, if ${\bf E}_{\rm fits}[\Delta_{\chi^2}^{(l)}] > 0$ then this indicates that under--fitting is occurring, i.e. in the current context a lack of flexibility in the PDF parameterisation. On the other hand, if ${\bf E}_{\rm fits}[\Delta_{\chi^2}^{(l)}]< 0  $ then this is acceptable as long as the absolute magnitude is sufficiently small. We can see from the Table that this is certainly true, and indeed the total value is very similar to that found by NNPDF in their analysis~\cite{NNPDF:2021njg}, albeit there with a different underlying fit and testing dataset.

In summary, we have considered a range of statistical estimators in both PDF and data space, and confirmed that in the context of the current (self--consistent) closure test that the PDF uncertainties are faithfully represented by the MSHT20 fixed parameterisation fit with textbook $T^2=1$ uncertainties, and in particular that there  is very limited evidence of any significant degree of under--fitting due to parameterisation inflexibility. In the following section we will enlarge this discussion to consider the case were the pseudodata, or indeed real dataset, are not completely self--consistent, and motivate the need for a tolerance in that context.

\section{The Role of the Tolerance}\label{sec:tol}

\subsection{General Remarks}\label{subsec:tolgen}

Having set up the global closure test above, where so far exact data and theory consistency has been imposed by construction, it is  interesting to consider a closure test but where dataset inconsistencies are injected into the fit. We will do this in the following section, but before doing so it is instructive to make some general remarks about the tolerance, and the relationship between the Hessian and Monte Carlo replica approach, within the context of a simple toy example.

We in particular consider the simplest possible case of a fit to two measured values, $D_{i}$ ($i=1,2$), of  a single observable $O$, with true value $D_0$. The fit theoretical prediction, $t$, which we assume to be otherwise unconstrained,  simply corresponds to the best fit value of the observable, $O$, that comes from this pair of measurements. We will assume that the $D_i$  have the same experimental uncertainty, $\sigma$, in which case we have 
\be\label{eq:t0}
t_0 = \frac{1}{2}\left(D_1+D_2\right)\;,
\ee
i.e. the best fit theory value is  given by the average of the two data values. This is assumed for simplicity, but we could readily take the errors to be different, in which case this would instead be a suitable weighted average of the two measured values, and the discussion below would be qualitatively unchanged. Similarly, generalising to the case of more than one datapoint in each dataset would lead the basic result unchanged; we will briefly discuss this at the end of the section. 

Writing $t=t_0 + \Delta t$, it is straightforwards to show that the $\Delta \chi^2$ as we deviate from this best fit value is simply given by
\be
\Delta \chi^2 = \frac{2 \Delta t^2}{\sigma^2}\;,
\ee
i.e. it  is {\it independent} of  the specific value of $t_0$ and  the particular values of the $D_i$. This is of course completely consistent with the underlying statistics of the measurement, namely that the $\Delta \chi^2=1$ error is given by 
\be\label{eq:deltat}
\Delta t=\pm \frac{\sigma}{\sqrt{2}}
\ee
as we would expect. In particular, the $D_{1,2}$ are given by  
\be
D_{1,2} =  D_0+\sigma\delta_{1,2}\;,
\ee
where are $\delta_{1,2}$ are normally distribution with unit variance, i.e. the $D_{1,2}$  are sampled from a normal distribution with true value $D_0$ and error $\sigma$. In this case, if the experiments corresponding to $D_1$ and $D_2$ are repeated multiple times, then $D_0$ will lie within the corresponding $t_0 \pm |\Delta t|$  with $1\sigma$  ($\sim 68\%$) frequency. Equivalently, the distribution of the fit $t_0$ over multiple repeated experiments will be centred on $\left\langle t_0 \right\rangle = D_0$ with standard deviation given by  \eqref{eq:deltat}. In other words, provided the underlying errors are faithful, and the measured $D_i$ are due to an underlying distribution that is statistically consistent with the true value $D_0$, then the $\Delta \chi^2=1$ error in the fit value of $t_0$ correctly indicates the expected 68\% C.L. consistency of this with the underlying true data value $D_0$.

The above discussion then immediately tells us what will happen if there is some inconsistency or tension in the measured data. We can in particular imagine that we instead have
\be\label{eq:atent}
D_{1,2} = a_{1,2} + (D_0+\sigma\delta_{1,2})\;,
\ee
where the $\delta_{1,2}$ are drawn from the same normal distribution as before, i.e. for simplicity we assume that the statistical error $\sigma$ is correctly known, though one could generalise to the case where this is not true without adding to the discussion below. However, we now introduce the $a_{1,2}$ as constant offsets that represent the (unknown) sources of inconsistency in the two measurements. Namely, they are non--zero in generating the measured values of $D_{1,2}$, but they are not accounted for in the fit, which still (now incorrectly) assumes that the $D_{1,2}$ are representative of the underlying  true value $D_0$ with statistical uncertainty $\sigma$. In this case, the best fit value of $t_0$ would be given by the same average as in \eqref{eq:t0}, which we can write explicitly as
\be
t_0 =  \frac{1}{2}\left(a_1+a_2\right)+D_0+ \frac{\sigma}{2}\left(\delta_1+\delta_2\right)\;,
\ee
 and applying the $\Delta \chi^2=1$ criterium would give exactly the same error \eqref{eq:deltat} as before; as this is independent of the particular value of $t_0$ it will clearly be independent of the $a_i$. This will however no longer be statistically consistent with the underlying true value, $D_0$. Indeed, if the experiments are repeated multiple times we will have
 \be\label{eq:ai}
 \left\langle t_0 \right\rangle = \frac{1}{2}\left(a_1+a_2\right) + D_0\;,
 \ee 
 and so for non--zero $a_{1,2}$ the average value of $t_0$ will be offset from the true value of $D_0$ in a manner that is not accounted for by the quoted uncertainty \eqref{eq:deltat}. 
 In other words, the $\Delta \chi^2=1$ uncertainty on the fit value of $t_0$ does not account at all for any  deviation between the measured $D_i$ and the true value $D_0$ or any tension between the measured $D_i$. The corresponding error is in particular identical to the case of exact data/theory consistency above. 
 
 The above results can be understood rather intuitively. Namely, the uncertainty \eqref{eq:deltat} corresponds to the statistical uncertainty on the average   \eqref{eq:t0} of the two measured values $D_{1,2}$ that is purely due to their statistical errors $\sigma_{1,2}=\sigma$. In the case of complete data/theory consistency ($a_{1,2}=0$) this is the correct statistical uncertainty on the fit to the true value $D_0$, in the frequentist sense described above. If we introduce some inconsistency ($a_{1,2} \neq 0$) then  \eqref{eq:deltat} still correctly corresponds to the statistical uncertainty, due to the errors $\sigma_i=\sigma$, on the average of the two measured values $D_{1,2}$, which will indeed be consistent with \eqref{eq:ai} at 68\% confidence according to the error given by \eqref{eq:deltat}. It is simply that this average is no longer representative of the true value $D_0$ due to the $a_{1,2} \neq 0$ offset, and as the above discussion makes clear, the quoted $\Delta \chi^2=1$ uncertainty is completely unrelated to that disagreement.

 If we consider instead the MC replica approach to uncertainty propagation, this should be completely consistent with the discussion above, though it is worth verifying this for complete clarity. In this approach, we generate $N_{\rm rep}$ pseudodata replicas of the two data points. Assuming complete data/theory consistency for now, these will be given by
\be
D_{1,2}^j =  \sigma \delta^j_{1,2} +D_{1,2}\;,
\ee
where the index $j$ labels the $j$th replica, the $\delta^j_{1,2}$ are as before drawn from a normal distribution with unit variance, and $D_{1,2}$ (without the $j$ superscript) correspond to the measured data values, as defined above \eqref{eq:t0}. The fit to the $j$th replica will give
\be\label{eq:tj}
t^j = \frac{1}{2}\left(D_1^j+D_2^j\right)\;,
\ee
for which we have
\be
t_0 = \left \langle t^j \right\rangle  = \frac{1}{N_{\rm rep}} \sum_j t^j  =  \frac{1}{2}\left(D_1+D_2\right)\;,
\ee
and 
\be \label{eq:deltamc}
\Delta t = \left(\frac{1}{N_{\rm rep}-1}  \sum_j  (t^j - t_0)^2\right)^{1/2} = \frac{\sigma}{2}\left(\frac{1}{N_{\rm rep}-1}  \sum_j  (\delta_j^1+\delta_j^2)^2\right)^{1/2}= \pm \frac{\sigma}{\sqrt{2}}\;, 
\ee
in the large $N_{\rm rep}$ limit. These are consistent with the results above, as we would of course expect. We can in particular again see that the uncertainty on $t$ is again independent on the particular value of $t_0$. So indeed error propagation through MC replicas leads to exactly the same issue as before, and for exactly the same reasons\footnote{We note that NNPDF4.0 fit, which applies this error treatment, does however propose a mechanism to assess the consistency of each dataset with the bulk of the data included in the fit and discard inconsistent datasets from the global fit, in an effort to avoid the need for any enlarged error definition.}.

While the discussion above has focused on a simple toy example, it should be clear that the issue will continue to occur in a genuine global PDF fit, in the presence of tensions between different datasets and between data and theory. In particular, the key issue will remain as above, namely that the best fit PDF will correspond to a fit to a suitable average of the data entering the fit, analogous to \eqref{eq:ai}, though in a much more complicated and indirect manner and now weighted by the corresponding data uncertainties. However, the PDF uncertainties evaluated using a $\Delta \chi^2=1$ criterion, or completely equivalently MC replica generation, will only correspond to the uncertainty on this average due to the quoted data errors (and any theoretical errors also included in the fit). In particular, as demonstrated above, if the underlying datasets are in tension or there is otherwise some underlying data/theory inconsistency, then the fit to this average will not in general coincide with the `true' PDF we wish to extract within the  $\Delta \chi^2=1$ uncertainty.

A proposed resolution to the above issue, or at least improvement to the situation, has been discussed at length in the literature~\cite{Collins:2001es,Pumplin:2001ct,Martin:2009iq,Pumplin:2009sc,Watt:2012tq,Kovarik:2019xvh}, and necessitates an enlargement of the textbook  $\Delta \chi^2=1$ criterion. Focusing briefly on the `dynamic' tolerance  developed in~\cite{Martin:2009iq}, it is instructive to examine how this would enter the toy model discussed above. Here, the uncertainty range is set by a hypothesis testing criterion, whereby all datasets entering the fit should be suitably described within their 68\% C.L. limits around the best fit value. To be precise, we require that
\be
\chi^2_i < \xi_{68}\left( \frac{ \chi^2_{i,0}}{\xi_{50}}\right)\;,
\ee
where $\chi^2_i$ corresponds to the fit quality to dataset $i$, and $\chi^2_{i,0}$ the corresponding value at the global best fit. The $\xi$ represent the appropriate 50th and 68th percentiles of the $\chi^2$ distribution for the $N_{\rm dat}^i$ degrees of freedom of the dataset $i$; we have $\xi_{50}\sim N_{\rm dat}$ for sufficiently large $N_{\rm dat}$. The factor of $\chi^2_{i,0}/\xi_{50}$ is introduced as in general the value of $\chi^2_{i,0}$ may depart rather significantly from a textbook `good' fit quality, and this renormalisation corrects for that.

If we assume for concreteness that $D_1 < D_2$ then the upper limit on $t_0$,  which we denote $\Delta t_+$, will be set by $D_1$, in which case we have
\be
\frac{(D_1 - t_0 - \Delta t_+)^2}{\sigma^2} = \frac{ \xi_{68}}{ \xi_{50}}\frac{(D_1 - t_0 )^2}{\sigma^2}\;.
\ee
 Rearranging, we find we pick up a contribution of order
 \be\label{eq:deltatp}
  \Delta t_+  \propto D_1-D_2 \sim a_1-a_2\;,
 \ee
 in the presence of dataset tensions. Thus, the dynamic tolerance will result in an uncertainty that is in general larger than the $\Delta \chi^2=1$ criterion, but which is representative of the spread in the datasets entering the fit, due to the non--zero values of the $a_i$. 
 
We note that in the MC replica approach such a spread  can in principle be accounted for by a suitable modification of the method for generating and sampling the MC replicas. In the current toy approach this could e.g. correspond to dataset 1 being highly weighted in the fit for some of the replica generation, and vice--versa for dataset 2 in other cases. A more general approach to this question has also been discussed recently in~\cite{Yan:2024yir}. We also note that the impact of training and validation on the corresponding uncertainties has not been considered in the above discussion, though it is considered in e.g.~\cite{Hunt-Smith:2022ugn}.

Finally, we note that the precise value of the overall constant of proportionality in \eqref{eq:deltatp} is not particularly relevant here, and indeed the toy example of two datasets with a single datapoint each is rather far from the case of a global PDF fit. It is therefore instructive to briefly consider a generalisation of our toy model to the case where the two datasets now each have $N/2$ datapoints generated according to \eqref{eq:atent}, i.e. with each individual datapoint having the same uncertainty $\sigma$ as before, and with a constant offset $a_{1,2}$ between the two datasets. 

One thing this will clarify is the extent to which one can have $a_i$ that are sufficiently non--zero to cause issues with the representative nature of the evaluated $\Delta \chi^2=1$ uncertainties without leading to significant deterioration in the overall fit quality. As discussed in~\cite{Collins:2001es}, the fact that this can occur is due to the fundamental difference between the hypothesis testing criterion, whereby a fit to a given dataset is good provided the $\chi^2$ lies within $\pm \sqrt{2N}$ of $N$, and the parameter fitting criterion discussed above, which applies the much stronger  $\Delta \chi^2=1$ constraint. Therefore, we can expect deviations in the fit quality due to non--zero values of $a_i$ that are still well within the $\pm \sqrt{2N}$ range of acceptable values, but which lead to deviations that are well outside the $\Delta \chi^2=1$ uncertainty estimate on $t$.

To show this in the context of our toy model, one can readily show that \eqref{eq:deltat} generalises to give
\be\label{eq:deltatn}
\Delta t=\pm \frac{\sigma}{\sqrt{2N}}\;.
\ee
Taking the large $N$ limit for simplicity, we have
\be
t_0 = \frac{1}{2}\left(a_1+a_2\right) + D_0\;,
\ee
and
\be
\chi^2_0 = N\left(1+\frac{1}{2\sigma^2}(a_1-a_2)^2\right)\;.
\ee
Requiring that the deviation from ideal fit quality $N$ be less than $\sqrt{2N}$ gives
\be\label{eq:a1ma2}
|a_1-a_2| < (8N)^{1/4} \cdot \Delta t\;,
\ee
i.e. for reasonable values of $N$ the deviation can be significantly larger than the textbook $\Delta \chi^2 =1$ uncertainty \eqref{eq:deltatn} without giving a particularly poor fit quality; for $N=10-1000$ the prefactor  ranges from $\sim 3-10$. This is of note as, dependent on the signs of the $a_i$ in the current example, if the L.H.S of~\eqref{eq:a1ma2} is such a factor larger than $\Delta t$, then this can result in a deviation in the best fit value, $t_0$, of this order. 

Turning to the dynamic tolerance, in the same large $N$ limit we again find for non--zero $a_{1,2}$ that $ \Delta t_+$ picks up a contribution that is proportional to $a_1-a_2$, as required. It is also interesting to consider the case where there is in fact no underlying tension, i.e. $a_{1,2}=0$. In this case we find that
\be
\Delta t_+ = \left(\frac{ \xi_{68}}{ \xi_{50}}-1\right)^{1/2}\sigma \propto \frac{\sigma}{N^{1/4}}\;,
\ee
where we have used that $\xi_{50} \sim N$ and $\xi_{68} - \xi_{50} \propto \sqrt{N}$. This highlights a potential issue with the dynamic tolerance (and indeed any enlarged tolerance) criteria, namely that in the absence of any dataset tensions it does not reproduce the underlying $\Delta \chi^2=1$ uncertainty, and in particular the scaling of the uncertainty reduction with $N$ is more gradual than the textbook $\sim 1/\sqrt{N}$ case. We note however that this is not in itself inconsistent, and the dynamic tolerance procedure assumes  by definition that we are not in this situation, and there is much evidence that in a real global fit we are not. 

\subsection{Global Closure Tests and the Tolerance}\label{subsec:tolclos}

Having set the general problem up in the context of a toy model, we now consider some example cases where we perform a closure test but now explicitly construct these so that the theory used to generate the pseudodata and that used to perform the fit are not consistent. This will provide arguably a more accurate model of a genuine PDF fit, where 100\% consistency between data and theory and different datasets is  not always achieved.

We begin by dividing the NNPDF4.0 global dataset into two subsets, in one case containing all low energy Drell Yan (DY) and DIS data, as well as the HERA collider DIS data, and in the other case all hadron collider data from the Tevatron and LHC. The former subset has 3294 datapoints and the latter 1332, however while the hadron collider dataset is smaller in terms of the number of datapoints it contains some rather constraining high precision data from the LHC in particular. Hence this provides a (roughly) comparable division of the global dataset in terms of constraining power, even if clearly there are some regions of $x$ and some PDFs that are more or less constrained by either subset.The kinematic coverage of these subsets can be read off from Fig. 2.1 of~\cite{NNPDF:2021njg} and we can e.g. see that the HERA data extend to rather lower $x$ than the hadron collider data. However, as we shall see, the dominant differences between the fits to these subsets that we will describe below lie in regions of PDF parameter space that are rather well constrained by both, such as on the quark sector at intermediate $x$. We use the same MSHT parameterisation described in Section~\ref{sec:gen}, with the same number of free parameters. These may be more flexible than is required for fitting these reduced datasets with the required level of precision, but for the purposes of the current comparison this is not a significant issue.

Having chosen this division, we now generate the two subsets of pseudodata  with two different underlying input sets. The basic idea here is to model the tendency in a real global fit for a given subset of the data to prefer a somewhat different set of PDFs to  other subsets, beyond the expectations of statistical fluctuations alone. Of course in reality the `true' underlying PDF set would be the same in both cases, and it would be inconsistencies in the actual data and/or theory that lead to differing result in the fit. Nonetheless, without injecting further inconsistencies into the closure test, and simply modifying the input PDF sets in this way, we can effectively model the more complex situation that would occur in reality, although we will show an example of instead modifying the  data rather than the input PDF set at the end of the section. We perform a fluctuated closure test, with the pseudodata fluctuations performed consistently at the level of the full global dataset (i.e. using the global covariance matrix) with a given input set before selecting the subset to be used. 

To be precise, we take the same NNPDF4.0 (perturbative charm) set as before, and now consider also the HERAPDF2.0 NNLO set~\cite{H1:2015ubc} as an input. We choose this as it is sufficiently different from the NNPDF4.0 set to clearly demonstrate the issue of dataset inconsistencies\footnote{We  note that the CT family of global sets, e.g. CT18~\cite{Hou:2019efy}, cannot straightforwardly be used in this study, as this requires the PDFs at $Q_0=1$ GeV, whereas the CT PDFs are only made available at a higher starting value of $Q_0$.}. We will show the case that the low energy DY/DIS + HERA pseudodata are generated by the HERAPDF2.0 set, and the hadron collider data by the NNPDF4.0 set, and for one particular fluctuated dataset, but we have confirmed that the basic conclusions are unchanged if this choice is swapped and/or a different random number seed is used for the pseudodata fluctuations. An example of this is shown in Appendix~\ref{app:inconsistentadd} where equivalent plots to Figs.~\ref{fig:tol_g}--~\ref{fig:tol_sigmachw} are shown, but with a different set of pseudodata fluctuations.

\begin{figure}
\begin{center}
\includegraphics[clip,scale=0.6]{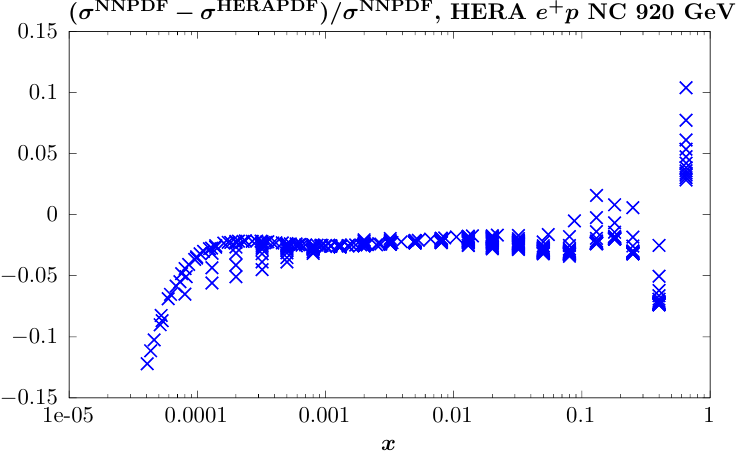}
\includegraphics[clip,scale=0.6]{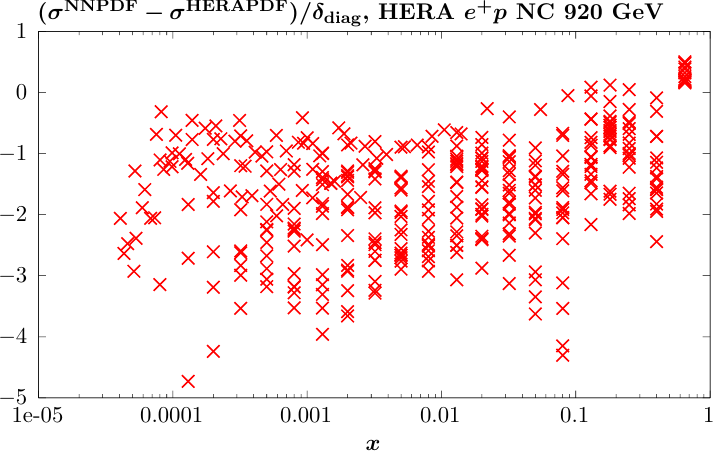}
\includegraphics[clip,scale=0.6]{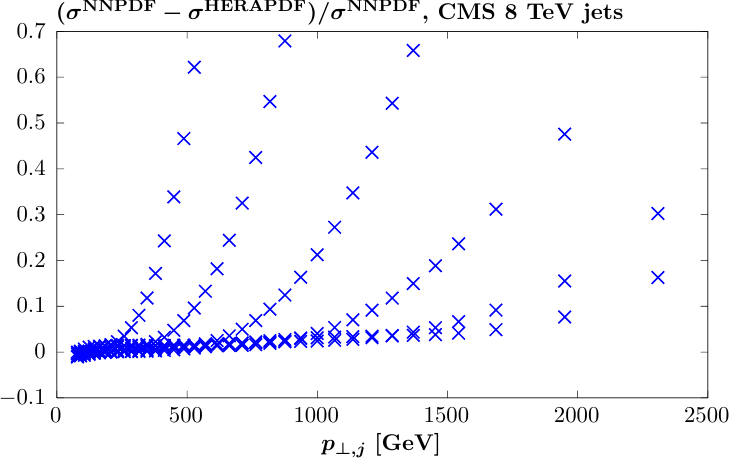}
\includegraphics[clip,scale=0.6]{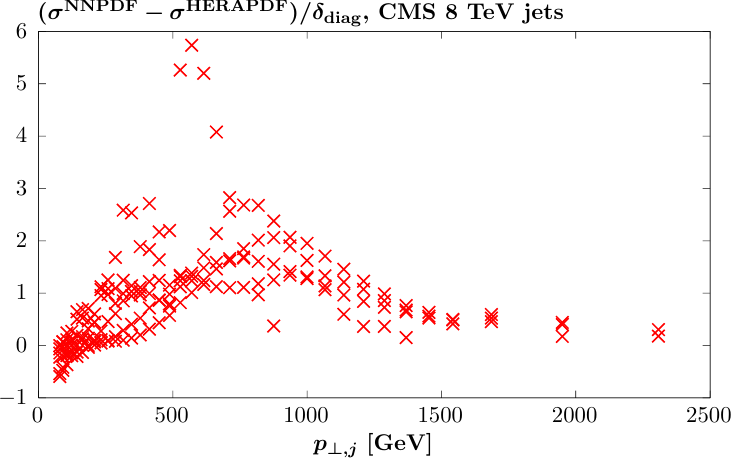}
\caption{\sf Normalized cross section differences between the NNPDF4.0 and HERAPDF2.0 unfluctuated predictions  for (top) the HERA $e+p$ 920 GeV NC~\cite{H1:2015ubc}  and (bottom) the CMS 8 TeV inclusive jet~\cite{CMS:2016lna} data. In the left figures these are normalized to the NNPDF4.0 prediction, while in the right they are normalized to the square root of the diagonal element, $\delta_{\rm diag}$, of the experimental covariance matrix.}
\label{fig:nnpdf_herapdf_sigs}
\end{center}
\end{figure}

\begin{figure}
\begin{center}
\includegraphics[trim={0 0.3cm 0 0.3cm},clip,scale=0.6]{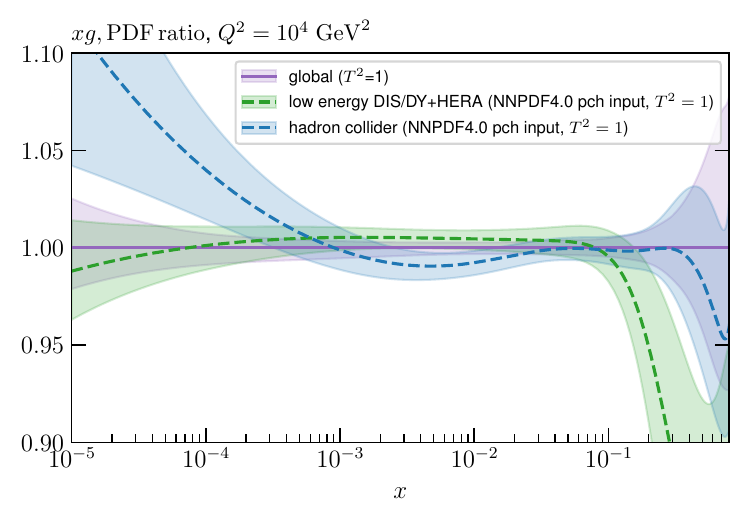}
\includegraphics[trim={0 0.3cm 0 0.3cm},clip,scale=0.6]{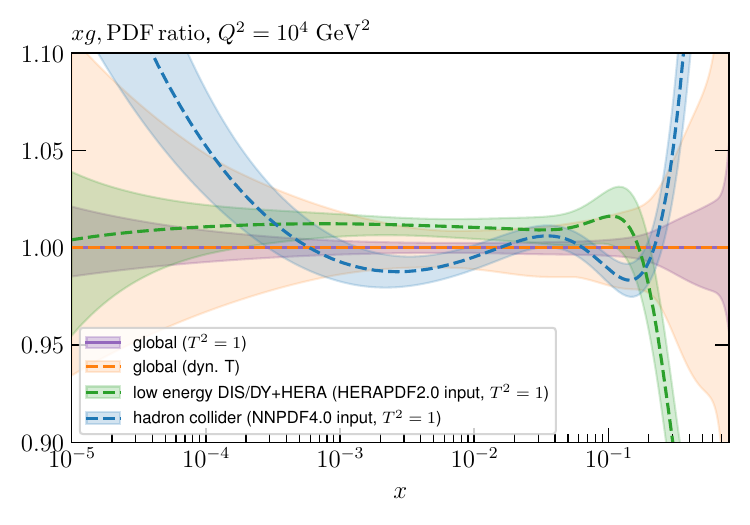}
\includegraphics[trim={0 0.3cm 0 0.3cm},clip,scale=0.6]{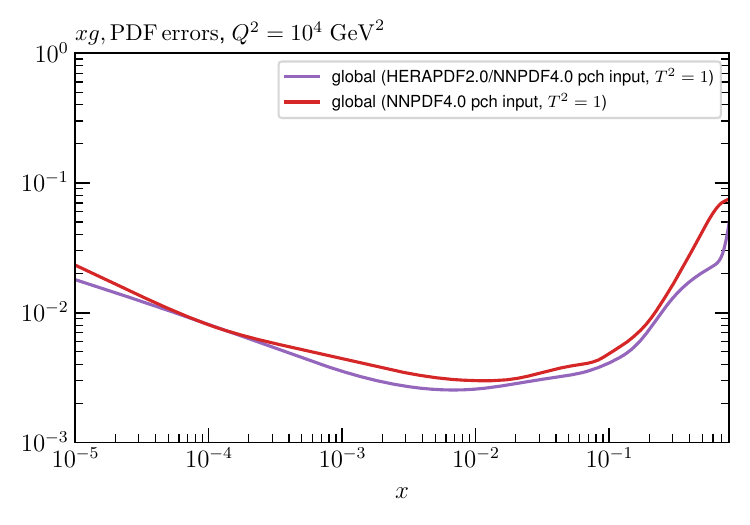}
\includegraphics[trim={0 0.3cm 0 0.3cm},clip,scale=0.6]{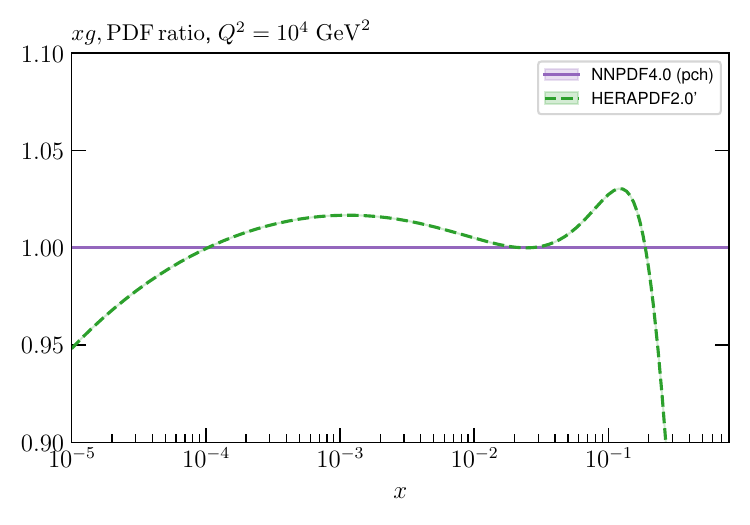}
\caption{\sf The gluon PDF at $Q^2=10^4 \, {\rm GeV^2}$  that results from a fluctuated closure test fit, using the MSHT20 parameterisation, to the NNPDF4.0 global dataset and to the low energy DIS/DY + HERA, and hadron collider datasets, such that the combination of these corresponds to the global case. Results are shown as a ratio to the respective global fits, and in the top left plot  the input pseudodata are all generated with the NNPDF4.0 (perturbative charm) as input, while in the top right plot the HERAPDF2.0 set is instead used for the low energy DIS/DY + HERA pseudodata. In both cases the global fit corresponds to a fit to the corresponding combination of these two subset. In the bottom left plot the corresponding PDF uncertainties for the two global closure tests fits  in the top plots are shown. In the bottom right plot the ratio of the central HERAPDF2.0'   set to the NNPDF4.0 (pch)  set is shown. All uncertainties are evaluated with $T^2=1$ with the exception of the global fit, where the result of applying a dynamic tolerance procedure is also shown.}
\label{fig:tol_g}
\end{center}
\end{figure}

\begin{figure}
\begin{center}
\includegraphics[trim={0 0.3cm 0 0.3cm},clip,scale=0.6]{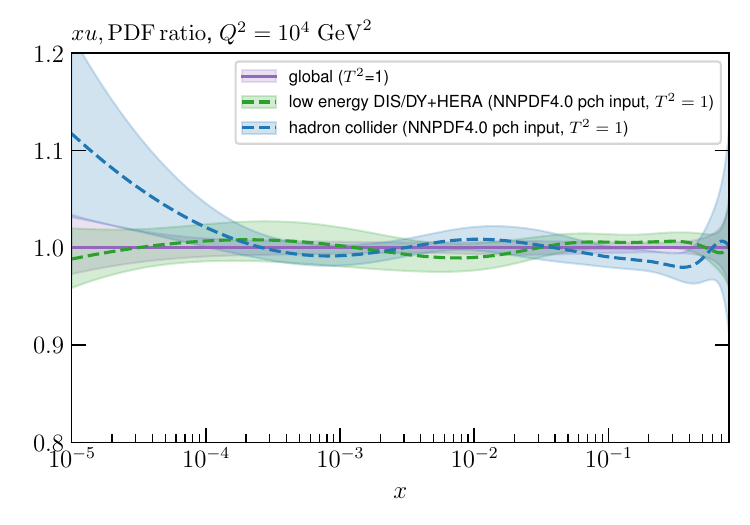}
\includegraphics[trim={0 0.3cm 0 0.3cm},clip,scale=0.6]{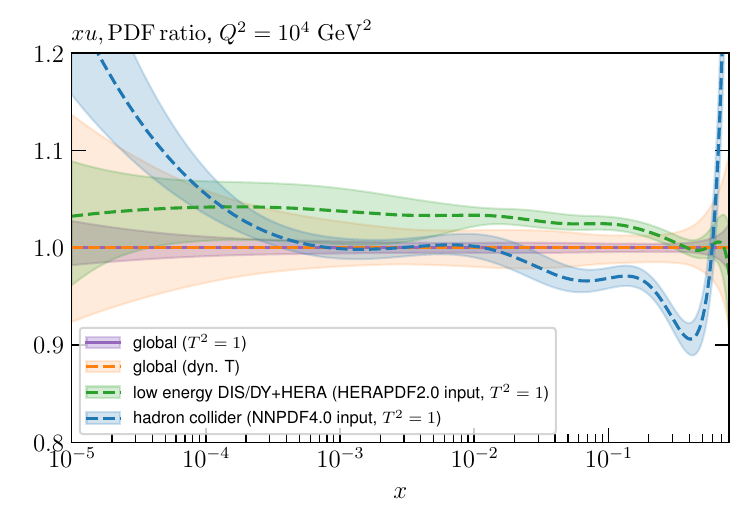}
\includegraphics[trim={0 0.3cm 0 0.3cm},clip,scale=0.6]{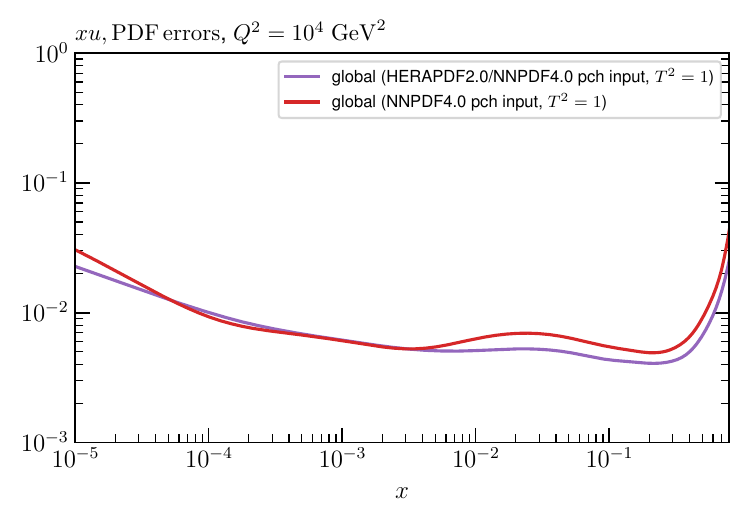}
\includegraphics[trim={0 0.3cm 0 0.3cm},clip,scale=0.6]{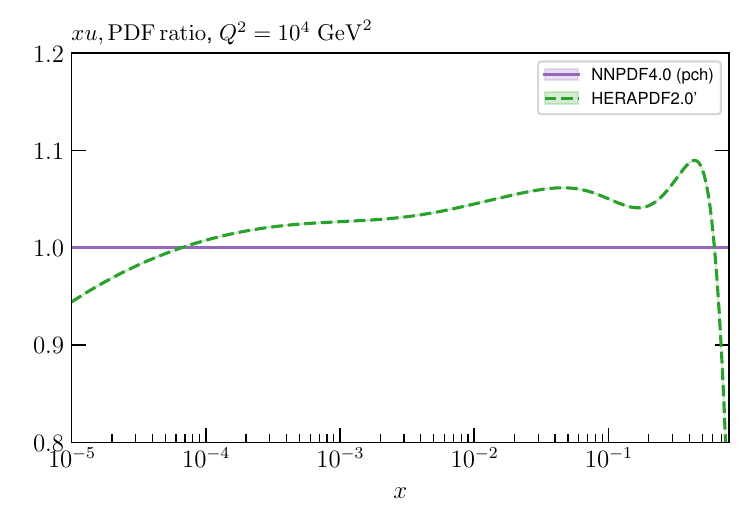}
\caption{\sf As in Fig.~\ref{fig:tol_g} but for the up quark.}
\label{fig:tol_u}
\end{center}
\end{figure}

\begin{figure}
\begin{center}
\includegraphics[trim={0 0.3cm 0 0.3cm},clip,scale=0.6]{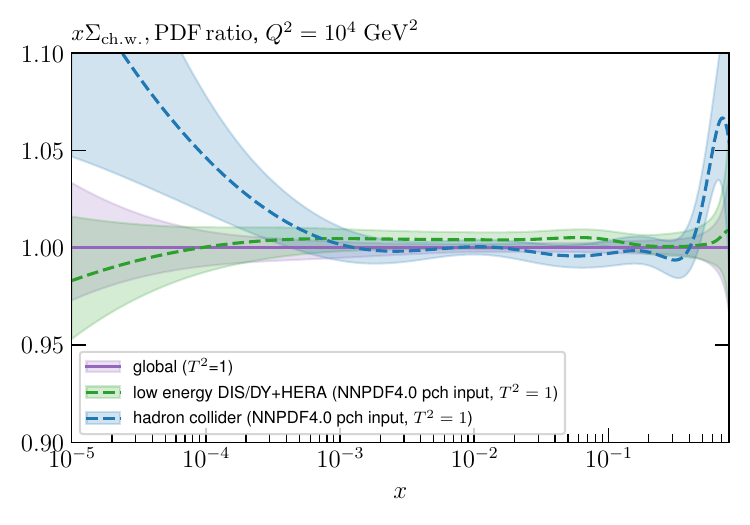}
\includegraphics[trim={0 0.3cm 0 0.3cm},clip,scale=0.6]{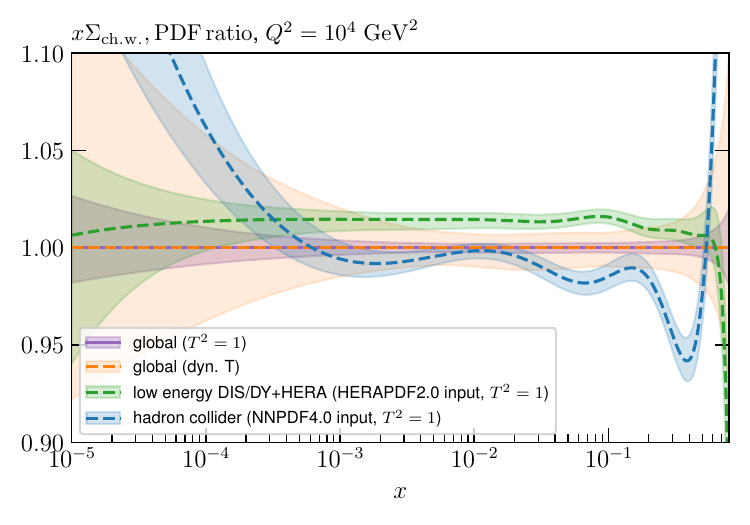}
\includegraphics[trim={0 0.3cm 0 0.3cm},clip,scale=0.6]{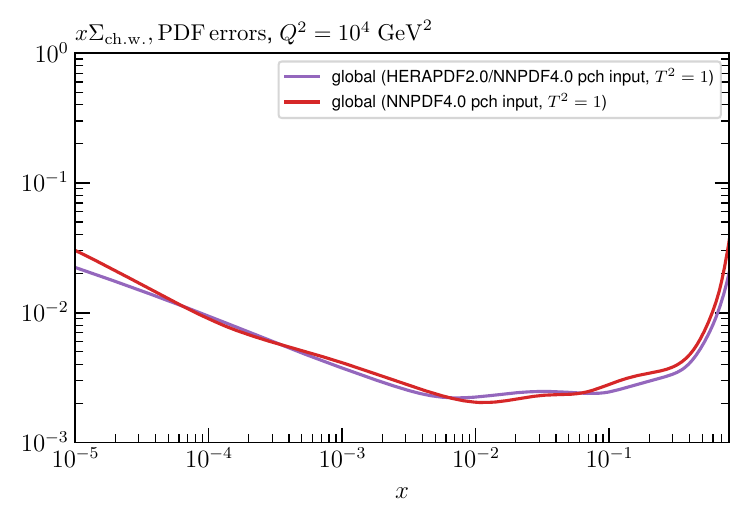}
\includegraphics[trim={0 0.3cm 0 0.3cm},clip,scale=0.6]{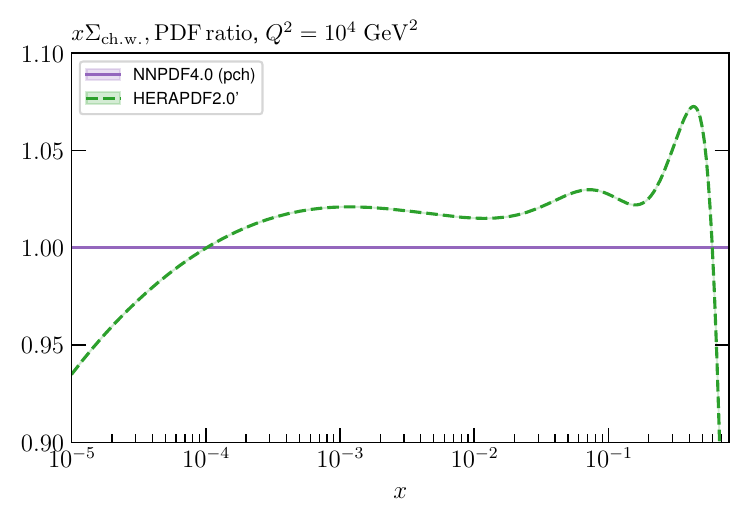}
\caption{\sf As in Fig.~\ref{fig:tol_g} but for the charge weighted quark singlet.}
\label{fig:tol_sigmachw}
\end{center}
\end{figure}

To give an idea of the difference that these two input sets lead to at the level of observable quantities, in Fig.~\ref{fig:nnpdf_herapdf_sigs} we show normalized cross section differences between the NNPDF4.0 and HERAPDF2.0 unfluctuated predictions  for (top) the HERA $e+p$ 920 GeV NC~\cite{H1:2015ubc}  and (bottom) the CMS 8 TeV inclusive jet~\cite{CMS:2016lna} data.   In the left figures these are normalized to the NNPDF4.0 predictions, and we can see clear fractional deviations at both low and high $x$ for the HERA data, and deviations at high jet $p_\perp$ (i.e. high $x$) for the jet data. These effects are driven by the differences in the PDFs, in these cases principally the gluon, at the relevant scales of the data. In the right plots the differences are instead normalized to the square root of the diagonal element, $\delta_{\rm diag}$, of the experimental covariance matrix, to give some indication of the level of this with respect to the size of the experimental uncertainties. This provides a clearer picture of the impact at the level of the fit quality; for example in the jet case we can see that while some of the higher $p_\perp$ points provide the largest fractional deviation in the pure predictions, the larger experimental uncertainties tend to wash this out. The largest impact is in the forward rapidity bin, i.e. the left most line of points in the bottom left figure. More broadly, at the level of the $\chi^2$ per point that comes from the prediction for the unfluctuated pseudodata generated with one set, using the other, we find this is $\sim 0.1,0.9$ and $0.5$ for the HERA, hadron collider and fixed--target data, respectively. This is consistent with expectations that the biggest difference should lie within data beyond HERA, and from the LHC in particular.

The fit results are shown in Figs.~\ref{fig:tol_g}--~\ref{fig:tol_sigmachw} for four choices of PDF, namely the gluon, up quark, up antiquark and charge weighted quark singlet. These are taken to be representative, and in certain $x$ regions there is an evident, but not too extreme deviation between the underlying PDF inputs. In each case, we show: (top left) the ratio of the fits to the two subsets to the fit to the combined global set when these are both consistently generated with the NNPDF4.0 (perturbative charm) set, i.e. as in the closure tests in the previous sections; (top right) the ratio of the fits to the two subsets to the fit to the combined global set, but where the HERAPDF2.0 set is now used to generate the DY/DIS + HERA pseudodata; (bottom left) the PDF uncertainties for the global pseudodata fits in the upper plots; (bottom right) the ratio of the HERAPDF2.0' input set to the NNPDF4.0 input, to serve as an indication of any expected tension between the results in the upper right figure. In all cases the PDF uncertainties correspond to the $T^2=1$ criterion, with the exception of the global fit in the top right plots, where the result of a dynamic tolerance procedure is also shown\footnote{For the dynamic tolerance uncertainties shown here, and elsewhere in this paper, to be precise we treat all ATLAS and CMS datasets that are from the same measurement as the same dataset, while also combining all HERA inclusive DIS and HERA heavy flavour measurements into two combined datasets, consistent with the treatment in the MSHT approach. We also exclude all datasets with fewer than 5 points from the error evaluation; while in general this has very little effect, it can be the case that a dataset with e.g. 1 datapoint (of which there are a handful in the default NNPDF4.0 dataset) can have an unrealistically large effect on the PDF uncertainty, due to the fact that the dynamic tolerance procedure is more applicable to datasets with a sufficient number of points.}.

Here the prime on the HERAPDF2.0' indicates that this is the result of taking the input set at $Q_0=1$ GeV and evolving to  $Q^2=10^4 \, {\rm GeV^2}$ using the NNPDF public code, as strictly speaking this is what is effectively used in the generation of the pseudodata. This differs a small amount from the default HERAPDF2.0 set due to e.g. differing quark masses in the evolution and so on. We also emphasise again that the global fit shown in the top right figure is to the pseudodata formed of the combination of the two (inconsistent) subsets that are produced with differing PDF inputs. 

Starting with the fit quality, we find that the result of the fit to the global, inconsistent dataset is $\chi^2/N_{\rm pt} \sim 1.036$, which is less than $2\sigma$ above $1$, i.e. while showing some mild deviation from ideal behaviour is clearly not so far from it to be considered a bad fit, not least in the context of global PDF fits where the fit quality is in general significantly worse. This is due to the fact that, while the input PDFs show some disagreement for certain PDFs in certain regions of $x$, the fit quality will remain good for many datasets (and within many datasets) which are not sensitive to this. The net effect of this will be to obscure the disagreement that is present in certain cases, when the global fit quality alone is considered. For the fit to the consistent pseudodata, generated with the NNPDF4.0 (perturbative charm) input set, we find $\chi^2/N_{\rm pt} \sim 0.995$, which is clearly in very good agreement with the expectation value of unity. Looking in more detail, while for the consistent fit we find that the $\chi^2/N_{\rm pt}$ for the fixed--target, HERA and hadron collider datasets is $\sim 1.00, 0.96$ and $1.02$, for the inconsistent case these deteriorate to $\sim 1.05, 0.98$ and $1.07$ per point. There is therefore some trend for the hadron collider and fixed--target data to be worse fit, although these values remain within $2\sigma$ of unity. Those datasets with the largest deviations are certain LHCb and ATLAS charged and neutral current DY data, with various NMC, SLAC and NuTeV datasets showing some reasonable deterioration.

Considering first the top left plots, which show the result of fitting to completely self--consistent pseudodata, we can see broad consistency between the fits to the two subsets in comparison to the global fit. There are as expected some regions of $x$ where the fit to a given subset disagrees with the global fit outside the quoted 68\% uncertainties, but this is to be expected given these are $1\sigma$ uncertainties alone, and indeed the agreement is almost always at the $\sim 2\sigma$ level. In the bottom left plots we can see that the uncertainties for the global fits are very close to identical, up to differences due to statistical fluctuations and the fact that the underlying HERAPDF2.0' set is somewhat simpler than the NNPDF4.0, leading to slightly smaller uncertainties in a closure test generated with this input. There is certainly no evidence at all that the $T^2=1$ uncertainty in the inconsistent case is any larger, completely consistent with the discussion in the previous section.

Turning now to the top right plots, we can see that clear differences emerge. Starting with the gluon, as indicated in Fig.~\ref{fig:tol_g} (bottom right) the largest disagreement between the underlying HERAPDF2.0' and NNPDF4.0 sets is at high $x$, where the former gluon is considerably softer. This is matched by the fits in the top right plot, with the low energy DY/DIS + HERA case (generated with HERAPDF2.0') being significantly softer than the hadron collider case (generated with NNPDF4.0). The global result lies somewhere between the two, as we would expect, but these are evidently far from being consistent within the quote uncertainties. 


Once the dynamic tolerance is applied to the global fit in the top right plot, however, we can see that the consistency is greatly improved. There is still some tendency for the subset fits to lie outside even this uncertainty band at the highest $x$ values, which is completely consistent with the underlying dynamic tolerance procedure; namely, the upper error band will be set by the fact that at some point the gluon becomes inconsistent with the pseudodata in the low energy DY/DIS + HERA  subset in a manner that is not guaranteed to overlap with the fit to the hadron collider pseudodata. In principle this might indicate some further room for further refinement of the procedure, but it should be emphasised that the consistency is still greatly improved with respect to the $T^2=1$ case. We note that the subset fits only show the $T^2=1$ uncertainties, whereas in a genuine fit the dynamic tolerance would also be applied here (and would correspondingly increase the PDF uncertainties). However, in the current set up the individual subsets are completely self--consistent, and hence the $T^2=1$ criterium is the most appropriate. It should also be emphasised that at high $x$ we are entering the extrapolation region for the gluon PDF, where it is less well constrained by either datasets.

Considering the up quark shown in Fig.~\ref{fig:tol_u}, a distinct difference in pulls between the two dataset is observable in the $x\gtrsim 0.01$ region in the top right plot, driven again by the difference in the underlying input sets shown in the bottom right plot. The individual results are clearly multiple $\sigma$ away from the global fit in some regions, indicating significant tension between the results, which is again not matched by any increase in the PDF uncertainty in the global fit. Indeed this provides a particularly clear illustration of the issue discussed in the previous section, namely that the global fit corresponds to a suitably weighted average of the fits to the two subsets, but that the $T^2=1$ uncertainty is not related to the spread between them when some tension is present. When the dynamic tolerance is instead used, the consistency is clearly greatly improved, even if again at high $x$ there some indication of a residual (but significantly reduced) tension between the global and hadron collider results. We note that the input to the fit is cross section observables, rather than the PDFs directly, and these depend on different combinations of PDFs in different $x$ regions in a non--trivial manner. There is therefore no strict requirement for the global fit result to lie between the individual fits, as is indeed observed not to occur in the low $x$ region of the top left figure.

Finally, for the charge weighted quark singlet in~Fig.~\ref{fig:tol_sigmachw}, the trend follows to a large extent that of the up quark, but is  rather more marked due to the fact that this PDF combination is more equally constrained by both datasets. Now, the difference in pulls  between the two dataset is observable in the $x\gtrsim 0.001$ region, and again it is only when the dynamic tolerance procedure is applied to the global fit that overall consistency is achieved, with the exception again of the highest $x$ region for the hadron collider result.

\begin{figure}
\begin{center}
\includegraphics[scale=0.6]{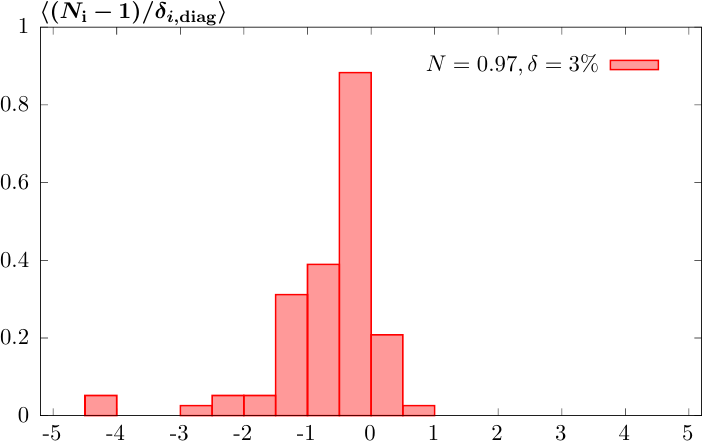}
\includegraphics[scale=0.6]{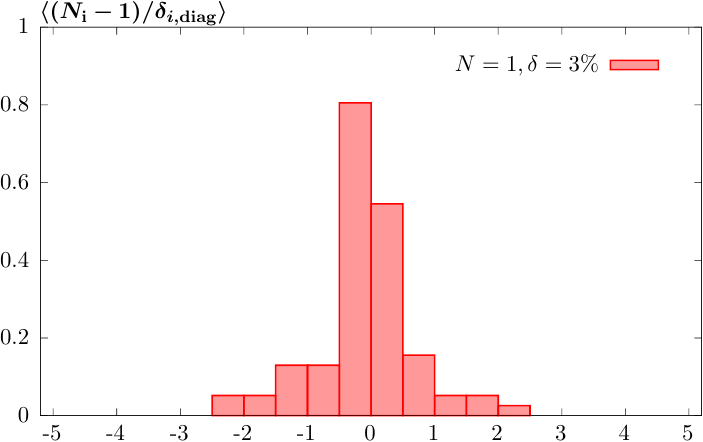}
\includegraphics[scale=0.6]{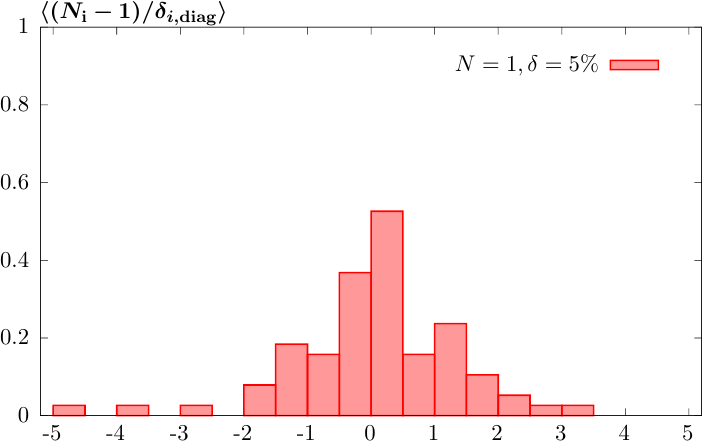}
\caption{\sf The average value of the ratio $(N_i-1)/\delta_{i,{\rm diag}}$ for each dataset $i$, where $N_i$ is the corresponding normalization factor, calculated as described in the text, and $\delta_{i,{\rm diag}}$ is the square root of the diagonal element experimental covariance matrix for each datapoint in the dataset.}
\label{fig:tol_renorm_norms}
\end{center}
\end{figure}

\begin{figure}
\begin{center}
\includegraphics[scale=0.6]{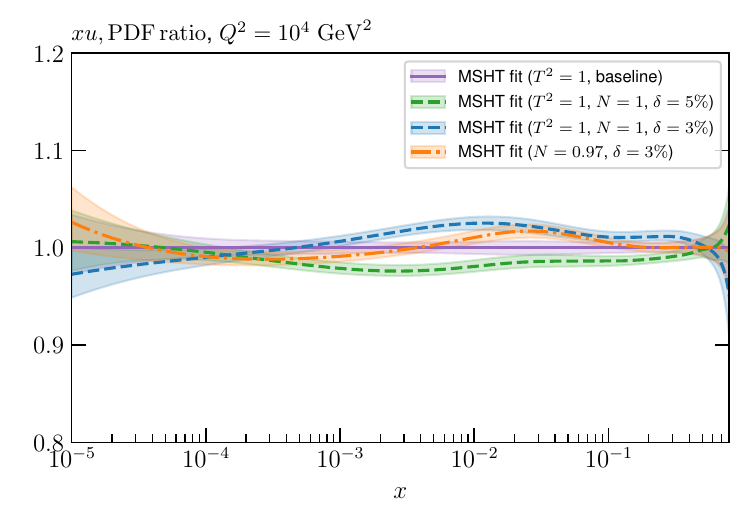}
\includegraphics[scale=0.6]{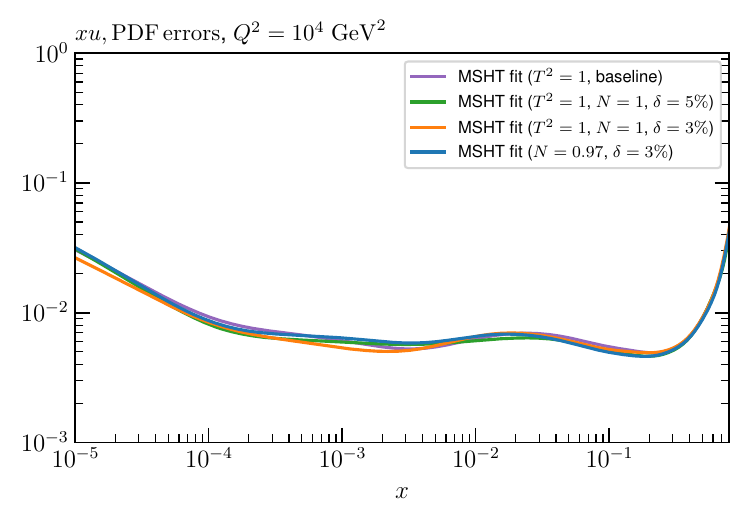}
\includegraphics[scale=0.6]{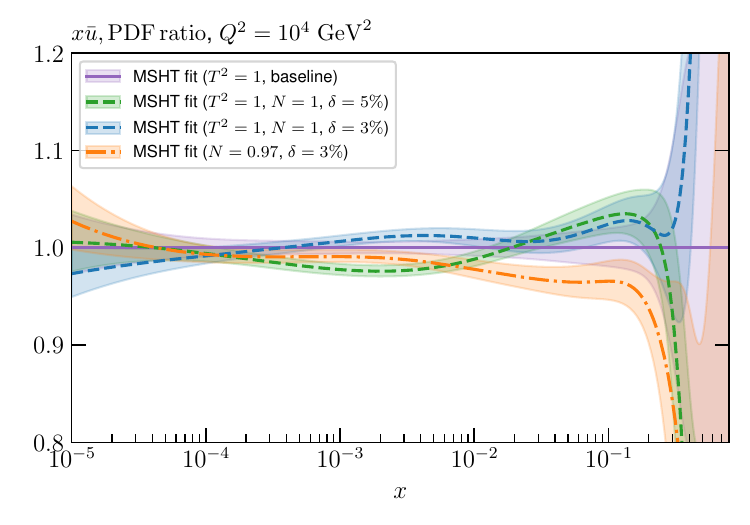}
\includegraphics[scale=0.6]{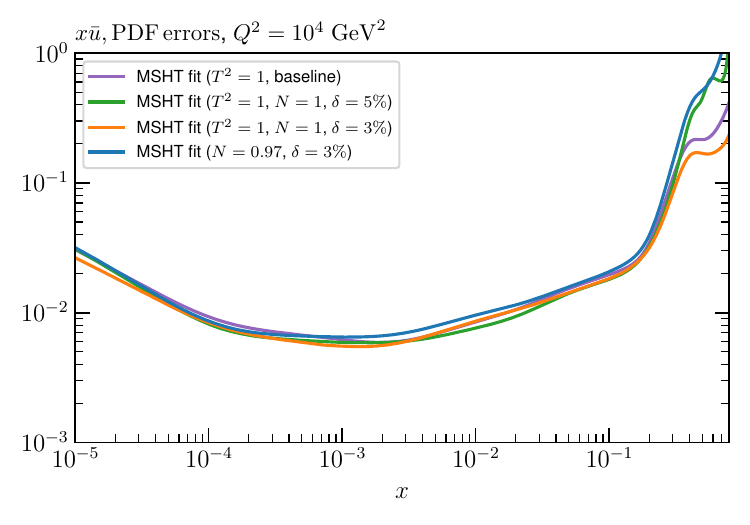}
\includegraphics[scale=0.6]{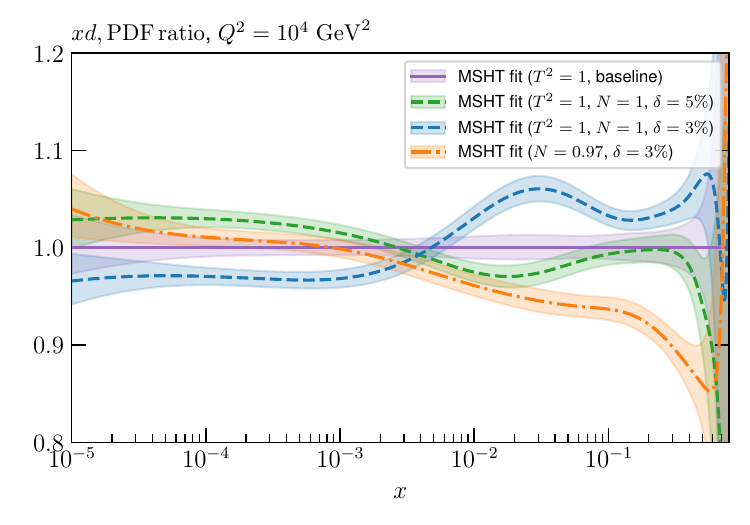}
\includegraphics[scale=0.6]{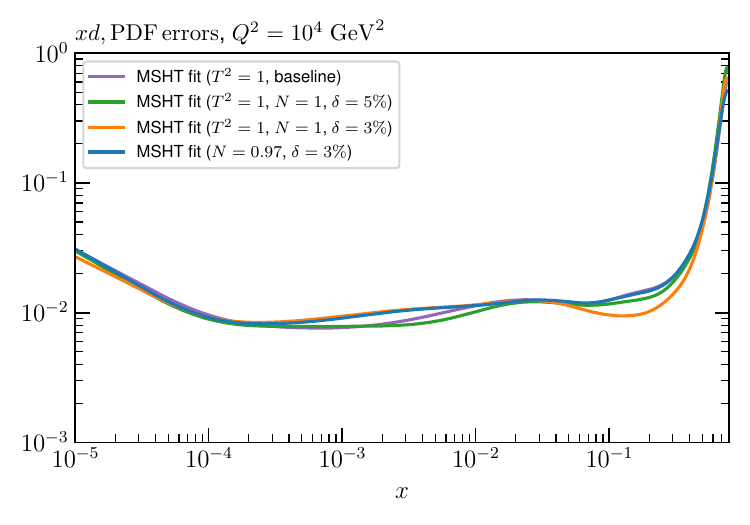}
\includegraphics[scale=0.6]{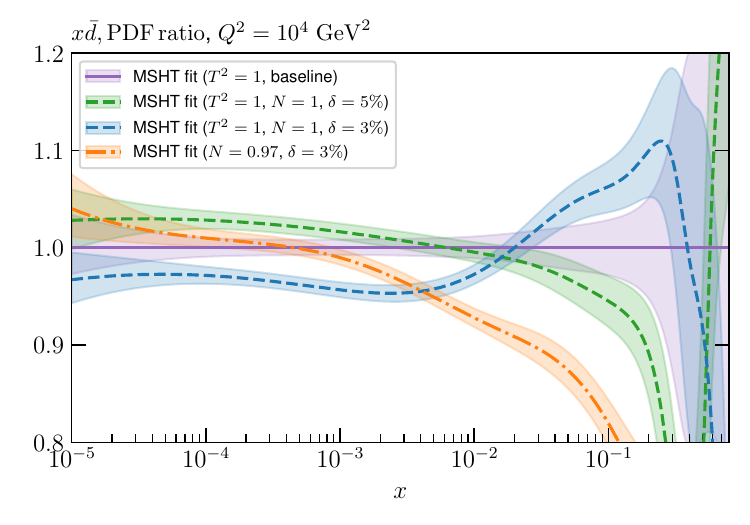}
\includegraphics[scale=0.6]{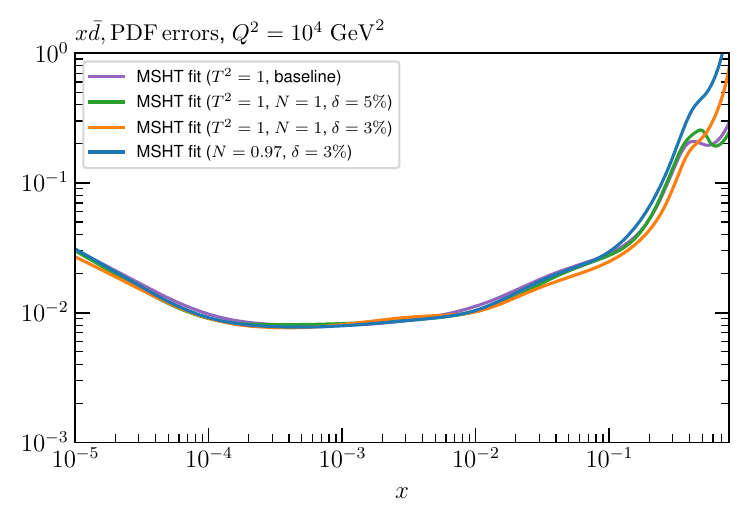}
\caption{\sf A selection of PDFs at $Q^2=10^4 \, {\rm GeV^2}$  that results from a fluctuated closure test fit, using the MSHT20 parameterisation, to the NNPDF4.0 global dataset. As well as the baseline case, we also show the results of fits where inconsistencies are included in the global dataset, on a dataset by dataset basis. In particular, the normalization of each dataset is fluctuated with a standard deviation given by the parameter $\delta$ and average given by $N$. The ratio of the PDFs to the baseline are shown in the left plots, while the corresponding PDF uncertainties are given in the right plots.}
\label{fig:tol_renorm}
\end{center}
\end{figure}

We next briefly consider an alternative closure test, where the fluctuated  pseudodata are now consistently generated with the same NNPDF4.0 (perturbative charm) input across the entire global dataset, but we now explicitly shift the normalisation of the datasets entering the fit by a given amount, in order to model inconsistencies between datasets and/or the theory used to describe them. In particular, the normalization of each dataset is randomly fluctuated with a standard deviation  $\delta$ and average  $N$. For these we take three choices for concreteness, driven by the requirement that the fit quality is not unrealistically poor, namely $\delta=(5\%,3\%,3\%)$ and $N=(1,1,0.97)$, respectively. To be precise, we treat each dataset as given by the default NNPDF labelling convention individually, such that e.g. each subset of the HERA inclusive data, and for multiple LHC datasets the $W^+$ vs. $W^-$ or differing rapidity regions of the same DY measurement, are shifted independently. This therefore to some extent will also allow for inconsistencies within the same measurements as well as across them to be modelled.

For the corresponding fit qualities we find  $\chi^2/N_{\rm p} \sim 1.36,1.18,1.10$ for the $\delta=(5\%,3\%,3\%)$ ($N=(1,1,0.97)$) cases, respectively. These  therefore demonstrate significant deviation away from the expected fit quality, given a $1\sigma$ deviation corresponds to $\sim 0.02$ per point. Nonetheless, with the exception of the rather more extreme first case, these are not larger than the fit qualities that tend to be observed in genuine NNLO global PDF fits. In a little more detail, in Fig.~\ref{fig:tol_renorm_norms} we show the average value of the ratio $(N_i-1)/\delta_{i,{\rm diag}}$ for each dataset $i$, where $N_i$ is the corresponding normalization factor, and $\delta_{i,{\rm diag}}$ is the square root of the diagonal element experimental covariance matrix for each datapoint in the dataset. This therefore provides some indication of the level of deviation from the correct normalization that these fluctuations induce, although given this measure omits the contribution from off--diagonal terms in the covariance matrix, i.e. uncertainty correlations, it is only intended to serve as a rough guide. Nonetheless we can see that in the $\delta=3\%$, $N=(1,0.97)$, cases in particular the normalization shifts are in the majority of cases lower than or comparable to the average (diagonal) experimental uncertainty, and hence do not correspond to particularly dramatic shifts in absolute terms. For the $\delta=5\%$ case there are a few more significant outliers, consistent with the poorer fit quality in this case.

The results are shown in Fig.~\ref{fig:tol_renorm} for a representative selection of PDFs, namely the $u$, $\overline{u}$, $d$ and $\overline{d}$ quarks. Clear deviations are observed at the level of the $T^2=1$ uncertainties in all three cases, with the results sometime lying well away from the baseline (self--consistent) fit. In the right plots the corresponding PDF uncertainties are given, which again emphasises the point that, up to minor fluctuations, the PDF uncertainty is independent of whether the underlying dataset is self--consistent, or has some tension present in it. For simplicity we do not consider the result of applying a dynamic tolerance here, but given this roughly coincides with a fixed $T^2=10$ tolerance, i.e. to a factor of $\sim 3$ increase in PDF uncertainty, it is clear that this would largely resolve any tensions between these fits and the self--consistent baseline.

In summary, we have considered in this section two different closure tests where the pseudodata and the theory used to perform the fits are by construction not fully consistent, but where the corresponding fit qualities are nonetheless acceptable by the standards currently achieved in global PDF fits. We have shown in all cases that the PDF uncertainty is essentially independent of whether there is self--consistency in the closure test or not. The net result of this is that if $T^2=1$ uncertainties alone are used, then the corresponding PDF uncertainties are no longer representative for the case that the underlying closure test includes inconsistencies in the data and/or theory. However, if an enlarged tolerance criterion is instead applied, such as the dynamic tolerance of the MSHT collaboration, then the situation is greatly improved and the PDF uncertainties are much more in line with what is required. We note that these results are exactly as expected from the discussion in the previous section.  Given we have already verified the minimal role that parameterisation flexibility plays in the context of the self--consistent closure tests of Sections~\ref{sec:closureun} and~\ref{sec:closure}, this is clearly a result that would also apply for other suitably flexible methods of parameterising the PDFs, such as in the NNPDF approach. Finally, while the size of the above effects is so large in the chosen examples that a detailed quantitative analysis is not essential, it would clearly be useful in future studies to apply a more quantitative approach, including applying different statistical estimators and a larger set of fluctuated fits.

\section{Full Global PDF Fits}\label{sec:fullfit}

\subsection{Perturbative Charm}\label{sec:pcharm}

\begin{table}
\begin{center}
  \scriptsize
  \centering
   \renewcommand{\arraystretch}{1.4}
\begin{tabular}{Xrccc}\hline 
&NNPDF4.0 pch&MSHT fit&MSHT fit (w positivity)
\\ \hline
NMC $\sigma^{{\rm NC,} p}$ (204)~\cite{NewMuon:1996fwh}  & 349.2 (1.71) & \textcolor{blue}{317.1 (1.55)} & \textcolor{blue}{337.2 (1.65)}\\
BCDMS $F_2^{p}$ (333)~\cite{BCDMS:1989qop} & 497.6 (1.49) & \textcolor{blue}{471.6 (1.42)} & \textcolor{blue}{483.2 (1.45)}\\
NuTeV $\sigma^{\overline{\nu}}_{\rm CC}$ (37)~\cite{NuTeV:2001dfo} & 28.2 (0.76) & \textcolor{red}{32.5 (0.88)} & \textcolor{red}{37.2 (1.00)}\\
{\bf DIS Fixed--Target (1881)} & {\bf 2076.1 (1.10)} &\textcolor{blue}{{\bf  2029.2 (1.08)}} & \textcolor{blue}{{\bf 2063.3 (1.09)}}
\\ \hline
E886 $\sigma^{p}$ (NuSea) (89)~\cite{NuSea:2003qoe} & 109.0 (1.23) & \textcolor{red}{120.7 (1.36)} & \textcolor{red}{118.4 (1.33)}\\
E906 $\sigma^{d}/2\sigma^p$ (SeaQuest) (6)~\cite{SeaQuest:2021zxb} & 5.55 (0.93) & \textcolor{blue}{4.37 (0.73)} & \textcolor{blue}{3.54 (0.59)}\\
{\bf DY Fixed--Target (195)} & {\bf 190.4 (0.98)} &\textcolor{red}{{\bf  201.8 (1.04)}} & {\bf 198.6 (1.02)}
\\ \hline
NC $e^- p$ 575 GeV (254)~\cite{H1:2015ubc} & 265.5 (1.05) & \textcolor{blue}{254.5 (1.00)} & \textcolor{black}{259.0 (1.02)}\\
NC $e^+ p$ 820 GeV (70)~\cite{H1:2015ubc} & 83.7 (1.20) & \textcolor{black}{87.0 (1.24)} & \textcolor{blue}{76.1 (1.09)}\\
NC $e^+ p$ 920 GeV (377)~\cite{H1:2015ubc} & 576.5 (1.53) & \textcolor{blue}{536.1 (1.42)} & \textcolor{black}{574.0 (1.52)}\\
NC, c (47)~\cite{H1:2018flt} & 125.8 (2.68) & \textcolor{black}{130.4 (2.77)} & \textcolor{blue}{120.8 (2.57)}\\
NC, b (26)~\cite{H1:2018flt} & 71.0 (2.73) & \textcolor{blue}{64.1 (2.47)} & \textcolor{black}{67.9 (2.61)}\\
{\bf HERA DIS (1145)} & {\bf 1698.4 (1.48)} &\textcolor{blue}{{\bf  1650.2 (1.44)}} & \textcolor{blue}{{\bf 1671.5 (1.46)}}
\\ \hline
D0 $W$ muon asymmetry (9)~\cite{D0:2014kma} & 16.2 (1.79) & 14.2 (1.58) & \textcolor{blue}{13.8 (1.53)}\\
ATLAS $W,Z$ 7 TeV ($\mathcal{L}=4.6\,{\rm fb}^{-1}$) (61)~\cite{ATLAS:2016nqi} & 119.9 (1.97) & \textcolor{blue}{101.1 (1.66)} & \textcolor{blue}{100.2 (1.64)}\\
ATLAS high--mass DY 2D 8 TeV (48)~\cite{ATLAS:2016gic} & 54.4 (1.13) & \textcolor{red}{59.3 (1.24)} & \textcolor{red}{59.8 (1.25)}\\
CMS electron asymmetry 7 TeV (11)~\cite{CMS:2012ivw} & 7.67 (0.70) & \textcolor{red}{11.0 (1.00)} & \textcolor{red}{11.1 (1.01)}\\
CMS DY 2D 7 TeV (110)~\cite{CMS:2013zfg}  & 156.3 (1.42) & \textcolor{blue}{143.0 (1.30)} & \textcolor{blue}{145.4 (1.32)}\\
CMS $W$ rapidity 8 TeV (22)~\cite{CMS:2015hyl} & 26.1 (1.18)& \textcolor{blue}{22.2 (1.01)} & \textcolor{blue}{22.1 (1.00)}\\
LHCb $Z\to ee$ (17)~\cite{LHCb:2012gii} & 28.8 (1.70) & \textcolor{blue}{25.2 (1.48)} & \textcolor{blue}{25.7 (1.51)}\\
LHCb $W,Z\to \mu$ 7 TeV (29)~\cite{LHCb:2015okr} & 45.7 (1.58) & \textcolor{blue}{40.6 (1.40)} & \textcolor{black}{43.9 (1.51)}\\
LHCb $W,Z\to \mu$ 8 TeV (30)~\cite{LHCb:2015mad} & 43.7 (1.46) & \textcolor{blue}{34.3 (1.14)} & \textcolor{blue}{36.1 (1.20)}\\
LHCb $Z\to \mu\mu$ 13 TeV (16)~\cite{LHCb:2016fbk} & 22.7 (1.42) & \textcolor{blue}{17.7 (1.11)} & \textcolor{blue}{18.9 (1.18)}\\
LHCb $Z\to ee$ 13 TeV (15)~\cite{LHCb:2016fbk} & 28.8 (1.92) & \textcolor{blue}{24.4 (1.63)} & \textcolor{black}{26.6 (1.78)}\\
{\bf Collider DY (576)} & {\bf 794.7 (1.38)} &\textcolor{blue}{{\bf 727.5 (1.26)}} & \textcolor{blue}{{\bf 741.6 (1.29)}}
\\
\hline
ATLAS incl. jets 8 TeV, $R=0.6$ (171)~\cite{ATLAS:2017kux} & 137.7 (0.81) & \textcolor{blue}{129.8 (0.76)} & \textcolor{black}{137.8 (0.81)}\\
ATLAS dijets 7 TeV (90)~\cite{ATLAS:2013jmu} & 242.5 (2.69) & \textcolor{blue}{235.2 (2.61)} & \textcolor{black}{240.6 (2.67)}\\
{\bf LHC Jets (500)} & {\bf 823.9 (1.65)} &\textcolor{black}{{\bf 813.2 (1.63)}} & \textcolor{black}{{\bf 826.5 (1.65)}}\\
\hline
ATLAS $W^\pm + $ jet 8 TeV (30)~\cite{ATLAS:2017irc} & 43.0 (1.47) & \textcolor{red}{48.4 (1.61)} & \textcolor{red}{47.8 (1.59)}\\
{\bf LHC $V+$ Jets (122)} & {\bf 136.6 (1.12)} &\textcolor{black}{{\bf 137.7 (1.13)}} & \textcolor{black}{{\bf 139.5 (1.14)}}\\ \hline
{\bf Isolated Photon (53)} & {\bf 39.3 (0.74)} &\textcolor{black}{{\bf 41.3 (0.78)}} & \textcolor{black}{{\bf 40.7 (0.77)}}\\ \hline
{\bf Top quark  (81)} & {\bf 82.7 (1.02)} &\textcolor{black}{{\bf 82.0 (0.78)}} & \textcolor{black}{{\bf 83.9 (1.04)}}
\\ \hline \hline 
{\bf Global, $\mathbf{t_0}$ (4626)}  &{\bf 5928.3 (1.282)}& \bf{5736.7 (1.240)} &\bf{5837.8 (1.262)}\\
\hline
{\bf Global, exp. (4626)} &{\bf 5543.7 (1.198)}& {\bf 5380.0 (1.163) }&{\bf 5470.7 (1.18)}\\
\hline
\end{tabular}
\end{center}
\caption{\sf $\chi^2$ values for the NNPDF4.0  (perturbative charm) fit and the MSHT fits to the NNPDF dataset/theory settings. The fit quality for the different major subsets the constitute the global dataset are given in bold, and  above each subtotal the fit qualities for individual experiments in these subsets where the difference with respect to the NNPDF4.0 cases is roughly larger than $\pm 0.5 \sigma = \pm \sqrt{N_{\rm pts}/2}$ for either MSHT fit is shown. When these differences are less than $-0.5\sigma$ the result is highlighted in blue, while the result is highlighted in red when it is greater than $0.5\sigma$. Both the absolute $\chi^2$ and the per point value in brackets, is given in all cases, while the number of points is indicated in brackets next to the dataset description.  For the total $\chi^2$ both the experimental and $t_0$ definitions are shown, while in all other cases only the latter definition is used. Results with and without positivity imposed are shown for the MSHT fit, though we note that the most appropriate comparison with the NNPDF4.0 result is with this imposed.}
\label{tab:chi2_pcharm}
\end{table}

In the previous sections we have considered pure closure tests, namely where pseudodata are generated with an input PDF set, which is then  fit with the MSHT20 parameterisation. We now apply exactly the same approach as in these tests, but instead fit the real data corresponding to the same NNPDF4.0 dataset; we will refer to these as `MSHT fits' for brevity. To be precise, we in fact now fix two PDF parameters in order to improve the stability of the fit in relatively unconstrained regions. Namely, as in MSHT20 we fix the low $x$ power of the strangeness to be the same as that of the sea in order to avoid potentially unphysical behaviour at  low $x$ and low $Q^2$ in the quark flavour decomposition. As we will discuss below, this is in fact also effectively done by the NNPDF collaboration, through the $T_8$ integrability that they impose. We also fix the low $x$ power of the strangeness asymmetry in order to improve the speed of convergence of the $\chi^2$ minimisation, although in this case this parameter could be left free without affecting the eventual results of the fit in anything but a very minor manner. 

We recall in particular that both the data in the fit, but also the theory settings, are completely  identical to those used in the corresponding NNPDF fit. The comparison is therefore completely like--for--like at the level of the full global PDF fit, with the only difference being due to the PDF parameterisation. Here, as we will discuss, there are various prior (i.e. non--data) constraints on the PDFs in the NNPDF fit, due to positivity and integrability as well as the training/validation split in the minimisation itself. Hence the PDF `parameterisation' is understood to denote also those additional constraints, and we will consider the impact of them as well in what follows.

We start by considering the case of a purely perturbatively generated charm PDF. The first question to explore is how the fit quality itself compares between the fit with the MSHT parameterisation and in the NNPDF fit quality. The latter is as usual given by the value of the fit quality of the central NNPDF4.0 replica to the data and we have verified that the results found by taking the public NNPDF4.0 (perturbative charm) set as an input is consistent with the value quoted in~\cite{NNPDF:2021njg}\footnote{To be precise, the quoted value in~\cite{NNPDF:2021njg} of $1.18$ per point is incorrect~\cite{Nocera}, and the correct value is instead  $1.198$ as given in~\cite{Ball:2022qks} and as we find in Table~\ref{tab:chi2_pcharm} below.}. 

The $\chi^2$ fit qualities are shown in Table~\ref{tab:chi2_pcharm}, for both the NNPDF4.0 public fit and the default MSHT fit, as well as the MSHT fit with positivity imposed, as will be described below. The global fit qualities are given, as well as the fit qualities for the different major subsets that constitute the global dataset. Within each subset, the fit quality for the individual experiments in these subsets where the difference with respect to the NNPDF4.0 case is roughly larger than $\pm 0.5 \sigma = \pm \sqrt{N_{\rm pts}/2}$ for either MSHT fit, is shown. When these differences are less than $-0.5\sigma$ the result is highlighted in blue, while the result is highlighted in red when it is greater than $0.5\sigma$. We note that fit qualities for certain individual datasets (e.g. a given HERA DIS subset) do not include the contribution from the correlation of this with other datasets, where relevant, and so are only to be take as a reasonable guide. The $\chi^2$ values correspond to those in the $t_0$ definition, which is also the figure of merit used in the fit itself,  in order to avoid the d'Agostini bias~\cite{Ball:2009qv}. However for the global fit quality we also show the experimental definition, for the sake of comparison. We note it is the latter definition that is used when fit qualities are quoted in~\cite{NNPDF:2021njg} for the NNPDF results. The $\chi^2$ for the experimental definition is somewhat lower than in the $t_0$ case, consistent with expectations in the simplest examples when the d'Agostini bias enters. In terms of the comparison between the different fits, however, the results are qualitatively similar irrespective of which definition is used for the final fit results (although not, we emphasise, necessarily for the figure of merit used in the fit, for which we only use the $t_0$ definition).  

The first observation is that the global fit quality for the MSHT fit is significantly better, by just over $190$ points in absolute value, or $\sim 0.04$ per point, than in the nominal NNPDF4.0 fit; for the experimental definition this is slightly less at $\sim 144$ points, but still comparable in size. In terms of the data subsets, the most significant improvements are in the fixed--target DIS, HERA DIS and collider DY data, which improve by $\sim 0.02, 0.04$ and 0.12 per point, respectively. For the fixed--target DY dataset, on the other hand, there is a moderate deterioration by $\sim 0.06$ per point, although the fit quality remains good in this case. For the fixed--target data, there are moderate improvements in the fit quality for the NMC and BCDMS Neutral Current (NC) data, while there is a mild deterioration in the case of the NuTeV $\overline{\nu}$ charged current (CC) data. For the HERA DIS data, there is some improvement in the NC data at 575 and 920 GeV, in $e^- p$ and $e^+ p$ collisions, respectively, with the biggest improvement in the latter. There is also some improvement in the case of the  bottom quark NC data. For the collider DY data, a significant improvement is seen across a wide range of datasets (9 in total), with some deterioration seen in two datasets, namely  the ATLAS 2D DY data at 8 TeV, and the CMS electron asymmetry at 7 TeV. For the fixed target DY, the deterioration is driven by the E886 NuSea proton data.

We therefore observe improvements for datasets that are sensitive to the quark flavour decomposition, as well as structure function data sensitive to both low and high $x$. That is, the improved fit quality occurs in a broad range of datasets that are sensitive to a broad range of PDF space, and we will expect this to be reflected in rather different PDF sets resulting from the fit, as we will confirm below. Before doing so, we first consider the interpretation of this apparently surprising result. As discussed in e.g.~\cite{Ball:2022uon} there are  reasons to expect that the fit quality of the central replica set for a given NNPDF fit will not correspond to an absolute minimum of the $\chi^2$ fit quality. In particular, three potential reasons are considered in~\cite{Ball:2022uon} (end of Section 3), and which we consider here in increasing order of relevance. 

Two of these reasons, which are related, refer to the statistical properties of the PDF replica ensemble itself. These can be summarised by the statement that the replica ensemble does not have to (and in general will not)  be completely symmetric and Gaussian in $\chi^2$ space, in a manner that is assumed by construction in a Hessian fit. Given this, it is perfectly possible for the fit quality away from that of the average central replica, to be better than the central value.  Indeed, it is observed directly that the fit quality for certain non--central replicas can be better than that of the central replica. This improvement is however limited to being a handful of points in $\chi^2$ (see Fig.~3.1 of~\cite{Ball:2022uon}). Given the significant level of improvement observed in Fig.~\ref{tab:chi2_pcharm} and as we will see the rather large difference then seen at the level of the PDFs themselves, it seems clear that this is not a relevant factor here. More generally, it is clear that the value of the central PDF replica, and in particular its corresponding fit quality to the data in the fit, is a meaningful measure of the quality of the overall fit, as represented by the replica ensemble, in matching the data, and indeed it is used in this manner by the NNPDF collaboration themselves. Therefore, we should expected any differences due to the above effects to be minor, and well within PDF uncertainties.

Turning to the potentially more relevant reason, for  NNPDF the fit quality  to any given pseudodata replica (and therefore of the central replica) does not correspond to the absolute minimum $\chi^2$ due to the stopping procedure that is implemented in order to avoid overfitting. In broad terms, the data are divided into a training and a validation set, and the fit is performed to the training dataset alone, with the minimisation stopped when the $\chi^2$ for the validation data begins to deteriorate (see~\cite{NNPDF:2021njg} for a detailed description of the precise implementation of this procedure). This is in principle essential when performing a PDF fit with highly flexible neural network (the NNPDF default case has 763 free parameters in total~\cite{NNPDF:2021njg}) in order to avoid overfitting, that is fitting the statistical noise in the data in a manner that does not produce reliable results; although we note that to some extent this validation will be automatically achieved by the fact that the global dataset contains multiple overlapping individual datasets that constrain similar PDFs in similar $x$ regions.  

The default MSHT parameterisation that we use has 52 free parameters, that is over an order of magnitude fewer, and hence in principle there should be rather limited scope for any overfitting. We also note that the improvement in the fit quality occurs across a wide range of datasets, e.g. in the case of the DY data for various individual datasets, each of which constrain overlapping regions of PDF parameter space.  This would arguably to a large extent exclude overfitting in the sense of fitting noise in any given individual dataset(s). Moreover, the fluctuated closure tests in Section~\ref{sec:closure} are performed with a pseudodataset that is derived from the same underlying data/theory as enters the fit here. Here, if the MSHT parameterisation allowed for a significant degree of overfitting, we might expect to see evidence of this occurring in the fits to the pseudodata replicas, but this is not observed, with the corresponding PDF uncertainties that result being found to be consistent with 68\% C.L. expectations. On first consideration, it therefore appears unlikely that this issue should be the relevant factor here, in the sense that the fit with the MSHT parameterisation (for which not training/validation is performed) might be overfitting. We will however reconsider this question in more detail in Section~\ref{sec:overfit}.

However, related to the above point is that there are further constraints imposed in the NNPDF fit that effectively form a prior PDF likelihood, and which we have not so far accounted for. One aspect of this is PDF integrability, which relates to the behaviour of the quark flavour decomposition at low $x$. This is discussed in Appendix~\ref{app:integrability}, and is found to not be a significant cause of any difference. Another more relevant factor is that positivity at the level of both certain observables and the PDFs is imposed by NNPDF, see~\cite{Candido:2020yat,Candido:2023ujx} for the motivation of this. The precise implementation of this in the NNPDF case is described in detail in Section 3.1.3 of~\cite{NNPDF:2021njg}. Positivity is in particular imposed at the level of various physical observables as well as directly on the PDFs at $Q^2=5\, {\rm GeV^2}$ for a fixed grid of 10 $x$ points logarithmically spaced between $5 \cdot 10^{-7}$ and $0.1$ and 10 points linearly spaced between $0.1$ and $0.9$. Focusing on the PDF positivity for concreteness, a $\chi^2$ penalty of
 \be\label{eq:pos}
\Lambda_k \cdot \left[- f(x_i,Q^2=5\, {\rm GeV^2})\right]
\ee
is imposed for each $x_i$ point, if the PDF is negative, with $\Lambda_k=10^6$  by the end of the fit, and something similar for the physical observables. A suitable smooth dependence between this and the positive PDF case is applied, where the contribution is negligible for positive PDFs, see~\cite{NNPDF:2021njg} for more details. 

These positivity requirements are available as part of the public NNPDF code, and hence we can also  apply them to the fit with the MSHT parameterisation. In practice, given the lower degree of flexibility of the MSHT parameterisation, taking  $\Lambda_k=10^6$ is too stringent a requirement, leading to rather unstable fit results; if e.g. even a single PDF is negative and $O(10^{-4})$ in size at the highest $x_i$ value of 0.9, this can begin to overwhelm the fit and the fixed MSHT parameterisation may not be sufficiently flexible to avoid this, even though it is well outside the data region. We therefore take a value of $\Lambda_k=10^3$  (for both PDF and physical observables) in the final fit iteration, though we will monitor the extent to which any negativity remains after this. As we will see, the principle effect of imposing positivity, namely on the low $x$ gluon, is already achieved with such a constraint. 

The global fit quality in this case is also shown in Table~\ref{tab:chi2_pcharm}, and we can see that this is rather higher than the default fit, but still $\sim 90$ points lower than the NNPDF4.0 case. The remaining positivity penalty (not included in the table value) is $\sim 5$ points in $\chi^2$, and is driven by the $\overline{d}$ PDF being negative but very small at $x=0.8$ and the $\sigma_{{\rm DY},d\overline{d}}$ observable similarly being negative at very forward rapidities. In principle, it may well be possible to further increase the stringency of the positivity requirement, but we judge this to be sufficient for our purposes, not least because the stability of a fixed parameterisation fit around the $\chi^2$ minimum begins to be called into question when such stringent positivity requirements are imposed.

The above results therefore indicate that PDF positivity is clearly playing some role in determining the eventual result of the fit. Indeed, if we examine the value of the $\chi^2$ penalty that corresponds to the output of the MSHT fit without positivity imposed then we find that this is significant, being $O(10^4)$ (with  $\Lambda_k=10^3$). By far the largest effect is in the low $x$ gluon, which is negative at low scales. That this negativity is preferred by the fit to the data alone has been known for a long time, see e.g.~\cite{Martin:2009iq} for some early discussion, and is observed in  the MSHT and MSTW global PDF fits~\cite{Martin:2009iq,Harland-Lang:2014zoa,Bailey:2020ooq}, as well as in the HERAPDF fit to the HERA DIS data alone~\cite{H1:2015ubc}. We will discuss in more detail the impact of this negativity on the data, and the comparison to the fitted charm case, in the next section.

\begin{figure}
\begin{center}
\includegraphics[scale=0.6]{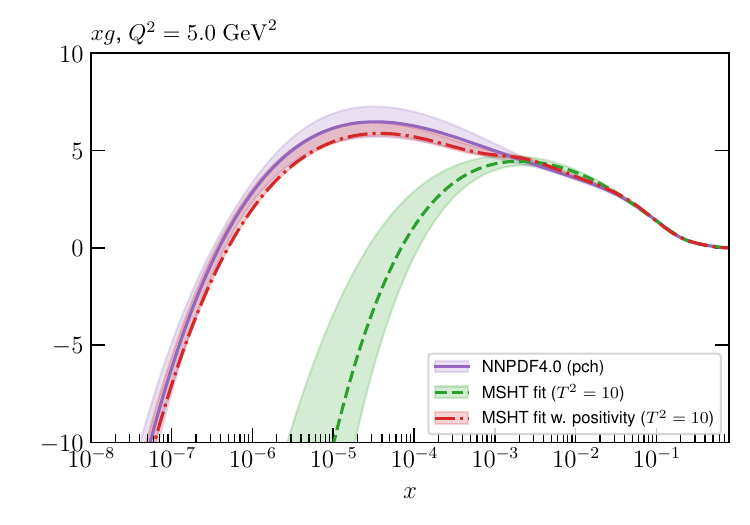}
\includegraphics[scale=0.6]{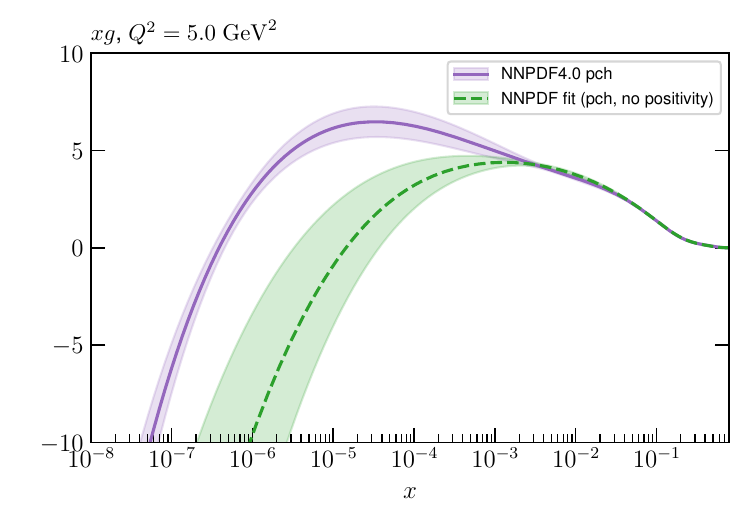}
\caption{\sf (Left) The gluon PDFs at $Q^2=5 \, {\rm GeV^2}$ that result from a  global PDF fit to the NNPDF4.0 (perturbative charm) dataset/theory setting, but using the MSHT20 parameterisation. Results with and without a positivity constraint applied, as described in the text, are shown. PDF uncertainties for the MSHT fits correspond to a fixed $T^2=10$ tolerance, and the NNPDF4.0 (perturbative charm) fit to the same dataset/theory settings is also shown. (Right) A comparison of the NNPDF4.0 (perturbative charm) public result to a fit within the NNPDF framework to the same dataset/theory settings, but without the positivity constraints imposed.}
\label{fig:fit_pch_dynT_gl}
\end{center}
\end{figure}

While this results in no negative observables in the fit itself, the prediction for the longitudinal structure function $F_L$ will become negative at low enough $x$ and $Q^2$, which is certainly unphysical. This is however in a region where the perturbative stability of the prediction is poor, as well as there being potential sensitivity to higher twist effects; a full account of low $x$ resummation effects would in particular be required (see e.g.~\cite{Martin:2009iq} for some early discussion and references and more recently \cite{Ball:2017otu,Bonvini:2017ogt,xFitterDevelopersTeam:2018hym}). Given this, it is arguably an open question as to whether it is better in a purely fixed order fit to obtain a negative $F_L$ outside the data region, which would eventually be cured by higher orders/higher twists and/or low $x$ resummation, or impose positivity on the gluon and achieve a worse fit quality to the existing DIS data. Certainly, the positivity arguments presented in~\cite{Candido:2020yat,Candido:2023ujx}  focus on the high $x$ region (see also~\cite{Collins:2021vke} for further relevant discussion), and in particular only apply where there is perturbative stability in the predicted result. 

In any case, the comparison in Fig.~\ref{fig:fit_pch_dynT_gl} (left) makes this apparently sharp distinction rather less clear. Here we show the resulting gluon PDFs at $Q^2=5$ ${\rm GeV}^2$, where positivity is imposed in the NNPDF fit. It is clear that the default NNPDF result (as well as the MSHT fit with positivity imposed) remains negative at low $x$, indeed precisely at $x$ values just below the lowest $x$ point where positivity is imposed, whereas when positivity is not imposed it is simply that this occurs at somewhat higher $x$. Thus, in either case some amount of negativity outside the data region appears to be preferred by the fit. 

In the right plot we show the result of performing a fit directly with the NNPDF framework, that is using the default NN architecture, but without imposing positivity. This is produced without further hyperparameter optimisation and hence the precise result may be treated with caution, though should be broadly representative. This gives an improvement in the fit quality of $\sim 100$  in the fit quality,  corresponding to a similar level of improvement in the fit quality for the MSHT (no positivity) fit in comparison to when positivity is imposed. We can see that this gives a gluon that is negative at lower $x$ and more in line with the MSHT result. 

\begin{figure}
\begin{center}
\includegraphics[scale=0.6]{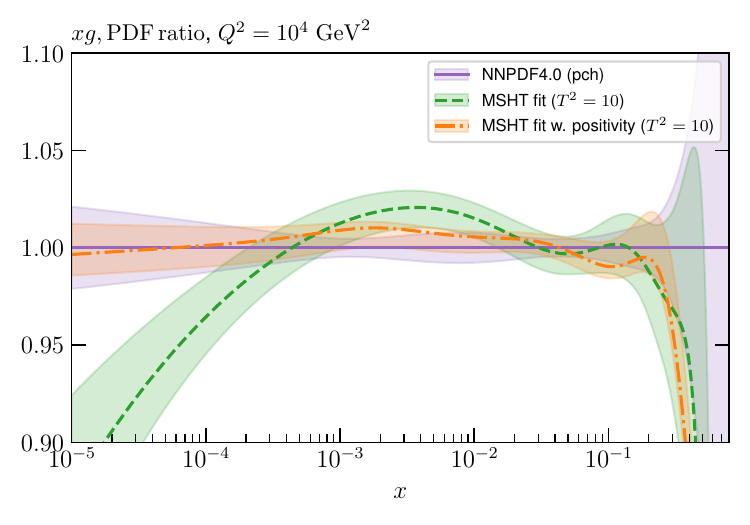}
\includegraphics[scale=0.6]{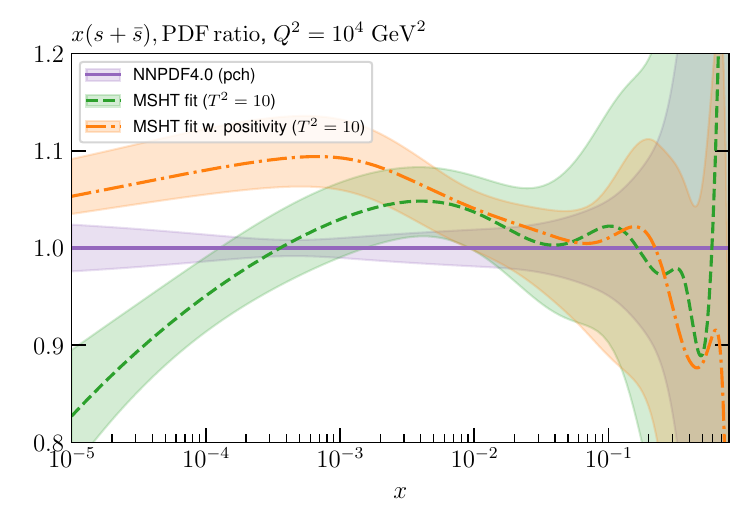}
\includegraphics[scale=0.6]{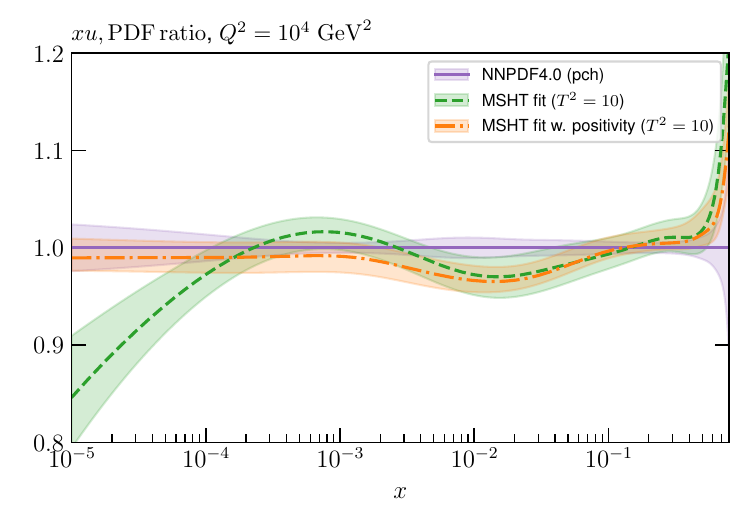}
\includegraphics[scale=0.6]{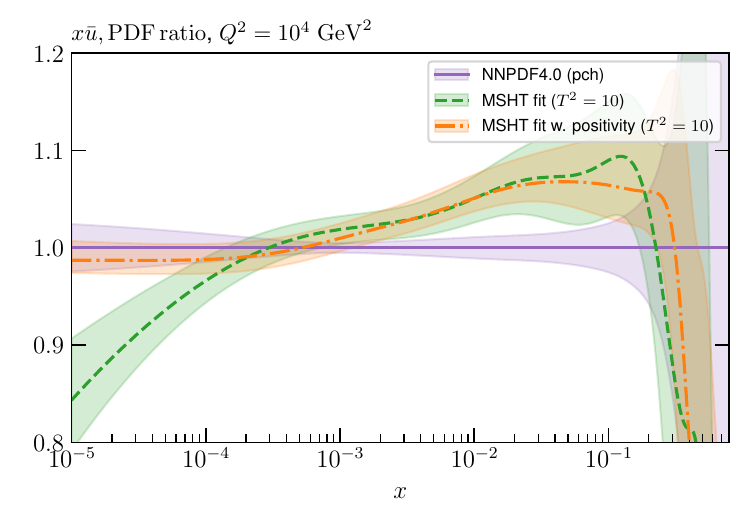}
\includegraphics[scale=0.6]{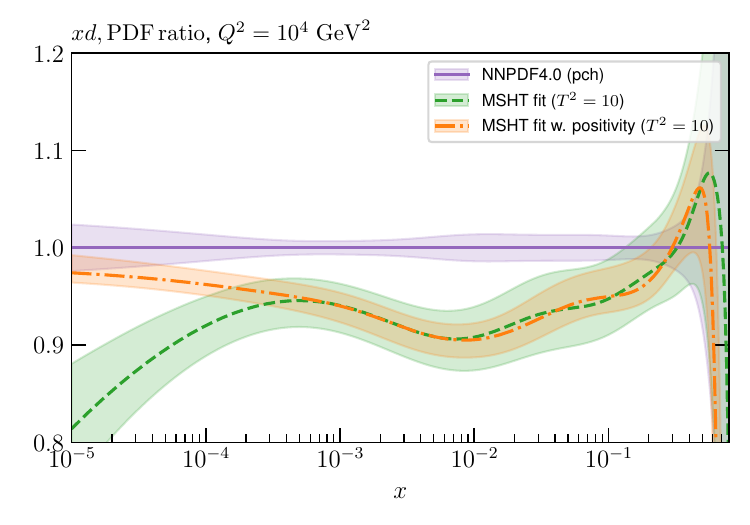}
\includegraphics[scale=0.6]{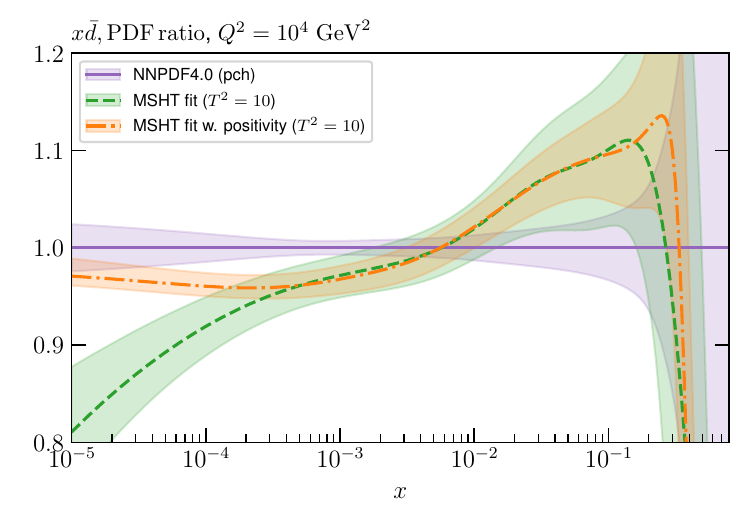}
\includegraphics[scale=0.6]{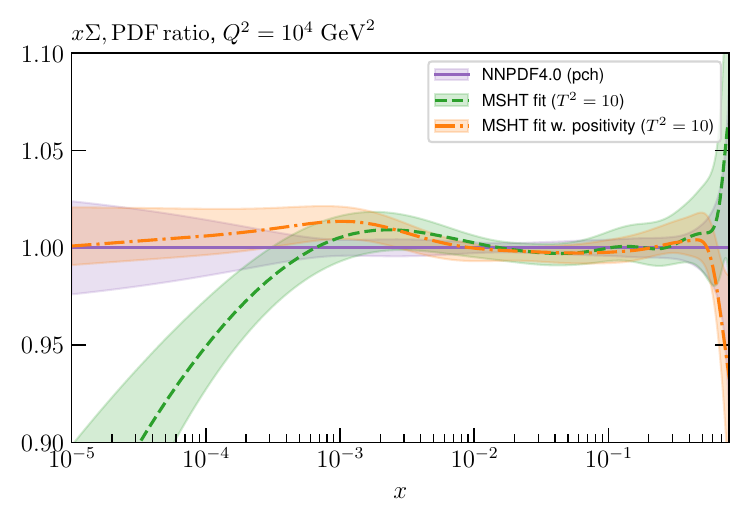}
\caption{\sf A selection of PDFs at $Q^2=10^4 \, {\rm GeV^2}$  that result from a  global PDF fit to the NNPDF4.0 (perturbative charm) dataset/theory setting, but using the MSHT20 parameterisation. Results with and without a positivity constraint applied, as described in the text, are shown. PDF uncertainties for the MSHT fits correspond to a fixed $T^2=10$ evaluation. Results are shown as a ratio to the NNPDF4.0 (perturbative charm) fit to the same dataset/theory settings.}
\label{fig:fit_pch_rat}
\end{center}
\end{figure}

Returning to Table~\ref{tab:chi2_pcharm}, in terms of the breakdown in the fit quality, the most significant difference with respect to the case without positivity imposed is in the fixed--target and HERA DIS data. In the latter case the difference is as expected focused on the low $x$ region, and is driven by the change in the gluon in this region when positivity is imposed. The fixed--target data on the other hand are not directly sensitive to this low $x$ region, but are indirectly via the momentum sum rule, and indeed are known to be in some tension with HERA DIS data due to this, see~\cite{Bailey:2020ooq} for a detailed discussion. Imposing positivity would appear to increase this tension further, resulting in some deterioration in the fit quality here. There is also some more moderate deterioration in the fit quality to various DY datasets, in particular those from LHCb, which is again consistent with the dominant change being at low $x$. However, overall the trend in the MSHT fit with positivity imposed with respect to the NNPDF4.0 fit is similar to that observed in the fit without positivity imposed, with some improvement observed in the description of the fixed--target DIS, HERA DIS and collider DY data, albeit somewhat milder, in particular for the fixed--target data, in comparison to the case without positivity. Most notable is the improved description of the collider DY data, with the ATLAS high precision $W,Z$ data~\cite{ATLAS:2016nqi} still being improved by $\sim 20$ points in $\chi^2$ relative to the NNPDF4.0 perturbative charm fit, which makes up almost half of the improvement of the collider DY data.

Considering now the impact on the PDFs, these are shown in Fig.~\ref{fig:fit_pch_rat}. For the MSHT fits we quote the errors with $T^2=10$, which are representative of the result of applying the dynamic tolerance criterion. We do not show results with this applied as in the case where positivity is imposed, it is not completely clear how to do so consistently, given in this case the $\chi^2$ deterioration away from the best fit value for some eigenvectors has a major contribution from the positivity penalty. Indeed, even taking a fixed tolerance, we can see for the case with positivity imposed that the PDF uncertainties are rather smaller than in the result without positivity, and can be rather asymmetric\footnote{In fact, given how strict the positivity requirement \eqref{eq:pos} is we find that behaviour of the global $\chi^2$ around the minimum can be strongly non--Gaussian, calling into question the applicability of the Hessian approach to evaluating PDF uncertainties in this case. The PDF uncertainties for the fit with positivity imposed can therefore only be taken as a rough guide.}. 

In broad terms, we can see that while there are regions of  agreement between the NNPDF4.0 result and the MSHT fits, there are also distinct differences. A significant difference across all PDF sets is at low $x$, without positivity applied to the MSHT fit, in which case these lie up $\sim 10-20\%$ below the NNPDF result\footnote{A qualitatively consistent trend is observed in the recent NNPDF study~\cite{Cruz-Martinez:2024cbz}, where a comparison between imposing the positivity constraint on the gluon and quark singlet at the default scale $Q^2=5$ ${\rm GeV}^2$ and a lower scale $Q^2=1$ ${\rm GeV}^2$  is presented.}. This effect is precisely driven by the negativity of the low $x$ and $Q^2$ gluon, and the corresponding reduction this leads to for all of the shown PDF combinations at higher $x$ scale due to evolution. Indeed, if we impose positivity, the MSHT result lies significantly closer to the NNPDF result at low $x$, for all PDFs. For the gluon PDF, this also improves the agreement in the intermediate $x$ region, such that the MSHT fit with positivity imposed results in a rather similar result for the gluon in the data region. The positivity constraint also has some impact on certain quark distributions at very high $x$ and lower scales, but these are not visible on the current plots. In Appendix~\ref{app:NNPDFnopos} we show the result of performing a fit within the NNPDF approach, using precisely the same settings as the NNPDF4.0 (perturbative charm) fit, but without the positivity constraints applied, and this confirms that in this case the NNPDF result at low $x$ matches that of the MSHT fit (without positivity) more closely.

\begin{figure}
\begin{center}
\includegraphics[scale=0.6]{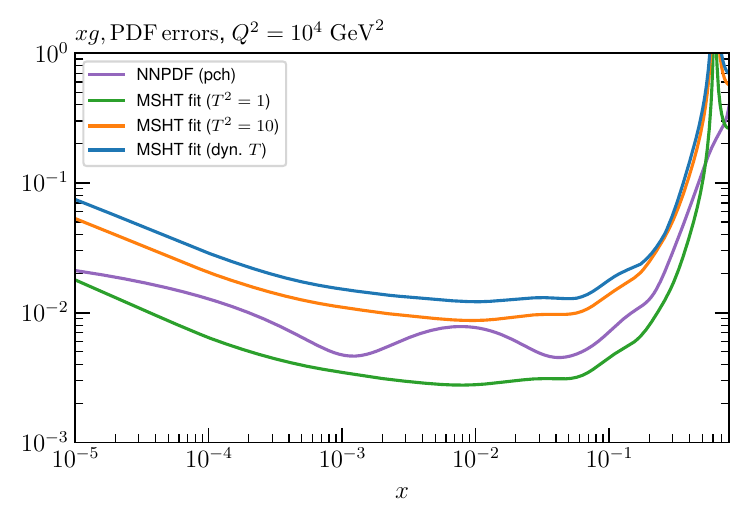}
\includegraphics[scale=0.6]{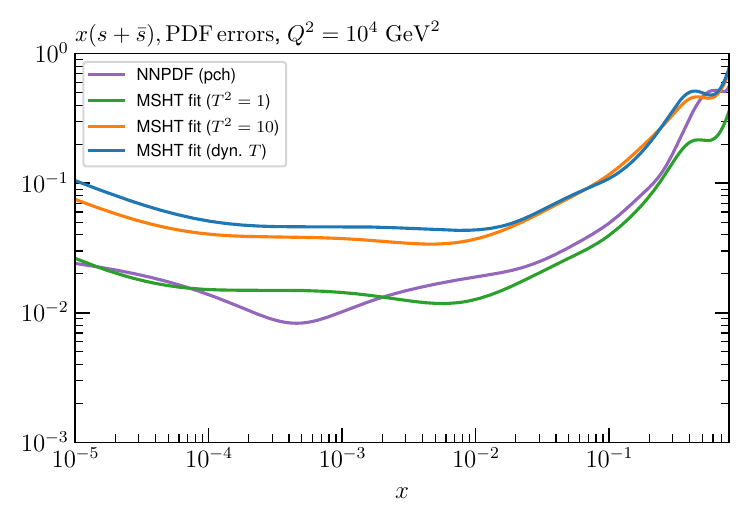}
\includegraphics[scale=0.6]{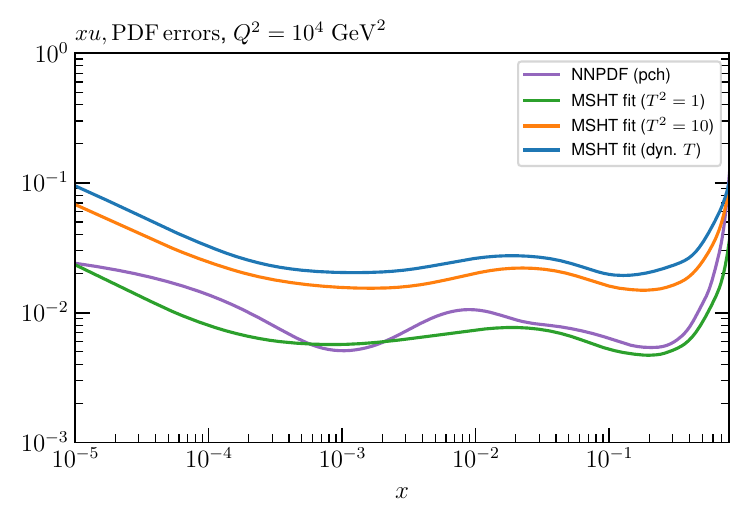}
\includegraphics[scale=0.6]{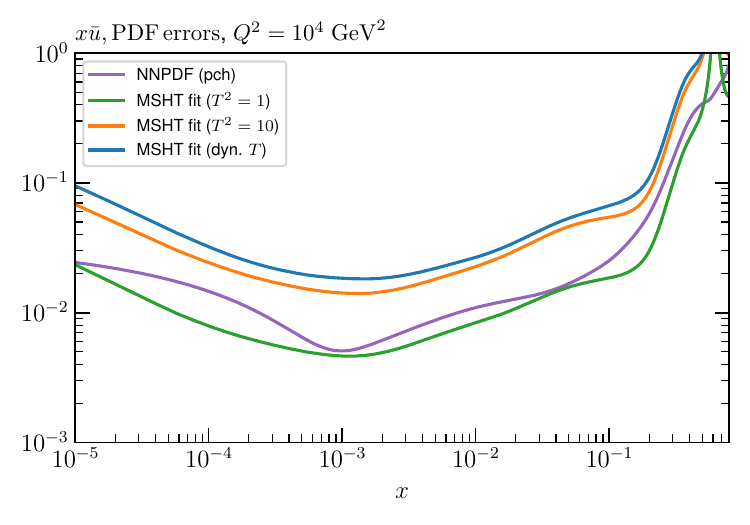}
\includegraphics[scale=0.6]{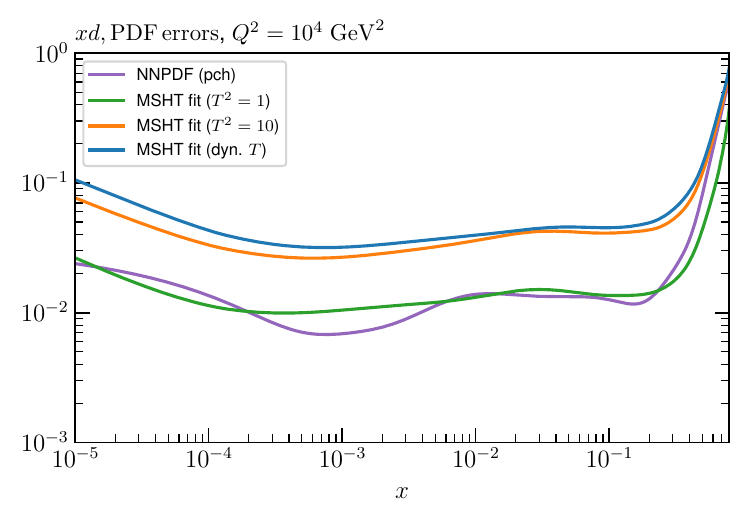}
\includegraphics[scale=0.6]{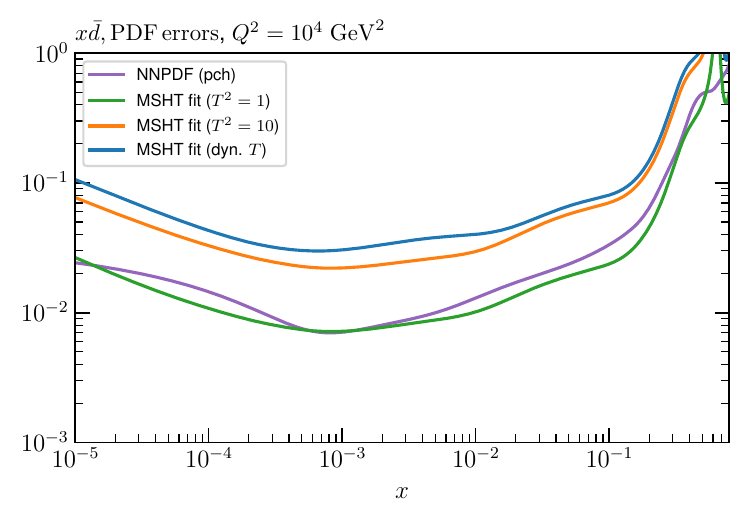}
\includegraphics[scale=0.6]{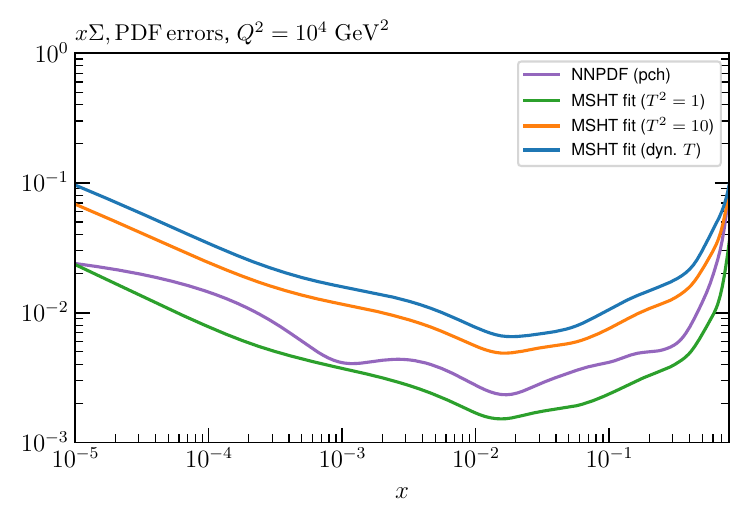}
\caption{\sf The PDF uncertainties at $Q^2=10^4 \, {\rm GeV^2}$  resulting from a  global PDF fit to the NNPDF4.0 (perturbative charm) dataset/theory setting, but using the MSHT20 parameterisation. Results  without a positivity constraint applied, as described in the text, are shown. PDF uncertainties for the MSHT fits correspond to dynamic and fixed tolerances are given as well as the corresponding uncertainty for the NNPDF4.0 (perturbative charm) fit to the same dataset/theory settings.}
\label{fig:fit_pch_errcomp_nopos}
\end{center}
\end{figure}

Even after imposing positivity, there remain however  significant differences in terms of the quark flavour decomposition, which is as we would expect largely insensitive to the positivity constraint away from the low $x$ region. That is, while there is broad stability in the overall quark singlet (and the charge--weighted singlet, not shown), there are non--negligible deviations in the individual $u,d,s$ quark and antiquark distributions. We note that while we show PDF uncertainties on the MSHT results for the sake of comparison, given the fit quality in the MSHT case is actually better than that of the NNPDF4.0 fit, the correct assessment of the overall consistency of the NNPDF PDFs with the MSHT result is between the central value of the MSHT and NNPDF PDFs, within the NNPDF errors. With this in mind, we can see that the differences are in many cases highly significant, with the MSHT result (either with or without positivity imposed) lying several standard deviations outside the NNPDF uncertainty band in many $x$ regions and for many PDFs. 
 
 It is unclear why this remaining difference, in either the fit quality or PDFs, is present once we have accounted for the differing treatment of PDF positivity, in particular for the low $x$ gluon. All we can straightforwardly establish is that the MSHT fit arrives at a somewhat better description of certain datasets, most notably the LHC DY data, and that this is achieved by  a rather distinct flavour decomposition in the intermediate  to high $x$ region to the NNPDF fit. The further difference at low $x$ is then explained by the differing treatment of PDF positivity. It is unclear what precisely in the NNPDF prior (i.e. the specific architecture, training/validation etc)  leads to this, though we will investigate the question of parameterisation flexibility further in Section~\ref{sec:overfit}. However, one potentially relevant observation is that, in~\cite{NNPDF:2021njg} (Fig. 8.6) we can see that qualitatively some of these differences in the $u$ and $d$ sector follow the difference that comes from using a flavour rather than an evolution basis, although the effect we see is clearly larger, and moreover the results in~\cite{NNPDF:2021njg} concern the fitted rather than perturbative charm.

This difference in the absolute values of the PDFs notwithstanding, it is also interesting to compare the size of the PDF uncertainties, as in Fig.~\ref{fig:glcl_errs} but now for a fit to the real data rather than in a closure test, for the MSHT result. The results are in fact very similar to this case, with the MSHT uncertainties with fixed tolerance being rather close to the uncertainties in the closure test. This is due to the fact that, as discussed in detail in Section~\ref{sec:tol}, the size of the PDF uncertainties is driven by the experimental (or in some recent fits, theoretical) uncertainties that are propagated through the fit, and is not determined by any underlying tensions between the datasets themselves, as will be present in a real fit but absent in a closure test (although we recall that the larger $T^2=10$ uncertainty is applied to reflect such tensions). There are some differences in detail, however, given the fits are not identical. 

In broad terms,  we therefore again find that at intermediate to higher $x$, for the quark flavour decomposition the NNPDF uncertainty is very similar in size  to (though on average very slightly larger than) the MSHT fit with $T^2=1$, with the most notable exception of the gluon and the related quark singlet, where the NNPDF uncertainty is rather larger, and somewhere between the $T^2=1$ and $T^2=10$ cases, though rather lower than the latter. In other words, in many cases the NNPDF uncertainty is indeed  roughly equivalent to taking a $T^2=1$ tolerance with the MSHT parameterisation in the data region, in the context of a global PDF fit. This uncertainty is in particular significantly lower than the result of taking $T^2=10$, and the dynamic tolerance criterion, which we now show; the latter is, as claimed, similar in size to the $T^2=10$, if slightly larger on average.

Finally, a comparison between the MSHT fit, without positivity, and the MSHT20~\cite{Bailey:2020ooq} result is given in Appendix~\ref{app:pdfwmsht}. Further discussion is given in the appendix, but here we simply note that a comparison is presented in Figs~\ref{fig:fit_pch_dynT_wMSHT} and~\ref{fig:fit_pch_Teq1_wMSHT} between the default MSHT20 result and the MSHT fit with a dynamic tolerance, and the cases where a fixed $T^2=1$ tolerance is instead used (in the MSHT20 case this is approximated by dividing the PDF uncertainties by a factor of 3). These fits differ only in the choice of datasets (and their treatments), and the underlying theory settings, while the fitting methodology is identical. Therefore, while differences in the underlying PDFs would be expected, and indeed are observed, these should remain compatible within them if the PDF uncertainties are representative. From this comparison, however, it is clear that there is strong statistical incompatibility between the two results if the textbook $T^2=1$ criterion is used. On the other hand, for the dynamic tolerance the compatibility is greatly improved, with no significant tension observed; although we note that a full account of the compatibility of these sets in this case would also require an evaluation of the correlation between them, which will certainly be non--zero given the similarly in many of the datasets in the fits.

\subsection{Fitted Charm}\label{subsec:fcharm}

Having considered a comparison to the NNPDF fit with perturbative charm, it is also instructive to account for the possibility of fitted charm, given this is the default treatment in the NNPDF fit. Here, the charm quark PDF is freely parameterised, rather than being determined from the other QCD partons via perturbatively calculated matching conditions (i.e. effectively via perturbative $g\to c\overline{c}$ splittings). As discussed in~\cite{Ball:2016neh,NNPDF:2021njg,Ball:2022qks}, this permits higher-order corrections to these perturbative matching conditions to be absorbed into the initial charm PDF, as well as allowing for a possible non-perturbative `intrinsic' charm component (see also~\cite{Hou:2017khm,Guzzi:2022rca} for studies by the CT collaboration and discussion of the definition of intrinsic charm and its process dependence). Procedurally, our fit works in exactly the same way as before, but now the  input can take an arbitrary set of PDFs defined at a higher value of $Q_0=1.65$ GeV ($ > m_c = 1.51$ GeV), rather than $Q_0=1$ GeV, i.e. the PDFs are parameterised at this higher scale, but otherwise using the same MSHT20 parameterisation as before. The modifications to the DGLAP evolution and the DIS cross section predictions are then consistently and automatically accounted for by the public NNPDF code, such that the PDF fit can be performed in a completely analogous manner to the previous case.

The charm PDF is therefore now freely parameterised at this scale, rather than being determined perturbatively in terms of the other PDFs. We do this by assuming that $c=\overline{c}$ for simplicity at input (see~\cite{NNPDF:2023tyk} for a study where this constraint is not imposed), while for $c_+ = c+\overline{c}$ we take the standard 6 Chebyshev parameterisation:
\be\label{eq:cheb6}
x c_+(x,Q_0) = A_{c_+} x^{\delta_{c_+}} (1-x)^{\eta_{c^+}}\left( 1+ \sum_{i=1}^6 a_{c,i}T_i(y(x))\right)\;.
\ee
This therefore represents the first PDF fit with fitted (i.e. freely parameterised) charm outside of a neural network approach to PDF parameterisation.

\begin{table}
\begin{center}
  \scriptsize
  \centering
   \renewcommand{\arraystretch}{1.4}
\begin{tabular}{Xrccc}\hline 
&NNPDF4.0&MSHT fit&MSHT fit (w positivity)
\\ \hline
BCDMS $F_2^{p}$ (333)~\cite{BCDMS:1989qop}  & 473.6 (1.42) & \textcolor{blue}{451.8 (1.36)} & \textcolor{blue}{453.9 (1.36)}\\
NuTeV $\sigma^{\overline{\nu}}_{\rm CC}$ (37)~\cite{NuTeV:2001dfo}  & 21.1 (0.57) & \textcolor{red}{33.9 (0.92)} & \textcolor{red}{35.0 (0.95)}\\
{\bf DIS Fixed--Target (1881)} & {\bf 2011.6 (1.07)} &\textcolor{black}{{\bf  2018.6 (1.07)}} & \textcolor{black}{{\bf 2015.3 (1.07)}}
\\ \hline
E886 $\sigma^{p}$ (NuSea) (89)~\cite{NuSea:2003qoe} & 105.3 (1.18) & \textcolor{red}{112.5 (1.26)} & \textcolor{black}{110.0 (1.24)}\\
E906 $\sigma^{d}/2\sigma^p$ (SeaQuest) (6)~\cite{SeaQuest:2021zxb}  & 5.72 (0.95) & \textcolor{blue}{3.33 (0.56)} & \textcolor{blue}{3.69 (0.62)}\\
{\bf DY Fixed--Target (195)} & {\bf 185.6 (0.95)} &\textcolor{black}{{\bf 192.1 (0.99)}} & {\bf 190.2 (0.98)}
\\ \hline
NC $e^+ p$ 920 GeV (377)~\cite{H1:2015ubc} & 518.6 (1.38) & \textcolor{blue}{506.0 (1.34)} & \textcolor{blue}{506.0 (1.34)}\\
CC $e^+ p$ (39)~\cite{H1:2015ubc} & 47.5 (1.22) & \textcolor{blue}{42.9 (1.10)} & \textcolor{black}{44.5 (1.14)}\\
NC, c (37)~\cite{H1:2018flt} & 82.8 (2.24) & \textcolor{black}{82.7 (2.24)} & \textcolor{red}{91.1 (2.46)}\\
{\bf HERA DIS (1145)} & {\bf 1575.6 (1.38)} &\textcolor{blue}{{\bf  1557.6 (1.36)}} & \textcolor{black}{{\bf 1565.8 (1.38)}}
\\ \hline
D0 $W$ muon asymmetry (9)~\cite{D0:2014kma} & 17.9 (1.99) &  \textcolor{blue}{15.2 (1.69)} & \textcolor{blue}{15.4 (1.72)}\\
CMS DY 2D 7 TeV (110)~\cite{CMS:2013zfg} & 146.2 (1.33) & \textcolor{blue}{138.6 (1.26)} & \textcolor{black}{140.7 (1.28)}\\
CMS $W$ rapidity 8 TeV (22)~\cite{CMS:2015hyl} & 26.2 (1.19) & \textcolor{blue}{22.7 (1.03)} & \textcolor{black}{23.3 (1.06)}\\
LHCb $W,Z\to \mu$ 7 TeV (29)~\cite{LHCb:2015okr} & 56.3 (1.94) & \textcolor{blue}{51.8 (1.78)} & \textcolor{black}{53.6 (1.85)}\\
{\bf Collider DY (576)} & {\bf 767.8 (1.33)} &\textcolor{blue}{{\bf 743.1 (1.29)}} & \textcolor{blue}{{\bf 754.4 (1.31)}}
\\
\hline
{\bf LHC Jets (500)} & {\bf 804.8 (1.61)} &\textcolor{black}{{\bf 796.7 (1.59)}} & \textcolor{black}{{\bf 797.6 (1.60)}}\\
\hline
ATLAS $W^\pm + $ jet 8 TeV (30)~\cite{ATLAS:2017irc} & 43.9 (1.46) & \textcolor{red}{48.2 (1.61)} & \textcolor{black}{47.3 (1.58)}\\
{\bf LHC $V+$ Jets (122)} & {\bf 136.1 (1.12)} &\textcolor{black}{{\bf 140.4 (1.15)}} & \textcolor{black}{{\bf 139.2 (1.14)}}\\ \hline
{\bf Isolated Photon (53)} & {\bf 41.9 (0.79)} &\textcolor{black}{{\bf 40.5 (0.76)}} & \textcolor{black}{{\bf 40.6 (0.77)}}\\ \hline
ATLAS $t\overline{t}$ $l+$ jets 8 TeV (8)~\cite{ATLAS:2015lsn} &25.9 (3.24) & \textcolor{red}{30.6 (3.82)} & 26.1 (3.26)\\
{\bf Top quark  (81)} & {\bf 85.0 (1.05)} &\textcolor{black}{{\bf 87.4 (1.08)}} & \textcolor{black}{{\bf 83.0 (1.02)}}
\\ \hline \hline 
{\bf Global, $\mathbf{t_0}$ (4616)}    &{\bf 5692.1 (1.233)}& {\bf 5645.2 (1.222)} &{\bf 5651.0 (1.224)}\\
\hline
{\bf Global, exp. (4616)} &{\bf 5354.1 (1.160)}& {\bf 5322.5 (1.153) }&{\bf 5341.5 (1.155)}\\
\hline
\end{tabular}
\end{center}
\caption{\sf $\chi^2$ values for the NNPDF4.0 fit   and the MSHT fits to the NNPDF dataset/theory settings, with fitted charm. The fit quality for the different major subsets the constitute the global dataset are given in bold, and  above each subtotal the fit qualities for individual experiments in these subsets where the difference with respect to the NNPDF4.0 cases is roughly larger than $\pm 0.5 \sigma = \pm \sqrt{N_{\rm pts}/2}$ for either MSHT fit is shown. When these differences are less than $-0.5\sigma$ the result is highlighted in blue, while the result is highlighted in red when it is greater than $0.5\sigma$. Both the absolute $\chi^2$ and the per point value in brackets, is given in all cases, while the number of points is indicated in brackets next to the dataset description.  For the total $\chi^2$ both the experimental and $t_0$ definitions are shown, while in all other cases only the latter definition is used. Results with and without positivity imposed are shown for the MSHT fit, though we note that the most appropriate comparison with the NNPDF4.0 result is with this imposed.}
\label{tab:chi2_fcharm}
\end{table}

Starting with the fit qualities, these are shown in Table~\ref{tab:chi2_fcharm}. We can see that the MSHT fit is again better than the NNPDF result, but by rather less $\sim 50$ points, rather than the $\sim 190$ points in the cases of the perturbative charm in Table~\ref{tab:chi2_pcharm}. Nonetheless, this is a non--negligible improvement in fit quality. Looking at the breakdown between different data subsets, the only particularly significant changes are now limited to the HERA DIS and collider DY data, both of which show moderate improvements. The difference in the latter case in particular indicates that again the quark flavour decomposition will play a role here.  For the fixed--target data, the fit quality is on average relatively stable even if, for certain individual datasets there are reasonable improvements and deteriorations in the fit quality. 

Interestingly, the role of positivity is now found to be almost negligible. The penalty~\eqref{eq:pos} that results from the fit with no positivity imposed is only $\sim 20$  (with  $\Lambda_k=10^3$), due to the $\overline{d}$ being slightly negative at high $x$, and $F_2^c$ being slightly negative at $x\sim 0.1$. After imposing positivity, these no longer occur, at the expense of a slightly worse fit quality, which remains improved with respect to the NNPDF4.0 result. The changes with respect to the case where positivity is not imposed are divided relatively evenly between the different data subsets, with the fit quality improving in some cases and deteriorating in others, in all cases by a handful of points in $\chi^2$.

\begin{figure}
\begin{center}
\includegraphics[scale=0.6]{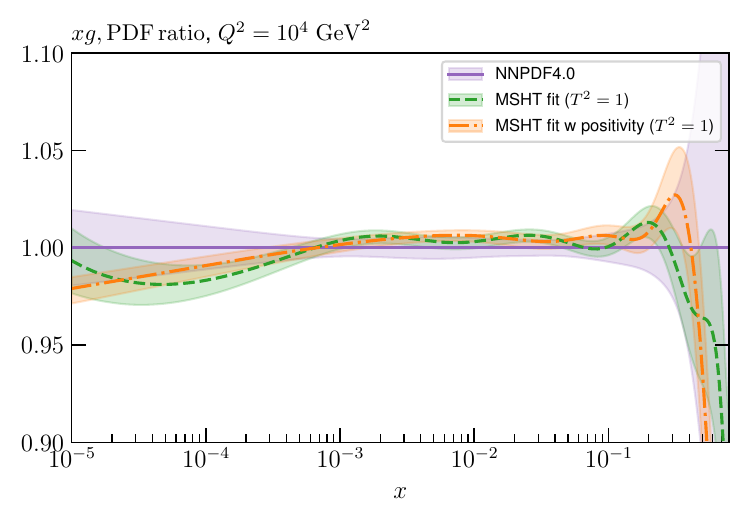}
\includegraphics[scale=0.6]{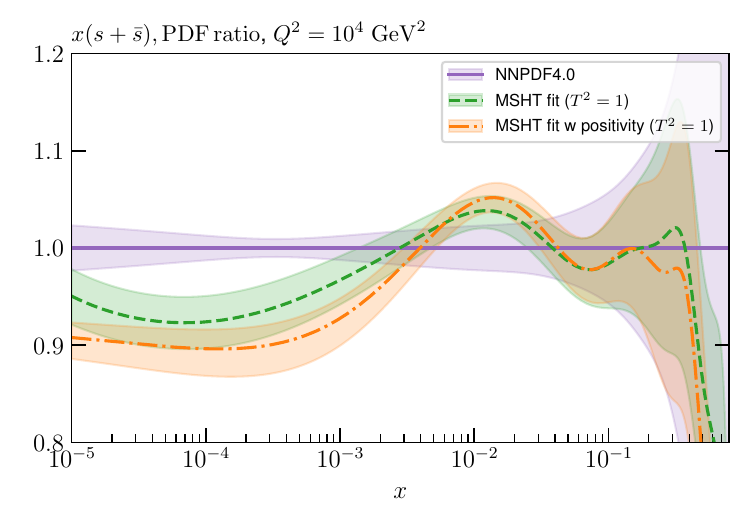}
\includegraphics[scale=0.6]{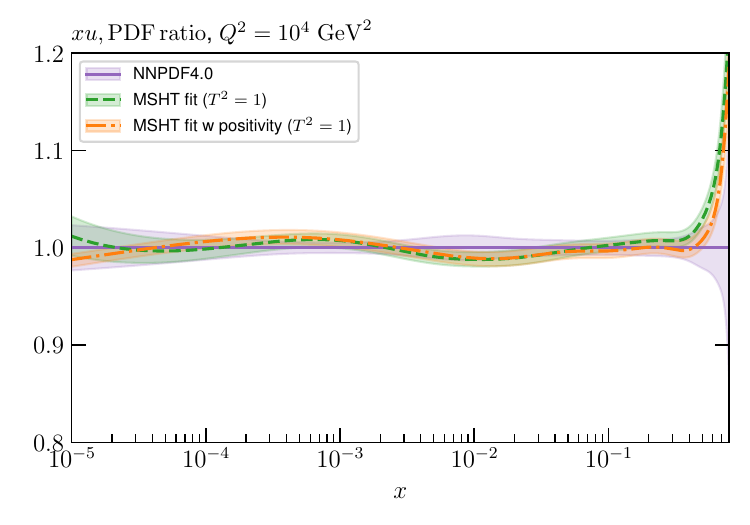}
\includegraphics[scale=0.6]{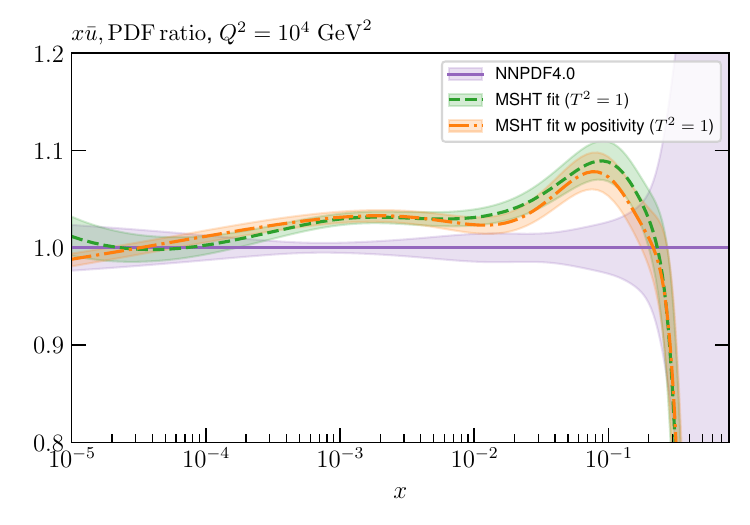}
\includegraphics[scale=0.6]{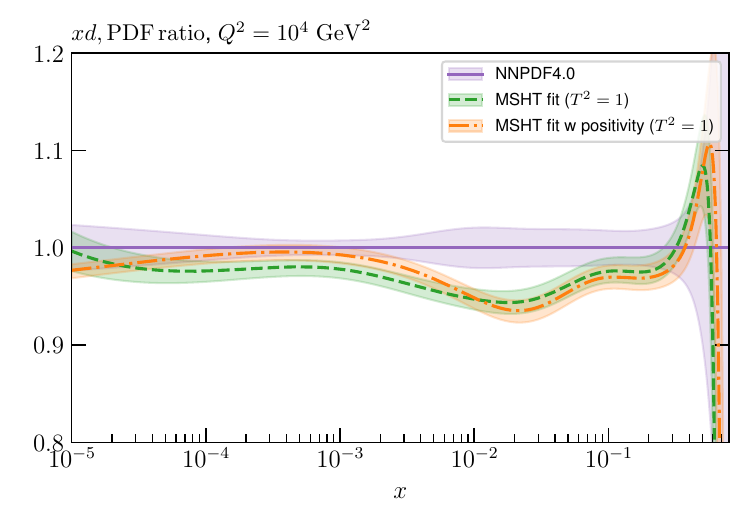}
\includegraphics[scale=0.6]{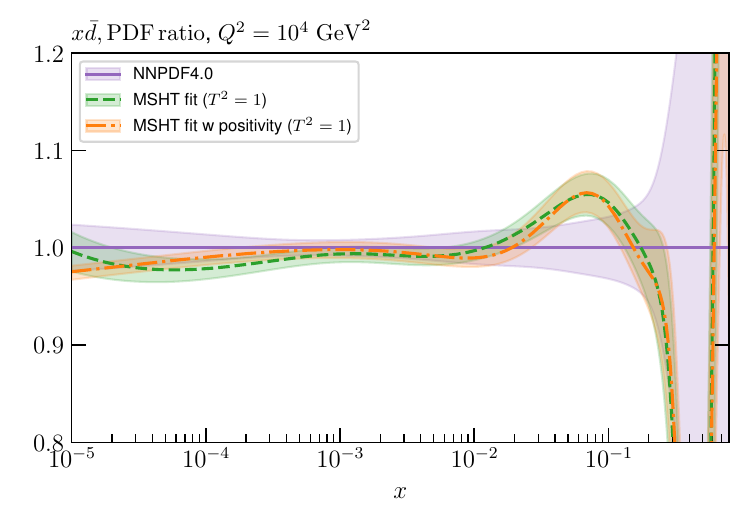}
\includegraphics[scale=0.6]{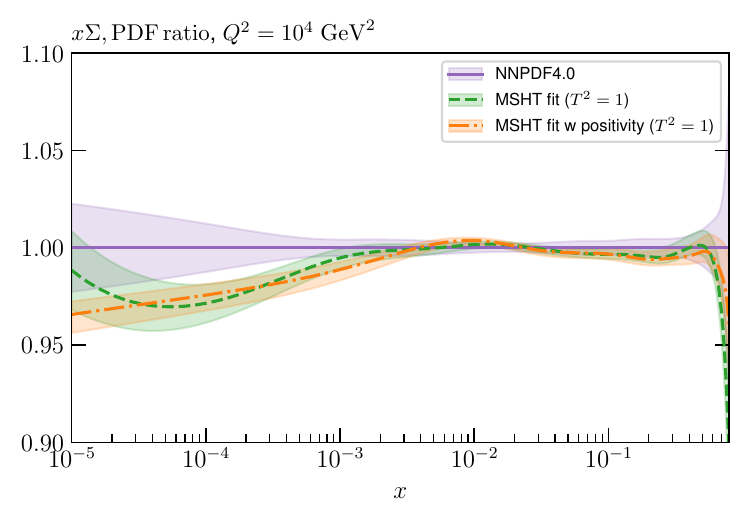}
\includegraphics[scale=0.6]{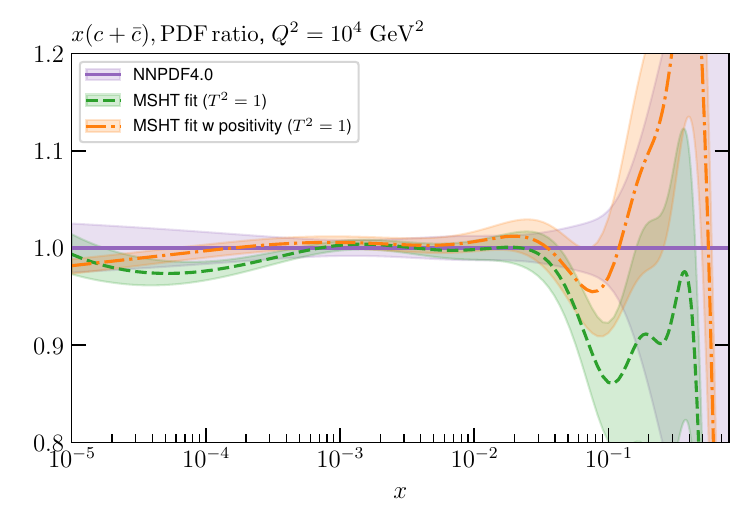}
\caption{\sf  A selection of PDFs at $Q^2=10^4 \, {\rm GeV^2}$  that result from a  global PDF fit to the NNPDF4.0 dataset/theory (fitted charm) setting, but using the MSHT20 parameterisation. Results with and without a positivity constraint applied, as described in the text, are shown. PDF uncertainties for the MSHT fits correspond to a fixed $T^2=1$ tolerance. Results are shown as a ratio to the NNPDF4.0  fit to the same dataset/theory settings.}
\label{fig:fit_fitch_Teq1}
\end{center}
\end{figure}

Turning now to the PDF comparison, this is shown in Fig.~\ref{fig:fit_fitch_Teq1} for the case of $T^2=1$ uncertainties, which we show in order to demonstrate more clearly the difference with respect to the NNPDF case. We recall  that, given the fit quality in the MSHT cases is actually better than that of the NNPDF4.0 fit, the more appropriate assessment of the overall consistency of the NNPDF result with the MSHT case is between the central value of the MSHT  and  NNPDF PDFs, within the NNPDF errors. The first observation is that, in comparison to the perturbative charm case shown in Fig.~\ref{fig:fit_pch_rat}, the agreement between the MSHT fit and the NNPDF result is greatly improved, but that nonetheless some significant differences with respect to the NNPDF uncertainty  remain. For certain PDFs, such as the gluon and up quark, there is near complete consistency, and indeed the most important areas of disagreement relate to the quark flavour decomposition, as we might expect given the discussion of the fit quality. For e.g. the $\overline{u}$, $d$ and strangeness in certain $x$ regions, the PDFs from the MSHT fit remain several standard deviations away from the NNPDF case, with respect to the NNPDF uncertainties.

The difference between the cases with and without positivity imposed is also rather small,  as we would expect given the small difference in fit qualities seen in Table~\ref{tab:chi2_fcharm}. Of particular note is the behaviour at low $x$, for which we do not see any significant deviation between the two cases, with no trend for the PDFs without positivity imposed to lie below those with positivity imposed, as was the case for the perturbative charm fit shown in Fig.~\ref{fig:fit_pch_rat}; we will return to this issue below. Indeed, more generally the two sets of PDFs are completely compatible even within the reduced $T^2=1$ uncertainties. The charm quark at $x\sim 0.1$ is observed to be somewhat higher when positivity is imposed, as a result of the $F_2^c$ positivity constraint in this region. In Appendix~\ref{app:NNPDFnopos} we show the result of performing a fit within the NNPDF approach, using precisely the same settings as the NNPDF4.0 fit, but without the positivity constraints applied, and find trends consistent with this, namely limited changes in the PDFs in general and an increase in the charm quark around $x\sim 0.1$, again to avoid the negativity of the charm structure function in this region at $Q^2=5$ ${\rm GeV}^2$. We note that this will  have an impact on any interpretation of these results in terms of the statistical significance of any (positive) intrinsic charm in the high $x$ region. This is arguably problematic, given one would expect the significance of any preference from the data for such a positive intrinsic charm component to be independent of such a positivity requirement via a hypothetical $F_2^c$ observable.

\begin{figure}
\begin{center}
\includegraphics[scale=0.6]{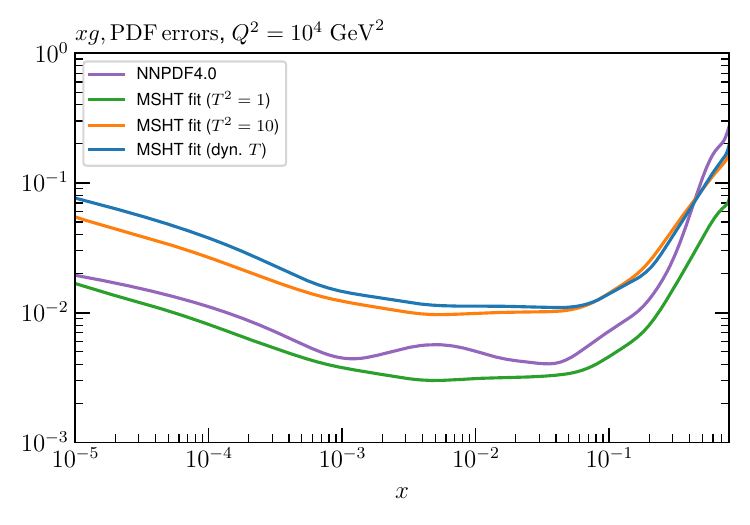}
\includegraphics[scale=0.6]{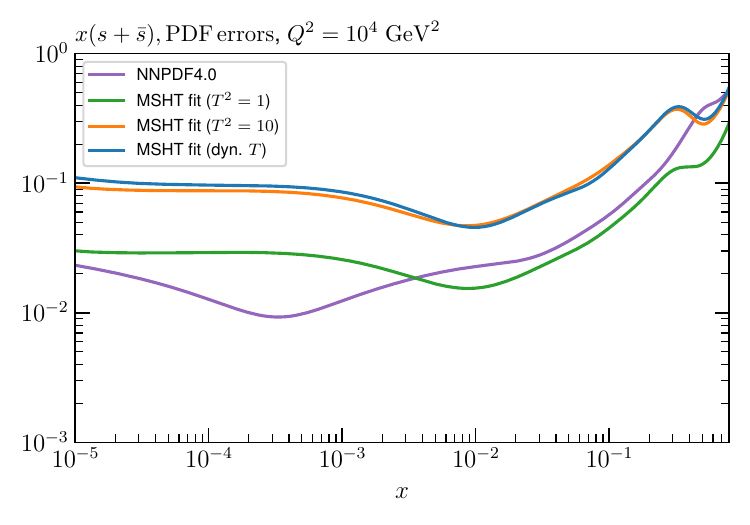}
\includegraphics[scale=0.6]{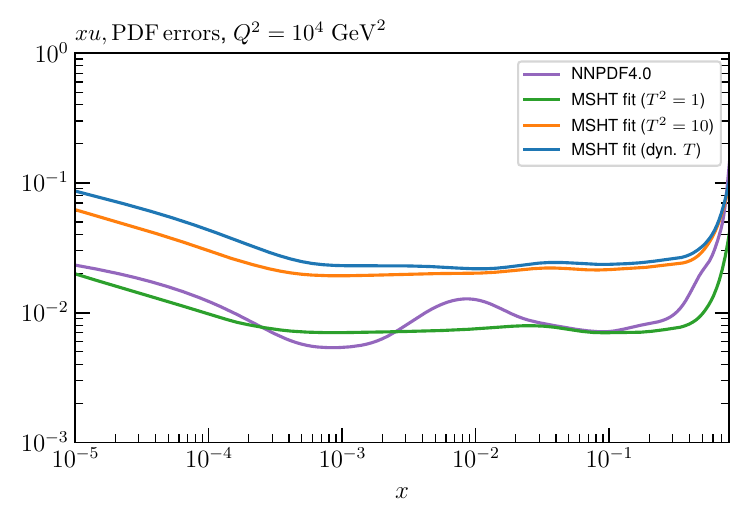}
\includegraphics[scale=0.6]{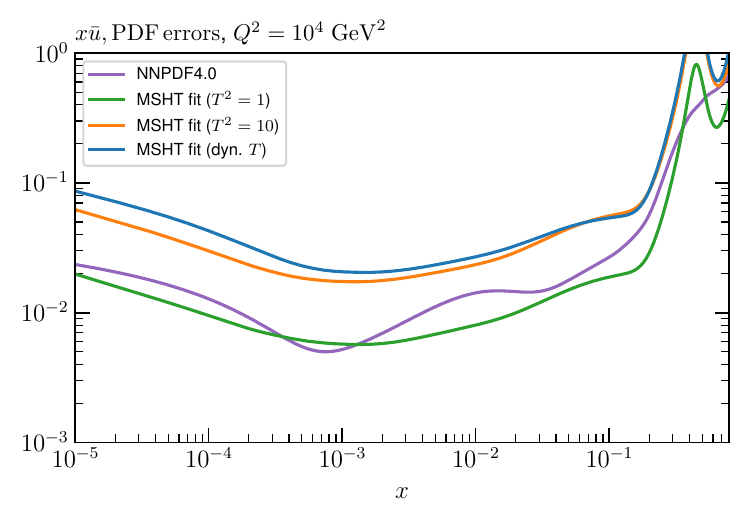}
\includegraphics[scale=0.6]{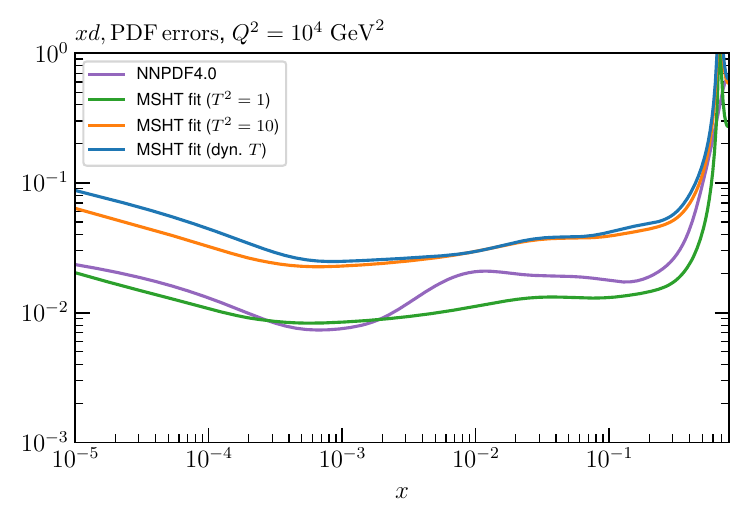}
\includegraphics[scale=0.6]{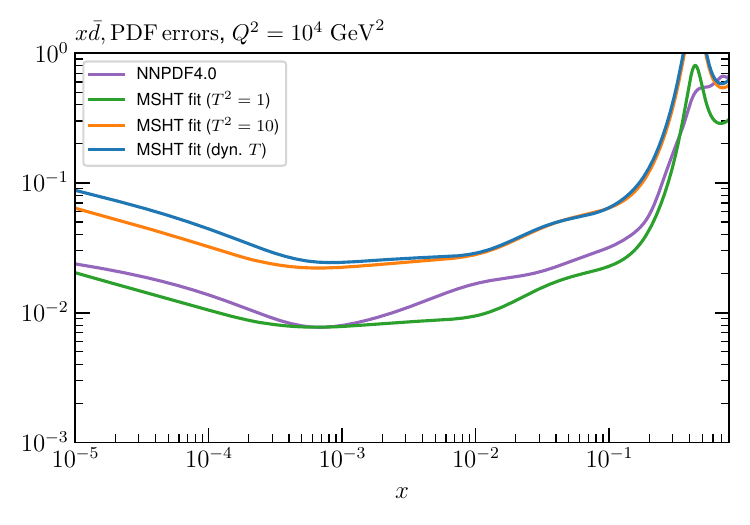}
\includegraphics[scale=0.6]{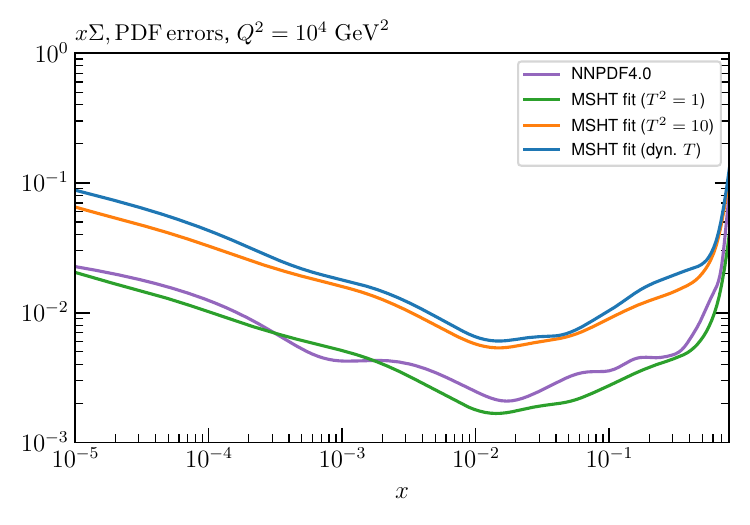}
\includegraphics[scale=0.6]{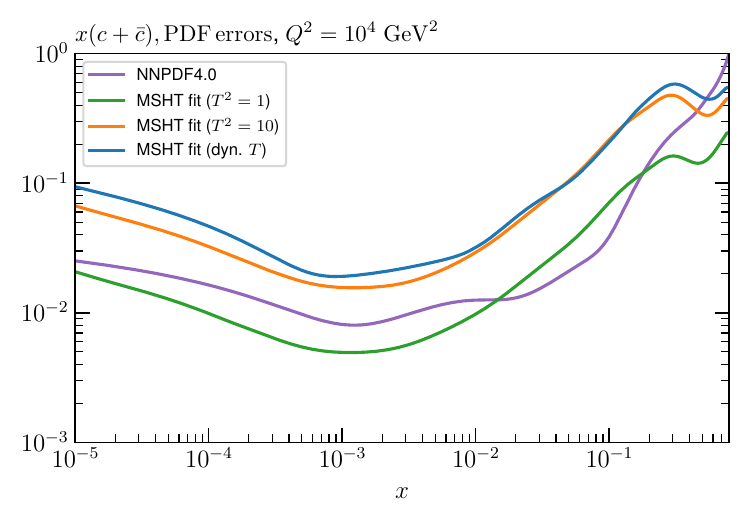}
\caption{\sf The PDF uncertainties at $Q^2=10^4 \, {\rm GeV^2}$  resulting from a  global PDF fit to the NNPDF4.0 dataset/theory setting (with fitted charm), but using the MSHT20 parameterisation. Results  without a positivity constraint applied, as described in the text, are shown. PDF uncertainties for the MSHT fits correspond to dynamic and fixed tolerances are given as well as the corresponding uncertainty for the NNPDF4.0  fit to the same dataset/theory settings.}
\label{fig:fit_fitch_errcomp_pos}
\end{center}
\end{figure}

Turning to the uncertainties, these are shown in Fig.~\ref{fig:fit_fitch_errcomp_pos}, and we can see that once again the NNPDF uncertainties are broadly of order (though on average very slightly larger than) the $T^2=1$ ones, with the main exception of the gluon and quark singlet. However, interestingly here the gluon uncertainty is somewhat closer to (though still somewhat larger than) the  $T^2=1$ uncertainty than in the perturbative charm case.

\begin{figure}
\begin{center}
\includegraphics[scale=0.6]{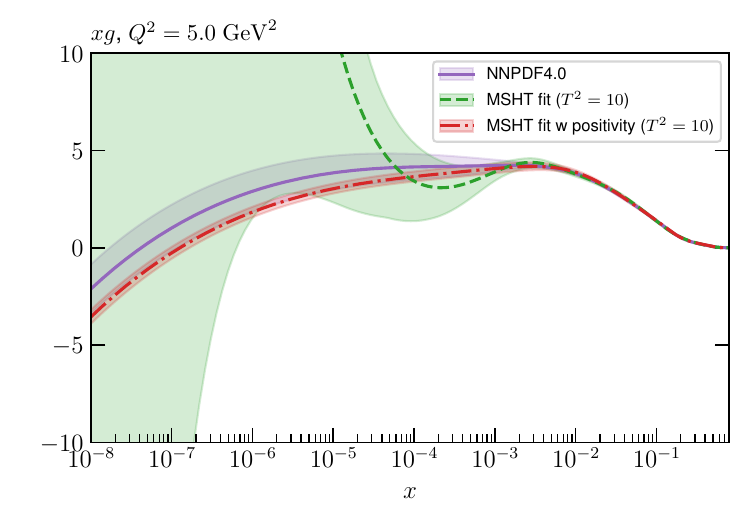}
\includegraphics[scale=0.6]{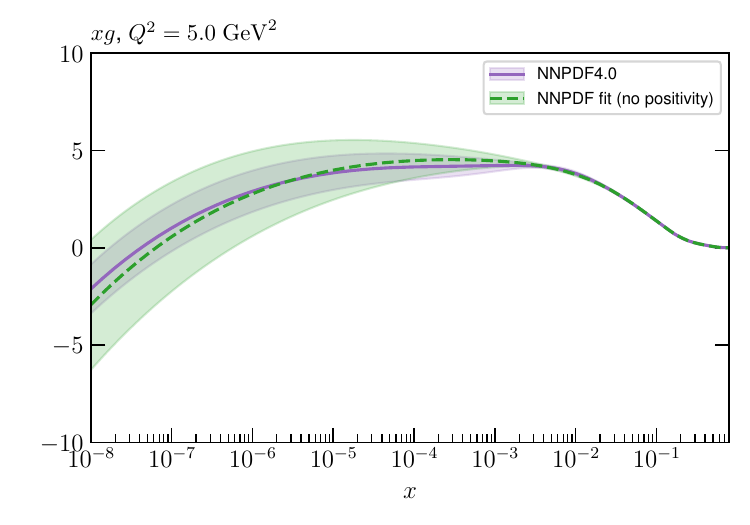}
\caption{\sf The gluon PDF at $Q^2=5 \, {\rm GeV^2}$ that results from a  global PDF fit to the NNPDF4.0 (fitted charm) dataset/theory setting, but using the MSHT20 parameterisation. Results with and without a positivity constraint applied, as described in the text, are shown. PDF uncertainties for the MSHT fits correspond to a fixed $T^2=10$ tolerance, and the NNPDF4.0 fit to the same dataset/theory settings is also shown. (Right) A comparison of the NNPDF4.0 public result to a fit within the NNPDF framework to the same dataset/theory settings, but without the positivity constraints imposed.}
\label{fig:fit_fch_gluon}
\end{center}
\end{figure}

We next return to the question of PDF positivity, in particular of the gluon at low $x$, and the observed differences between the perturbative and fitted charm cases.  First, to highlight the difference with respect to the perturbative charm case, we  show the gluon at  $Q^2=5$ ${\rm GeV}^2$ in Fig.~\ref{fig:fit_fch_gluon} for the NNPDF4.0 set as well as the MSHT fits. We can see that the NNPDF4.0 result now falls much more gently at low $x$ in comparison to the perturbative charm case, with only a mild tendency to be slightly negative at very low $x\sim 10^{-8}$ values. The central value of the MSHT fit without positivity imposed is on the other hand in fact increasing at low $x$, with no negativity occurring at all, although within the extremely large PDF uncertainties in this region that do allow for negative values. When positivity is imposed, the negative region of the PDF uncertainty band is excluded, and the result lies much closer to the NNPDF case. We recall however that the size of the PDF uncertainties in the Hessian approach when positivity is imposed cannot be interpreted too literally, given the lack of Gaussian behaviour around the $\chi^2$ minimum this results in. We also find that if a somewhat different parameterisation of the gluon is taken, as in the form of~\eqref{eq:cheb6} than the behaviour of the low $x$ gluon without positivity imposed lies rather closer to the NNPDF result, i.e. without a trend to increase at low $x$. It is equally true if we perform a fit within the NNPDF framework, but remove positivity, that the result lies close to the default result, as shown in the right plot. Therefore, in summary it is clear that there is no comparable trend for the low $x$ gluon to prefer to be negative at low scales that occurs in the perturbative charm fit, either with the NNPDF4.0 or the MSHT parameterisation. 

\begin{figure}
\begin{center}
\includegraphics[scale=0.6]{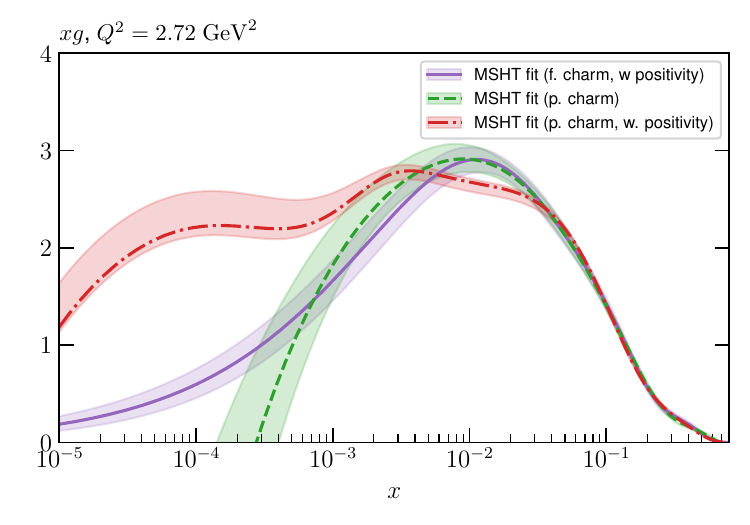}
\includegraphics[scale=0.6]{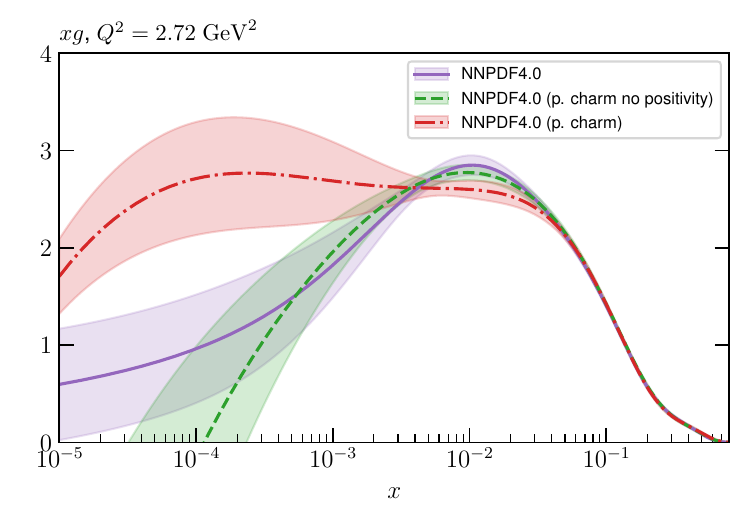}
\caption{\sf (Left) The gluon PDF at $Q=1.65 \, {\rm GeV}$ that results from a  global PDF fit to the NNPDF4.0 dataset/theory setting, with fitted and perturbative charm, and using the MSHT20 parameterisation.  MSHT PDF uncertainties correspond to a fixed $T^2=10$ tolerance. (Right) The gluon  PDF at $Q=1.65 \, {\rm GeV}$ for the  NNPDF4.0 fits, with fitted and perturbative charm.}
\label{fig:fit_gluon}
\end{center}
\end{figure}

To shed light on the question of why this trend is so different between the fitted and perturbative charm cases, in Fig.~\ref{fig:fit_gluon} we show the resulting gluon PDFs at low scale $Q=1.65 \, {\rm GeV}$ (i.e. the input scale for the fitted charm case) for the fitted charm case, with positivity imposed, as well as the perturbative charm case with and without positivity imposed. In the left plots the MSHT fits are shown, while in the right the NNPDF fits; we show both to demonstrate that the impacts on the gluon of these  choices is rather similar between the two fits. In the latter case the perturbative charm fit without positivity imposed comes from performing a fit directly with the NNPDF public code,  while the other two results simply correspond to the relevant public releases. For the perturbative charm fit, the negativity of the low $x$ and $Q^2$ gluon occurs in a region of $x$ that is largely outside the data region, but is  seen to allow for a somewhat larger gluon around $x\sim 10^{-2}$, due to the  small (at $Q^2=5$, per mille level) negative contribution to the momentum sum rule this provides at low scales. Both this effect, and more directly the impact of the reduced gluon at lower $x$ on the relevant DIS cross sections are observed to match the HERA DIS better, namely by providing a reduction in the NC cross section at up to $x\sim 10^{-4}$ and an increase at up to $x\sim 10^{-3}$.  Imposing positivity, on the other hand, leads to a lower gluon around  $x\sim 10^{-2}$, but also somewhat unexpected behaviour at lower $x$, with a larger gluon (which as we have seen in Fig.~\ref{fig:fit_pch_dynT_gl} becomes negative at low enough $x$) in comparison to the fitted charm case. In terms of the comparison between the MSHT and NNPDF fits, it is clear that the trends of the three fit results are rather similar, indicating that parameterisation is not playing a significant role in generating these differences.

\begin{figure}
\begin{center}
\includegraphics[scale=0.6]{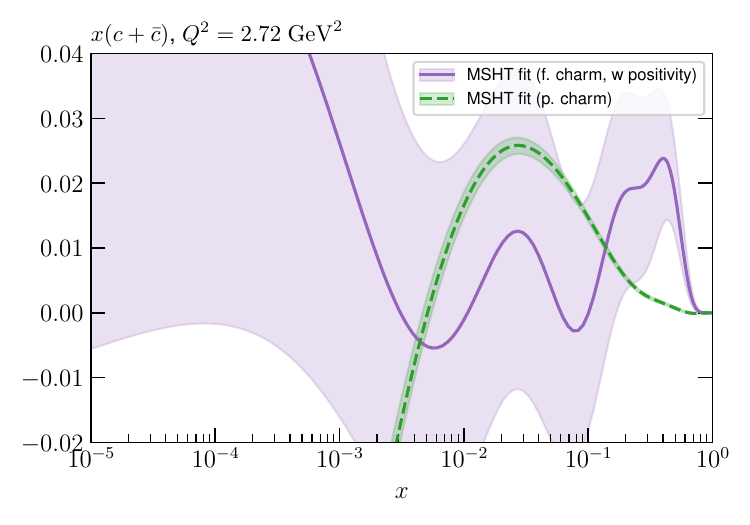}
\includegraphics[scale=0.6]{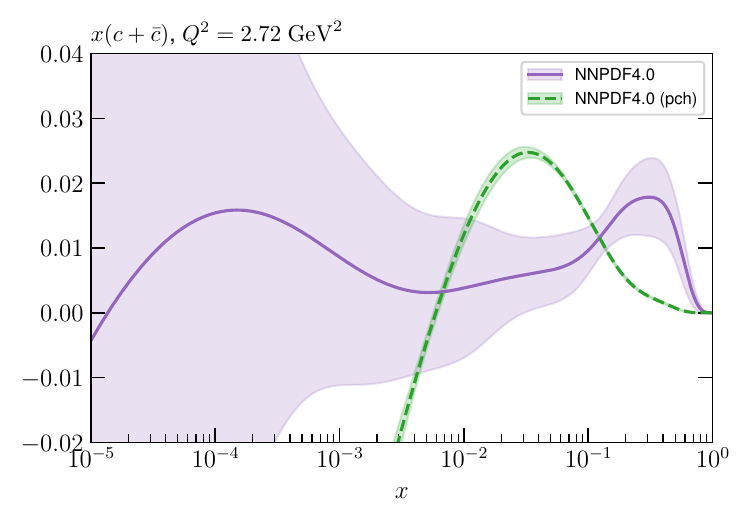}
\caption{\sf (Left) The charm quark PDF at $Q=1.65 \, {\rm GeV}$ that results from a  global PDF fit to the NNPDF4.0 dataset/theory setting, with fitted and perturbative charm, and using the MSHT20 parameterisation.  MSHT PDF uncertainties correspond to a fixed $T^2=10$ tolerance. (Right) The charm quark PDFs at $Q=1.65 \, {\rm GeV}$ for the  NNPDF4.0 fits, with fitted and perturbative charm.}
\label{fig:fit_charm}
\end{center}
\end{figure}

For the fitted charm result, a rather similar increase in the gluon  around  $x\sim 10^{-2}$ and at $Q=1.65 \, {\rm GeV}$ is observed with respect to the perturbative charm case with positivity imposed. However here, as seen by comparing Figs.~\ref{fig:fit_pch_dynT_gl} and~\ref{fig:fit_fch_gluon}, there is a much milder trend for the gluon to be negative at low $x$, although it it still decreasing in a manner that is distinct from the quark sector. Instead, it is now the freedom for the charm PDF to be fitted that allows the DIS data as well as other data sensitive to the quark flavour decomposition, most notably LHC DY measurements, to be fit well but with a reduction in the overall momentum fraction carried by the down--type $d$ and $s$ quarks/antiquarks. This can be seen in Appendix~\ref{app:pdf_pch_vs_fch}, Fig.~\ref{fig:fit_pch_vs_fch} where a comparison of the MSHT fitted and perturbative charm fits is shown. The strangeness distribution, in particular, is suppressed in the fitted charm case (as is also observed in the NNPDF fits~\cite{NNPDF:2021njg}). This results in an overall reduction in the momentum fraction carried by the quark sector (despite the fact that the charm quark momentum fraction is increased in the fitted charm case), and hence allows the gluon to be increased in this region. Indeed, also at rather higher $x\gtrsim 0.05$, in the fitted charm case the gluon momentum fraction is increased, which may explain why the LHC jet data are somewhat better described in this fit, see Appendix~\ref{app:fch_vs_pch} for a detailed set of comparisons, in both the MSHT and baseline NNPDF fits.

Thus the freedom allowed by fitting the charm PDF allows for a modification in the quark flavour decomposition with respect to the perturbative charm case that permits a somewhat larger momentum fraction carried by the gluon, as is preferred by the data in the fit. In Fig.~\ref{fig:fit_charm} we show the resulting charm  PDF at $Q=1.65$ GeV, for both the MSHT and baseline NNPDF fits. We also show for comparison the result of the perturbative charm fits, at the same scale. While in both cases there is a  peak towards the high $x\gtrsim 0.1$ region, it is clear that the modification of the charm quark with respect to the perturbative charm case is not limited to this region. The  trend for the fitted charm quark at low $x$ to be significantly larger than the (negative) perturbative charm case (which is a consequence of the to matching conditions) is driven by the fit to the HERA charm structure function data, which at low $x$ tends to lie above the result of the purely perturbative charm fits (both MSHT and NNPDF). A reduction in the fitted charm is also preferred in the $x\sim 0.01-0.1$ region (albeit within the $T^2=10$ MSHT uncertainties) which is again driven by the details of the fit to the HERA charm structure function data and also likely driven to some extent by the fit to LHC DY data. 
We note this reduction indicates that there will be a negative intrinsic, or more precisely fitted, charm contribution in this region\footnote{This provides a good but not exact indication as the perturbative charm distribution corresponds to the result of a different fit; to be precise we must evaluate the charm quark PDF in the 3--flavour scheme of the same set, as described in~\cite{Ball:2022qks}.}, as is observed in~\cite{Ball:2022qks}\footnote{We note that the negativity of the intrinsic charm component around  $x\sim 0.01-0.1$ becomes less statistically significant once MHOUs are included at  NNLO, but is found to be statistically significant at approximation ${\rm N}^3$LO in the most recent NNPDF analysis~\cite{NNPDF:2024nan}.}. Similarly, the enhancement at low $x$ will correspond to a significant positive intrinsic charm component, which is inconsistent with the valence--like models that are commonly used for this, see~\cite{Ball:2022qks,Brodsky:2015fna}. Note that this significant component of 
fitted charm at very low $x$, as well as the relative decrease compared to perturbative charm near $x=0.01$, could well be mimicking the effect of missing higher orders/small-$x$ resummation, which are not known very precisely for heavy flavours for $Q^2$ in the region of the heavy quark mass. Note, however, that a correction to the prediction for HERA charm structure function data can just as easily come from missing higher order corrections to the hard cross section as from the matching conditions for the charm PDF.
 
\begin{figure}
\begin{center}
\includegraphics[scale=0.6]{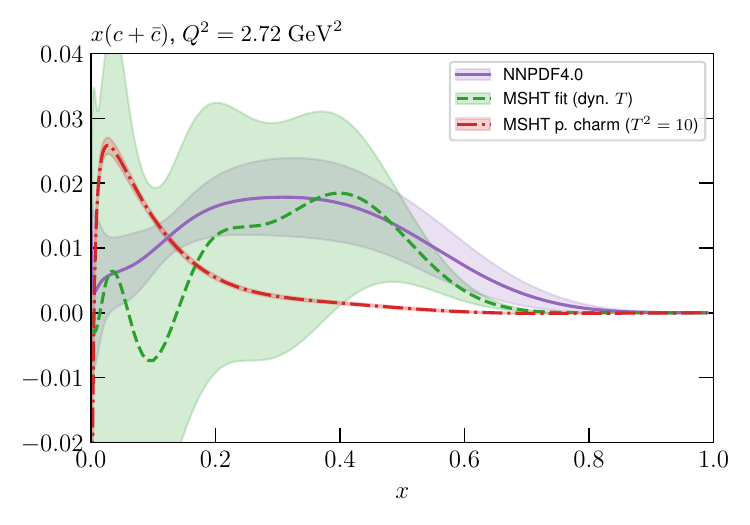}
\caption{\sf The charm quark PDF at $Q=1.65 \, {\rm GeV}$ that results from a  global PDF fit to the NNPDF4.0 dataset/theory setting, with fitted and perturbative charm, and using the MSHT20 parameterisation. The MSHT fit with fitted charm and the dynamic tolerance criterium applied for the PDF uncertainty is shown.}
\label{fig:fit_charm_highx}
\end{center}
\end{figure}
 
 In terms of the MSHT fit to the charm PDF, it is not the aim here to present a precise analysis of its implications for the intrinsic charm content of the proton. However, here we make a few observations. First, we can see in Fig.~\ref{fig:fit_charm} that the MSHT fit qualitatively matches the behaviour of the NNPDF results in terms of the difference between the perturbative and fitted charm. That is, at low $x$ the fitted charm exceeds the perturbative significantly (albeit increasing rather more steeply at low $x$ in comparison to the NNPDF result), at intermediate $x$ it lies below it, and at high $x$ there is a distinct peak structure. It is interesting to observe that in the intermediate $x$ region, the MSHT fitted charm qualitatively tracks the peaked shape of the perturbatively generated charm (within large uncertainties) in a manner that is not seen  in the NNPDF fit. 

 In Fig.~\ref{fig:fit_charm_highx} we focus on the high $x$ region, where we show the fitted charm quark PDF at $Q=1.65$ GeV, for  the NNPDF4.0 default result and the MSHT fit, with fitted and perturbative charm. The MSHT fitted charm result does not have positivity imposed and the dynamic tolerance criterion is applied. To give an estimate of the implications of this for the intrinsic charm content of the proton, we simply observe that at higher $x$ the perturbative charm PDF is very close to zero, and hence to good approximation we can simply identify the deviation from zero of the fitted charm PDF as the statistical significance for intrinsic charm. While the result without positivity imposed will correspond in some cases to $F_2^c$ cross sections at  $Q^2=5\,{\rm GeV^2}$ that are negative, in particular at the lower end of the uncertainty band, these will occur by construction in regions where the  intrinsic charm is negative. Hence we are free to use this result as a suitable evaluation of the statistical significance of positive intrinsic charm at high $x$, as dictated by the fit to the data alone (recalling also that no other PDFs exhibit significant negativity in this fit), bearing in mind that some of the lower region of the uncertainty band should in principle be excluded by the additional constraint of positivity. Indeed, arguably from the point of view of evaluating the statistical significance of any positive intrinsic charm this is the more appropriate approach to take; certainly if this is modified significantly by directly imposing positivity of $F_2^c$ at the level of the fit then this would indicate some caution is needed in the interpretation. With this in mind, it seems clear that the significance of intrinsic charm would be less in the MSHT fits, which given the observations above about the in general larger uncertainties implied by the MSHT tolerance, is not surprising (a similar conclusion is arrived at in~\cite{Guzzi:2022rca} for related reasons). On the other hand, if positivity is imposed as in Fig.~\ref{fig:fit_charm} this significance becomes somewhat larger, if still less than that of NNPDF, though here we recall that the PDF uncertainties in the Hessian approach should be treated with caution, given the breakdown in Gaussian behaviour that occurs around the minimum when positivity is imposed. 

\section{The Role of Parameterisation Flexibility}\label{sec:overfit}

It has been observed in the previous sections, see in particular Tables~\ref{tab:chi2_pcharm} and~\ref{tab:chi2_fcharm}, that the fits to exactly the same dataset and theory setting as in the NNPDF4.0 case, but applying the fixed MSHT20 parameterisation, in fact have a better global fit quality to the data entering the fit. In the perturbative charm fit, this is in part explained by the impact of positivity, while the impact of integrability constraints have been assessed in Appendix~\ref{app:integrability} and found to be small. Accounting for these, there remains a improvement of $\sim 90$ (40) point in $\chi^2$ in the perturbative (fitted) charm cases, i.e. $\sim 0.02$ (0.01) per point. More significantly, these result in PDFs that, as seen in Figs.~\ref{fig:fit_pch_rat} and~\ref{fig:fit_fitch_Teq1}, in many cases lie outside the quoted PDF uncertainty bands for the NNPDF fits. 

The remaining most obvious source of  difference in the fit methodologies then relates to the question of overfitting, namely that in the NN approach the data are divided into training and validation sets, and the $\chi^2$ for the validation sets (which are not fit to) is monitored in order to avoid fitting noise in the data with the very flexible input NN. Indeed, it is argued in~\cite{Ball:2022uon} that overfitting is the cause of the improved fit qualities observed in~\cite{Courtoy:2022ocu} when suitable linear combination of the NNPDF4.0 replicas are form via the so--called `hopscotch' scans, although this is not demonstrated with certainty.

In the current case the default MSHT parameterisation has 52 free parameters and so on the face of it appears unlikely that this would result in a significant degree of overfitting, but certainly this is a question that is worth exploring further. To examine this, in this section we present the result of a number of closure test and real fits with modified MSHT parameterisation containing both more and fewer Chebyshev polynomials, and examine the sensitivity of the results to these. In particular, we recall the basic form of the MSHT parameterisation:
\be\label{eq:cheb6_gen_rep}
x f(x,Q_0) = A x^{\delta} (1-x)^{\eta}\left( 1+ \sum_{i=1}^n a_{i}T_i(y(x))\right)\;,
\ee
for which we have $n=6$ by default for all distributions other than the gluon, for which a differing parameterisation is by default taken as in \eqref{eq:gluonpar}, and the $s_-$, which is more restricted. A straightforward check is then to repeat our analysis, but with $n$ set to less or greater than 6. In the cases of the gluon we will instead change the number of Chebyshev polynomials in the first term in \eqref{eq:gluonpar} (i.e. away from the default of 4), while we do not modify the  $s_-$, for simplicity.

\begin{figure}
\begin{center}
\includegraphics[scale=0.6]{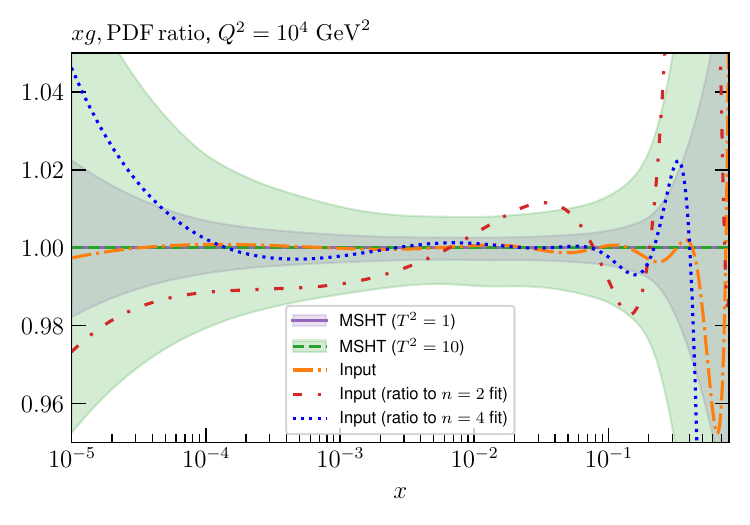}
\includegraphics[scale=0.6]{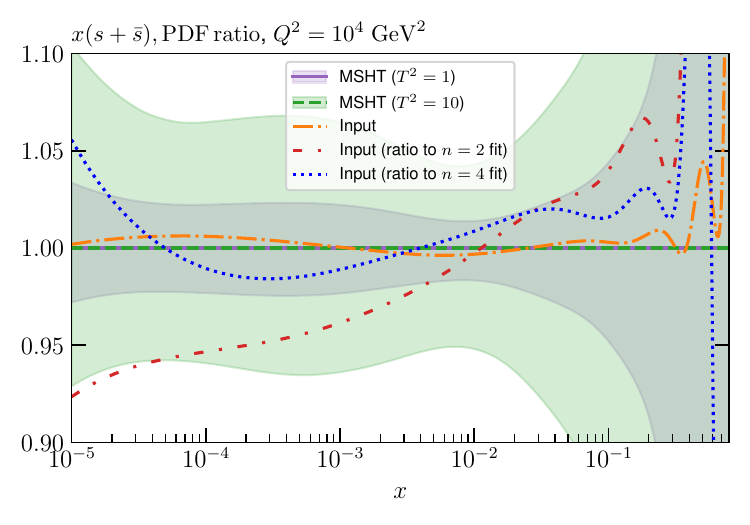}
\includegraphics[scale=0.6]{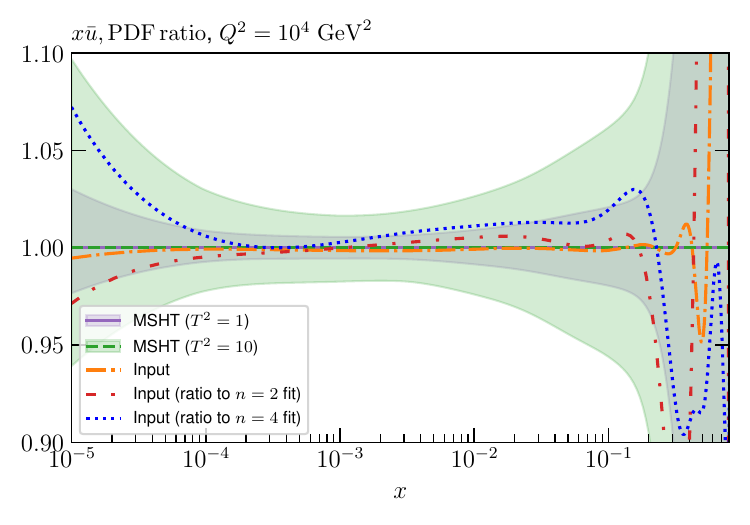}
\includegraphics[scale=0.6]{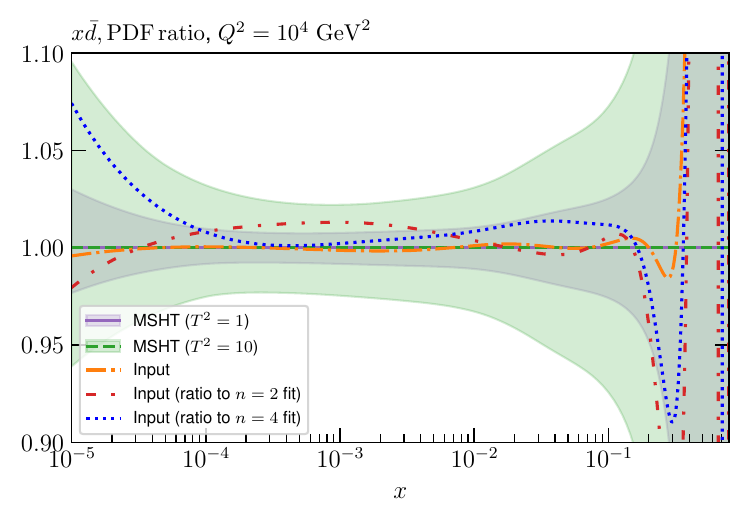}
\includegraphics[scale=0.6]{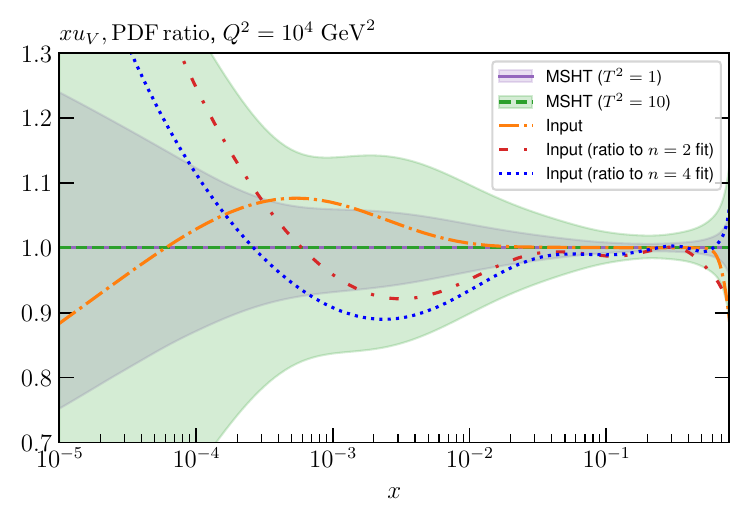}
\includegraphics[scale=0.6]{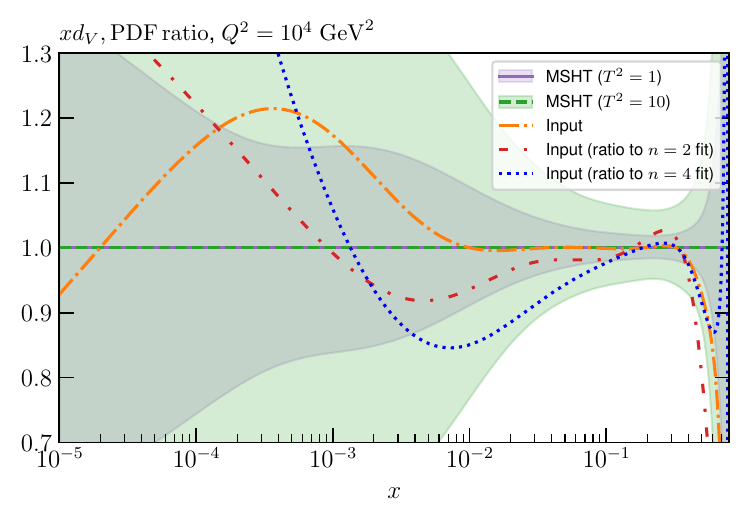}
\includegraphics[scale=0.6]{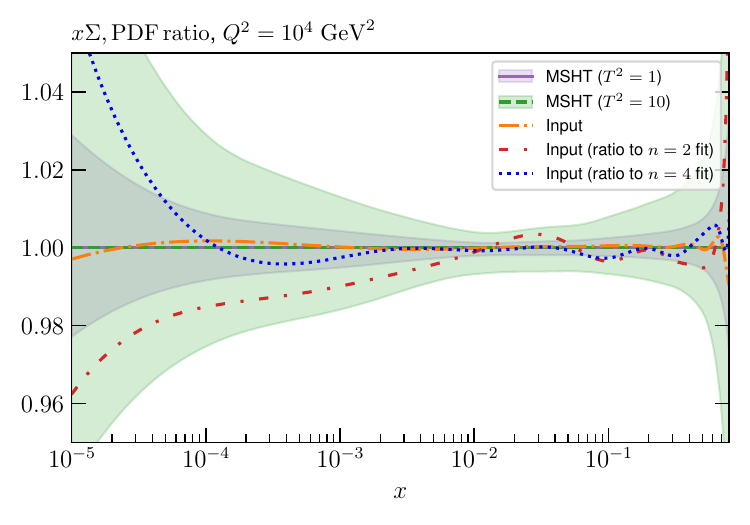}
\caption{\sf As in Fig.~\ref{fig:glcl_rat} but also showing the ratio of the NNPDF4.0 (perturbative charm) input to the fit result with the number of Chebyshev polynomials fixed to 2 and 4 (i.e. reduced by 4 and 2, respectively, for each PDF in comparison  to the default).}
\label{fig:glcl_cheb24}
\end{center}
\end{figure}

We start by repeating exactly the same unfluctuated closure test described in Section~\ref{sec:closureun}, but now restricting the MSHT parameterisation to have $n=2,4$ (i.e. for the gluon $n=0,2$ in the first term). While the default $n=6$ parameterisation has 52 free parameters in the current closure tests, this restricts the number of free parameters to be 28 40, respectively.  We note it is quite common in the literature for fits with similar number of free parameters to be performed: for example, the latest global CT18 fit~\cite{Hou:2019efy}  has 29 free parameters, i.e. very similar to the $n=2$ case here, though we note the parameterisation basis itself is different\footnote{Note also that the 29 parameters have been chosen as representative after an investigation of fit quality obtained from a range of about 250 fits with different bases, and a sometimes larger, i.e. 35-40, number of free parameters.}. In the case of the HERAPDF2.0~\cite{H1:2015ubc} set there are 14 free parameters, ABMP16~\cite{Alekhin:2017kpj} has 25, 
 while for the ATLASpdf21 set~\cite{ATLAS:2021vod} has 21, and so these are significantly fewer than even the $n=2$ case, although these are of course not all  global PDF fits. It is also common practice in a range of LHC experimental analyses to assess the impact of a given new dataset on PDFs via a parameterisation that is identical or very close to the HERAPDF2.0 case. 
 
In terms of the fit quality, we recall that we find $\chi^2=2.4$ for the 4627 points for the default $n=6$ case. For the  $n=2$ and 4 Chebyshev fits the results are as expected worse, with $\chi^2 = 103.8$ and 69.0, respectively, that is 0.022 and 0.015 per point, as opposed to 0.0005 in the default case. These deteriorations are relatively significant, and certainly larger than the difference in the number of free parameters, which is 12 (24) fewer in the $n=4$ (2) case. The results for the PDFs are given in Fig.~\ref{fig:glcl_cheb24}, which shows the same ratio of the NNPDF4.0 (pch) input to the default MSHT fit as in Fig.~\ref{fig:glcl_rat}, but in addition the ratio of the input set to the $n=2, 4$ Chebyshev fits. The $T^2=1$ and 10 uncertainties for the default fit are given, for comparison. We also consider the case of extending the MSHT parameterisation to have $n=7$ and 8, i.e. 64 and 76 free parameters, respectively. The fit quality is only observed to improve marginally, with  $\chi^2=2.3$ and 2.1, respectively, while the impact on the matching between the input and fits is similarly mild, and is not shown here for brevity. 

We can clearly see that the precision with which the $n=2,4$ fits match the input set is significantly worse than in the default case. For the $n=4$ fit, while there are regions of good agreement, e.g. for the gluon at intermediate $x$, in general the agreement between the input and the fit is only towards the edge of the $T^2=1$ uncertainty bands, and sometimes (notably at lower $x$) outside it. For the $n=2$ fit the agreement is as expected worse again, and in some cases even at the level or outside of the $T^2=10$ uncertainty bands, notably for the gluon and strangeness, but also for other distributions. It is therefore clear that these parameterisations do not have sufficient flexibility to match the input set at the level of precision corresponding to the $T^2=1$ PDF uncertainty. Given this uncertainty is representative of the overall experimental precision in current state--of--the art global PDF fits, we can therefore also conclude that these more limited parameterisations are insufficient to match this. The $n=5$ case is not shown for clarity, but we note that this in rather better agreement with the input than the $n=4$ case, as we would expect, but that the matching is still rather less good than the default $n=6$ fit in certain regions.

In this closure test setting the $T^2=1$ uncertainties are therefore not faithful representations of the PDF uncertainties for these restricted $n=2,4$ fits, in the sense that if a fluctuated closure test were performed we would not expect the fit to agree with the input within these uncertainties at 68\% confidence. These results then indicate that an enlarged tolerance would  certainly be required here, and by extension in a real PDF fit, due to parameterisation inflexibility alone, in a manner that is not apparent for the more flexible default MSHT parameterisation. For the $n=2$ parameterisation, it is even the case in some regions that the enlarged $T^2=10$ tolerance would not be sufficient, given the agreement between the input set and fit can be at the edge of or even beyond these uncertainties. 

As described above, in the CT18 analysis~\cite{Hou:2019efy},  in the default parameterisation there are 29 free parameters, in line with the $n=2$ case, albeit with a different underlying parameterisation. The above results therefore indicates that this number of free parameters would be insufficient to provide a faithful representation of a global fit of the sort performed by CT18. Indeed, this point is acknowledged in this analysis, where an enlarged uncertainty band is formed by performing multiple fits with differing forms of the parameterisation, and numbers of free parameters (in some cases more than 29) and ensuring that the nominal PDF uncertainty covers these. However, as the above results show, the $n=2$ MSHT fit would result in a significant loss of accuracy at the level of this closure test and hence we would expect at the level of a genuine fit. Enlarging the PDF uncertainties to account for this loss of accuracy may in the end ensure that the PDF uncertainty is representative, but only at the cost of an unnecessary loss of precision that could be avoid by taking a more flexible parameterisation. 

In~\cite{Hou:2019efy} it is argued that increasing the number of free parameters beyond $\sim 29$ would result in overfitting of statistical noise in the data, and destabilisation of the fits. We will discuss this further below, but note here that while in the context of the unfluctuated global fit overfitting is by construction less of an issue, the fact that the $n=2$ (and $n=4$) fits provide a rather poor description of the input set at the level of the $T^2=1$ uncertainties (which are due to the experimental errors on the data) would indicate that overfitting is unlikely to occur at this point, at least in the context of the MSHT parameterisation. That is, overfitting would only be expected to occur in the case of a parameterisation that has sufficient flexibility to provide a good description of the underlying data, and this is not seen to be the case for the $n=2,4$ fits. As further evidence for this point in the case of the default $n=6$ fit, we recall that we do also perform fluctuated closure tests in Section~\ref{sec:closure}, where overfitting could occur more readily, and find no evidence that the PDF uncertainties are not faithful, as would be expected to occur if there a significant amount of overfitting occurring to the fluctuated pseudodata replicas\footnote{We note that evidence for overfitting is found in~\cite{Kovarik:2019xvh}, see Fig.~4, when the number of free parameters extends beyond $\sim 30$. This is however to the CT14HERA2 dataset, for which the precision and extent of data entering the fit is more in line with the older MMHT14 fit~\cite{Harland-Lang:2014zoa}, where indeed fewer free PDF parameters, 37 rather than the 52 used in MSHT20, are required. In addition the fit is performed at NLO, which to some extent is expected to be less stable than a NNLO fit. A  parameterisation in terms of Chebyshev polynomials is also not used, which we recall is motivated in part by the avoidance of overfitting. There are therefore various potential reasons for this apparently different result, though a detailed analysis of this would be useful in future work.}.

\begin{table}
\begin{center}
  \scriptsize
  \centering
   \renewcommand{\arraystretch}{1.4}
\begin{tabular}{Xrccccccc}\hline 
$n=2$ & $n=3$&$n=4$&$n=5$&{\bf $\mathbf{n=6}$ (default)}&$n=7$&$n=8$
\\ \hline
+215.9 (0.047)&+147.5 (0.032) &+128.3 (0.028)&+71.1 (0.015)&{\bf  5736.7 (1.240)} &-4.3 (0.0009) &-7.2 (0.0016)\\
\hline
\end{tabular}
\end{center}
\caption{\sf Fit qualities for  a range of MSHT fits to the NNPDF dataset/theory settings, with perturbative charm. The value of $n$ indicates the  number of Chebyshev polynomials used in the parameterisation of each parton set, as described in the text, with $n=6$ being the default MSHT20 parameterisation, with 52 free parameters. The relative number of free parameters is then given by the change in $n$ with respect to this, multiplied by 6. The absolute global $\chi^2$ for the default case (which is the same as in Table~\ref{tab:chi2_pcharm}) is given, while the change in $\chi^2$ with respect to this is given in all other cases. All results correspond to the $t_0$ definition.}
\label{tab:chi2_cheb}
\end{table}

We next turn to the impact of modifying the number of free parameters on the full global fit, starting with the perturbative charm case. In Table~\ref{tab:chi2_cheb} we show the impact on the fit quality of reducing the number of free parameters, now with $n=2$ to 5 (i.e. 28 to 46 free parameters), and also of increasing them, with $n=7,8$ (i.e. 58 to 64 free parameters). A clear trend for an increase in the $\chi^2$ as we decrease $n$ is observed, as we would expect, and indeed the deterioration is  larger than in the closure test: the $\chi^2$ increases by a factor of $\sim 2$ more for the $n=2,4$ cases considered there. This is perfectly possible, as the underlying best fit distribution will be different in the baseline $n=6$ case, and bearing in mind the fact that the pseudodata in the closure test case are not fluctuated. Moreover, the underlying nature of the real data in the fit, with the known departures from statistical consistency between data and theory, and between datasets may be more sensitive to parameterisation inflexibility. 

On the other hand, when the number of free parameters is increased we actually observe a relatively limited improvement in the fit quality. Some caution may be needed in the interpretation of these $n=7,8$ results, however as given the increasing number of free parameters the possibility that e.g. a saddle point, rather than the true global minimum of the $\chi^2$, has been found in the minimisation becomes more likely. Therefore it is possible that some further improvement in the fit quality may be achievable here, although clearly no trend for any significant further improvement is seen.

We recall that in the MSHT20 analysis~\cite{Bailey:2020ooq}, the impact of increasing the number of free parameters from 36 (corresponding to the MMHT14 parameterisation\footnote{In~\cite{Harland-Lang:2014zoa} this is described as being 37, but in this definition the low $x$ power of the strangeness asymmetry is considered to be free, when it is in fact fixed to a sensible value, as is done here. Accordingly, labelling this as a fixed parameter the number of free parameters is 36.})  to 52, i.e. by 16 was found to improve the fit quality by $\sim 73$ points. Given the 36 free parameter case lies in terms of the number of free parameters between the  $n=3$ and 4 fits in Table~\ref{tab:chi2_cheb}, this is a somewhat milder, but qualitatively similar change; given the underlying fits are rather different, this is perfectly consistent. 

\begin{figure}
\begin{center}
\includegraphics[scale=0.6]{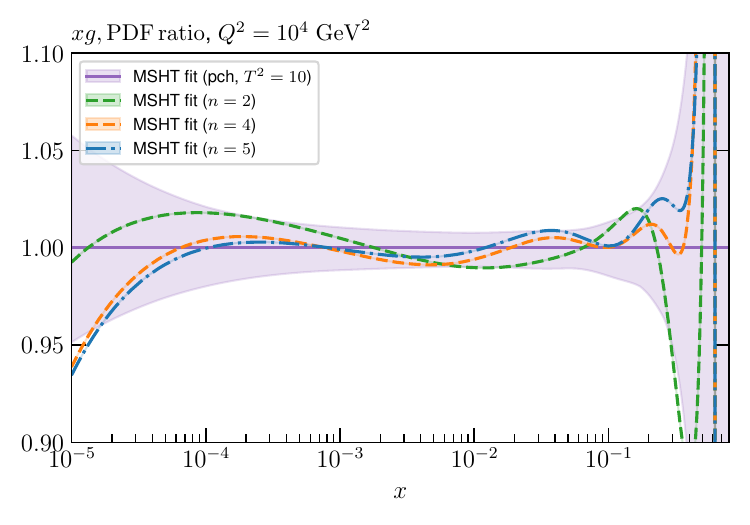}
\includegraphics[scale=0.6]{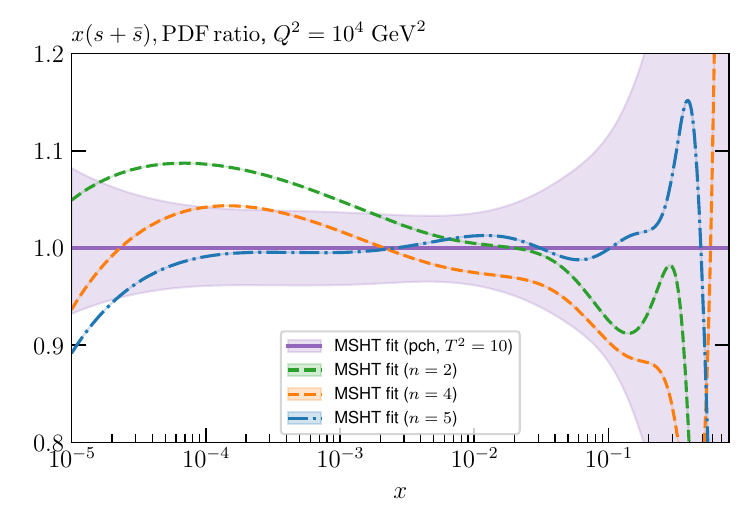}
\includegraphics[scale=0.6]{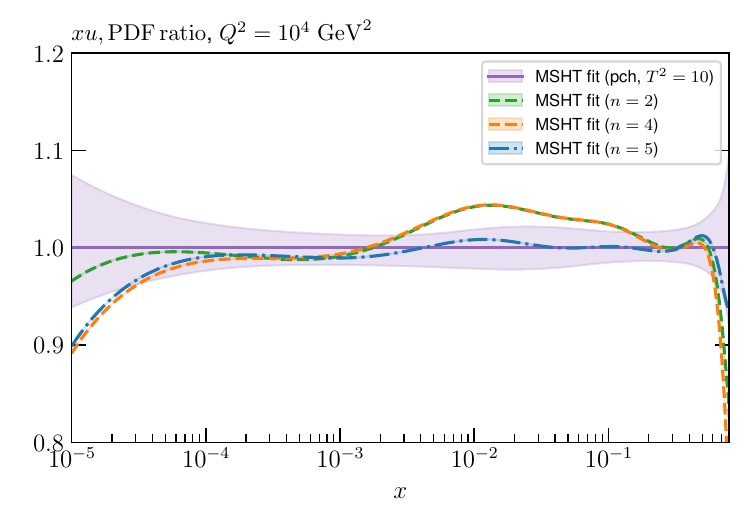}
\includegraphics[scale=0.6]{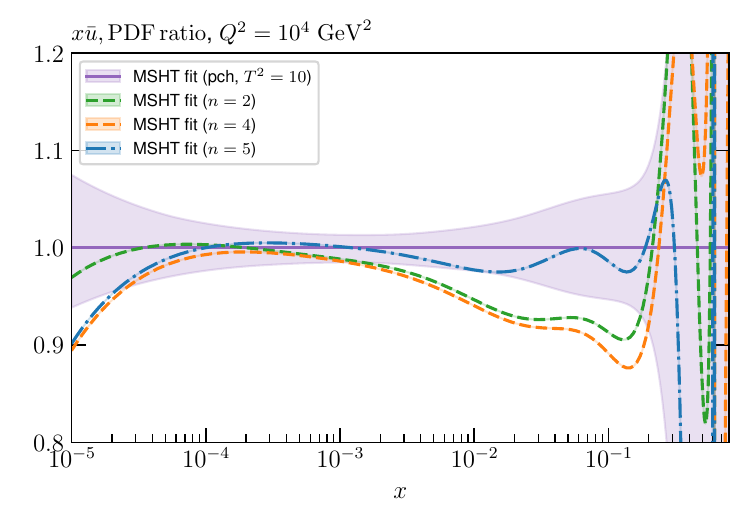}
\includegraphics[scale=0.6]{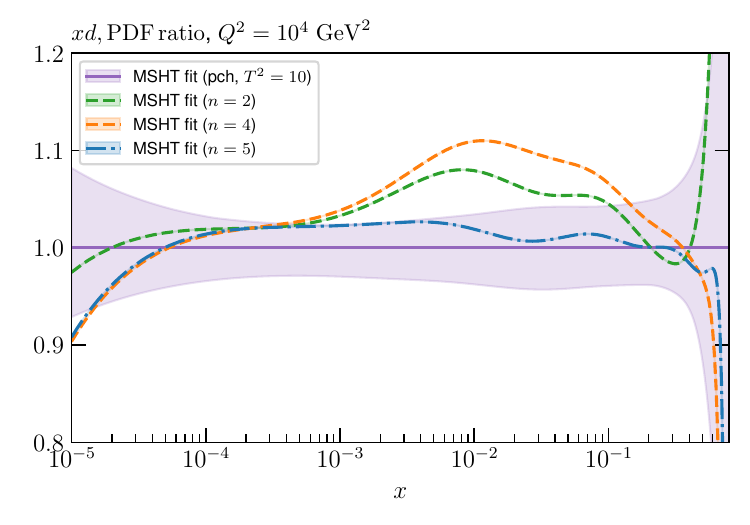}
\includegraphics[scale=0.6]{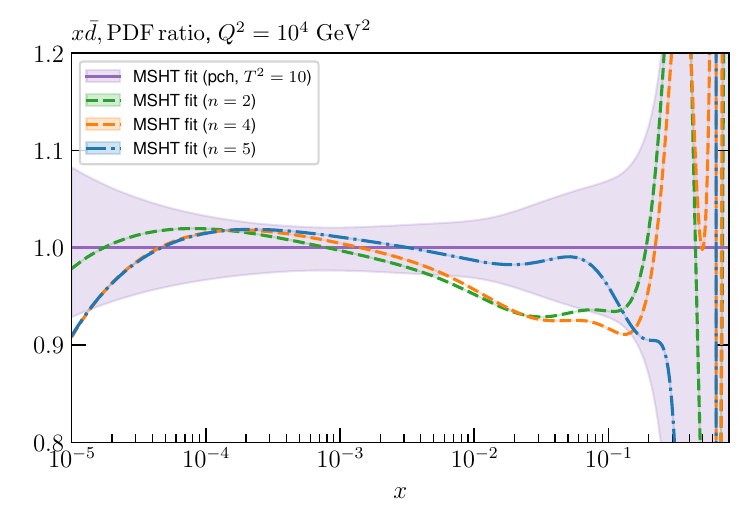}
\includegraphics[scale=0.6]{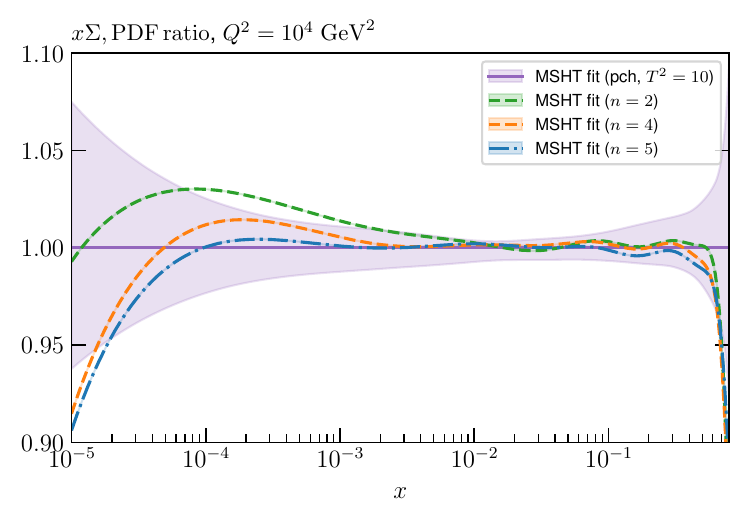}
\caption{\sf A selection of PDFs at $Q^2=10^4 \, {\rm GeV^2}$  that result from a  global PDF fit to the NNPDF4.0 dataset/theory (perturbative charm) setting, but using the MSHT20 parameterisation. The default result, and fits with the number of Chebyshev polynomials fixed to 2, 4 and 5 (i.e. decreased by 4, 2 and 1, respectively, for each PDF in comparison to the default)  are shown. Results are shown as a ratio to the default fit, with $T^=2=10$ uncertainties indicated.}
\label{fig:fit_pch_cheb245}
\end{center}
\end{figure}

\begin{figure}
\begin{center}
\includegraphics[scale=0.6]{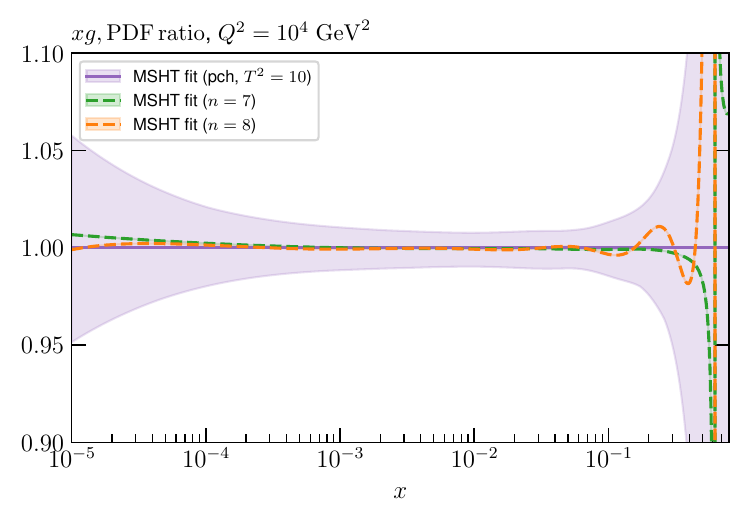}
\includegraphics[scale=0.6]{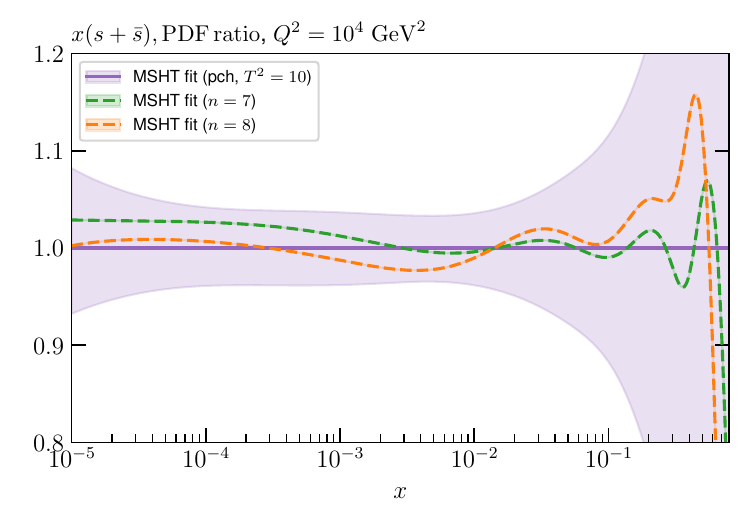}
\includegraphics[scale=0.6]{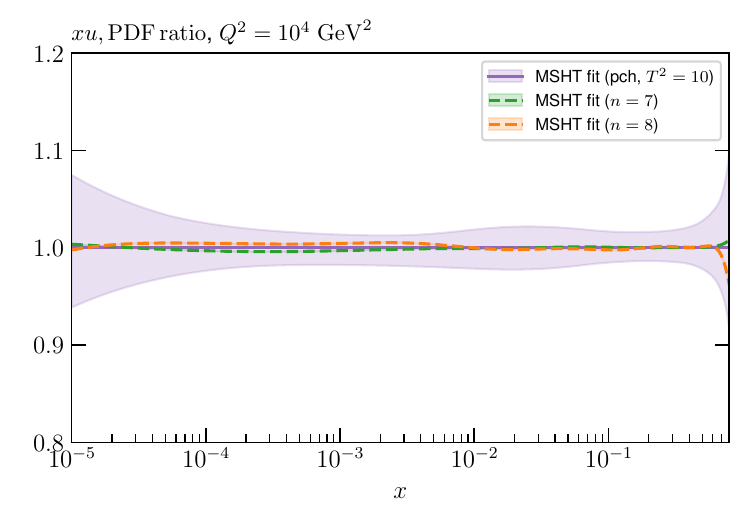}
\includegraphics[scale=0.6]{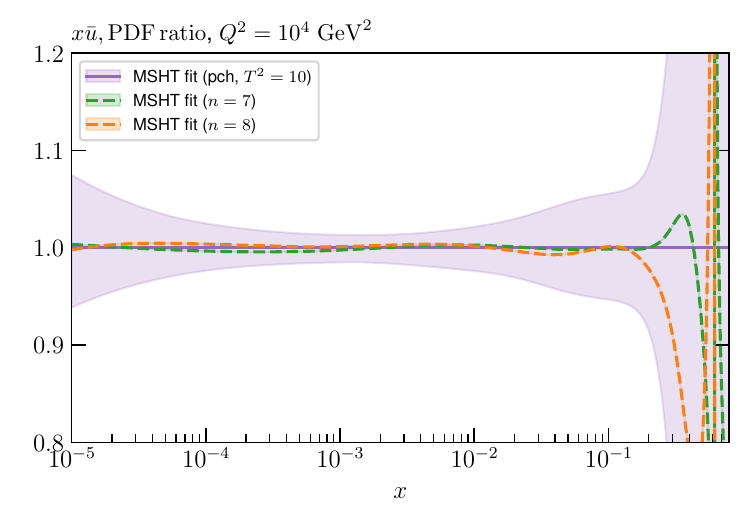}
\includegraphics[scale=0.6]{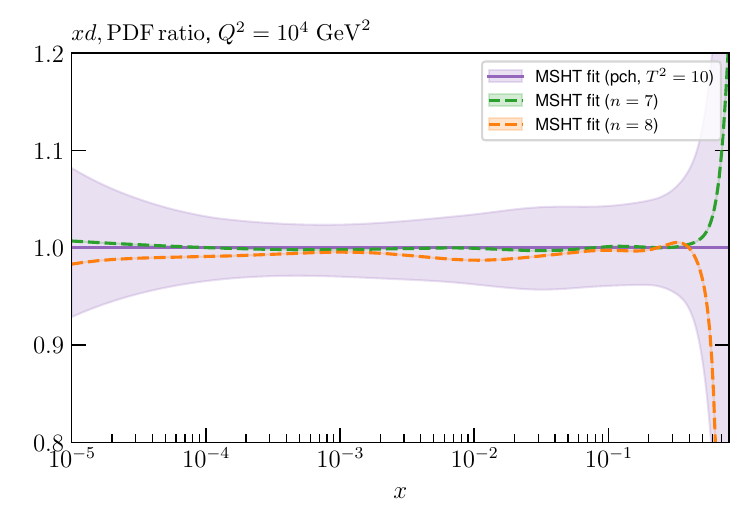}
\includegraphics[scale=0.6]{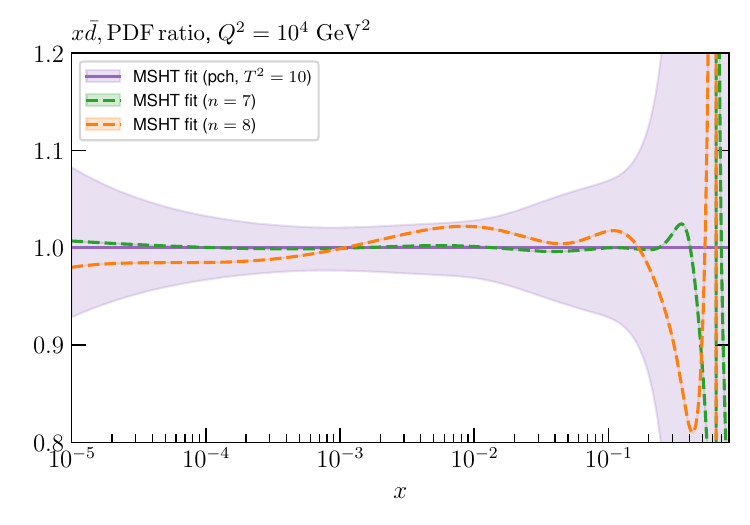}
\includegraphics[scale=0.6]{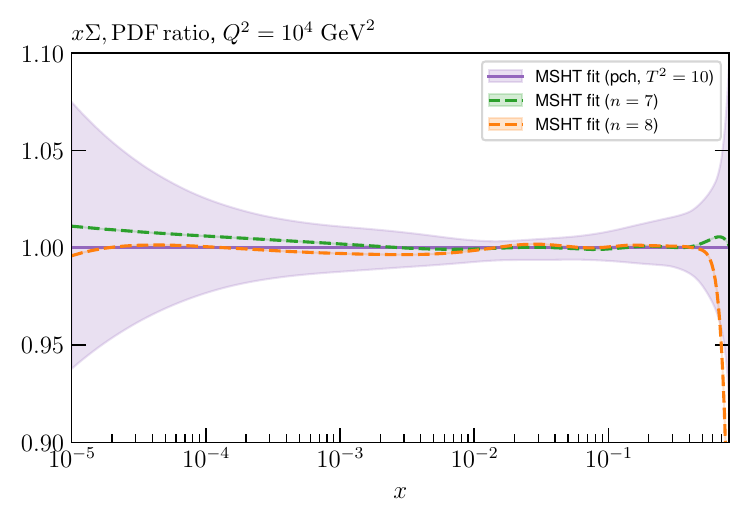}
\caption{\sf A selection of PDFs at $Q^2=10^4 \, {\rm GeV^2}$  that result from a  global PDF fit to the NNPDF4.0 dataset/theory (perturbative charm) setting, but using the MSHT20 parameterisation. The default result, and fits with the number of Chebyshev polynomials fixed to 7 and 8 (i.e. increased by 1, and 2, respectively, for each PDF in comparison to the default)  are shown. Results are shown as a ratio to the default fit, with $T^=2=10$ uncertainties indicated.}
\label{fig:fit_pch_cheb78}
\end{center}
\end{figure}

The impact on the PDFs are shown in Figs.~\ref{fig:fit_pch_cheb245} and~\ref{fig:fit_pch_cheb78}, which give the ratio of the $n=2,4,5$ and $n=7,8$ fits, respectively, to the baseline at $Q^2=10^4$ ${\rm GeV}^2$ (with $T^2=10$ uncertainties given) for a range of PDFs. Only the central values are given for the fits with $n\neq 6$, but we note that there is a moderate but non--negligible trend observed for a reduction in PDF uncertainty with reducing  $n$, such that in the $n=2$ case these are in many regions a factor of $\sim 1.5-2$ smaller. This point will be relevant when considering other public PDF fits with more restricted parameterisations.

Starting with Fig.~\ref{fig:fit_pch_cheb245}, we can see a clear trend for the $n=2$ fit but also the $n=4$ to disagree with the baseline result at a level that is comparable to and often even larger than the quoted $T^2=10$ uncertainties.  Indeed, this level of disagreement is greater than that observed in the case of a closure test as shown in Fig.~\ref{fig:glcl_cheb24} for the $n=2$ fit, consistent with the larger deterioration in fit quality observed. This difference is somewhat larger in the case of the quark flavour decomposition, though is also non--negligible for the gluon and quark singlet in the $n=2$ case in particular. It is therefore clear that these fits are not compatible with the baseline, even within the $T^2=10$ uncertainties of this fit, and this strongly suggests that performing fits with such a restricted number of parameters will lead to unrepresentative results. 

For the $n=5$ case the agreement with the baseline result is significantly better, although far from perfect in some regions. Therefore we can see that there is some trend towards stability as we increase $n$, with the $n=5$ case providing a better but imperfect representation of the baseline. Related to this, and turning now to  Fig.~\ref{fig:fit_pch_cheb78}, we can see that the impact of increasing $n$ further to 7 and 8 is in general very limited. There are some PDFs where the change as $n$ is increased is more visible, most notably for the strangeness and $\overline{d}$, although these changes are always within the $T^2=10$ uncertainties. Broadly then we can see an encouraging degree of stability, with limited changes in the PDFs as $n$ is increased above the baseline value of 6, but on the other hand a more significant effect if $n$ is reduced. The choice of $n=6$ for the MSHT20 fit~\cite{Bailey:2020ooq} was motivated by the original study of~\cite{Martin:2012da} and in particular the observation that this choice allows a fit with sub--percent level precision, it being expected that this would be required by the increasingly high precision LHC data now entering the fit. Our results here clearly support this expectation quantitatively.

\begin{figure}
\begin{center}
\includegraphics[scale=0.6]{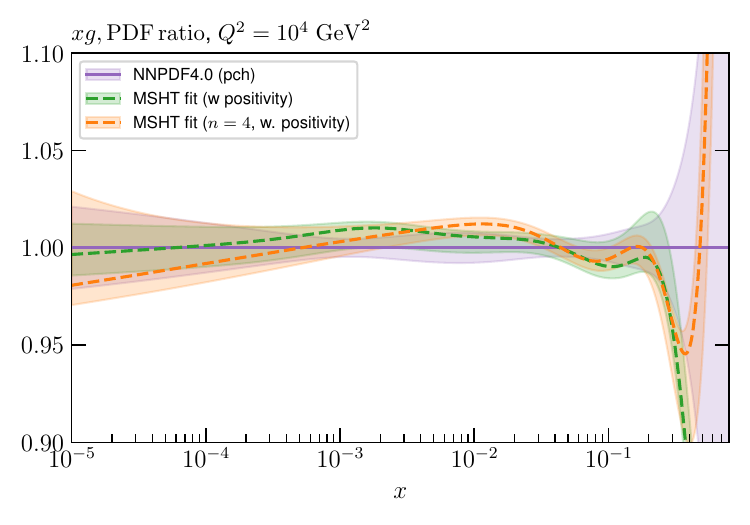}
\includegraphics[scale=0.6]{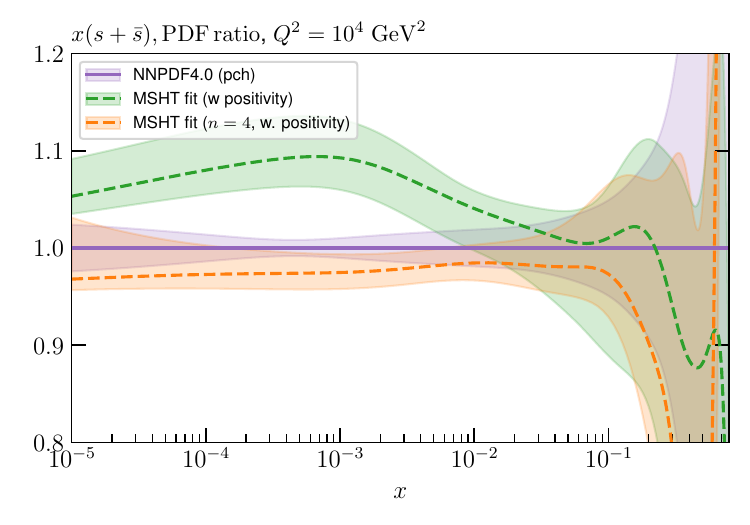}
\includegraphics[scale=0.6]{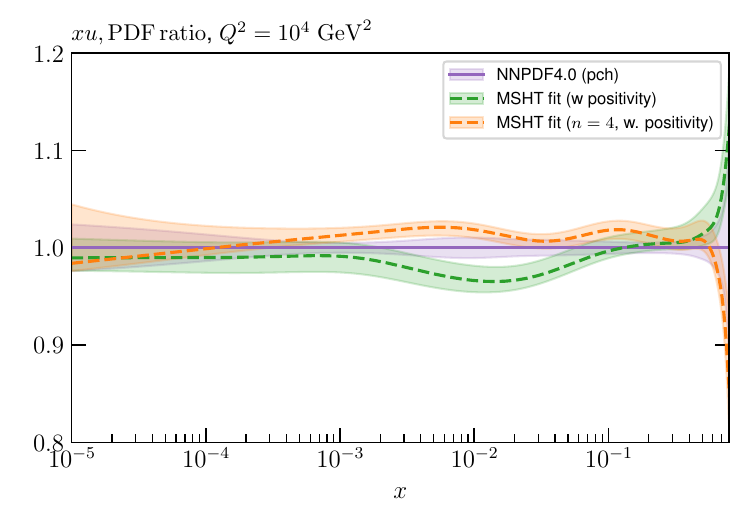}
\includegraphics[scale=0.6]{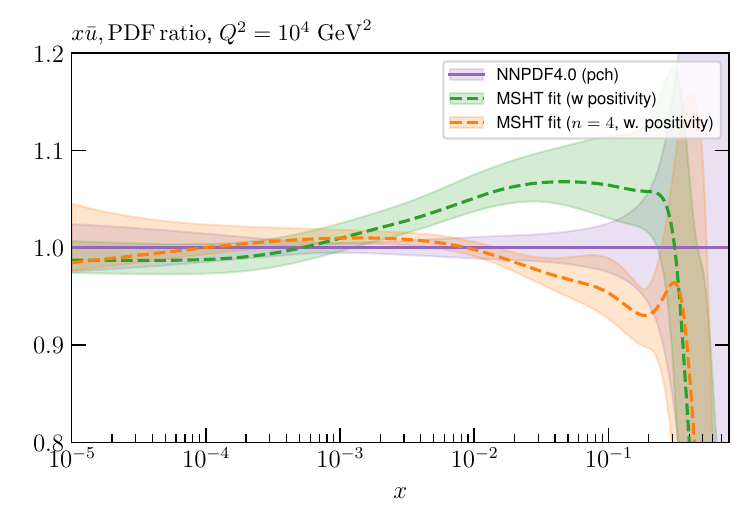}
\includegraphics[scale=0.6]{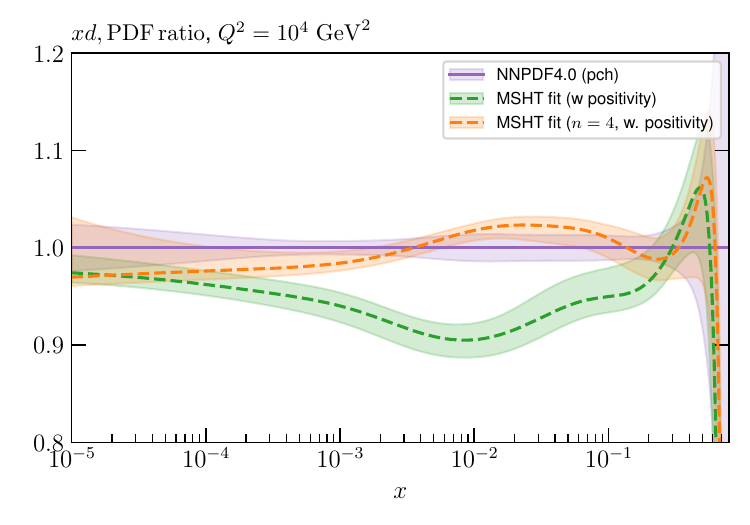}
\includegraphics[scale=0.6]{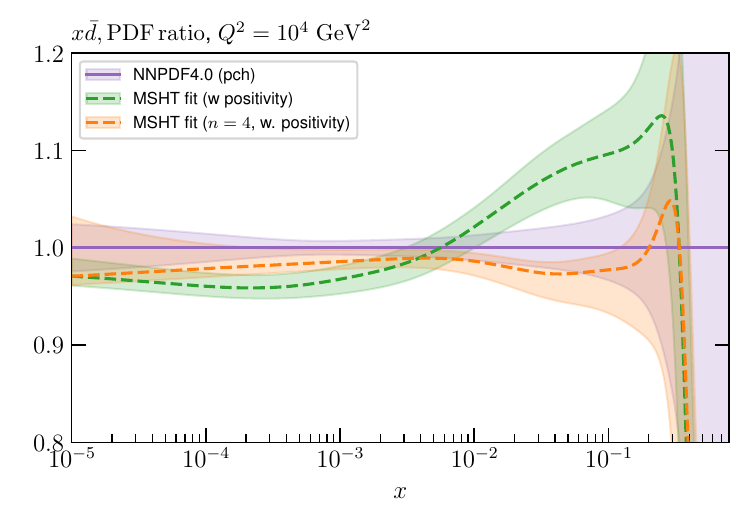}
\includegraphics[scale=0.6]{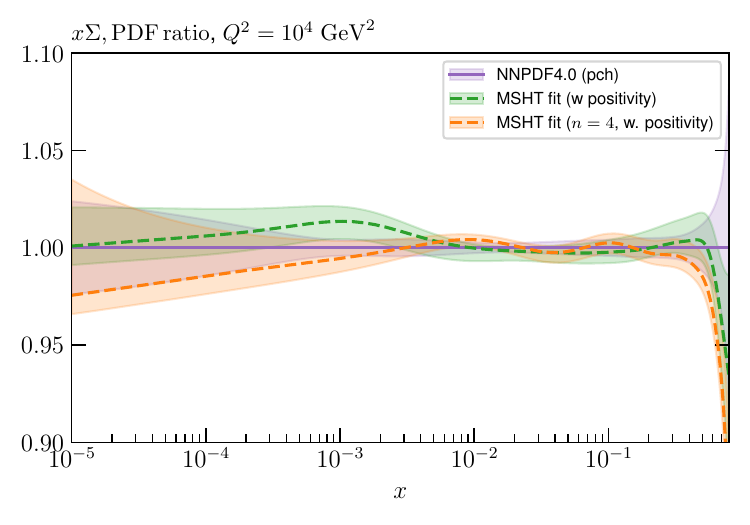}
\caption{\sf A selection of PDFs at $Q^2=10^4 \, {\rm GeV^2}$  that result from a  global PDF fit to the NNPDF4.0 dataset/theory (perturbative charm) setting, but using the MSHT20 parameterisation. The default result, without positivity applied, and a fit with the number of Chebyshev polynomials fixed to 4 (or more generally reduced by 2 for each PDF in comparison  to the default), and with and without a positivity constraint applied, are shown. PDF uncertainties for the MSHT fits correspond to $T^2=1$. Results are shown as a ratio to the NNPDF4.0  fit to the same dataset/theory settings.}
\label{fig:fit_pch_cheb4}
\end{center}
\end{figure}

\begin{figure}
\begin{center}
\includegraphics[scale=0.6]{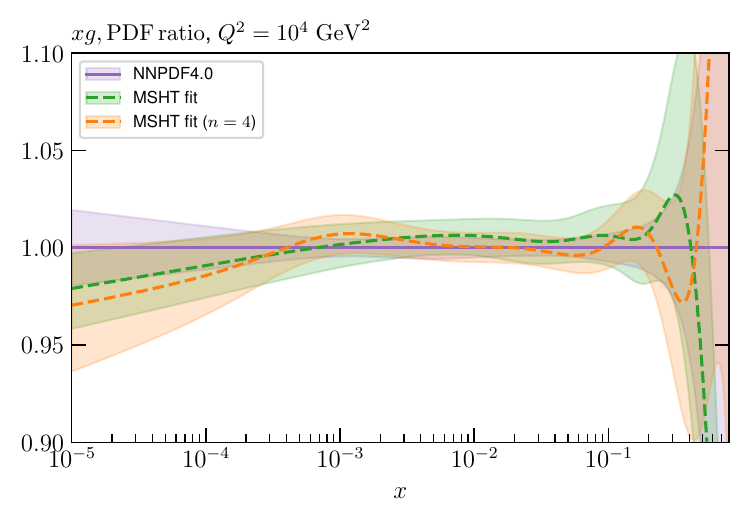}
\includegraphics[scale=0.6]{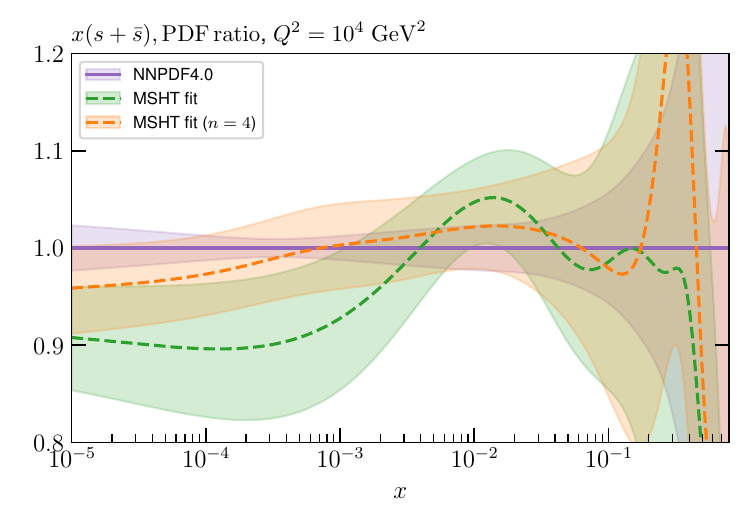}
\includegraphics[scale=0.6]{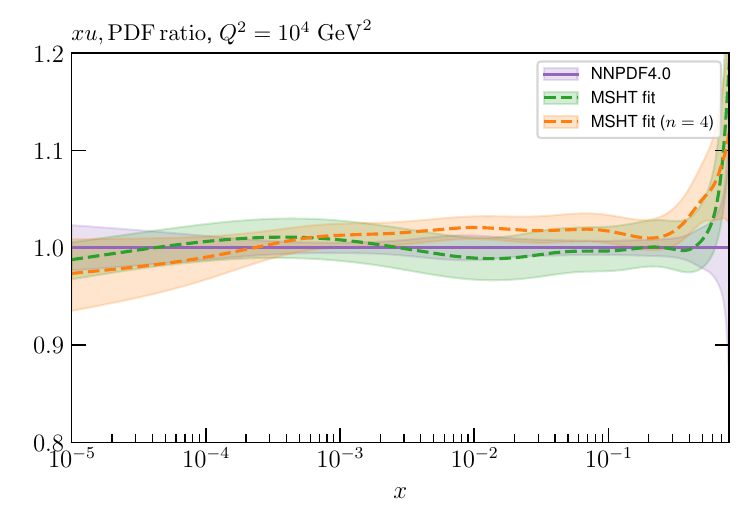}
\includegraphics[scale=0.6]{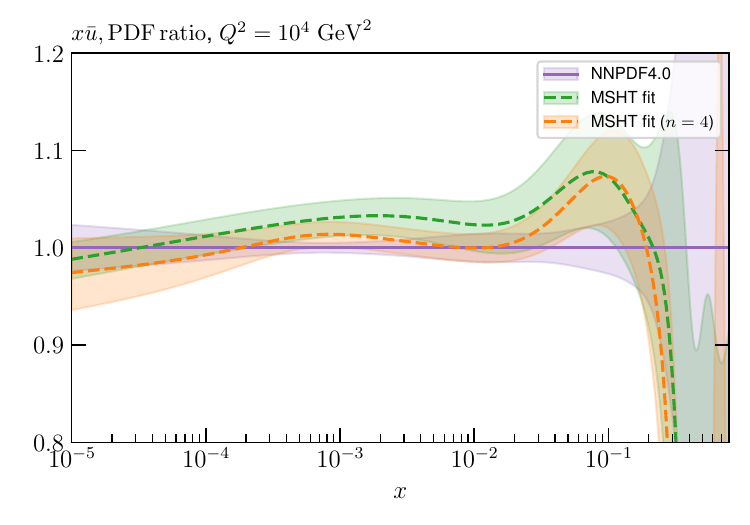}
\includegraphics[scale=0.6]{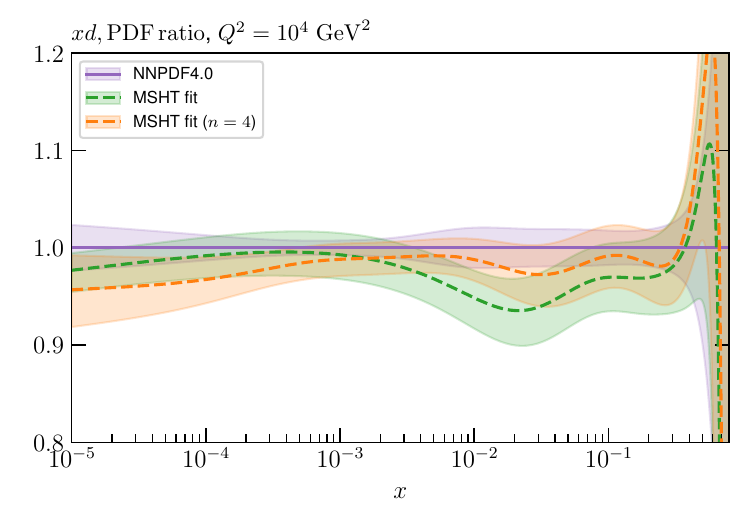}
\includegraphics[scale=0.6]{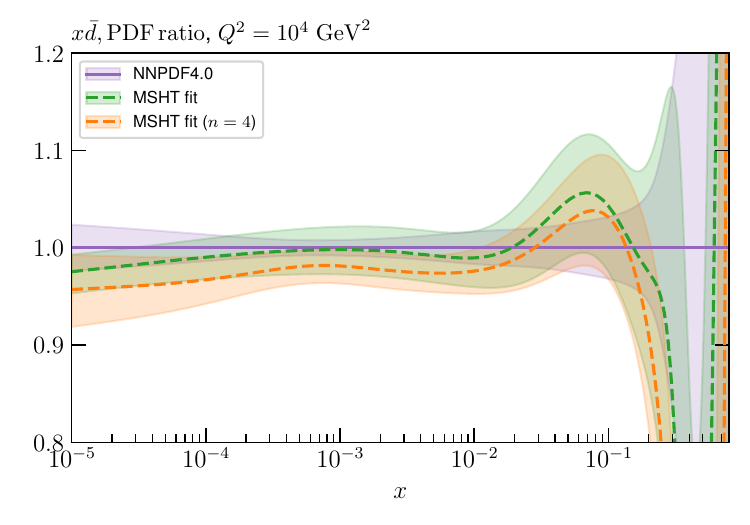}
\includegraphics[scale=0.6]{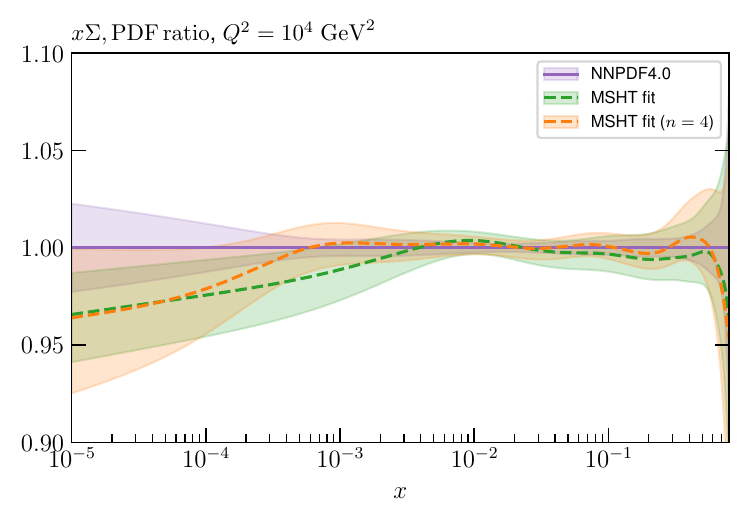}
\includegraphics[scale=0.6]{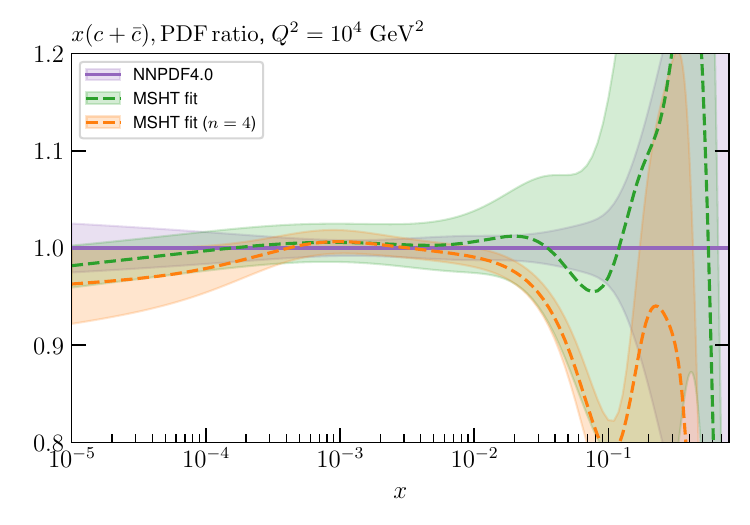}
\caption{\sf A selection of PDFs at $Q^2=10^4 \, {\rm GeV^2}$  that result from a  global PDF fit to the NNPDF4.0 dataset/theory (fitted charm) setting, but using the MSHT20 parameterisation. The default result, and a fit with the number of Chebyshev polynomials fixed to 4 (or more generally reduced by 2 for each PDF in comparison to the default), in both cases with a positivity constraint applied, are shown. PDF uncertainties for the MSHT fits correspond to $T^2=10$. Results are shown as a ratio to the NNPDF4.0  fit to the same dataset/theory settings.}
\label{fig:fit_fch_cheb4}
\end{center}
\end{figure}

Having discussed the overall trend with changing $n$, it is interesting to consider these results in light of the original comparison in Section~\ref{sec:pcharm} to the NNPDF4.0 fit. From Table~\ref{tab:chi2_pcharm} we can see that once positivity is imposed the fit quality remains $\sim 90$ points better for the baseline MSHT fit, while we can see in Table~\ref{tab:chi2_cheb} that a similar level of difference is seen between the MSHT fits with $n=4,5$ and the baseline case. It is therefore interesting to compare directly the resulting PDFs to the NNPDF4.0 set. This is shown in Fig.~\ref{fig:fit_pch_cheb4} where we compare the $n=4$ fit, with positivity now imposed in order to be consistent with the NNPDF4.0 fit, to the NNPDF4.0 (perturbative charm) result, as well as the baseline MSHT fit, again with positivity imposed. Rather strikingly, we can see that the MSHT fit with $n=4$ is in significantly better agreement with the NNPDF4.0 fit. This result, along with the comparison of the fit qualities, is therefore suggestive that the NNPDF case may be characterised by a lower degree of variation or flexibility than the baseline MSHT set, and be more in line with the $n=4$ case; we will return to this issue later in the section. In terms of the fit quality, once positivity is imposed the $n=4$ result is actually $\sim 100$ worse than the NNPDF result, but this deterioration is arguably indicative of the difficulty the rather less flexible parameterisation has in satisfying such stringent positivity requirement on the PDFs at rather high $x$, in this case the $\overline{d}$.

We do not present a detailed analysis of the fitted charm case here for brevity, but in Fig.~\ref{fig:fit_fch_cheb4} we  show the same comparison as in Fig.~\ref{fig:fit_pch_cheb4} but for the fitted charm case, and in this cases without positivity imposed. The impact on the fit quality is to give a deterioration of $\sim 54$ points (i.e. very closely in line with the NNPDF4.0 fit quality), which is rather less than is observed in the perturbative charm fit, see Table~\ref{tab:chi2_cheb}. This is most likely in part driven by the fact that the number of free parameters in the baseline fitted charm case is higher, due to the treatment of the charm, with 61 rather than 52 free parameters. Hence the $n=4$ case has 49 free parameters in total, which may be still be sufficient to provide a reasonable description of the data. This effect is also observed in the PDFs, for which the change is rather more moderate. There is arguable some tendency though for the $\overline{u}$, $d$ and $s$ ($\overline{s}$) quarks to lie closer to the NNPDF4.0 case, but this is less clear for the $u$ and $\overline{d}$. A rather lower charm quark at high $x$ appears also to be preferred, though here the role of $F_2^c$ charm positivity is most significant.

We note that, for the reasons described above we do not consider increasing the value of $n$ to 7 and/or 8, as was investigated in the perturbative charm case, as at that this point the overall stability of the fit may become unclear. Similarly, we in fact find that for the reduced $n=2$ fit the charm parameterisation is not sufficiently flexible to give a suitable description of the fitted charm, with the resulting fit becoming rather unstable in order to compensate for this. We therefore do not show any results in this case either. 

We also note that in~\cite{Yan:2024yir} two measures of overfitting are discussed, namely the Akaike information criterion (AIC)~\cite{Akaike:1974vps} and  Bayesian information criterion (BIC)~\cite{Schwarz:1978tpv}, whereby the improvement in fit quality with an increase in the number of free parameters, $N_{\rm par}$, for a given number of datapoints, $N_{\rm pt}$, is judged in comparison to the increase in $N_{\rm par} \ln N_{\rm pt}$ and $2 N_{\rm par}$, respectively. In the context of the unfluctuated closure tests, the improvement in the $\chi^2$ also corresponds to an improvement in the BIC, for which the addition of $2 N_{\rm par}$ corresponds to 48 and 24 in the $n=2,4$ cases, while the $\chi^2$ improves by 103.8 and 69, respectively. On the other hand, for the BIC we find that $N_{\rm par} \ln N_{\rm pt}$ (which gives $\sim 200$ and 100, for the $n=2,4$ cases, respectively) is larger than these improvements, and so according to that criterion using the $n=6$ parameterisation would not correspond to a relevant improvement in the fit. However, looking at Fig.~\ref{fig:glcl_cheb24} it is clear that at the level of the PDFs the $n=2,4$ fits are significantly worse at representing the input set, which calls into question to use of the BIC in this context. In the full fit, as indicated in Table~\ref{tab:chi2_cheb}, the improvement in the AIC is again evident in going from $n=2,4$ to $n=6$, while for the BIC it is marginal. For the $n=7,8$ case there is clearly no improvement in either measure.

Finally, it is useful to consider in a little more detail the form of the fit PDFs themselves, given the evidence we have found above that the MSHT baseline fits may actually be representative of a higher degree of flexibility than the NNPDF4.0 central sets. To this end, we consider the so--called PDF kinetic energy discussed in~\cite{Ball:2022uon}, defined as
\be
{\rm KE} = \sqrt{1+\left(\frac{{\rm d}}{{\rm d}\ln x} xf(x,Q^2)\right)^2}\;,
\ee
such that this variable integrated over $\ln x$ gives the arc--length of the curve of the PDF as a function of $\ln x$. This is then in essence is a measure of the amount of variation in a curve, given for any two fixed endpoints in $x$ a PDF with a larger value of the arc--length will by definition vary more between the points. Of course for any given comparison between two PDF sets the values at fixed endpoints in $x$ are not the same, and hence this interpretation requires some care. Indeed in~\cite{Ball:2022uon} the kinetic energy is discussed in terms of the context of the `wiggliness' of the PDFs, with a larger local (or average) value of the KE corresponding to more of this property, however as we will see even a relatively smooth change in the PDF such that it increases or decreases in a certain region can result in a larger KE with respect to a PDF set that exhibits less change. This point should be kept in mind in the comparison which follow.

\begin{figure}
\begin{center}
\includegraphics[scale=0.6]{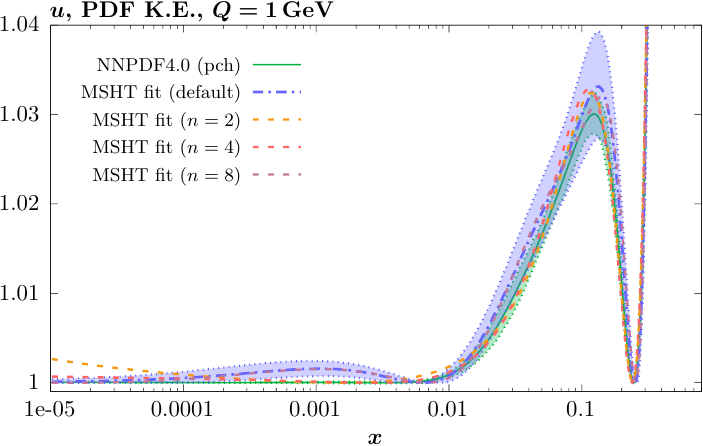}
\includegraphics[scale=0.6]{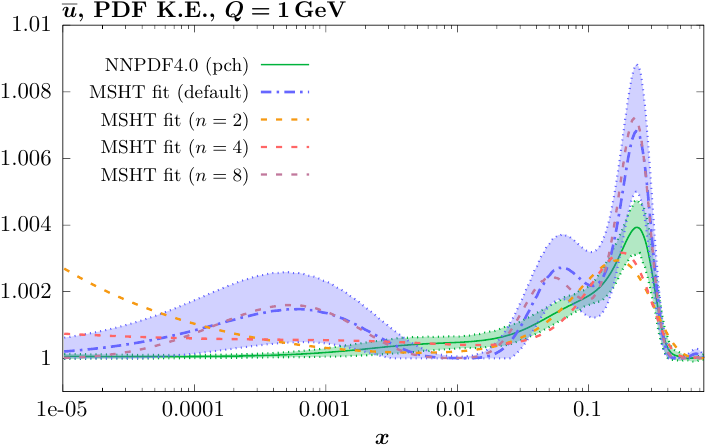}
\includegraphics[scale=0.6]{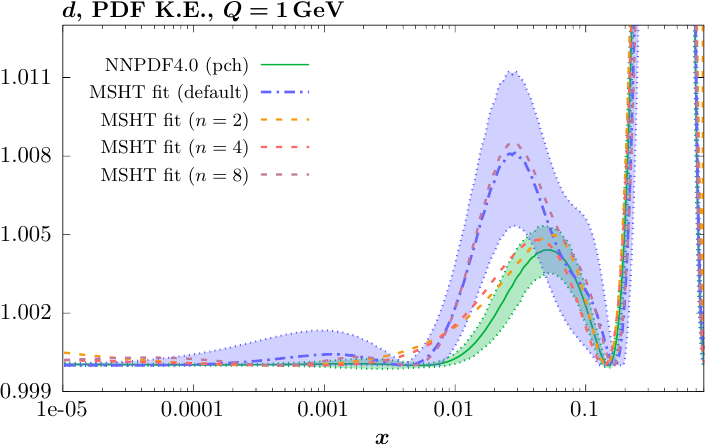}
\includegraphics[scale=0.6]{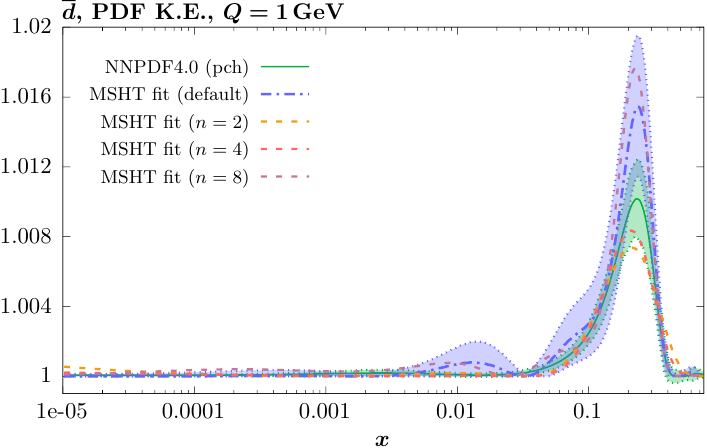}
\includegraphics[scale=0.6]{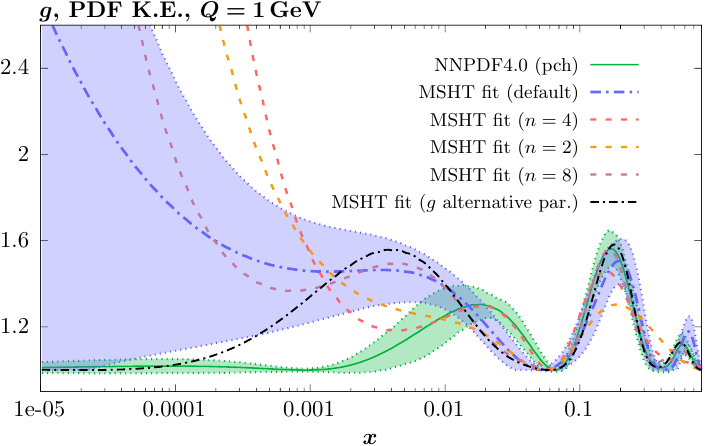}
\includegraphics[scale=0.6]{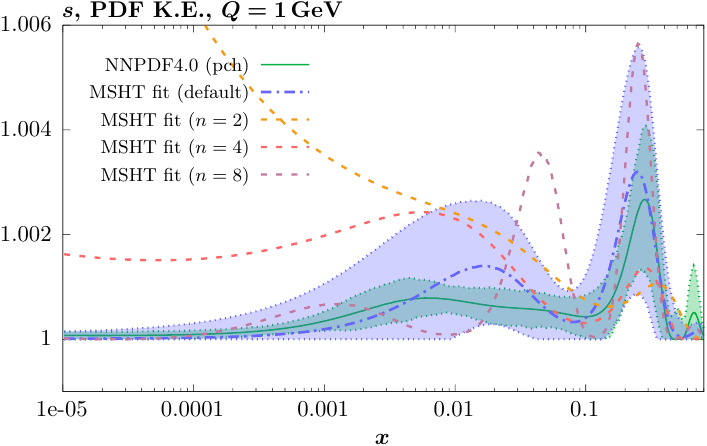}
\caption{\sf The PDF Kinetic energy, defined in the text, for a selection of PDFs at $Q=1 \, {\rm GeV}$  that result from a  global PDF fit to the NNPDF4.0 dataset/theory (perturbative charm) setting, but using the MSHT20 parameterisation. The default result, and fits with the number of Chebyshev polynomials fixed to 2, 4 and 8 (i.e. in the first cases reduced  by 4 and 2 and in the last increased by 2, respectively, for each PDF in comparison to the default)  are shown. For the gluon, the result without an explicit second term in the parameterisation, of the form given in \eqref{eq:cheb6_gen}, is also shown.}
\label{fig:arclength_pch}
\end{center}
\end{figure}

In Fig.~\ref{fig:arclength_pch} we show the kinetic energy for a selection of PDFs at the input scale $Q_0=1$ GeV for the perturbative charm fit. We compare the default MSHT fit with the NNPDF4.0 (pch) result, while also plotting the central values for the $n=2,4,8$ cases. Care should be taken to note the $y$ scales on these plots, which differs significantly between partons in order to highlight the principle features. For the default fit, the $T^2=10$ uncertainties are shown, which consistent with the discussion above tend to be rather larger then the NNPDF4.0 uncertainties. Broadly speaking, we can see that in various regions of $x$ there are certain peaking structures in the KE, the most significant of which occur in similar fashions in the NNPDF and MSHT fits. In some cases these are essentially guaranteed by the sum rules, e.g. for the $u$ and $d$ quark we can see sizeable peaks (beyond the upper limit of the presented $y$ axes) in the high $x$ region, driven by the non--zero up and down valence structure of the proton PDFs, which results in  turn overs of the PDFs in this region. Other peaks are simply preferred by the fit to the data, e.g. in the $\overline{u},\overline{d},s$ and gluon at higher $x$. 

\begin{figure}
\begin{center}
\includegraphics[scale=0.6]{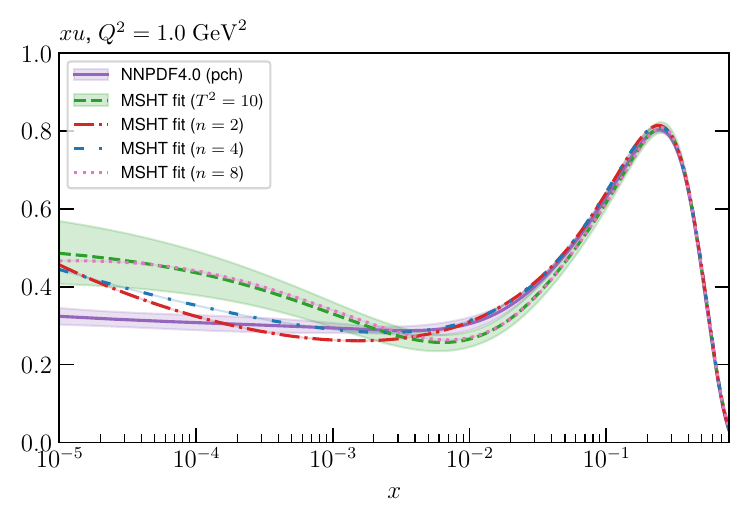}
\includegraphics[scale=0.6]{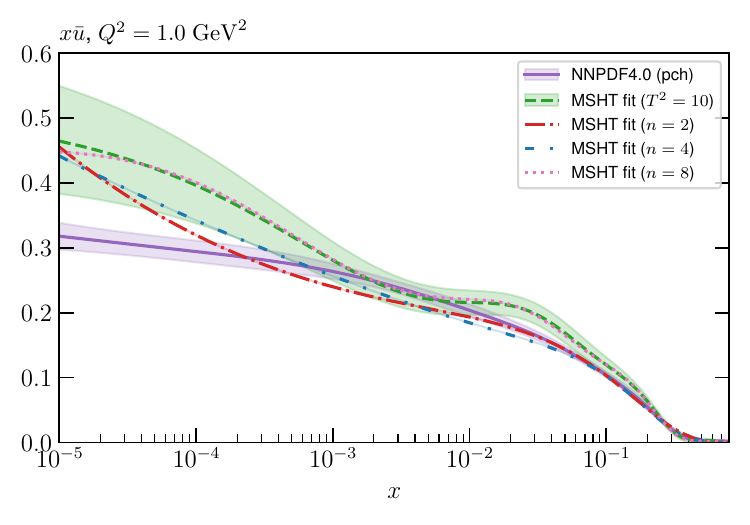}
\includegraphics[scale=0.6]{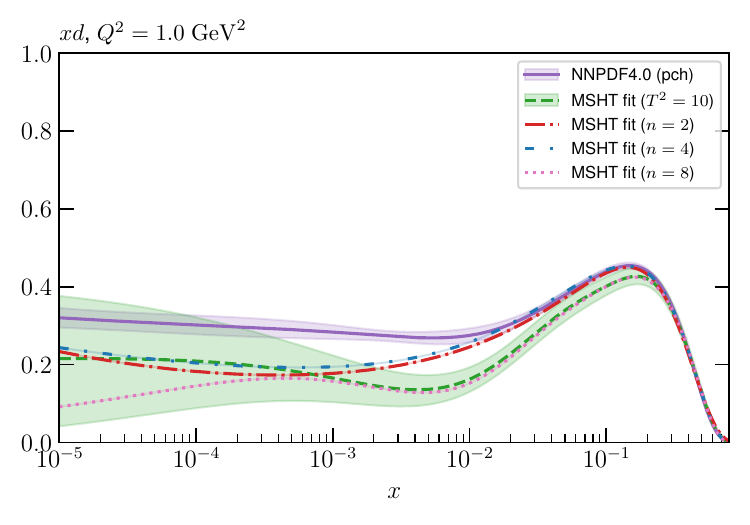}
\includegraphics[scale=0.6]{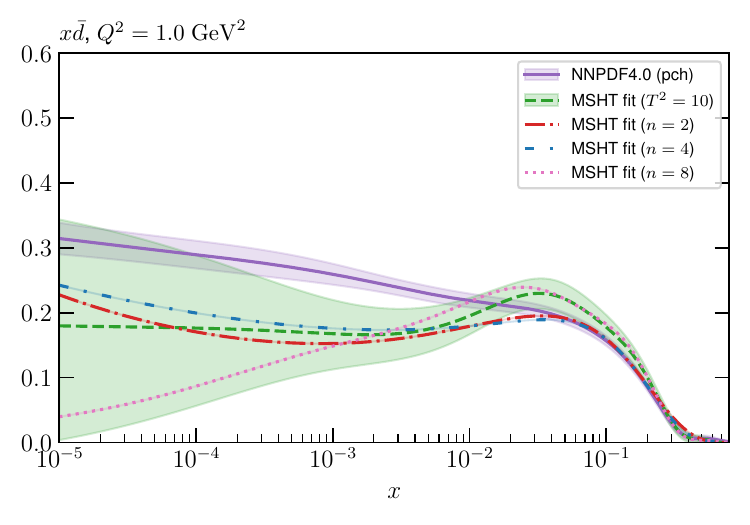}
\includegraphics[scale=0.6]{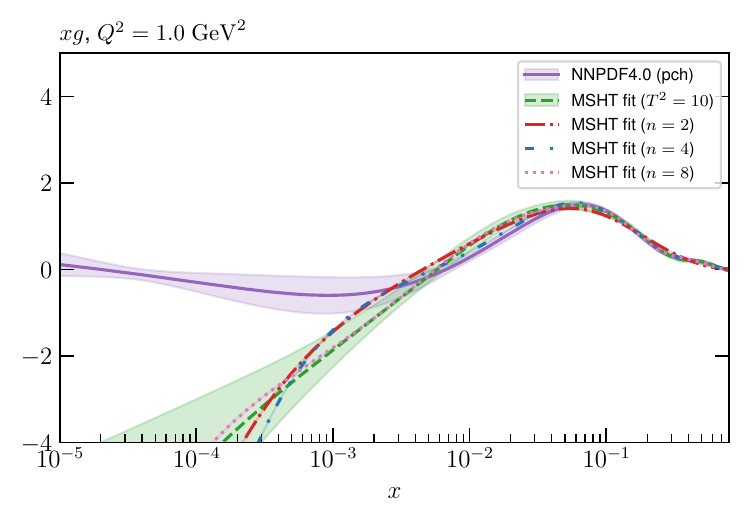}
\includegraphics[scale=0.6]{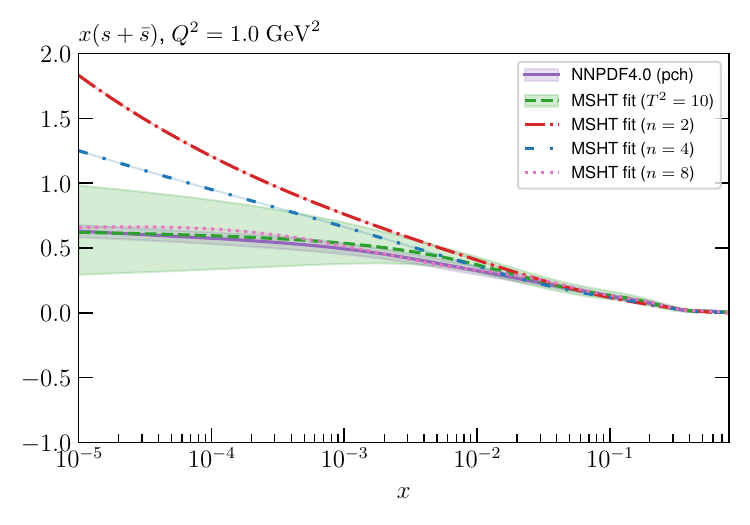}
\caption{\sf  A selection of PDFs at $Q=1 \, {\rm GeV}$  that result from a  global PDF fit to the NNPDF4.0 dataset/theory (perturbative charm) setting, but using the MSHT20 parameterisation. The default result, and fits with the number of Chebyshev polynomials fixed to fixed to 2, 4 and 8 (i.e. in the first cases reduced  by 4 and 2 and in the last increased by 2, respectively, for each PDF in comparison to the default)   are shown.  The NNPDF4.0 (pch) set is also shown.}
\label{fig:fit_pch_q0}
\end{center}
\end{figure}

We note that there is no evidence for a particularly larger KE in general, and indeed the arc--length integrated over the displayed $x$ regions agree at the sub permille level in all cases, with the exception of the gluon in the perturbative charm fit, where the behaviour at low $x$ leads to a $\sim 10\%$ increase with respect to the NNPDF4.0 PDFs. However, there are some regions of increased structure in the MSHT default fits. In particular, the $\overline{u}$ shows moderate local increase above the NNPDF result around $x\sim 0.05$ and again at lower $x$. Similarly for the $d$, the peak in the KE at $x\sim 0.03$ is somewhat larger in the MSHT case. The strangeness at large $x$ is one exception, where the NNPDF result exhibits some peaking not seen in the MSHT fit. For the low $x$ gluon the rather striking behaviour at low $x$ is driven by the turn over of the gluon as it become negative, which as discussed above is not present in the NNPDF fit, where positivity is imposed. While these effects appear relatively prominent when the results are plotted in terms of the KE, in Fig.~\ref{fig:fit_pch_q0} we show the corresponding PDFs at $Q_0=1$ GeV, and we can see that in general there is no particularly obvious lack of smoothness for these PDFs. The one exception is arguably the $\overline{u}$, though this not dramatic. 

In terms of the fits with modified number of Chebyshevs, we can clearly see that for the $n=2,4$ cases these match the KE of the NNPDF set more closely. For example, the additional peaking structure in the $\overline{u}$ at  $x\sim 0.05$  is absent, and the peak in the $d$ at $x\sim 0.03$  is much more in line with the NNPDF case. It therefore seems  clear for this perturbative charm case that the NNPDF fit, while initially deriving from a significantly more flexible NN parameterisation, provides an inherent degree of flexibility that is more in line with the restricted $n=2,4$ cases then the default $n=6$ MSHT fit,  once the NNPDF methodology (training/validation, post--selection etc) has been applied to produce the PDF prior. This lower degree of variation is also evident in Fig.~\ref{fig:fit_pch_q0}, at the level of the PDFs, where again the $n=2,4$ results are more in line with (though not in perfect agreement with) the NNPDF case.  For the $n=8$ case  this is largely in line with the baseline result, with the most significant exception of some additional features in the strangeness at intermediate $x$ seen in Fig.~\ref{fig:arclength_pch}; however here differences are also observed in the $n=2,4$ cases.

\begin{figure}
\begin{center}
\includegraphics[scale=0.6]{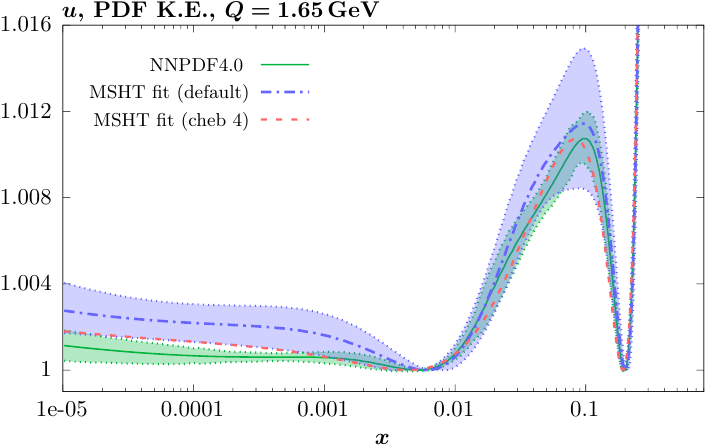}
\includegraphics[scale=0.6]{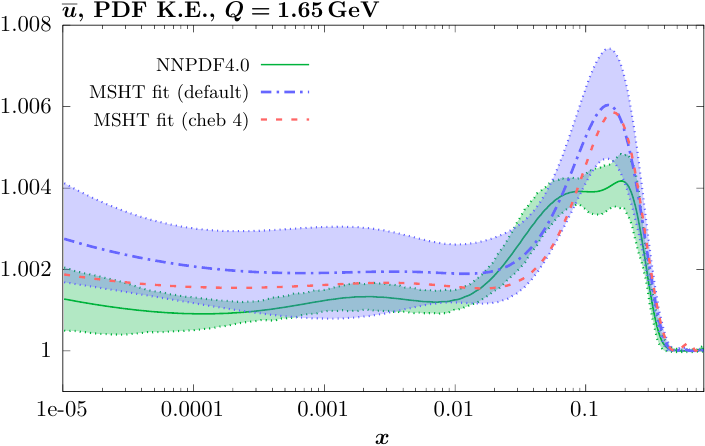}
\includegraphics[scale=0.6]{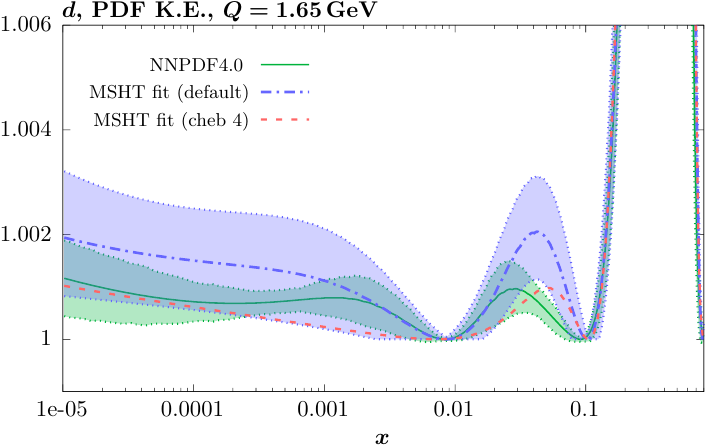}
\includegraphics[scale=0.6]{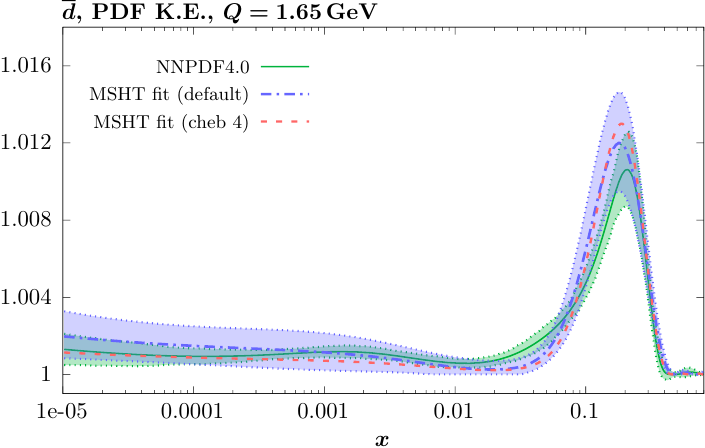}
\includegraphics[scale=0.6]{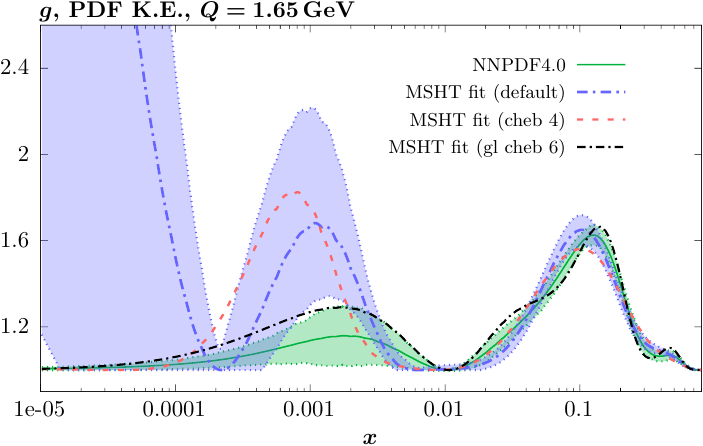}
\includegraphics[scale=0.6]{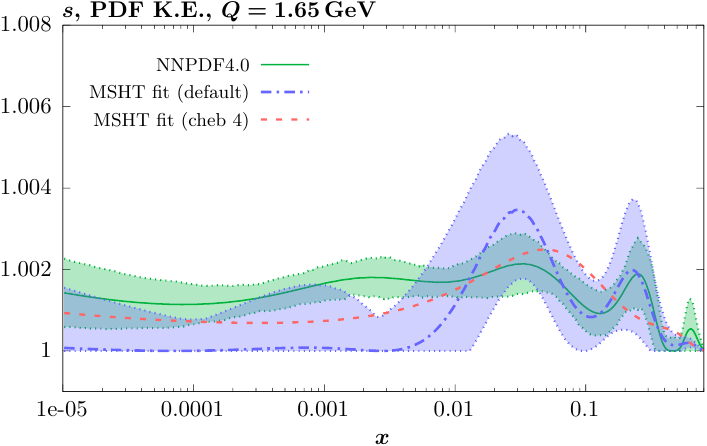}
\includegraphics[scale=0.6]{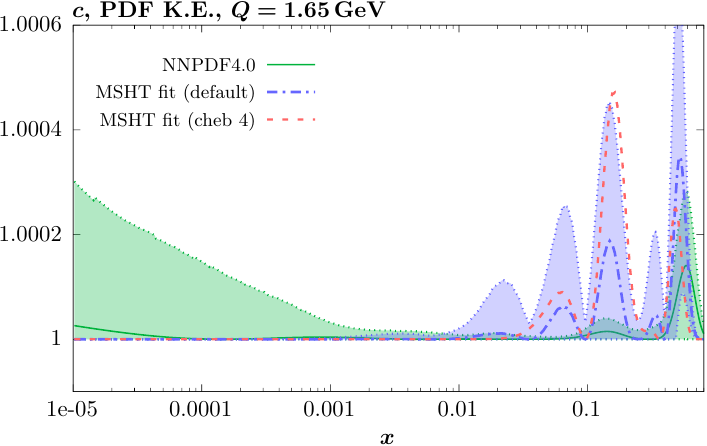}
\caption{\sf  The PDF Kinetic energy, defined in the text, for a selection of PDFs at $Q=1.65 \, {\rm GeV}$  that result from a  global PDF fit to the NNPDF4.0 dataset/theory (fitted charm) setting, but using the MSHT20 parameterisation. The default result, and fits with the number of Chebyshev polynomials  fixed to 2, 4 and 8 (i.e. in the first cases reduced  by 4 and 2 and in the last increased by 2, respectively, for each PDF in comparison to the default)  are shown. For the gluon, the result without an explicit second term in the parameterisation,  of the form given in \eqref{eq:cheb6_gen}, is also shown.}
\label{fig:arclength_fch}
\end{center}
\end{figure}

Therefore, while the default MSHT parameterisation does result in some local regions of larger PDF KE, these are in general moderate and do not provide any particularly strong evidence that a significant degree of overfitting is occurring. Moreover, we emphasise that in Fig.~\ref{fig:arclength_pch} (and in Fig.~\ref{fig:arclength_fch} below for the fitted charm case) there are clear regions of locally increased kinetic energy in both the MHST and NNPDF4.0 fits, at a similar level. For example, the increases in the up quark case around $x\sim 0.1$ or the gluon around $x\sim 0.2$ are clearly preferred by both fits. With this in mind, it is arguably not clear to what extent one can identify the `correct' amount of variation in the kinetic energy. 

\begin{figure}
\begin{center}
\includegraphics[scale=0.6]{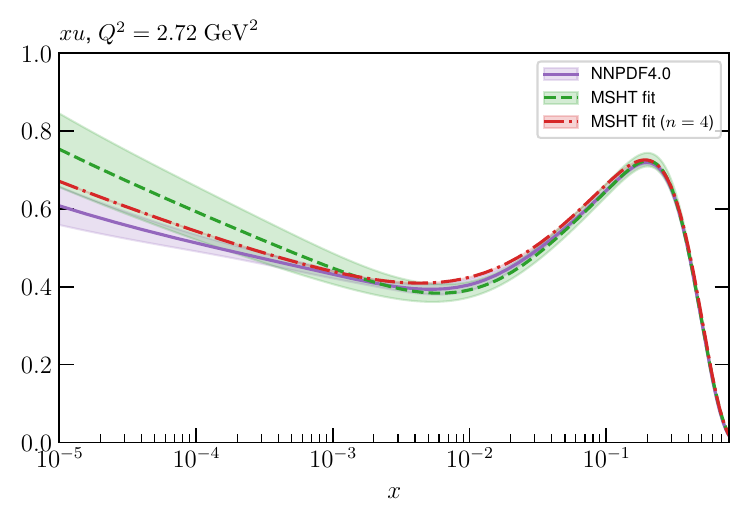}
\includegraphics[scale=0.6]{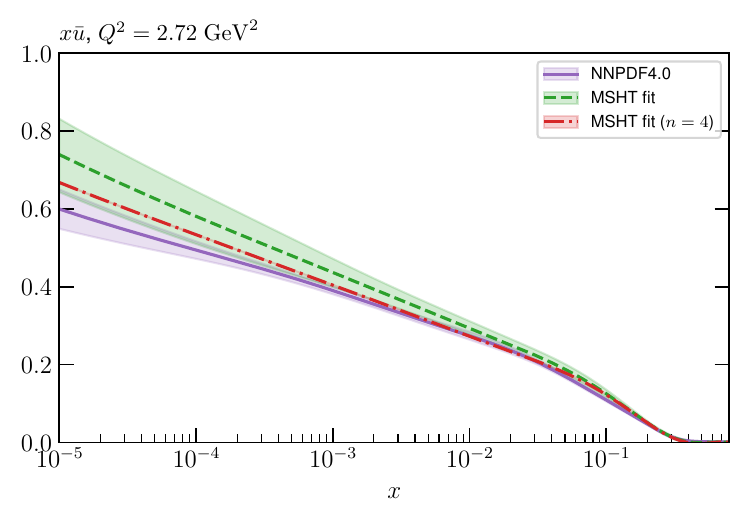}
\includegraphics[scale=0.6]{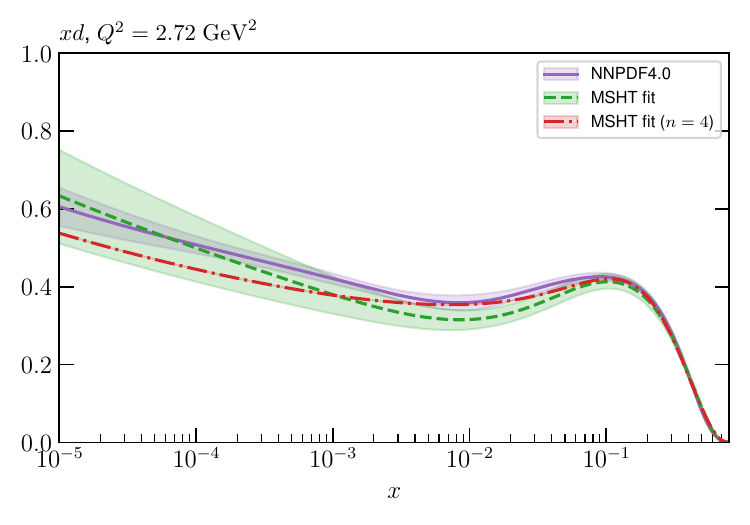}
\includegraphics[scale=0.6]{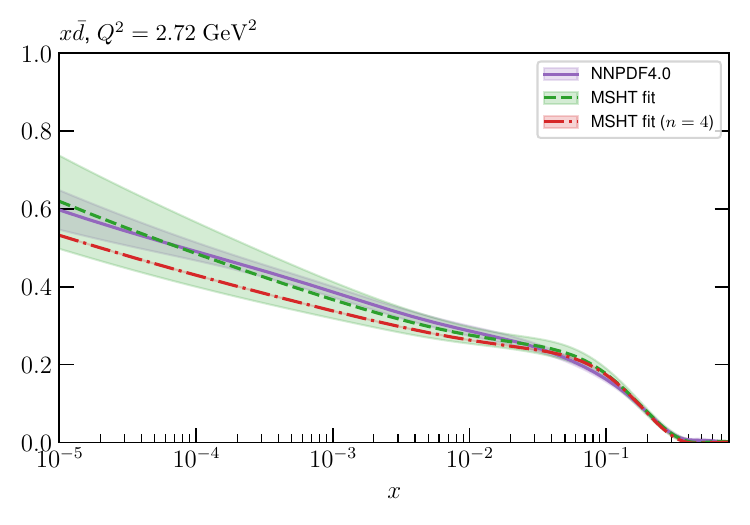}
\includegraphics[scale=0.6]{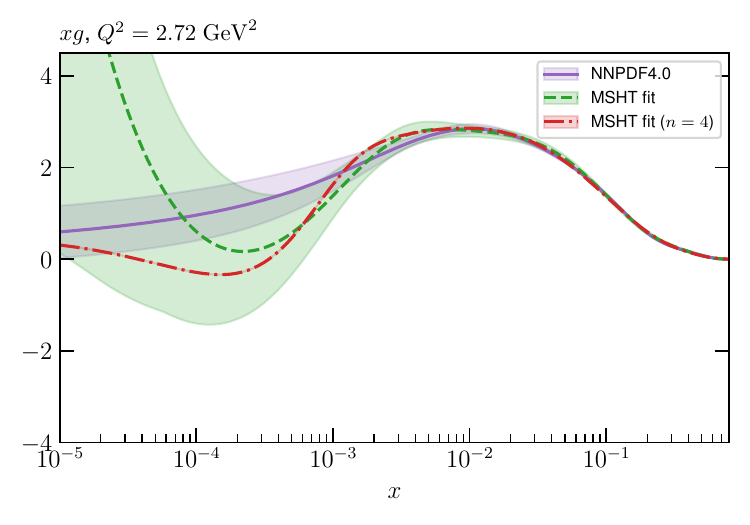}
\includegraphics[scale=0.6]{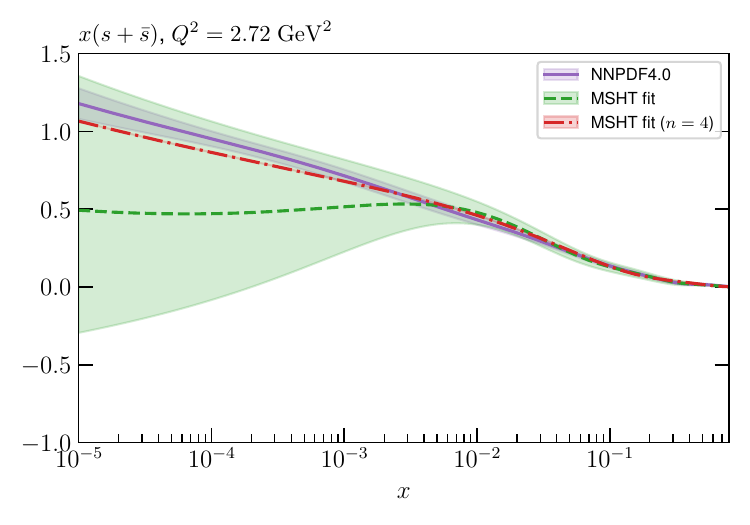}
\includegraphics[scale=0.6]{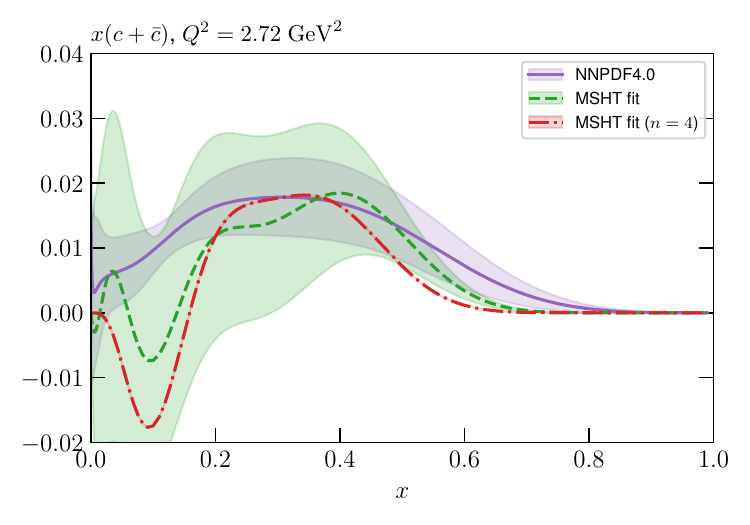}
\caption{\sf  A selection of PDFs at $Q=1.65 \, {\rm GeV}$  that result from a  global PDF fit to the NNPDF4.0 dataset/theory (fitted charm) setting, but using the MSHT20 parameterisation. The default result, and fits with the number of Chebyshev polynomials fixed to 2, 4 and 8 (i.e. in the first cases reduced  by 4 and 2 and in the last increased by 2, respectively, for each PDF in comparison to the default)   are shown. The NNPDF4.0 set is also shown.}
\label{fig:fit_fch_q0}
\end{center}
\end{figure}

We next turn to the fitted charm case, for which the KE is shown in Fig.~\ref{fig:arclength_fch}. Here, the difference between the baseline MSHT and NNPDF is less visible, in line with the closer fit qualities in this case. There is some mild indication of a larger peak in the $\overline{u}$ are $x\sim 0.1$, and the rather different behaviour of the low $x$ gluon also results in evident differences in this region, albeit within large PDF uncertainties in the MSHT case. The one exception is the charm quark, where some more prominent peaking structure at intermediate to high $x$ is seen, again within reasonable PDF uncertainties.  In terms of the fit with $n=4$ the differences with respect to the baseline are generally moderate, but not negligible.

The PDFs at input $Q_0=1.65$ GeV are shown in Fig.~\ref{fig:fit_fch_q0}, with results in line with these findings, i.e. no significant degree of additional variation in the gluon and light quark PDFs. For the charm, a dominant factor here is that we have not imposed positivity in the case of the MSHT fit; if this is imposed the dip around $x\sim 0.1$ is no longer present and the KE is correspondingly lower. Indeed we have verified that performing a NNPDF4.0 fit directly, but with $F_2^c$ positivity removed, results in a similar dip structure in this region. In terms of the large KE at lower $x$, we can see in Fig.~\ref{fig:fit_charm} that the peak around $x=0.03$ may be expected, given the corresponding peak in the purely perturbative component in this region.

In summary we have seen clear evidence, at the level of both closure test and fit to the real data that restricted parameterisations with fewer Chebyshev polynomials present in the input PDFs are not sufficient to describe the data entering current global PDF fits with sufficient precision and accuracy. This is achieved by the default $n=6$, for which there is no strong evidence of overfitting, or of the fit quality and description changing dramatically if further free parameters are included. In terms of the comparison to NNPDF, it would appear the resulting PDFs are more in line with the restricted $n=4$ parameterisation (i.e. with 12 fewer free parameters then the baseline case of 52), with the default MSHT fit exhibiting a moderately increased degree of variation for certain PDFs, in the perturbative charm case. For fitted charm, this distinction is less evident, perhaps because of the additional freedom allowed by fitting the charm PDF.

 In the future, it would be interesting to extend these studies to the case of fluctuated closure tests, to e.g. provide a more quantitative assessment of the (lack of) faithfulness of the fits with restricted parameterisations in that context. Beyond this, the question of overfitting clearly deserves further attention: a more detailed evaluation of this, for example by considering a partition of the data into training and validations sets, will be the focus of a future study.

\section{Cross Section Benchmarks}\label{sec:cross}

Having compared the MSHT and NNPDF fits at the level of the fit quality and PDFs, it is useful to also compare the corresponding predictions for some relevant LHC observables. In this section we therefore show some benchmark cross section results. In particular, we use \texttt{n3loxs}~\cite{Baglio:2022wzu} to calculate the Higgs boson production cross section in gluon fusion, and the $Z, W^\pm$ production cross sections, both at 14 TeV. The theoretical settings are as described in~\cite{Cridge:2023ryv}.

\subsection{Unfluctuated Closure Tests and Restricted Parameterisations}

\begin{figure}
\begin{center}
\includegraphics[scale=0.6]{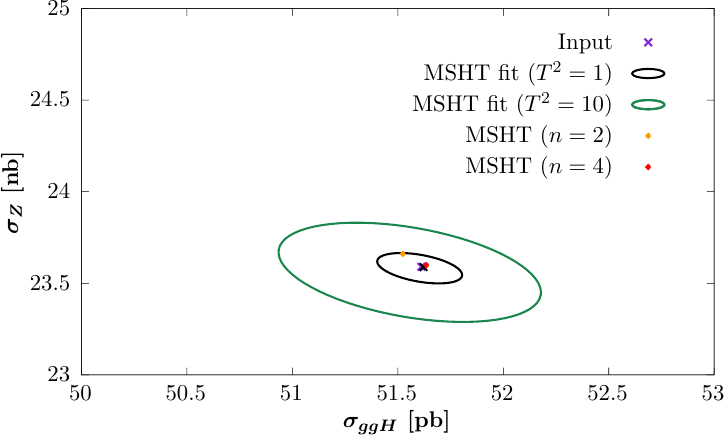}
\includegraphics[scale=0.6]{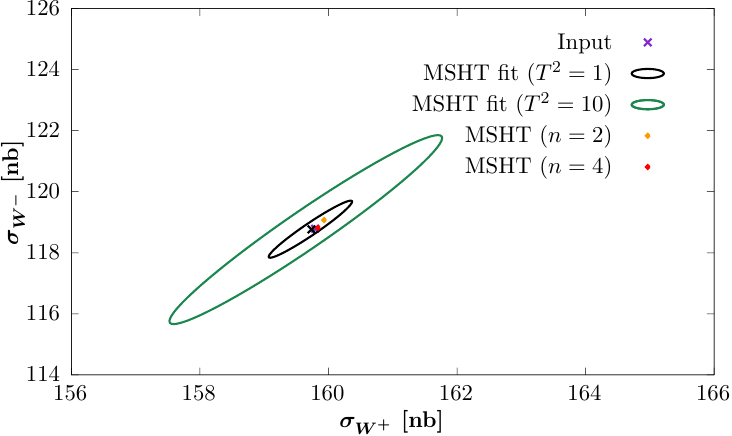}
\caption{\sf Cross section predictions for Higgs production via gluon fusion and on--peak $Z$ production, calculated as described in the text. Results shown for the unfluctuated closure test with $T^2=1$ and $T^2=10$ errors, and the central value from the input NNPDF set is  shown (the difference is barely visible from the predicted central value). Also shown  are the central values of the MSHT fits with restricted $n=2,4$ parameterisations.}
\label{fig:cs_L0}
\end{center}
\end{figure}

First, we show in Fig.~\ref{fig:cs_L0} the result of using the unfluctuated closure test. In the left plot the $Z$ boson and $ggH$ production cross sections are shown, with the corresponding PDF uncertainty ellipses for the MSHT fits, with both  $T^2=1$ and $T^2=10$. As usual, the corresponding axes of the ellipses indicate the size of the PDF uncertainties, while the angle indicates the correlation between the two predictions. In the right plot the $W^\pm$ results are shown.  The difference between the  input and unfluctuated closure test result is as expected almost invisible on the plot, and much less than the $T^2=1$ (or 10) uncertainties, providing further support for the minor role of parameterisation flexibility at the level of such a closure test, and in the default MSHT20 parameterisation. 

We also give the results of taking a restricted parameterisation as described in Section~\ref{sec:overfit}, with $n=2,4$ Chebyshevs, corresponding to 28 and 40 free parameters in total. While the $n=4$ case in fact shows rather good agreement with the input at the level of these cross sections, for $n=2$ we can see that the agreement reaches the edge of the $T^2=1$ uncertainty. This is in fact a rather closer matching then is always seen at the level of the PDFs shown in Fig.~\ref{fig:glcl_cheb24}, which is perhaps unsurprising given these cross sections involve sums over various PDF combinations and integrals over their $x$ dependence, that may be rather better constrained than a given individual PDF. It is in particular of note that there is much DY cross section data entering the fits.

\subsection{Impact of PDF Positivity}

\begin{figure}
\begin{center}
\includegraphics[scale=0.6]{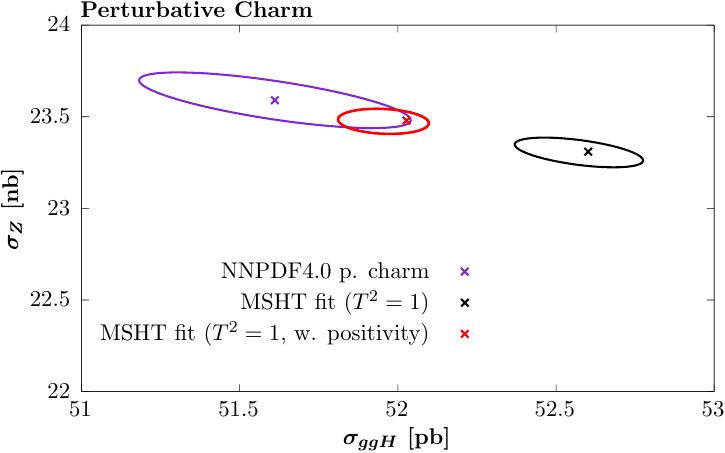}
\includegraphics[scale=0.6]{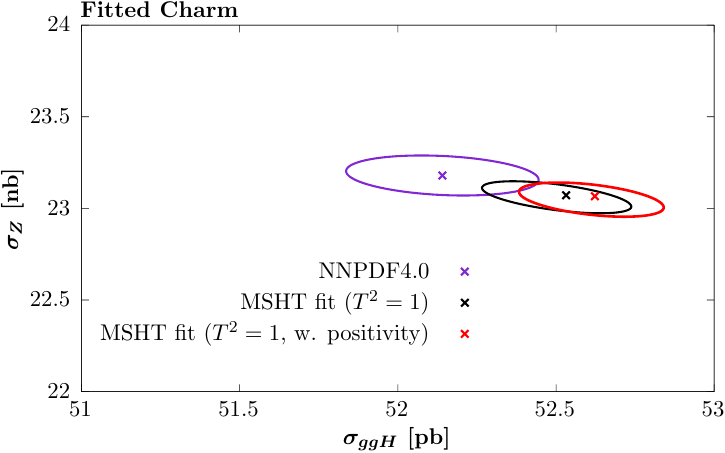}
\includegraphics[scale=0.6]{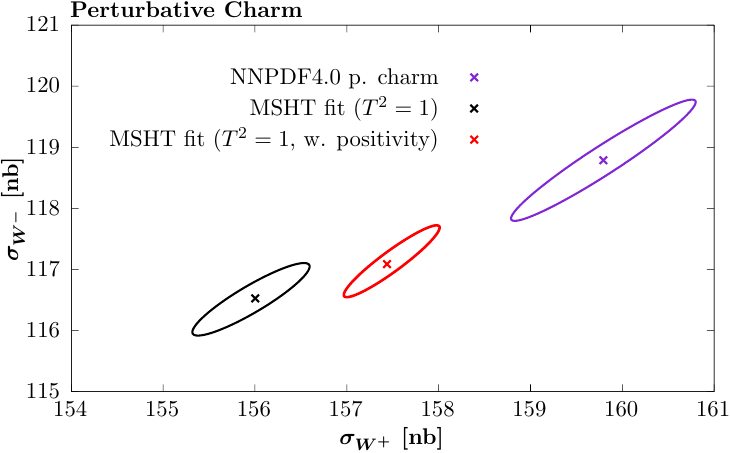}
\includegraphics[scale=0.6]{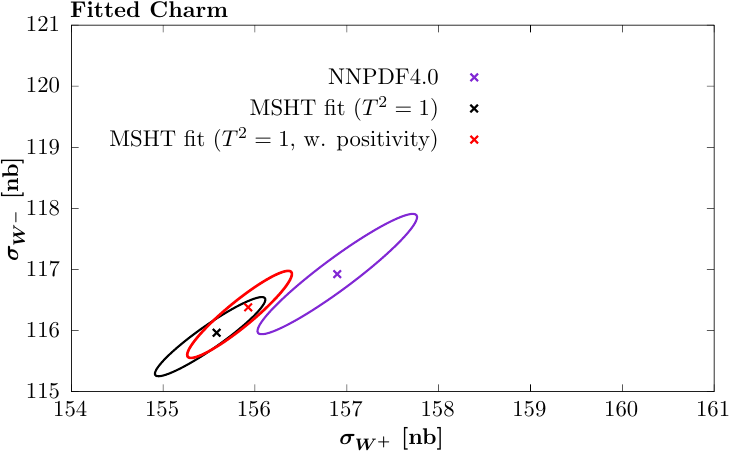}
\caption{\sf Cross section predictions for Higgs production via gluon fusion and on--peak $Z, W^\pm$ production, calculated as described in the text. Results shown for the (left) perturbative and (right) fitted charm cases, and with the MSHT fits to the NNPDF dataset/theory settings with $T^2=1$ uncertainties, both with and without positivity imposed. The corresponding NNPDF4.0 results are also shown.}
\label{fig:cs_fits_poscomp}
\end{center}
\end{figure}

Next, in Fig.~\ref{fig:cs_fits_poscomp}  we show the results for the MSHT fits to the real data entering the NNPDF fits, with perturbative and fitted charm in the left and right hand figures, respectively. The MSHT fit results are shown with $T^2=1$ uncertainties for clarity of the comparison. In the perturbative charm case we can see that there is as expected rather poor agreement between the NNPDF prediction and the MSHT fits without positivity. The Higgs cross section is larger in the MSHT fit, in line with the discussion relating to Fig.~\ref{fig:fit_fch_gluon} where it was observed that perturbative charm fits without positivity of the low $x,Q^2$ gluon imposed allow a larger gluon at intermediate $x$. The $Z$ cross section is mildly suppressed, due to the differing flavour decomposition. Both the $W^\pm$ are also suppressed, due to the suppression in the $u,d$ and $\overline{d}$ observed in the relevant $x$ region in Fig.~\ref{fig:fit_pch_rat}, while for the $\overline{u}$ some enhancement is seen, leading to a milder suppression for the $W^-$.

When positivity is imposed the agreement is greatly improved, as we would expect given the discussion before. Of particular note is the improved agreement for the Higgs cross section, an effect which is again in line with the discussion relating to Fig.~\ref{fig:fit_fch_gluon}, although improved agreement is also seen in the $W^-$ case in particular. Nonetheless, some difference remains. We recall  in particular that, as described in the case of the PDF comparison,  given the fit quality in the MSHT cases is actually better than that of the NNPDF4.0 fit, the correct assessment of the overall consistency between the NNPDF and the MSHT results is between the central value of the MSHT and the NNPDF results, within the NNPDF errors. With this in mind, we can see that the central MSHT result for the Higgs and $Z$ case lies at the edge of the NNPDF uncertainty band, while for the $W^\pm$ it is over $2\sigma$ away. 

In the fitted charm case, the agreement between the MSHT and NNPDF results is  better, and the difference between the cases with and without positivity imposed smaller, as we would expect given the results in the previous sections. However, again given the fit quality is better in the MSHT fit, the relevant comparison is between the central value of the MSHT  and the NNPDF results, within the NNPDF errors alone. Here we can see that there are still clear differences, with the central MSHT prediction lying roughly $\sim 1-2\sigma$ away from the NNPDF central result, with respect to the NNPDF uncertainties.  

\subsection{PDF uncertainties and Role of Tolerance}

\begin{figure}
\begin{center}
\includegraphics[scale=0.6]{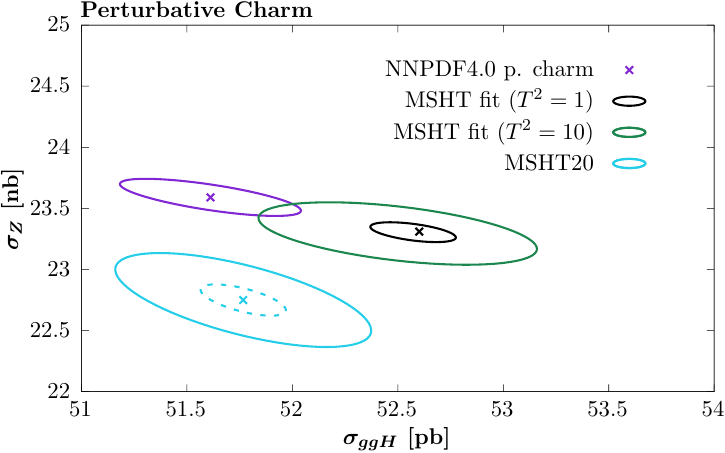}
\includegraphics[scale=0.6]{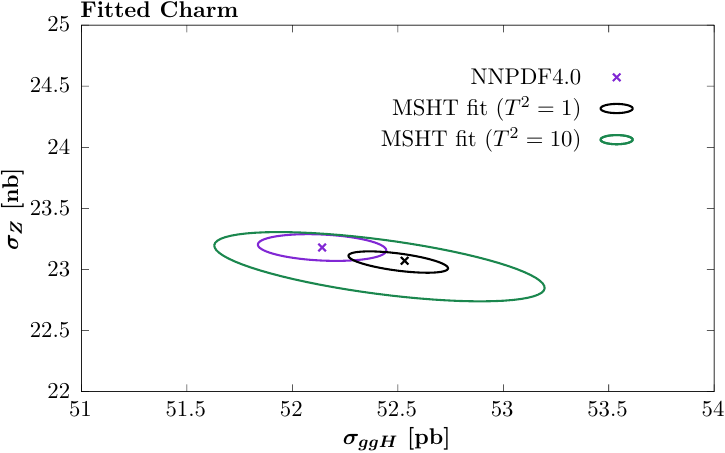}
\includegraphics[scale=0.6]{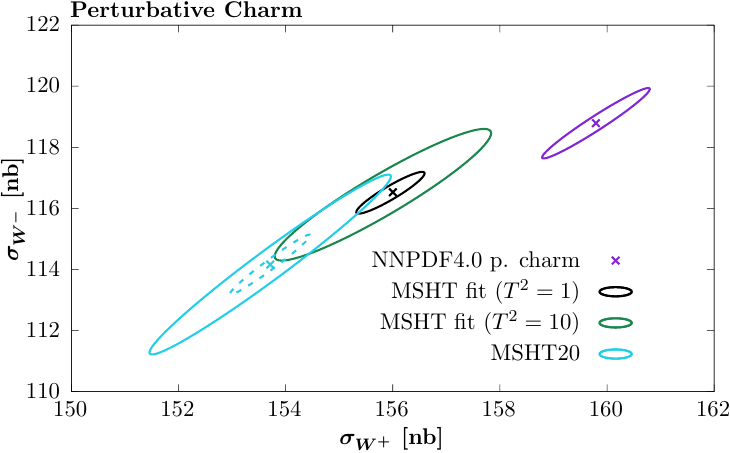}
\includegraphics[scale=0.6]{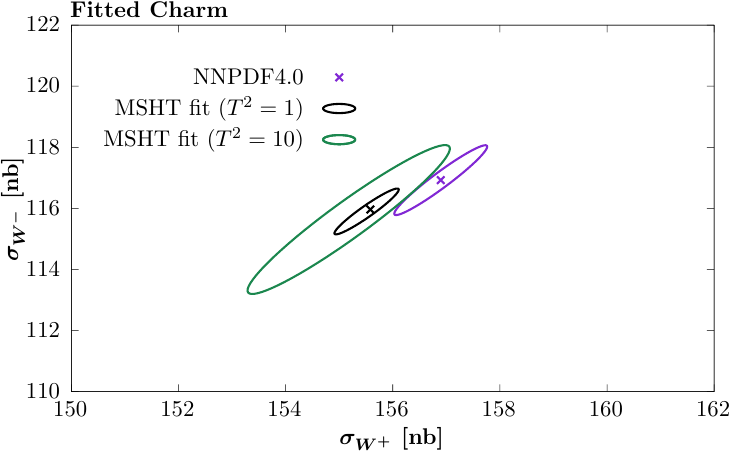}
\caption{\sf Cross section predictions for Higgs production via gluon fusion and on--peak $Z, W^\pm$ production, calculated as described in the text. Results shown for the (left) perturbative and (right) fitted charm cases, and with the MSHT fits to the NNPDF dataset/theory settings with both $T^2=1$ and 10 uncertainties. Positivity is not imposed in the MSHT fits. In the left plots the MSHT20 prediction is shown for comparison, where the dashed ellipse corresponds to dividing the uncertainty by a factor of 3, approximately corresponding to $T^2=1$ PDF errors. The corresponding NNPDF4.0 results are also shown.}
\label{fig:cs_fits}
\end{center}
\end{figure}

In Fig.~\ref{fig:cs_fits} we show the same cross section results as in Fig.~\ref{fig:cs_fits_poscomp} but now without positivity imposed, and also including the MSHT fit results with $T^2=10$ uncertainties. In terms of the size of the PDF uncertainties, for the perturbative charm fits the NNPDF result is found to lie somewhere between the $T^2=1$ and $T^2=10$ case, with the biggest departure from the MSHT $T^2=1$ case being in the Higgs cross section, as we would expect given the previous comparison of the PDF uncertainties. For fitted charm, the NNPDF result is clearly  closer to the $T^2=1$ case (though still somewhat larger than it) than for the perturbative charm fit, and is certainly significantly smaller than the $T^2=10$ case. 

For the sake of comparison we also show the result of the MSHT20 fit,  which applies the same parameterisation as the MSHT fits presented here, but with different theoretical and data inputs. The size of the uncertainty is in line with the MSHT $T^2=10$ case, as we would expect. In the perturbative charm case, comparing the MSHT20 and MSHT ($T^2=10$) results provides an indication of the level of difference between the MSHT20 and NNPDF4.0 fits due to the datasets included and theory settings alone. These are clearly playing an important role and interestingly some degree of tension is observed in the predicted $Z$ production cross section, though the agreement in the $W^\pm$ case is rather better. However, we also show in the dashed ellipse the case with the MSHT20 uncertainties scaled down by a factor of 3, approximately corresponding to $T^2=1$ PDF errors. From this, we can see that both comparisons provides further evidence of the need for a tolerance in the fit; clearly if $T^2=1$ uncertainties are used we would find significant tension between these predictions, which can only be due to the fact that different datasets are fit (or given datasets are treated differently) and theoretical inputs are used. 

\subsection{Comparison to Hopscotch study}

\begin{figure}
\begin{center}
\includegraphics[scale=0.6]{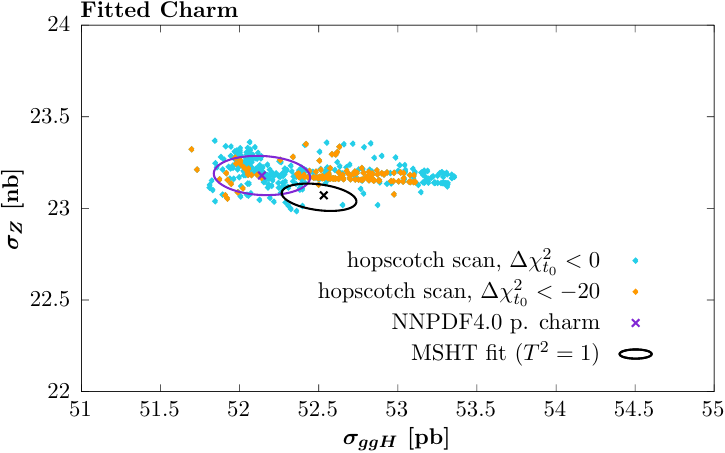}
\includegraphics[scale=0.6]{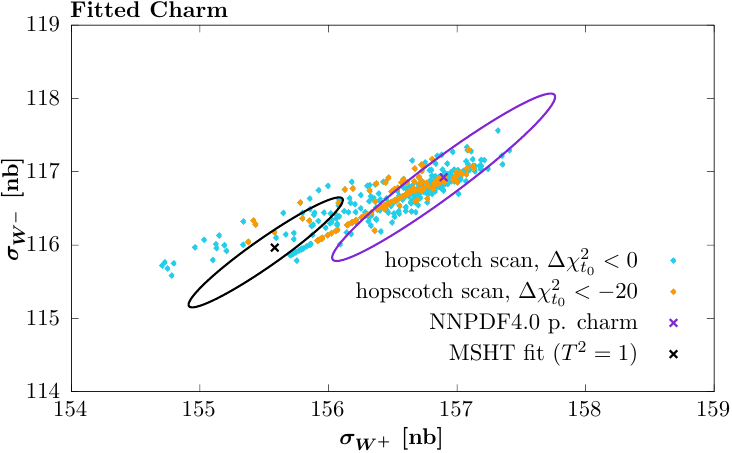}
\caption{\sf Cross section predictions for Higgs production via gluon fusion and on--peak $Z, W^\pm$ production, calculated as described in the text. The MSHT fits to the NNPDF dataset/theory settings, with  $T^2=1$ uncertainties, are compared to the hopscotch scans of~\cite{Courtoy:2022ocu} for the case that $\Delta \chi^2 < 0 $ and -20, in the $t_0$ definition. The corresponding NNPDF4.0 results are also shown.}
\label{fig:cs_fits_hop}
\end{center}
\end{figure}

In~\cite{Courtoy:2022ocu},  by taking a suitable linear combination of the NNPDF4.0 replicas via a so--called `hopscotch' scan, better fit qualities were found to the central data than for the the central NNPDF replica, and cross section predictions outside the quoted NNPDF uncertainty band were found. Taking the subset of the publicly available replicas~\cite{hopweb} that correspond to an improvement with respect to the $t_0$ definition of the fit quality, we compare these predictions against our fit results in Fig.~\ref{fig:cs_fits_hop}. As discussed in~\cite{Courtoy:2022ocu} the precise distribution of these replicas does not have a statistical interpretation but is due to the hopscotch scan methodology. Moreover,  the majority of these replicas correspond to changes in the fit quality that are significantly lower than the improvement observed in the MSHT fit\footnote{See also~\cite{Ball:2022uon} where a response to this analysis is presented by NNPDF; it is not the aim of the current paper to comment further on the discussion presented in these papers, which we simply highlight here.}. Nonetheless, it is interesting to examine to what extent these lie in a region of cross section space that is consistent with the MSHT best fit.

Comparing to these we observe a similar trend for larger Higgs production cross section to be larger. On the other hand,  for the $Z$ cross section no particular  trend is observed. For the $W^\pm$ cross sections there is again an interesting level of consistency between the general region of hopscotch replicas that lie outside of the NNPDF uncertainty band and the MSHT fit results. However, the majority of these that lie towards the MSHT fit region show rather milder improvements in fit qualities of less than $-20$.

\section{PDF Sets: Availability}\label{sec:availability}
To enable further studies and comparisons by the community we make a selection of PDF sets resulting from the MSHT fits to the NNPDF4.0 data and theory settings publicly available at the following website:
\\
\\
\href{https://www.hep.ucl.ac.uk/msht/NNPDFgrids.shtml}{https://www.hep.ucl.ac.uk/msht/NNPDFgrids.shtml}
\\
\\
The perturbative charm sets described in Section~\ref{sec:pcharm} are available via the following links:
\\
\\
\href{https://www.hep.ucl.ac.uk/msht/NNPDFgrids/XX.tar.gz}{MSHT\_NNPDF40input\_pch\_Teq1}
\\
\href{https://www.hep.ucl.ac.uk/msht/NNPDFgrids/XX.tar.gz}{MSHT\_NNPDF40input\_pch\_Teq10}
\\
\\
while the fitted charm sets described in Section~\ref{subsec:fcharm} are available via the following links:
\\
\\
\href{https://www.hep.ucl.ac.uk/msht/NNPDFgrids/XX.tar.gz}{MSHT\_NNPDF40input\_fch\_Teq1}
\\
\href{https://www.hep.ucl.ac.uk/msht/NNPDFgrids/XX.tar.gz}{MSHT\_NNPDF40input\_fch\_Teq10}
\\
\\
In both cases the PDF sets without positivity imposed, and with a fixed tolerance of $T^2=1$ and 10 are made available. We do not make the sets with positivity imposed available because, as discussed in earlier sections, the Hessian uncertainty is rather unstable in this case.

\section{Summary and Outlook}\label{sec:conc}

In recent years there has been a great deal of progress in the development of global PDF fits, as exemplified by the CT, MSHT and NNPDF fitting collaborations. These combine high precision theory, such that NNLO is the standard and now even approximate ${\rm N}^3$LO QCD precision is being accounted for, with a wealth of global measurements, in particular with much high precision data from the LHC. Combing these ingredients, we can hope to constrain proton structure with high levels of precision and accuracy, in order to match the ever increasing precision requirements of the LHC physics programme.

However, as well as differing in the data and the relevant theoretical ingredients that enter the fit, these global PDF analyses also differ rather significantly in the methodologies they apply. Moreover, these methodological effects alone are observed to have a significant impact on the resulting PDFs and, crucially, their uncertainties. Therefore, arriving at  more complete understanding of this source of difference is arguably as important to the LHC precision programme as the continued  progress being made within each fit. It has been the aim of this paper to begin to address this issue directly. To achieve this, it is certainly necessary that the fitting methodology of each individual collaboration  be tested as robustly as possible. However, it is also essential that a completely direct comparison between the different methodologies be made, in order to make real progress.

We have presented results that aim to address both of these directions of investigation. With the first issue in mind, we have presented the first full global closure test of a fixed parameterisation approach to PDF fitting, focusing on the MSHT20 case. We have found that the default MSHT20 parameterisation can reproduce the features of the input set in such a closure test to well within the textbook $\Delta \chi^2 =T^2=1$ uncertainties. This non--trivial result provides strong evidence that parameterisation inflexibility in the MSHT20 fit is not a significant issue in the data region, and hence that should not be a major contribution in the enlarged error definition (a so called `tolerance'), which is commonly applied in PDF fits. In the extrapolation region, most notably at very high $x$, some discrepancy is found, highlighting a potential limitation of the fixed parameterisation approach. Interestingly, if instead MC replica error generation is applied, but with the same fixed parameterisation, then the PDF uncertainties at high $x$ are potentially more representative.

With the second issue in mind, we have also presented the first completely like--for--like comparison between two global PDF fits, namely MSHT and NNPDF, where the only difference is guaranteed to be due to the fitting methodology, and PDF parameterisation in particular. To achieve this, we have made use of the public NNPDF fitting code to perform a PDF fit to exactly the same data and theory settings that are used in the NNPDF4.0 NNLO public release, but with the MSHT20 fixed parameterisation applied instead of a neural network (NN). Two fits, one with perturbative and one with fitted charm, have been presented; in the latter case this represents the first model independent fitted charm determination outside of the NN approach. Somewhat surprisingly we find that, despite the inherently larger flexibility  in the NN used in the NNPDF4.0 case, the fit with the MSHT parameterisation produces a moderately, but noticeably, better fit quality than the central NNPDF results, both with perturbative and fitted charm. More significantly, we have found that the resulting PDFs and various predicted benchmark cross sections  due to the public NNPDF4.0 releases are not compatible with these MSHT fits within the nominal NNPDF uncertainties.

The role of prior constraints, in particular the PDF positivity that is imposed by NNPDF, are considered and this is found to play an important role in the fit with perturbative charm with respect to the gluon at low $x$, but is not found to explain the difference entirely. The question of whether the MSHT fit may lead to overfitting (which is controlled against  in the NNPDF fits, given the more flexible NN architecture) is also considered by a close examination of the breakdown in the fit quality between datasets, the form of the underlying PDFs, and the impact of restricting and extending the number of free parameters in the fit. No particular evidence for this is found, although if on further analysis some evidence for  this were indicated, we note that this would only serve to further support the result that parameterisation inflexibility is not a major issue in the MSHT fit. Studying in detail the form of the resulting PDFs, we have found that the MSHT fit case appears to have moderately more flexibility associated with them, which therefore results in a better fit to the data being achieved. This is also supported by restricting the number of free parameters in the MSHT parameterisation; reducing this from 52 to 40 gives a result more closely in line with the NNPDF4.0 case, although it should be emphasised that this performs rather less well in the closure tests fits than the default MSHT parameterisation. 

Putting aside the reason for this difference in the underlying PDFs, we have also compared the corresponding PDF uncertainties and have indeed found that with respect to the quark flavour decomposition the NNPDF4.0 uncertainties are rather closely in line with the MSHT $T^2=1$ uncertainties, while the gluon and quark singlet being somewhat larger, but still significantly lower than the $T^2=10$ uncertainties, which are representative of the enlarged tolerance applied in e.g. MSHT20. In other words, the NNPDF4.0 fit, as a result of changes in methodology alone,  produces PDF uncertainties that are rather more closely in line with those of the fixed MSHT20 parameterisation if the textbook $T^2=1$ criterion is applied. Given it has long been argued by the MSHT collaboration, and others, that this $T^2=1$ criterion is not applicable in global PDF fits, this 
 at a minimum points to an inherent inconsistency between the approaches. In particular, the possibility that the MSHT approach is less accurate due to its fixed parameterisation and therefore requires an enlarged error definition due to this, has been largely ruled out by the results of this study, namely the successful closure tests and the improved fit quality that is seen with respect to NNPDF in the real fit. Therefore, either the NNPDF uncertainties are too aggressive (i.e. too small), or the MSHT uncertainties are too conservative (i.e. too large), or the truth lies somewhere in between.

We have formulated our conclusion in this rather agnostic way in order to state this key result  as clearly, and with as minimal additional interpretation, as possible. In this way, we believe it could serve as a motivation for further developments both on the fixed parameterisation side, through the refinement of the tolerance criterion that is applied, but also on the NN side, through a reconsideration of how the uncertainties are defined. However, it should be emphasised that there are very good reasons for including an enlarged tolerance in global PDF fits. In particular, the textbook $\Delta \chi^2=1$ criterion is applicable to the ideal scenario of complete statistical compatibility between the multiple datasets entering the fit, a completely faithful evaluation of the experimental uncertainties within each dataset, and theoretical calculations that match these exactly. There is a great deal of evidence that the first two situations do not hold in a PDF fit, while for the latter case it is of course well known that the fixed--order theoretical predictions are not exact (although as described above there has been recent progress in evaluating the uncertainty due to this~\cite{McGowan:2022nag,NNPDF:2024dpb,NNPDF:2024nan}). Indeed, at the level of the global fit quality, and the pulls of individual datasets in global fits, departures from textbook statistics are certainly evident. 

In the current study, we have also compared the result of the public MSHT20 fit with the MSHT fits presented here. In other words, these compare the difference due to the change in dataset and theory setting alone (keeping the fitting methodology fixed) in the resulting PDFs. We find broad consistency at the level of the PDFs and benchmark cross sections if an enlarged $T^2=10$ definition is used, but crucially not if  the $T^2=1$ definition is used, when evident significant tensions appear. This provides further support for the need for an enlarged error definition, as provided by the tolerance. Further to this, we have also presented a detailed discussion of the role of statistical incompatibilities within a toy model, and highlighted the fact that tensions between datasets are not reflected by increased PDF uncertainties if the textbook $T^2=1$ criterion is applied. These are then found to be unrepresentative, as they do account for any increased spread in the PDF error that should be present due to these tensions. We have then demonstrated this effect explicitly in the context of a set of global closure tests, where incompatibility between different datasets, or between data and theory, is included in the test.

Finally, we have as described above addressed the question of parameterisation flexibility from the point of view of restricting the number of free parameters to be less than the nominal 52 in the MSHT20 case. By restricting the PDF parameterisation to instead have 28 or 40 free parameters, we confirm that these more restricted parameterisations are insufficient to match the input of the global closure test within the $T^2=1$ (and even in some cases $T^2=10$) uncertainties. Hence, in such cases parameterisation inflexibility would clearly play a more significant role in requiring an enlarged tolerance, but this can be largely avoided by simply taking a suitably flexible parameterisation. This result should  be of particular relevance to those PDF analyses, such as CT and ATLASpdf, where more restricted parameterisation are taken by default.

We have in addition considered the evidence for intrinsic charm within the MSHT fit to the NNPDF4.0 data and theory. Given the discussion above about the increased uncertainties in the default MSHT fit (with the appropriate $T^2=10$ uncertainty definition) it is not surprising that the statistical significance of an intrinsic charm component at high $x$ is observed to be markedly lower than in the nominal NNPDF4.0 fit. However, a mild preference for this is evident, albeit one where the significance is rather increased by imposing positivity of the predicted $F_2^c$ structure functions at relatively low scale. The most significant improvement in the fit quality that comes from fitting charm, is observed to lie in the low to intermediate $x$ region, which indicates that the most statistically significant intrinsic charm, or more strictly fitted charm component may lie in this region, where it is even found to be negative at intermediate $x$. However, a full analysis of these effects and indeed precise interpretation of our results in terms of purely intrinsic charm has not been presented here, and is beyond the scope of this study.

In summary, we have presented in this study the first full global closure test of a fixed parameterisation approach to PDF fitting, focusing on the MSHT20 case, and the first completely like--for--like comparison between two global PDF fits, namely MSHT and NNPDF, where the only difference is guaranteed to be due to the fitting methodology, and PDF parameterisation in particular. The closure tests are successfully passed, and  we in particular find no evidence that parameterisation inflexibility plays a significant role in the uncertainty budget of the MSHT fit. At the level of the full fit, we find a moderate improvement in the MSHT fit quality with respect to the NNPDF4.0 central results, and a non--negligible difference in the corresponding PDFs and benchmark cross sections. At the level of the PDF uncertainties, the NNPDF4.0 case is found to be broadly in line with the MSHT result, but only if the textbook tolerance of $T^2=1$ of used, which is arguably unsuitable in the non--trivial environment of a global PDF fit. Indeed, in this study we have explicitly demonstrated the impact of including data/theory tensions in a global closure test and show that these have essentially no impact on the resulting PDF uncertainties. Hence, these are found to be unrepresentative, as they do not account for the increased spread in the PDF error due to these tensions. These results are completely in lines with first principles considerations, and should apply whether a NN or fixed parameterisation is used.

This indicates an inherent inconsistency between these approaches that clearly begs to be resolved if progress is going to made with respect to the LHC precision physics programme. On the NN side, these results indicate that a critical reassessment of the origin of the current PDF uncertainties in this approach, and extension to account for an effective tolerance may be required. On the fixed parameterisation side, refinement and reassessment of the current approaches to including such a tolerance is well motivated, and related to this further work towards an explicit account of the elements in a fit that contribute to such a tolerance in the fit.  Future studies and comparisons of the two approaches are planned, including providing the fixed parameterisation fitting code used in this study for  public use.  More broadly, progress has already been made and is ongoing on for example the inclusion of theoretical uncertainties from missing higher orders, and an estimation of the error on experimental systematic uncertainties, both of which will play a role in this. Through continued such work, and collaboration between groups that apply different fitting approaches we can hope to continue to make progress in high precision LHC era.

\section*{Acknowledgements}

We thank  Emanuele Nocera, Juan Cruz--Martinez and Zahari Kassabov for assistance in using the public NNPDF code. We thank  Aurore Courtoy for assistance in using the hopscotch scan PDFs. We thank Amanda Cooper--Sarkar, Stefano Forte and other members of the NNPDF collaboration for useful comments on the first arxiv version of the manuscript.

TC acknowledges that this project has received funding from the European Research Council (ERC) under the European Union’s Horizon 2020 research and innovation programme (Grant agreement No. 101002090 COLORFREE).  L. H.-L. and R.S.T. thank STFC for support via grant awards ST/T000856/1 and ST/X000516/1.

\appendix

\section{Subleading Eigenvectors and Parameter Fixing in Hessian PDF Uncertainties}\label{app:PDFparfixing}

In this appendix we demonstrate in more detail the role of subleading eigenvectors and the fixing of PDF parameters in the Hessian matrix when the PDF uncertainties are evaluated according to the eigenvector prescription described in Section~\ref{sec:gen}. In particular,  it has been demonstrated clearly in Section~\ref{sec:overfit} that a given degree of parameterisation flexibility is required in order to match the requirements of a given PDF fit. Nonetheless, when it comes to considering variations of these PDF parameters around the global $\chi^2$ minimum it is common for there to be  a certain amount of  redundancy between some of these PDF parameters, such that small changes in the values of some parameters can be largely compensated for by changes in other parameters. As a result of this high degree of correlation, the behaviour of certain PDF eigenvectors about the $\chi^2$ minimum can become highly non--quadratic. 

Further details are provided  in~\cite{Martin:2002aw,Martin:2009iq}, where a practical solution of fixing certain PDF parameters at their best fit values, when these exhibit a significant degree of correlation with other PDF parameters, is proposed. In this way, a set of more quadratic eigenvectors is arrived at, and a more stable application of the Hessian approach becomes possible. We emphasise  that this is not the same as starting with fewer free parameters to start with in the fit, e.g. setting higher Chebyshev coefficients to zero as in Section~\ref{sec:overfit}, and there is in particular no contradiction that fewer free parameters are required in the eigenvector evaluation stage than in the initial fit. Moreover, as discussed in~\cite{Martin:2002aw,Watt:2012tq}, this method of fixing certain PDF parameters provides PDFs uncertainties that are consistent with other methods of error propagation, such as Lagrange multiplier scans and the MC replica technique.

\begin{figure}
\begin{center}
\includegraphics[scale=0.6]{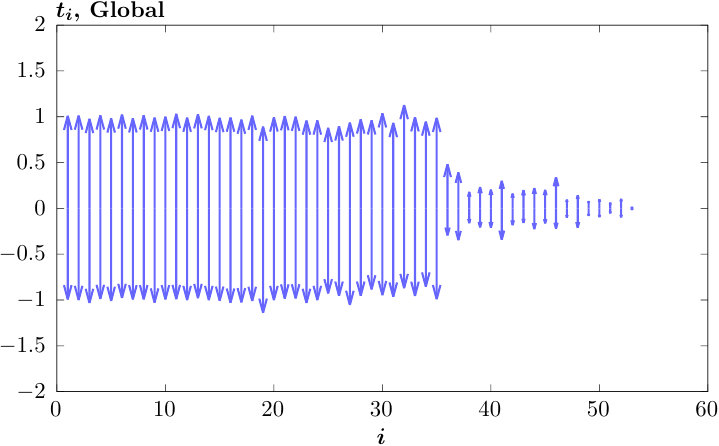}
\includegraphics[scale=0.6]{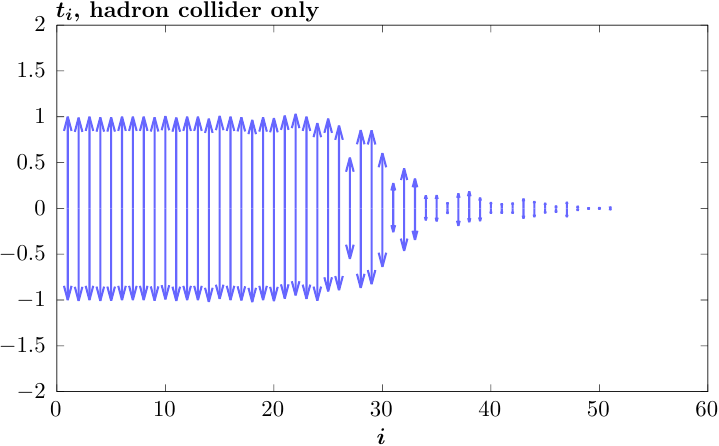}
\includegraphics[scale=0.6]{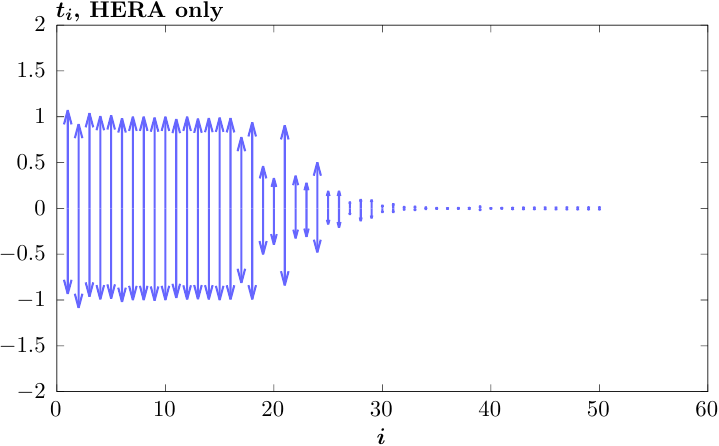}
\caption{\sf Values of the rescaled PDF eigenvector variations, $t_i$, defined in \eqref{eq:ti} for  unfluctuated closure test fits to the NNPDF4.0 global dataset, as well as the HERA only and hadron collider only subset. The $t_i$ are defined such that a $\chi^2$ variation of $T^2=1$ occurs for each eigenvector, and these are ordered by decreasing size of the eigenvalues, $\lambda_i$, defined in \eqref{eq:hesseigenvalue}. The positive/negative values correspond to the $\pm$ variations for each eigenvector.}
\label{fig:eigenvalues}
\end{center}
\end{figure}

Nonetheless, it is useful to return to this issue here, not least given the emphasis placed in this study on having a sufficiently flexible fixed parameterisation at the fit stage. We first show in Fig.~\ref{fig:eigenvalues} the values of the rescaled PDF eigenvector variations, $t_i$, defined in \eqref{eq:ti} for  unfluctuated closure test fits to the NNPDF4.0 global dataset, as well as the HERA only and hadron collider only subsets. The $t_i$ are defined such that a $\chi^2$ variation of $T^2=1$ occurs for each eigenvector, and these are ordered by decreasing size of the eigenvalues, $\lambda_i$. These are provided, in particular, without fixing any PDF parameters in the Hessian before performing the eigenvector evaluation\footnote{To be precise, these particular fits are performed with the low $x$ power of the strangeness free, but of the strangeness asymmetry fixed, and hence have 53 free parameters in total.}. In general, doing this can result in a Hessian matrix with negative eigenvalues, in which case these are discarded as they correspond effectively to directions where the global minimum has not been reached. This  does not occur in the full global case, but does for 2 (3) eigenvectors in the hadron collider (HERA) only fits. 

\begin{figure}
\begin{center}
\includegraphics[scale=0.6]{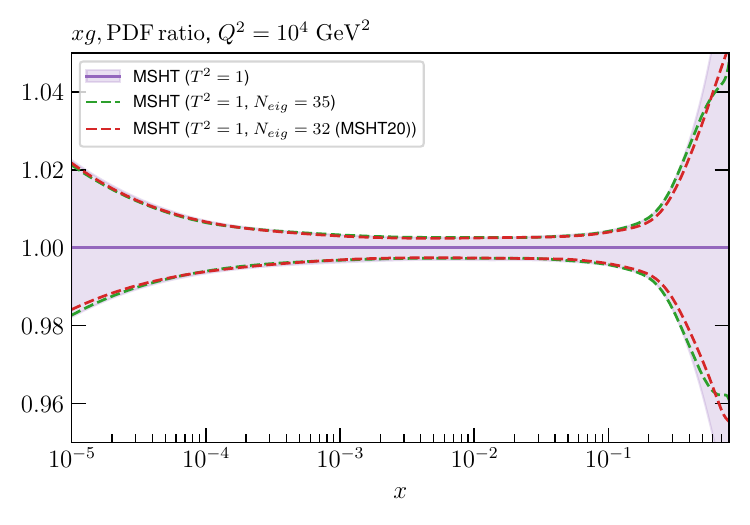}
\includegraphics[scale=0.6]{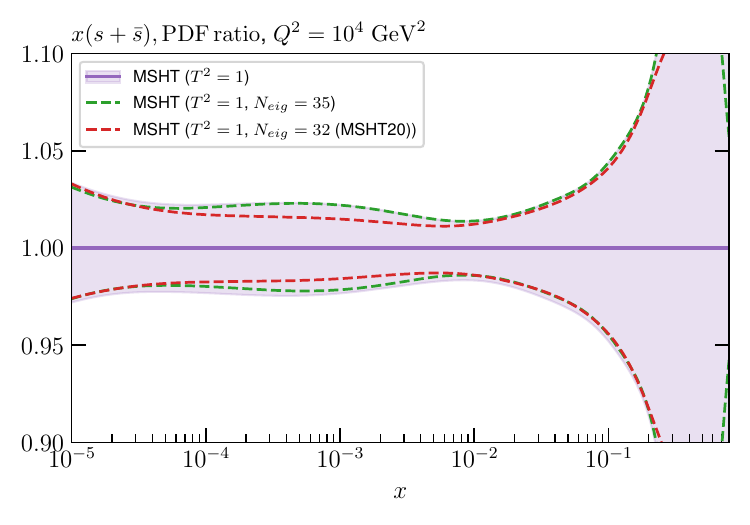}
\includegraphics[scale=0.6]{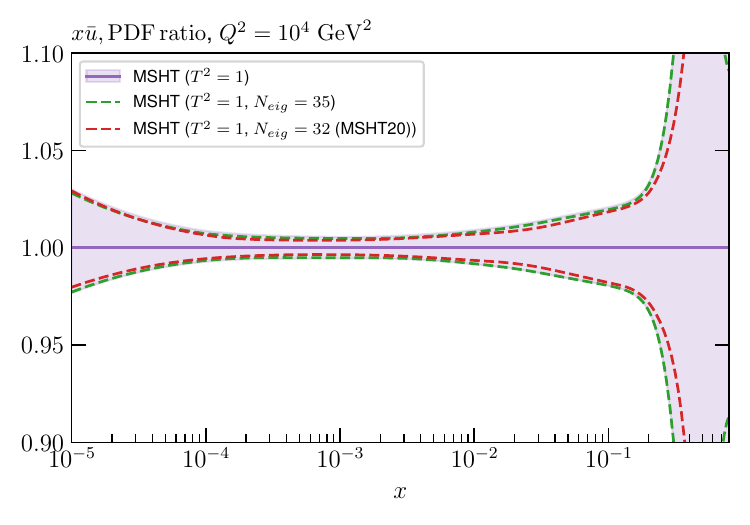}
\includegraphics[scale=0.6]{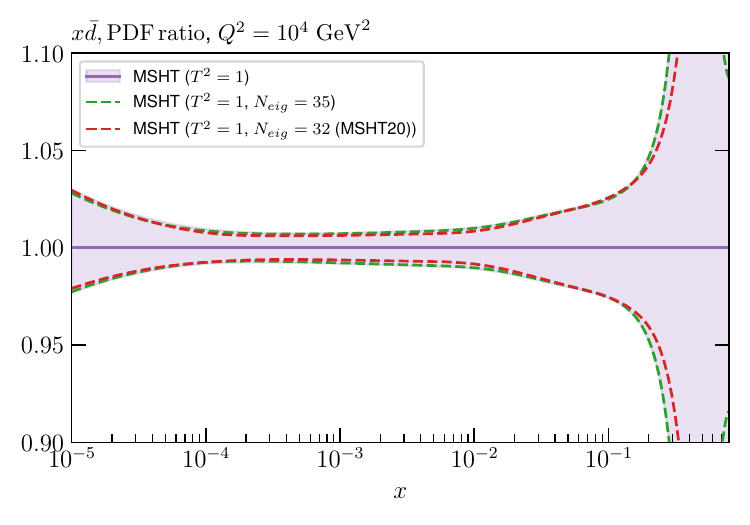}
\includegraphics[scale=0.6]{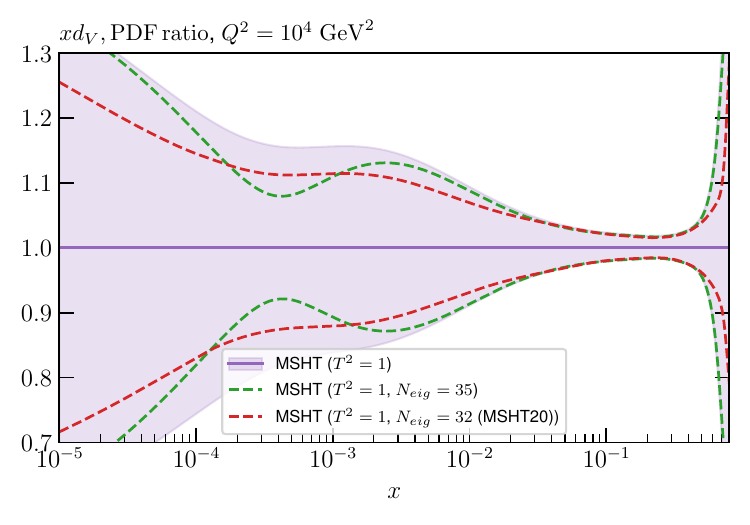}
\includegraphics[scale=0.6]{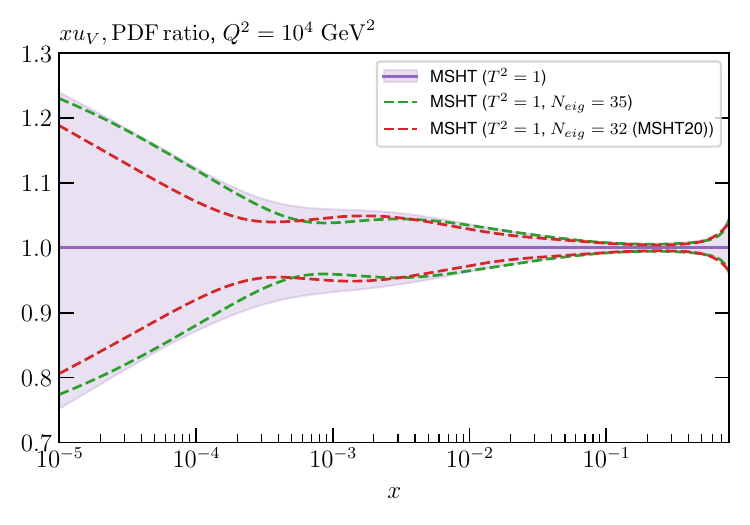}
\caption{\sf The $T^2=1$ PDF uncertainties that result from the  unfluctuated closure test fit to the NNPDF4.0 dataset, as shown in  Fig.~\ref{fig:glcl_rat}, but now also with the result of including only the first 35 eigenvectors (out of 53) shown, as ordered in decreasing size of the corresponding eigenvalue, and of fixing PDF parameters in the Hessian matrix such that only 32 are left free in the eigenvector evaluation. The latter case is indicated by the MSHT20 label.}
\label{fig:fit_neig}
\end{center}
\end{figure}

\begin{figure}
\begin{center}
\includegraphics[scale=0.6]{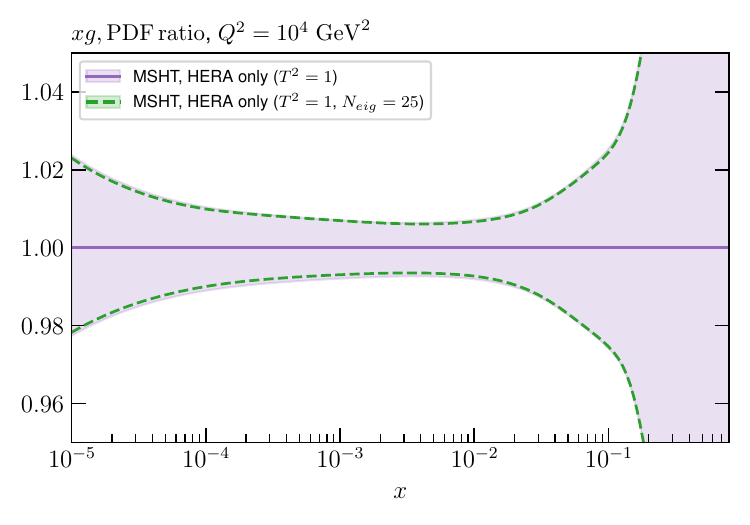}
\includegraphics[scale=0.6]{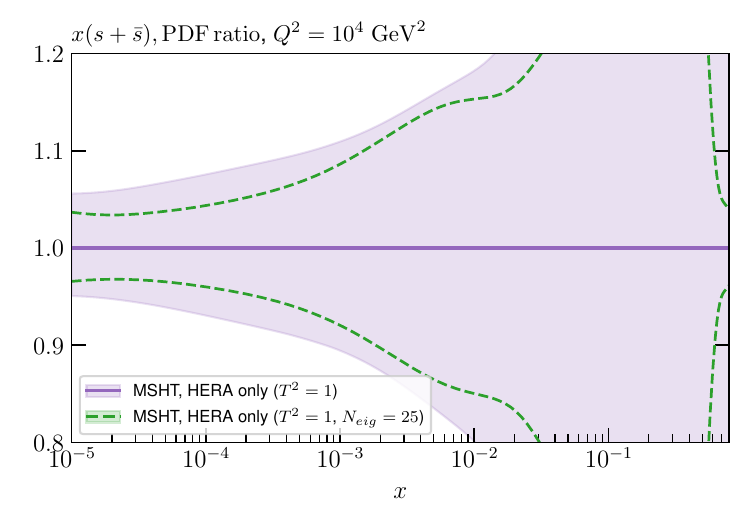}
\includegraphics[scale=0.6]{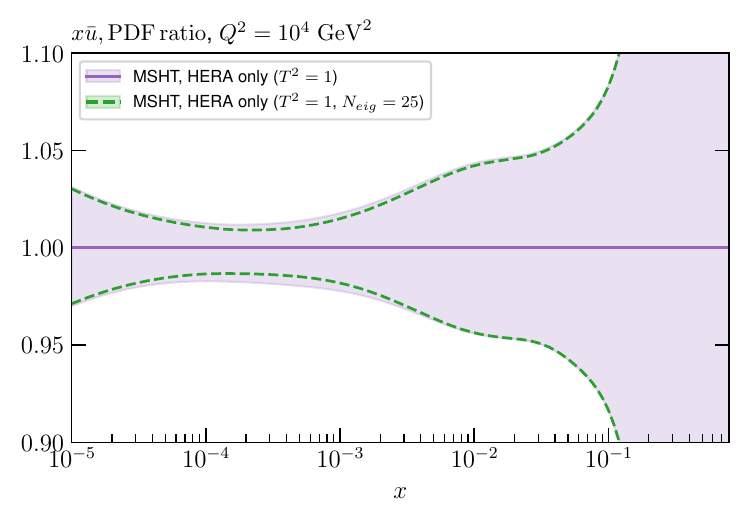}
\includegraphics[scale=0.6]{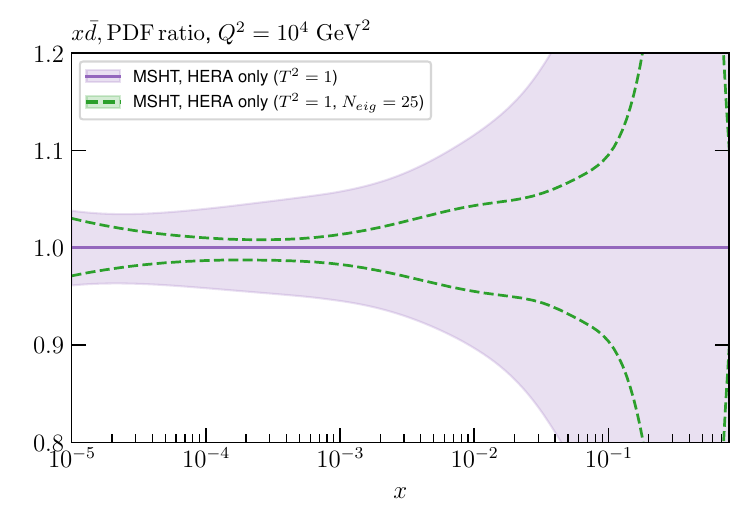}
\includegraphics[scale=0.6]{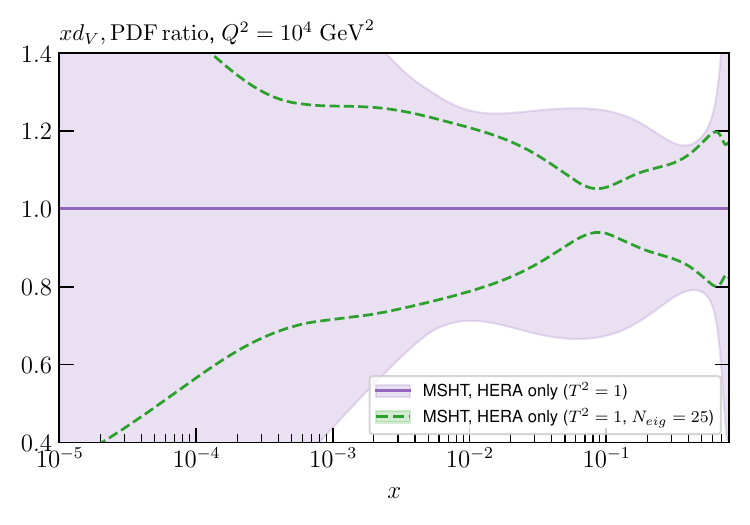}
\includegraphics[scale=0.6]{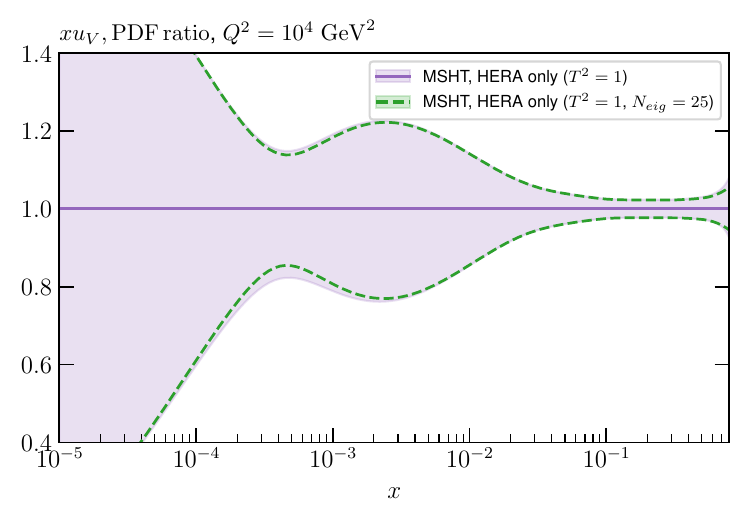}
\caption{\sf As in Fig.~\ref{fig:fit_neig} but for the HERA only fit. The result including only the first 20 eigenvectors (out of 54) is shown, while that of fixing PDF parameters is not given. Note the different $y$ scale in certain cases.}
\label{fig:fit_neig_HERAonly}
\end{center}
\end{figure}

Focusing on the global dataset shown in the top left figure, we can see that there is a noticeable trend for the first 35 eigenvectors to have $t_i \sim 1$ rather closely, indicating that these exhibit  good quadratic behaviour, but beyond this there is a rather distinct change in behaviour, with $t_i \ll 1$, indicating a higher degree of redundancy for these higher eigenvectors in the manner described above. It is notable that this is rather close to the number of eigenvectors that are produced in the MSHT20 fit~\cite{Bailey:2020ooq}, where 32 is arrived at through the process of PDF parameter fixing. For the more constrained fits, on the other hand, there is a clear trend for the onset of this $t_i \ll 1$ behaviour to occur for a lower number of eigenvectors, in particular at $\sim 20$ (30) for the HERA (hadron collider) only fits. This is of course exactly as we would expect: for these reduced fits a lower degree of parameterisation flexibility is required, and this translates into a higher degree of redundancy between PDF parameters. The negativity of certain eigenvalues for these reduced fits can be understood in the same light.

A further noticeable trend in all cases (with the exception of two eigenvectors in the HERA only fit) is that the trend with increasing eigenvalue is for $t_i$ to decrease rather close to monotonically beyond the point where $t_i \sim 1$. This is consistent with the interpretation given in~\cite{Martin:2002aw,Martin:2009iq}, namely that while  these correspond to those eigenvectors with smaller eigenvalues, such that according to \eqref{eq:eij} these are nominally less well constrained parameter directions, in reality this simple counting masks the fact that non--quadratic terms tend to dominate for these variations. As a result, the compensation between PDF parameters fails rather dramatically with only a small variation from the minimum, and the $\chi^2$ rises sharply. This results in rather low values of $t_i$. This is particularly true for the lowest eigenvalues, where the contribution to the PDF uncertainty becomes negligible, and the precise determination of $t_i$ less numerically stable.

We next turn to the impact of these higher eigenvectors, and of fixing PDF parameters in the Hessian matrix, on the PDF uncertainties themselves. This is shown in Fig.~\ref{fig:fit_neig} for the same unfluctuated global closure test discussed in Section~\ref{sec:closureun}. We consider the $T^2=1$ uncertainty for concreteness, although note that the comparison with $T^2=10$ leads to very similar conclusions. Here the $T^2=1$ uncertainty band corresponds to precisely that shown in Fig.~\ref{sec:closureun}, but we also indicate the size of the PDF uncertainty if instead of including all eigenvectors in the uncertainty evaluation only the first 35 are included, guided by the observation in Fig.~\ref{fig:eigenvalues}.

We can see that broadly the contribution to the PDF uncertainty that comes from these higher eigenvectors is  negligible, especially in the data region.  The only visible exception is in the down and up quark valence at relatively low $x$, and in certain distributions, such as the gluon, at very high $x$. Clearly, to very good approximation we could just account for the first 35 eigenvectors, for which quadratic behaviour is well observed, and get a realistic evaluation of the uncertainty in the Hessian approach in almost all cases. It is in particular encouraging that the contribution from the higher eigenvectors, where a breakdown in quadratic behaviour and the linear assumptions underlying linear error propagation is observed, is very small. 

If instead we fix the same 32 PDF parameters as in the MSHT20 fit we can see again this rather closely matches the result of including all eigenvectors, that is of not fixing any parameters. The PDF uncertainties are somewhat smaller in some cases in comparison to the 35 eigenvector results, most notably at higher $x$ and in strangeness at intermediate $x$, but these differences are relatively minor. This therefore validates the approach taken so far in the MSHT global PDF fits. We note in particular that as the underlying data being fit here and in the MSHT20 fits are far from identical, there is not strict requirement for the number of relevant eigenvector here, namely 35, to exactly match the number of parameters left free in the MSHT20 uncertainty evaluation, namely 32. In particular, a direct application of the parameter fixing approach to this differing dataset may well lead, upon inspection of the corresponding Hessian matrix and the correlations between different PDF parameters, to the fixing of fewer parameters, more closely in line with the 35 eigenvector result. 

We note that the above trend is not guaranteed to occur in all fits, and is in particular a feature of a more global PDF fit where all PDFs in the plotted regions are relatively well constrained by data. To demonstrate this, in Fig.~\ref{fig:fit_neig_HERAonly} we show the equivalent comparison for the HERA only fit, but where the number of eigenvectors is limited to 25 eigenvectors, again guided by Fig.~\ref{fig:eigenvalues}. For some PDFs in the plotted region, notably the gluon and up quark sector, there is again a close matching between the restricted and full eigenvector uncertainties. However for the strangeness and down quark sector there is a clear trend for the 25 eigenvector uncertainty to lie below the full result, most notably for the $\overline{d}$. These are the PDF combinations that are least well constrained by HERA data alone. In this case, some of the higher eigenvectors will be particularly sensitive to these PDF combinations, with the lower eigenvalues genuinely indicating that these PDF directions are rather poorly constrained by the data. By removing these, or simply restricting the parameterisation in the first place~\cite{H1:2015ubc}, we are therefore liable to underestimate the true uncertainty on these PDFs. We note that by including 25 eigenvectors, we are already accounting for some cases where the $t\sim T$ equivalence has broken down significantly, and hence if we took a more restrictive choice of e.g. 15 eigenvectors, demanding that the quadratic $t\sim T$ behaviour is always obeyed, then the corresponding PDF uncertainty would be smaller still in some cases.

Finally, we emphasise for concreteness that in all results shown in this paper, we include all (positive) eigenvectors in the PDF uncertainty evaluation. However, as shown above we could take an approach that more closely matches that taken in the MSHT20 family of fits, with little impact on the overall comparisons.

\section{Global Closure Tests with Inconsistent Inputs: Additional Results}\label{app:inconsistentadd}

In this section we show equivalent plots to Figs.~\ref{fig:tol_g}--~\ref{fig:tol_sigmachw} but for a different set of pseudodata fluctuations. While the specific results are of course different, the most important trends, namely the incompatibility between the subset fits with inconsistent inputs (in the up quark and charge weighted quark singlet at intermediate to high $x$ and in the gluon at high $x$) and the lack of overlap with the global fit with $T^2=1$ uncertainties, is clear.

\begin{figure}
\begin{center}
\includegraphics[trim={0 0.3cm 0 0.3cm},clip,scale=0.6]{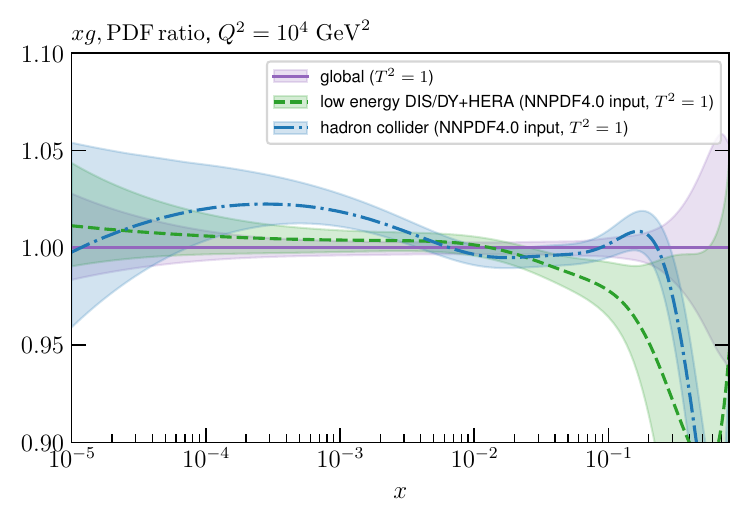}
\includegraphics[trim={0 0.3cm 0 0.3cm},clip,scale=0.6]{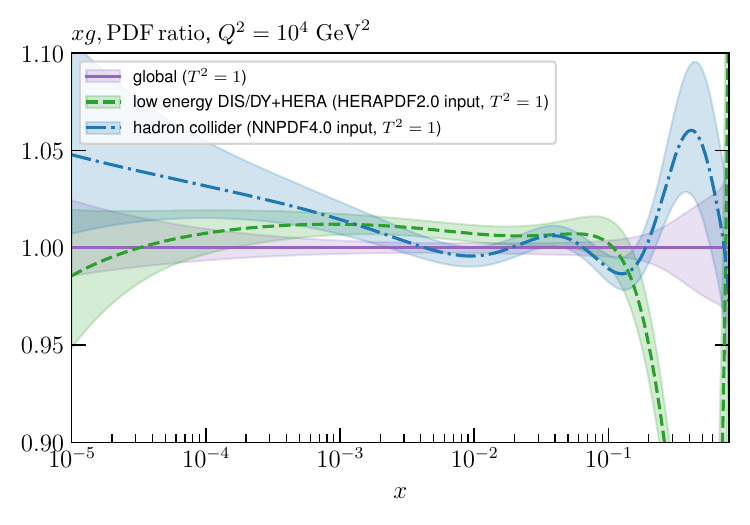}
\includegraphics[trim={0 0.3cm 0 0.3cm},clip,scale=0.6]{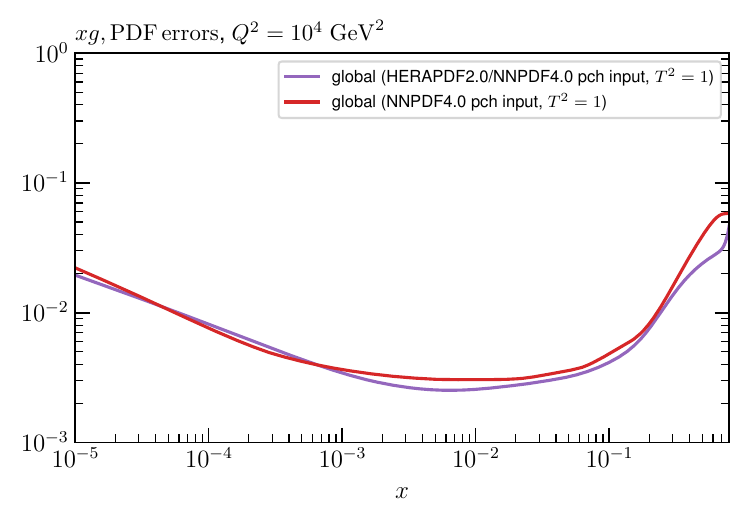}
\includegraphics[trim={0 0.3cm 0 0.3cm},clip,scale=0.6]{Figs/pdfplot-rat-L1_inputcomp2_g}
\caption{\sf As in Fig.~\ref{fig:tol_g} but for a different set of pseudodata fluctuations. The global fit result applying the dynamic tolerance is not shown here for simplicity, while the plot of PDF ratios is exactly as in Fig.~\ref{fig:tol_g}.}
\label{fig:tol_g_add}
\end{center}
\end{figure}

\begin{figure}
\begin{center}
\includegraphics[trim={0 0.3cm 0 0.3cm},clip,scale=0.6]{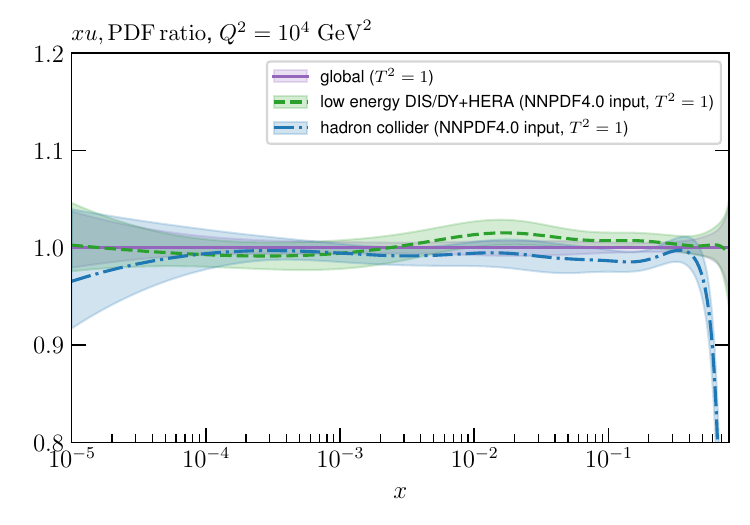}
\includegraphics[trim={0 0.3cm 0 0.3cm},clip,scale=0.6]{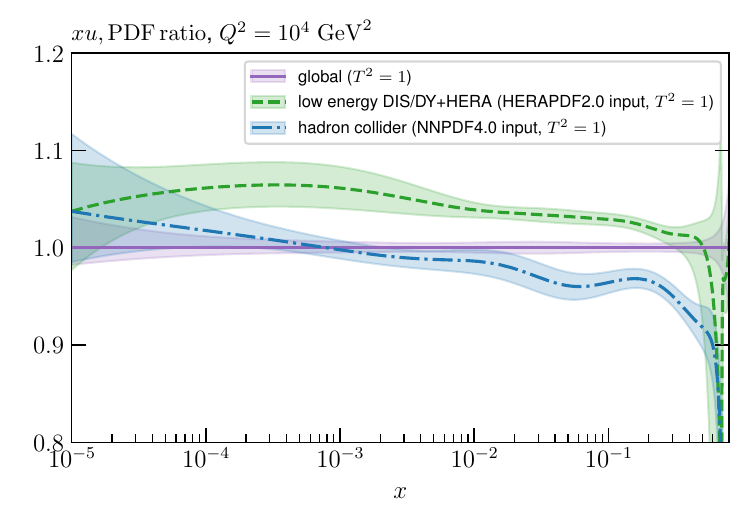}
\includegraphics[trim={0 0.3cm 0 0.3cm},clip,scale=0.6]{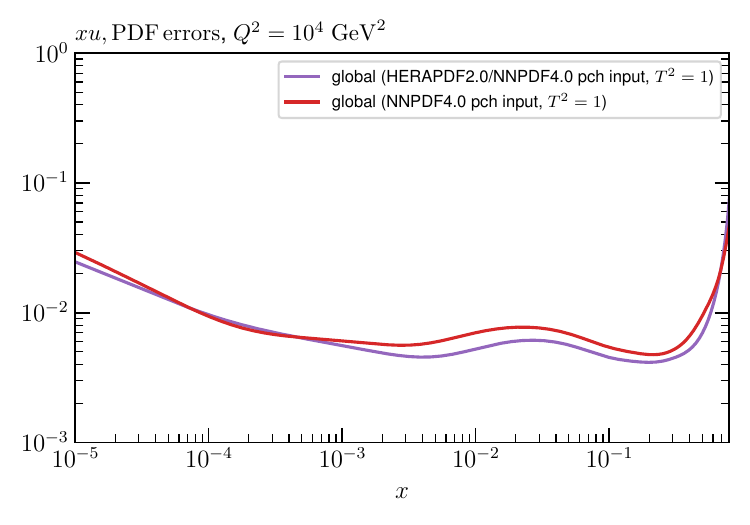}
\includegraphics[trim={0 0.3cm 0 0.3cm},clip,scale=0.6]{Figs/pdfplot-rat-L1_inputcomp2_u}
\caption{\sf As in Fig.~\ref{fig:tol_g_add} but for the up quark.}
\label{fig:tol_u_add}
\end{center}
\end{figure}

\begin{figure}
\begin{center}
\includegraphics[trim={0 0.3cm 0 0.3cm},clip,scale=0.6]{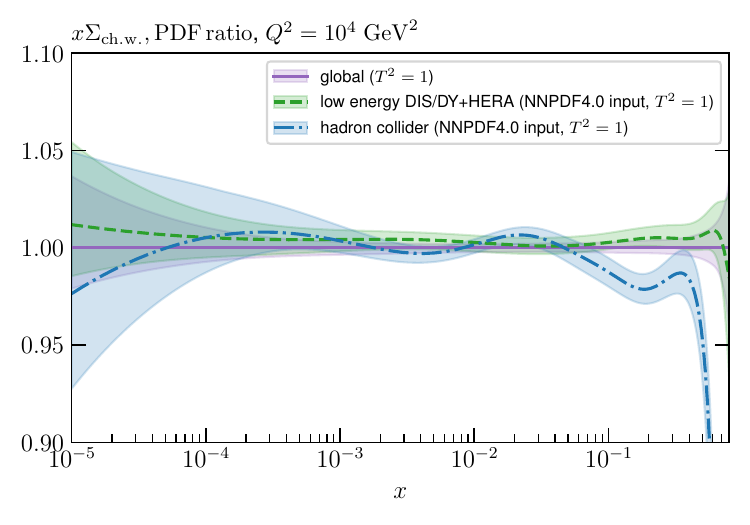}
\includegraphics[trim={0 0.3cm 0 0.3cm},clip,scale=0.6]{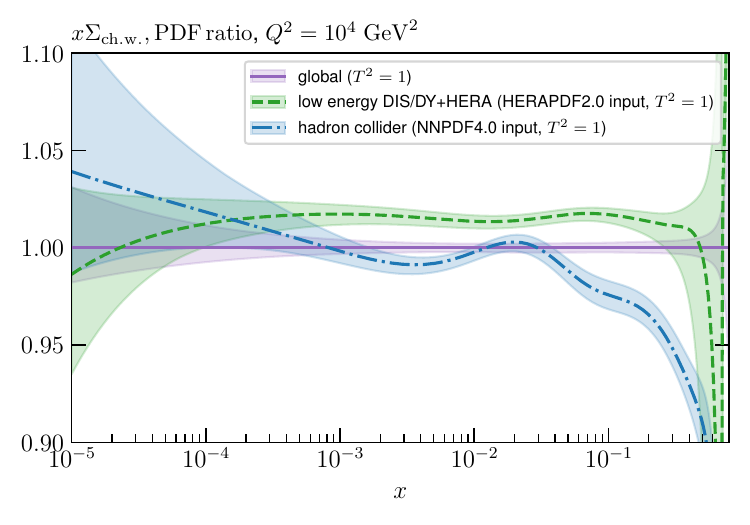}
\includegraphics[trim={0 0.3cm 0 0.3cm},clip,scale=0.6]{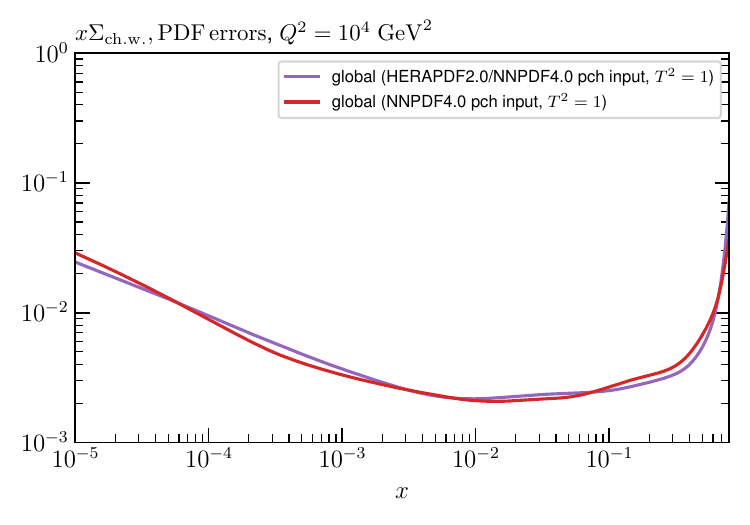}
\includegraphics[trim={0 0.3cm 0 0.3cm},clip,scale=0.6]{Figs/pdfplot-rat-L1_inputcomp2_sigamchw}
\caption{\sf As in Fig.~\ref{fig:tol_g_add} but for the charge weighted quark singlet.}
\label{fig:tol_sigmachw_add}
\end{center}
\end{figure}

\newpage

\section{Role of Integrability}\label{app:integrability}

In this appendix we discuss the impact of imposing integrability on the MSHT fits, as is done in the case of the NNPDF fits, see~\cite{NNPDF:2021njg}. In principle, this corresponds to requiring that 
\be
\lim_{x\to 0} xf(x,Q^2)\to 0 \;,
\ee
for all $Q^2$ for the relevant PDF combinations that we will consider below. In practice this is imposed by NNPDF at $Q^2=5$ ${\rm GeV^2}$ at certain low $x$ points $x_i \in \{10^{-9},10^{-8},10^{-7}\}$, with a $\chi^2$ penalty imposed to suppress larger values of $x_i f(x_i,Q^2)$. In the MSHT fit this can instead be imposed directly at the level of the PDF parameterisation (albeit at the somewhat lower value of the input scale $Q_0$). The integrability of the valence distributions is at low enough $x$  is guaranteed by the relevant sum rules, while numerically we have confirmed they satisfy the integrability criteria imposed by NNPDF very well. The remaining distributions where this is not by default imposed in the MSHT fits are for
\begin{align}
T_3 &= (u+\overline{u})-(d+\overline{d})\;,\\
T_8 &= (u+\overline{u} + d+\overline{d}) - 2(s+\overline{s})\;.
\end{align}
In the former case this would correspond to requiring that $\overline{d}/\overline{u} \to 1$ at low $x$, which while in general rather well satisfied is not required; this could be readily imposed by fixing the normalization of the $\overline{d}/\overline{u}$ distribution in terms of the coefficients and the known low $x$ behaviour of the Chebyshev polynomials. For $T_8$ while the low $x$ power of the strangeness is fixed to that of the sea, this is not sufficient to ensure that $T_8$ vanishes at low $x$, but this can be achieved by suitably fixing the normalization of the strangeness in terms of the normalization of the sea, and again the coefficients and the known low $x$ behaviour of the  Chebyshev polynomials in these distributions.

\begin{figure}
\begin{center}
\includegraphics[scale=0.6]{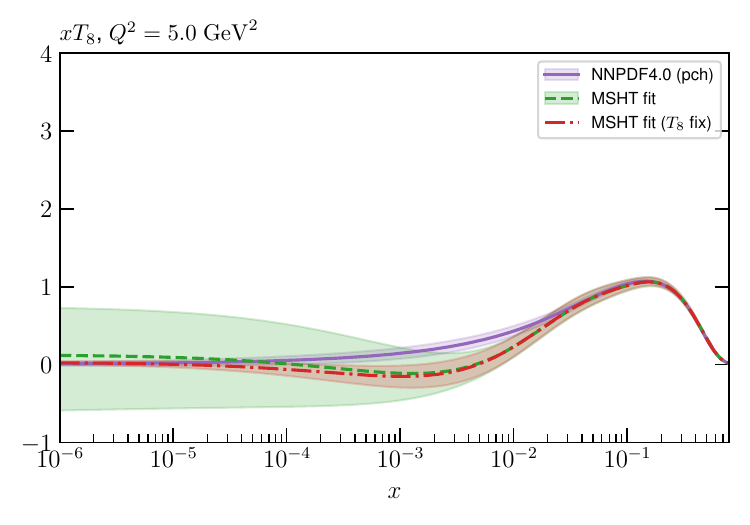}
\includegraphics[scale=0.6]{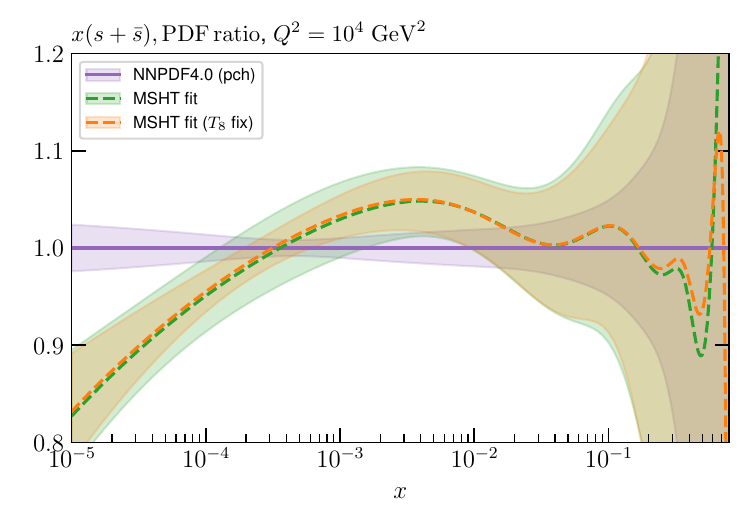}
\caption{\sf (Left) The $T_8$ distribution, defined in the text, at $Q^2=5$ ${\rm GeV}^2$, result from a MSHT fits to the NNPDF dataset/theory settings, with perturbative charm, with the low $x$ behaviour allowed to be free (the default), or fixed to vanish as in the NNPDF fit. (Right) The impact of fixing the low $x$ behaviour of the $T_8$ distribution on the strangeness, which is only distribution that shows any noticeable change, at  $Q^2=100$ ${\rm GeV}^2$. The NNPDF (pch) result is given in both cases for comparison and the MSHT uncertainties correspond to a fixed $T^2=10$ tolerance.}
\label{fig:t8fix_pch}
\end{center}
\end{figure}

\begin{figure}
\begin{center}
\includegraphics[scale=0.6]{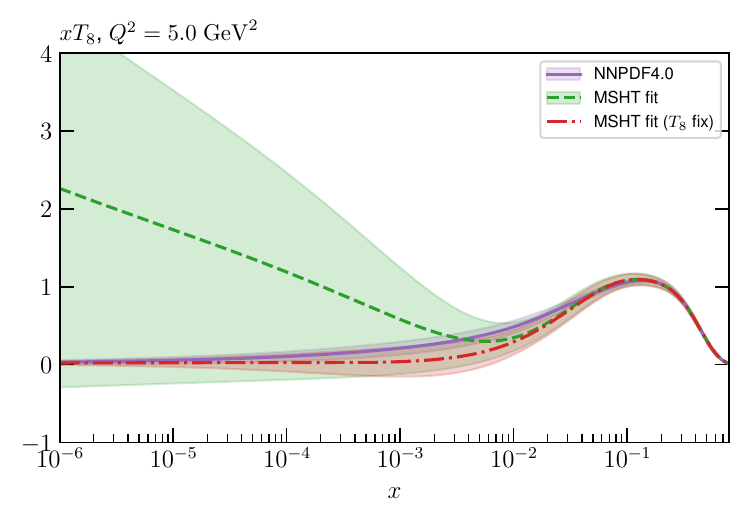}
\includegraphics[scale=0.6]{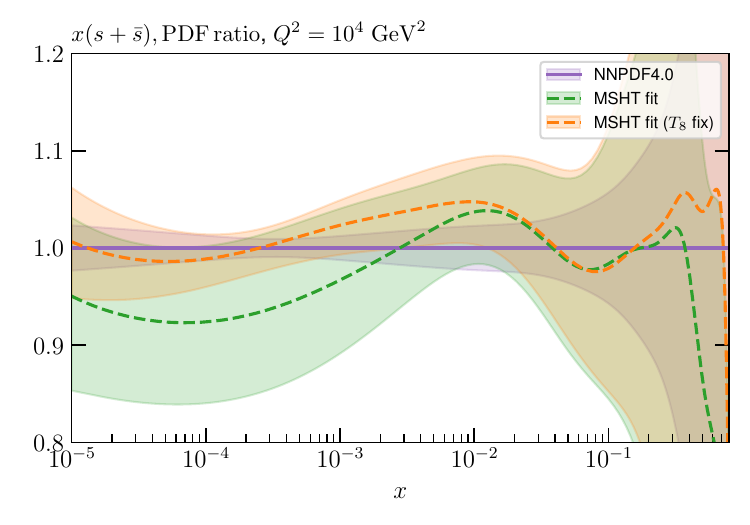}
\includegraphics[scale=0.6]{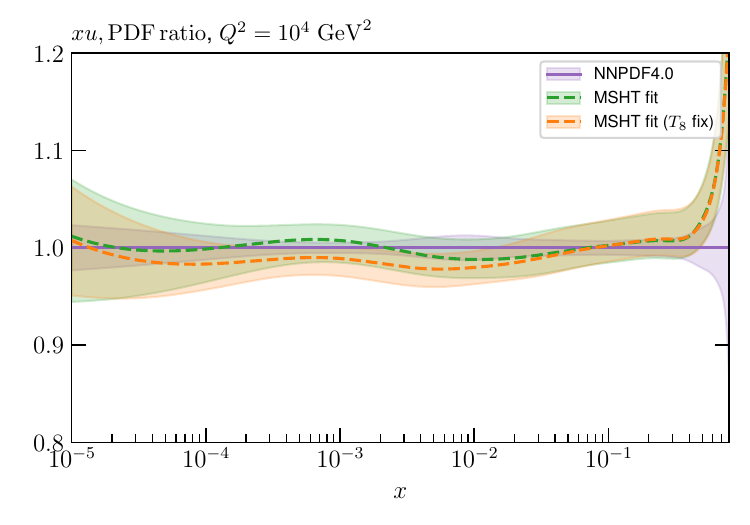}
\includegraphics[scale=0.6]{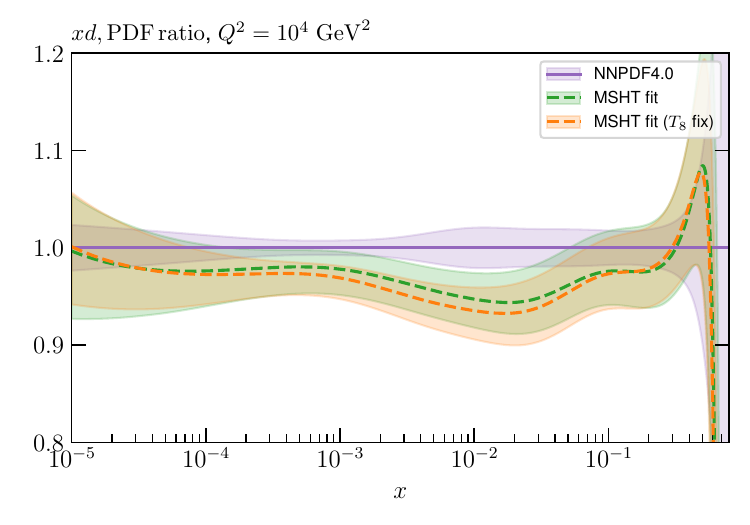}
\caption{\sf (Top) As in Fig.~\ref{fig:t8fix_pch} but for the fitted charm case. As well as the impact of fixing the low $x$ behaviour of the $T_8$ distribution on the strangeness at  $Q^2=100$ ${\rm GeV}^2$, the impact on the up and down quark distributions is shown.}
\label{fig:t8fix_fch}
\end{center}
\end{figure}

In general, as observed in\cite{NNPDF:2021njg} for the case of the MSHT20 fit, the central values of the $T_{3,8}$ distributions do not vanish as $x$ approaches zero, but these are consistent with zero within uncertainties. This is as we might hope for if there are general expectations for these to vanish, but this is not imposed directly, i.e. the fit itself prefers distributions that are consistent with zero. It is not the purpose of the current discussion to consider in detail the merits of imposing the constraints at the level of the fit, however we do note that in the case of $T_8$ in particular the larger strange quark mass is expected to lead to some deviation in this behaviour, as is clearly true when considering e.g. the rather more massive charm quark distribution, especially at the rather stringent level imposed by NNPDF. Hence it is an open question as to what extent it is sensible to impose this integrability, rather than observe to what extent it is (or is not) preferred by the fit. However, we note that not imposing this will clearly lead to more conservative PDF uncertainties in the low $x$ region.

Here, we focus for simplicity on the $T_8$ distribution, for which the MSHT fits tend to show larger deviations from zero at low $x$ than the $T_3$. Fixing the strangeness normalization so that this integrability is imposed at low $x$, we find that the fit quality in the perturbative charm case is almost indistinguishable from the case that this is allowed to be free, while for the fitted charm case the fit quality deteriorates by $\sim 3$ points. That is, in both fits the impact of this is very minor indeed on the fit quality. The impact on the PDFs is shown in Figs.~\ref{fig:t8fix_pch} and~\ref{fig:t8fix_fch}, both on $T_8$ at  $Q^2=5$ ${\rm GeV^2}$ but also on any relevant distributions where some change is seen at higher scale  $Q^2=100$ ${\rm GeV^2}$.  We can see that a non-zero value of $T_8$ is indeed permitted by both fits at $Q^2=5$ ${\rm GeV^2}$, but that these are consistent with zero within the representative $T^2=10$ uncertainties. 

Imposing integrability of $T_8$, in the perturbative charm case, we can see that $T_8$ vanishes at low $x$ as required. The impact on the PDFs in the data region is very small, consistent with the smaller impact on the fit quality. The strangeness distribution exhibits the only noticeable change, must notably at the level of the PDF uncertainties, which are rather smaller due to the additional constraint imposed. For the fitted charm case, the deviation of  $T_8$ is somewhat larger, and the impact of imposing integrability correspondingly a little more significant, consistent with the somewhat larger (though still small) impact on the fit quality. The impact on the low $x$ strangeness is clearly non--negligible, and brings the result closer in line with the NNPDF fit, with a notably smaller uncertainty. For the $u,d$ PDFs the central values at intermediate to low $x$ in fact lie somewhat further away from the NNPDF baseline. However all changes are within the MSHT uncertainties. We note however that integrability is imposed at the input scale $Q_0$, i.e. slightly lower than in the NNPDF case, and so this may enlarge its impact somewhat.

In summary, the impact of imposing integrability of $T_8$ at the level of the fit quality is very mild and in general small at the level of the PDFs. The biggest exception is for the strangeness in the fitted charm case, where imposing integrability has a clear impact on the PDF uncertainty at low $x$.

\section{Fit Quality: Perturbative vs. Fitted Charm}\label{app:fch_vs_pch}

In this appendix we compare the fit qualities between the  fits with purely perturbative  and fitted charm. In Tables~\ref{tab:chi2_fcharm_pcharm_nopos} and~\ref{tab:chi2_fcharm_pcharm_wpos} we show results for the MSHT fits to the NNPDF dataset/theory, with and without positivity imposed, respectively. In Table~\ref{tab:chi2_fcharm_pcharm_NNPDF} we show results for the default NNPDF4.0 fits.

\begin{table}[H]
\begin{center}
  \scriptsize
  \centering
   \renewcommand{\arraystretch}{1.4}
\begin{tabular}{Xrcc}\hline 
&Perturbative Charm&Fitted Charm
\\ \hline
SLAC $F_2^{p}$ (33)~\cite{Whitlow:1991uw}  & 25.4 (0.77) & \textcolor{red}{31.2 (0.94)} \\
BCDMS $F_2^{p}$ (333)~\cite{BCDMS:1989qop}  & 471.6 (1.42) & \textcolor{blue}{451.8 (1.36)} \\
BCDMS $F_2^{d}$ (248)~\cite{BCDMS:1989qop}  & 262.9 (1.06) & \textcolor{blue}{251.6 (1.01)} \\
{\bf DIS Fixed--Target (1881)} & 2029.2 (1.08) &\textcolor{blue}{{\bf  2018.6 (1.07)}} 
\\ \hline
E886 $\sigma^{p}$ (NuSea) (89)~\cite{NuSea:2003qoe}& 120.7 (1.36) & \textcolor{blue}{112.5 (1.26)} \\
{\bf DY Fixed--Target (195)} & {\bf 201.8 (1.04)} &\textcolor{black}{{\bf 192.1 (0.99)}} 
\\ \hline
NC $e^+ p$ 920 GeV (377)~\cite{H1:2015ubc} & 536.1 (1.42) & \textcolor{blue}{506.0 (1.34)} \\
NC, c (37)~\cite{H1:2018flt} & 117.7 (3.18) & \textcolor{blue}{82.7 (2.24)} \\
{\bf HERA DIS (1145/1135)} & 1650.2 (1.44) &\textcolor{blue}{{\bf  1557.6 (1.36)}} 
\\ \hline
LHCb $W,Z\to \mu$ 7 TeV (29)~\cite{LHCb:2015okr} & 40.6 (1.40) & \textcolor{red}{51.8 (1.78)} \\
LHCb $W,Z\to \mu$ 8 TeV (30)~\cite{LHCb:2016fbk} & 34.3 (1.14) & \textcolor{red}{41.1 (1.37)} \\
{\bf Collider DY (576)} &{\bf 727.5 (1.26)} &\textcolor{black}{{\bf 743.1 (1.29)}} 
\\
\hline
CMS incl. jets 8 TeV (185)~\cite{CMS:2016lna} & 353.2 (1.91) & \textcolor{blue}{332.7 (1.80) }\\
{\bf LHC Jets (500)} & {\bf 813.2 (1.63)} &\textcolor{blue}{{\bf 796.7 (1.59)}} \\
\hline
{\bf LHC $V+$ Jets (122)} & {\bf 137.7 (1.13)} &\textcolor{black}{{\bf 140.4 (1.15)}} \\ \hline
{\bf Isolated Photon (53)} & {\bf 41.3 (0.78)} &\textcolor{black}{{\bf 40.5 (0.76)}} \\ \hline
ATLAS $t\overline{t}$ $l+$ jets 8 TeV (8)~\cite{ATLAS:2015lsn} &23.8 (2.97) & \textcolor{red}{30.6 (3.82)}\\
{\bf Top quark  (81)} & {\bf 82.0 (1.01)} &\textcolor{black}{{\bf 87.4 (1.08)}} 
\\ \hline \hline 
{\bf Global, $t_0$ (4616)}    &{\bf 5736.7 (1.240)}& {\bf 5645.2 (1.222)} \\
\hline
{\bf Global, exp. (4616)} & {\bf 5374.3 (1.164)}& {\bf 5322.5 (1.153)}\\
\hline
\end{tabular}
\end{center}
\caption{\sf $\chi^2$ values for the MSHT fits to the NNPDF dataset/theory settings, without positivity imposed, and with and without fitted charm. The fit quality for the different major subsets the constitute the global dataset are given in bold, and  above each subtotal the fit qualities for individual experiments in theses subsets where there difference with respect to the perturbative charm case is roughly larger than $\pm 0.5 \sigma = \sqrt{N_{\rm pts}/2}$. When these differences are less than $-0.5\sigma$ the result is highlighted in blue, while the result is highlighted in red when it is greater than $0.5\sigma$. Both the absolute $\chi^2$ and the per point value in brackets, is given in all cases, while the number of points is indicated in brackets next to the dataset description.  For the total $\chi^2$ both the experimental and $t_0$ definitions are shown, while in all other cases only the latter definition is used. For the sake of comparison the same cuts as in the fitted charm case are applied for both fits.}
\label{tab:chi2_fcharm_pcharm_nopos}
\end{table}

\begin{table}[H]
\begin{center}
  \scriptsize
  \centering
   \renewcommand{\arraystretch}{1.4}
\begin{tabular}{Xrcc}\hline 
&Perturbative Charm&Fitted Charm
\\ \hline
NMC $\sigma^{{\rm NC,} p}$ (204)~\cite{NewMuon:1996fwh}  & 337.2 (1.65) & \textcolor{blue}{313.2 (1.54)}\\
SLAC $F_2^{p}$ (33)~\cite{Whitlow:1991uw} & 25.0 (0.76) & \textcolor{red}{31.8 (0.96)} \\
BCDMS $F_2^{p}$ (333)~\cite{BCDMS:1989qop}  & 483.2 (1.45) & \textcolor{blue}{453.9 (1.36)} \\
BCDMS $F_2^{d}$ (248)~\cite{BCDMS:1989qop}  & 260.9 (1.05) & \textcolor{blue}{250.1 (1.01)} \\
{\bf DIS Fixed--Target (1881)} & 2063.3 (1.10) &\textcolor{blue}{{\bf  2015.3 (1.07)}} 
\\ \hline
E886 $\sigma^{p}$ (NuSea) (89)~\cite{NuSea:2003qoe} & 118.4 (1.33) & \textcolor{blue}{110.0 (1.24)} \\
{\bf DY Fixed--Target (195)} & {\bf 198.6 (1.02)} &\textcolor{black}{{\bf 190.2 (0.98)}} 
\\ \hline
NC $e^- p$ 575 GeV (254)~\cite{H1:2015ubc} & 259.0 (1.02) & \textcolor{blue}{247.2 (0.97)} \\
NC $e^+ p$ 920 GeV (377)~\cite{H1:2015ubc} & 574.1 (1.52) & \textcolor{blue}{506.0 (1.34)} \\
NC, c (37)~\cite{H1:2018flt} & 109.8 (2.97) & \textcolor{blue}{91.1 (2.46)} \\
NC, b (26)~\cite{H1:2018flt} & 67.9 (2.61) & \textcolor{blue}{60.5 (2.33)} \\
{\bf HERA DIS (1145/1135)} & 1671.5 (1.46) &\textcolor{blue}{{\bf  1565.8 (1.37)}} 
\\ \hline
LHCb $W,Z\to \mu$ 7 TeV (29)~\cite{LHCb:2015okr} & 43.9 (1.28) & \textcolor{red}{53.6 (1.85)} \\
LHCb $W,Z\to \mu$ 8 TeV (30)~\cite{LHCb:2016fbk}  & 36.1 (1.20) & \textcolor{red}{43.1 (1.44)} \\
{\bf Collider DY (576)} &{\bf 741.6 (1.29)} &\textcolor{black}{{\bf 754.4 (1.31)}} 
\\
\hline
CMS incl. jets 8 TeV (185)~\cite{CMS:2016lna} & 354.6 (1.92) & \textcolor{blue}{334.7 (1.81) }\\
{\bf LHC Jets (500)} & {\bf 826.5 (1.65)} &\textcolor{blue}{{\bf 797.6 (1.60)}} \\
\hline
{\bf LHC $V+$ Jets (122)} & {\bf 139.5 (1.14)} &\textcolor{black}{{\bf 139.2 (1.14)}} \\ \hline
{\bf Isolated Photon (53)} & {\bf 40.7 (0.77)} &\textcolor{black}{{\bf 40.6 (0.77)}} \\ \hline
ATLAS $t\overline{t}$ $l+$ jets 8 TeV (8)~\cite{ATLAS:2015lsn} &23.8 (2.97) & \textcolor{red}{30.6 (3.82)}\\
{\bf Top quark  (81)} & {\bf 83.9 (1.04)} &\textcolor{black}{{\bf 83.0 (1.02)}} 
\\ \hline \hline 
{\bf Global, $t_0$ (4616)}    &{\bf 5724.0 (1.240)}& {\bf 5651.0 (1.224)}\\
\hline
{\bf Global, exp. (4616)} &{\bf  5463.1 (1.183)}& {\bf 5341.5 (1.155)} \\
\hline
\end{tabular}
\end{center}
\caption{\sf $\chi^2$ values for the MSHT fits to the NNPDF dataset/theory settings, with positivity imposed, and with and without fitted charm. The fit quality for the different major subsets the constitute the global dataset are given in bold, and  above each subtotal the fit qualities for individual experiments in these subsets where the difference with respect to the perturbative charm case is roughly larger than $\pm 0.5 \sigma = \sqrt{N_{\rm pts}/2}$. When these differences are less than $-0.5\sigma$ the result is highlighted in blue, while the result is highlighted in red when it is greater than $0.5\sigma$. Both the absolute $\chi^2$ and the per point value in brackets, is given in all cases, while the number of points is indicated in brackets next to the dataset description.  For the total $\chi^2$ both the experimental and $t_0$ definitions are shown, while in all other cases only the latter definition is used. For the sake of comparison the same cuts as in the fitted charm case are applied for both fits.}
\label{tab:chi2_fcharm_pcharm_wpos}
\end{table}

\begin{table}[H]
\begin{center}
  \scriptsize
  \centering
   \renewcommand{\arraystretch}{1.4}
\begin{tabular}{Xrcc}\hline 
&NNPDF4.0 (pch)& NNPDF4.0 
\\ \hline
NMC $\sigma^{{\rm NC,} p}$ (204)~\cite{NewMuon:1996fwh}  & 349.2 (1.71) & \textcolor{blue}{311.0 (1.52)}\\
BCDMS $F_2^{p}$ (333)~\cite{BCDMS:1989qop}  & 497.6 (1.49) & \textcolor{blue}{473.6 (1.42)} \\
BCDMS $F_2^{d}$ (248)~\cite{BCDMS:1989qop}  & 263.6 (1.06) & \textcolor{blue}{252.6 (1.02)} \\
NuTeV $\sigma^{\overline{\nu}}_{\rm CC}$ (37)~\cite{NuTeV:2001dfo} & 28.2 (0.76) & \textcolor{blue}{21.1 (0.57)}\\
{\bf DIS Fixed--Target (1881)} & 2076.1 (1.10) &\textcolor{blue}{{\bf  2011.6 (1.07)}} 
\\ \hline
{\bf DY Fixed--Target (195)} & {\bf 190.4 (0.98)} &\textcolor{black}{{\bf 185.6 (0.95)}} 
\\ \hline
NC $e^- p$ 575 GeV (254)~\cite{H1:2015ubc} & 265.5 (1.05) & \textcolor{blue}{250.6 (0.99)} \\
NC $e^+ p$ 920 GeV (377)~\cite{H1:2015ubc} & 576.5 (1.53) & \textcolor{blue}{518.6 (1.38)} \\
NC, c (37)~\cite{H1:2018flt} & 114.7 (3.10) & \textcolor{blue}{82.8 (2.24)} \\
NC, b (26)~\cite{H1:2018flt} & 71.0 (2.73) & \textcolor{blue}{62.0 (2.38)} \\
{\bf HERA DIS (1145/1135)} & 1698.4 (1.48) &\textcolor{blue}{{\bf  1575.6 (1.38)}} 
\\ \hline
ATLAS $W,Z$ 7 TeV ($\mathcal{L}=4.6\,{\rm fb}^{-1}$) (61)~\cite{ATLAS:2016nqi} & 119.9 (1.97) & \textcolor{blue}{104.5 (1.71)} \\
CMS DY 2D 7 TeV (110) & 156.3 (1.42) & \textcolor{blue}{146.2 (1.33)}\\
LHCb $Z\to ee$ (17)~\cite{LHCb:2012gii} & 28.8 (1.70) & \textcolor{blue}{22.4 (1.32)}\\
LHCb $W,Z\to \mu$ 7 TeV (29)~\cite{LHCb:2015okr} & 45.7 (1.58) & \textcolor{red}{56.3 (1.94)} \\
LHCb $Z\to \mu\mu$ 13 TeV (16)~\cite{LHCb:2016fbk} & 22.7 (1.42) & \textcolor{blue}{17.2 (1.08)} \\
LHCb $Z\to ee$ 13 TeV (15)~\cite{LHCb:2016fbk} & 28.8 (1.92) & \textcolor{blue}{24.6 (1.64)}\\
{\bf Collider DY (576)} &{\bf 794.7 (1.38)} &\textcolor{blue}{{\bf 767.8 (1.33)}} 
\\
\hline
CMS incl. jets 8 TeV (185)~\cite{CMS:2016lna} & 348.2 (1.88) & \textcolor{blue}{330.8 (1.79) }\\
{\bf LHC Jets (500)} & {\bf 823.9 (1.65)} &\textcolor{blue}{{\bf 804.8 (1.61)}} \\
\hline
{\bf LHC $V+$ Jets (122)} & {\bf 136.6 (1.12)} &\textcolor{black}{{\bf 136.1 (1.12)}} \\ \hline
{\bf Isolated Photon (53)} & {\bf 39.3 (0.74)} &\textcolor{black}{{\bf 41.9 (0.79)}} \\ \hline
{\bf Top quark  (81)} & {\bf 82.7 (1.02)} &\textcolor{black}{{\bf 85.0 (1.05)}} 
\\ \hline \hline 
{\bf Global, $t_0$ (4616)}    &{\bf 5917.1 (1.282)}& {\bf 5692.1 (1.233)}\\
\hline
{\bf Global, exp. (4616)} & {\bf 5538.2 (1.200)}&{\bf  5354.1 (1.160)}\\
\hline
\end{tabular}
\end{center}
\caption{\sf $\chi^2$ values for the default NNPDF fits and with and without fitted charm. The fit quality for the different major subsets the constitute the global dataset are given in bold, and  above each subtotal the fit qualities for individual experiments in these subsets where the difference with respect to the perturbative charm case is roughly larger than $\pm 0.5 \sigma = \sqrt{N_{\rm pts}/2}$. When these differences are less than $-0.5\sigma$ the result is highlighted in blue, while the result is highlighted in red when it is greater than $0.5\sigma$. Both the absolute $\chi^2$ and the per point value in brackets, is given in all cases, while the number of points is indicated in brackets next to the dataset description.  For the total $\chi^2$ both the experimental and $t_0$ definitions are shown, while in all other cases only the latter definition is used. For the sake of comparison the same cuts as in the Fitted charm case are applied for both fits.}
\label{tab:chi2_fcharm_pcharm_NNPDF}
\end{table}

\newpage

\section{Further PDF Comparisons}\label{app:PDFs}

\subsection{MSHT Fits with Perturbative Charm: Comparison to MSHT20}~\label{app:pdfwmsht}

\begin{figure}[H]
\begin{center}
\includegraphics[scale=0.55]{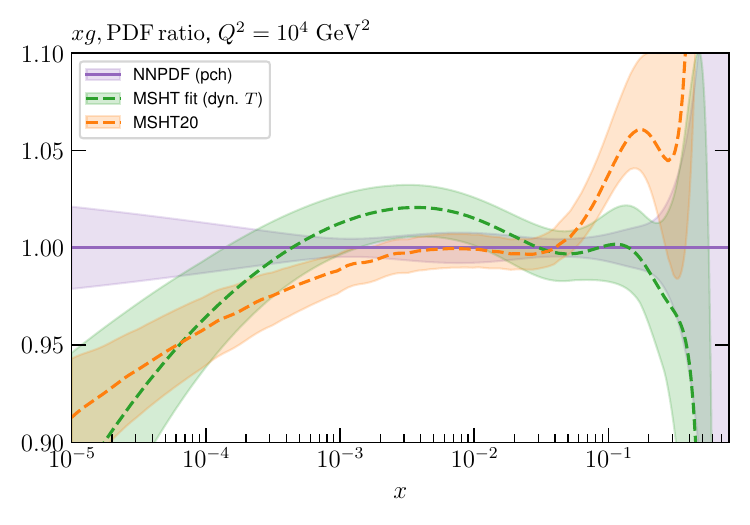}
\includegraphics[scale=0.55]{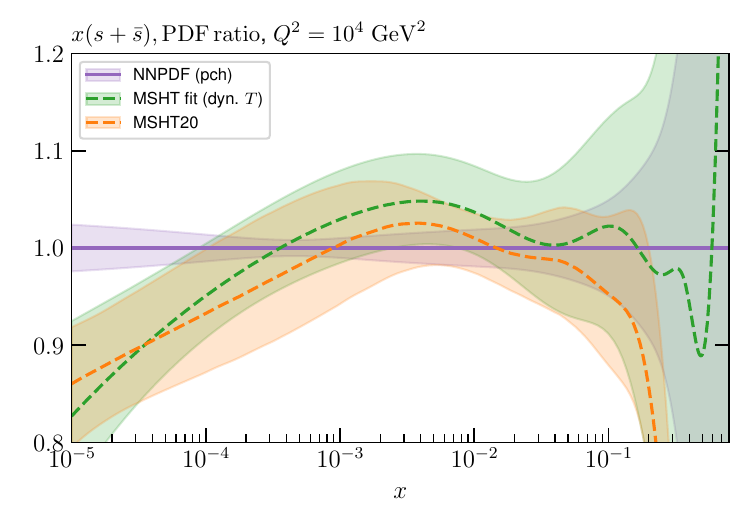}
\includegraphics[scale=0.55]{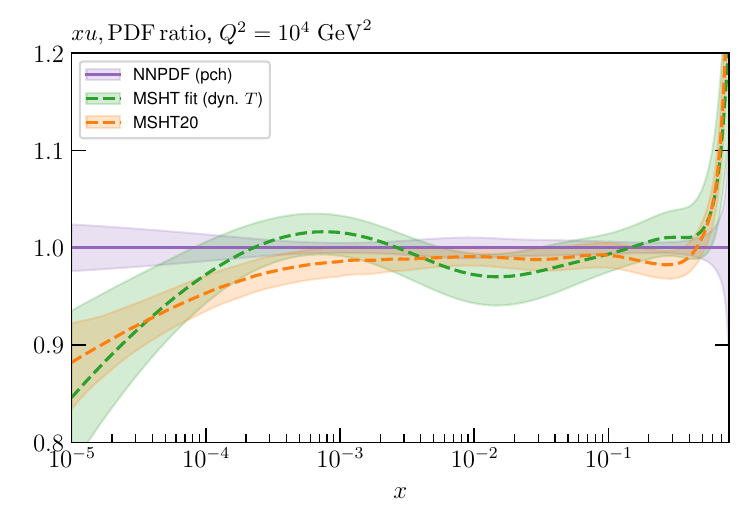}
\includegraphics[scale=0.55]{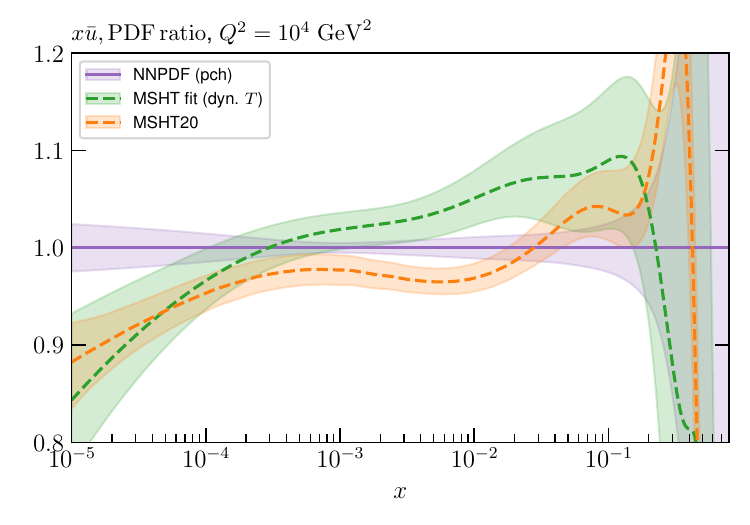}
\includegraphics[scale=0.55]{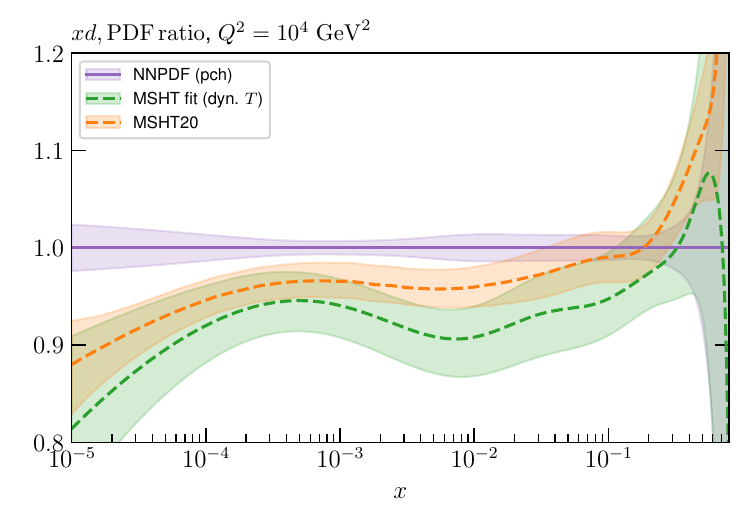}
\includegraphics[scale=0.55]{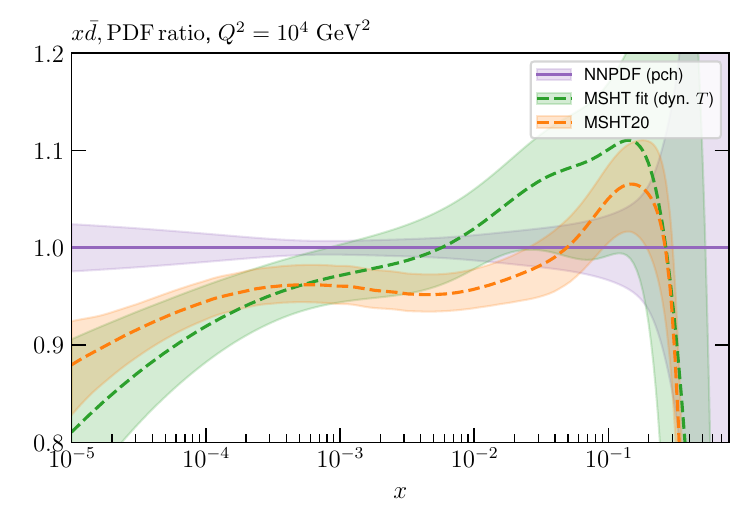}
\includegraphics[scale=0.55]{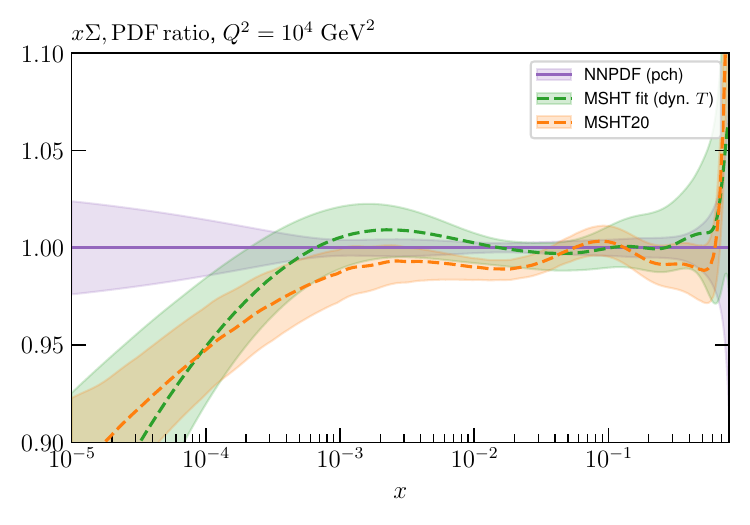}
\caption{\sf A selection of PDFs at $Q^2=10^4 \, {\rm GeV^2}$  that result from a  global PDF fit to the NNPDF4.0 dataset/theory (perturbative charm) setting, but using the MSHT20 parameterisation. This results of the MSHT fit with the dynamic tolerance criterium applied for the PDF uncertainty are shown, as well as the MSHT20  and the NNPDF (perturbative charm) PDFs, for comparison.}
\label{fig:fit_pch_dynT_wMSHT}
\end{center}
\end{figure}

In this appendix (Fig.~\ref{fig:fit_pch_dynT_wMSHT}) we compare the results of the public MSHT20 fit~\cite{Bailey:2020ooq} to the MSHT fit to the NNPDF4.0 data/theory settings, consistently with perturbative charm. The PDF parameterisation is therefore identical in the two cases, and the difference is due purely to the differing data and theory settings entering these fits. The NNPDF4.0 (perturbative charm) result is also shown for guidance.

As discussed further in the main body of the text, it is clear that there is strong statistical incompatibility between the two results if the textbook $T^2=1$ criterion is used. For the dynamic tolerance the compatibility is greatly improved, with no significant tension observed. On the other hand, even if these are broadly consistent within these uncertainties, there are differences between the underlying PDFs. We do not present a full analysis of this here, but note that for example: in the case of the high $x$ gluon, this is found~\cite{Bailey:2020ooq} to be rather sensitive to the precise choice and treatment of LHC data that are sensitive to this region, and which is known to be rather different between the NNPDF4.0 and MSHT20 fits; the difference in the down quark may be due in part to the differing treatments of deuteron corrections, which in NNPDF4.0 are by default centred on zero~\cite{Ball:2020xqw}, while being allowed to vary away from this in MSHT20~\cite{Bailey:2020ooq}; the treatment of the $D\to \mu$ branching ratio differs between the two groups, which enters into the neutrino--induced dimuon production cross section in DIS and has an important impact on the strangeness.

\begin{figure}[H]
\begin{center}
\includegraphics[scale=0.55]{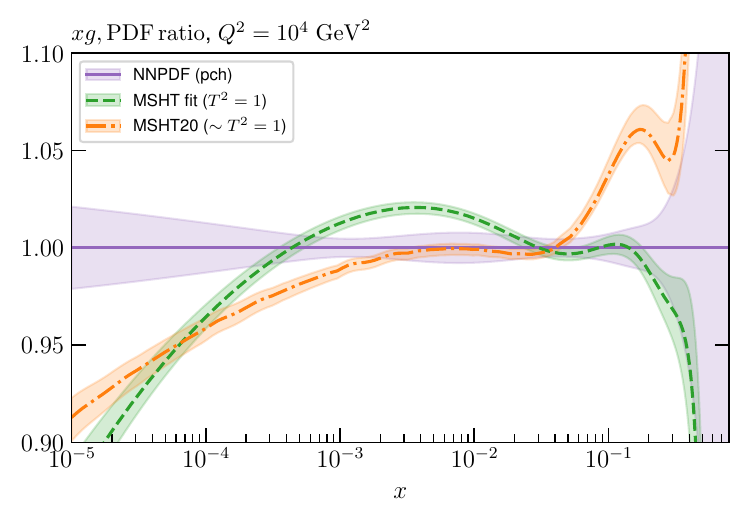}
\includegraphics[scale=0.55]{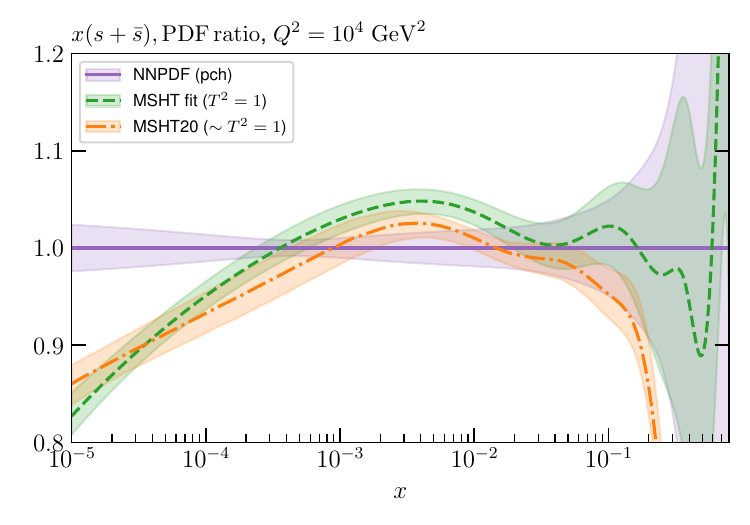}
\includegraphics[scale=0.55]{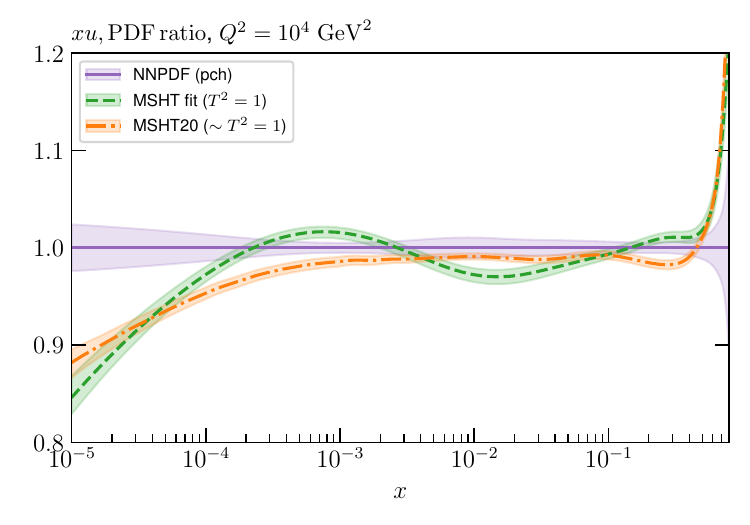}
\includegraphics[scale=0.55]{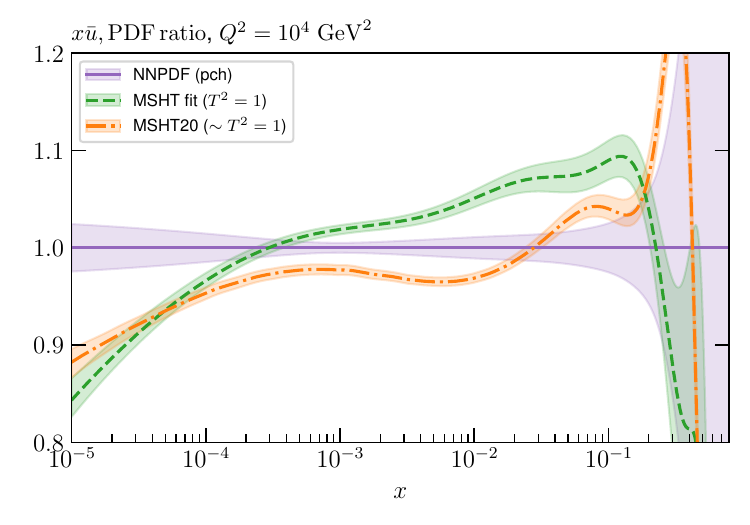}
\includegraphics[scale=0.55]{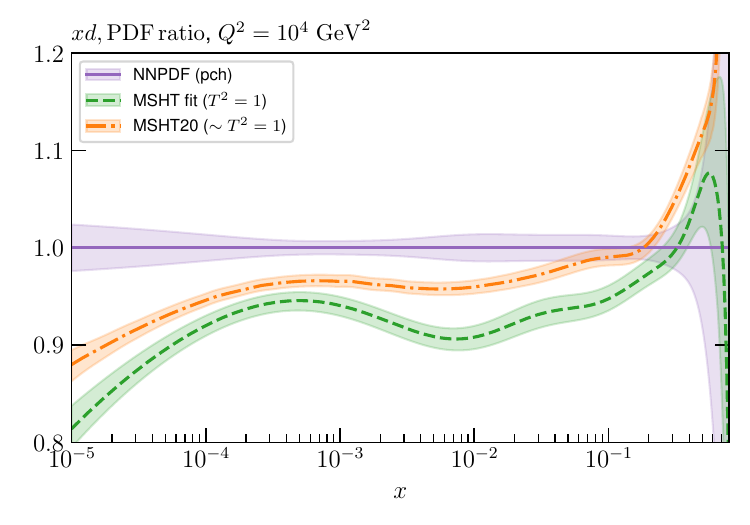}
\includegraphics[scale=0.55]{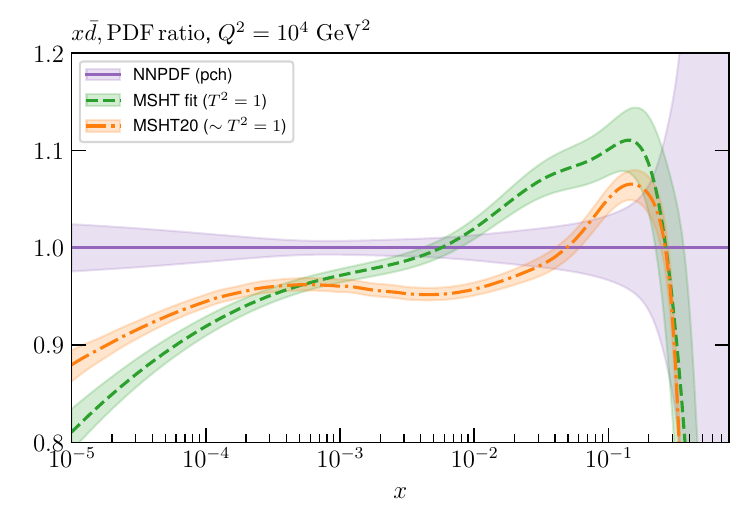}
\includegraphics[scale=0.55]{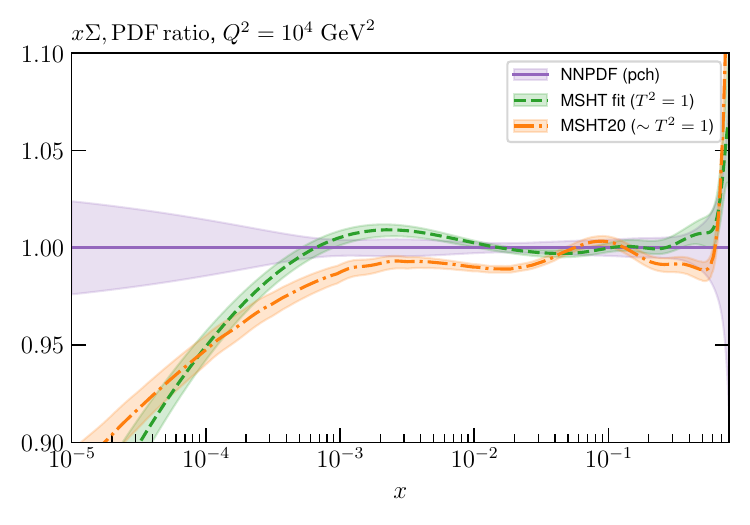}
\caption{\sf As in Fig.~\ref{fig:fit_pch_dynT_wMSHT} but the MSHT fit is shown with $T^2=1$ uncertainties, while the MSHT20 PDF uncertainties are divided by 3, approximately corresponding to $T^2=1$.}
\label{fig:fit_pch_Teq1_wMSHT}
\end{center}
\end{figure}

\subsection{MSHT Fits with Perturbative and Fitted Charm: Comparison}~\label{app:pdf_pch_vs_fch}

\begin{figure}[H]
\begin{center}
\includegraphics[scale=0.55]{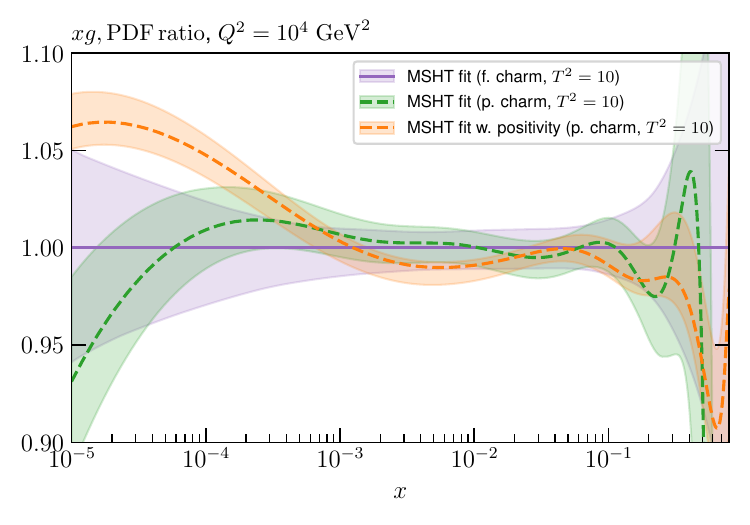}
\includegraphics[scale=0.55]{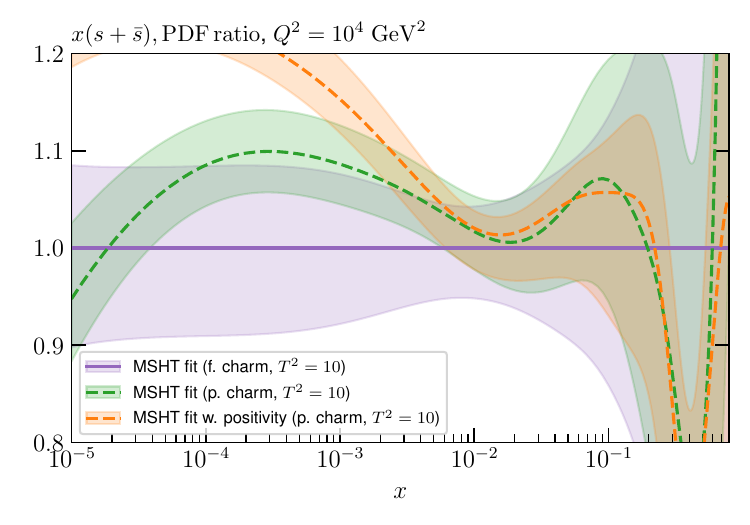}
\includegraphics[scale=0.55]{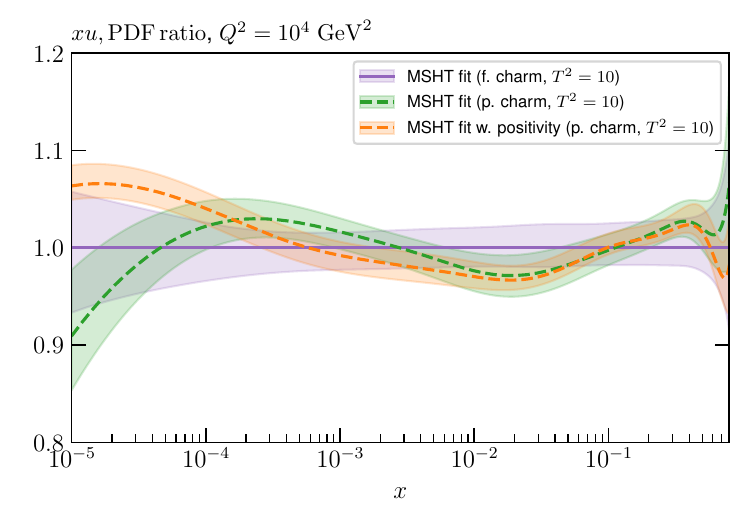}
\includegraphics[scale=0.55]{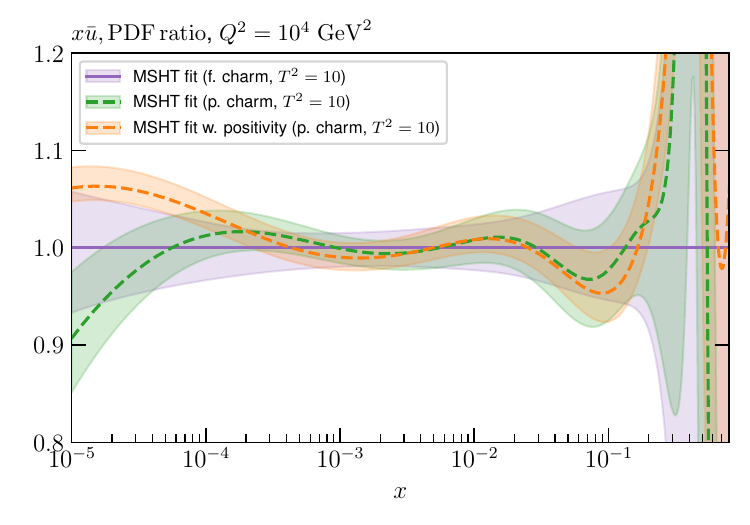}
\includegraphics[scale=0.55]{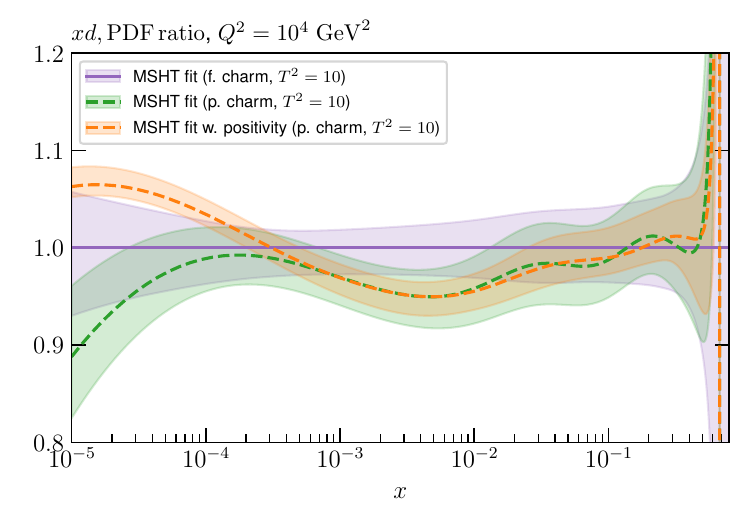}
\includegraphics[scale=0.55]{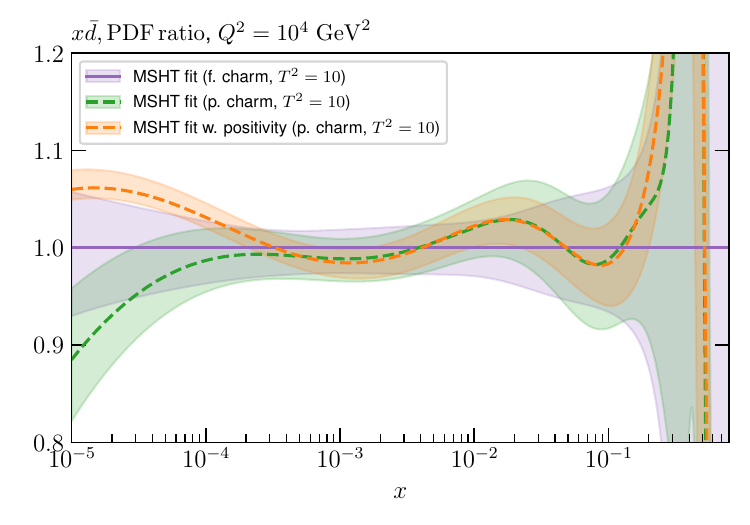}
\includegraphics[scale=0.55]{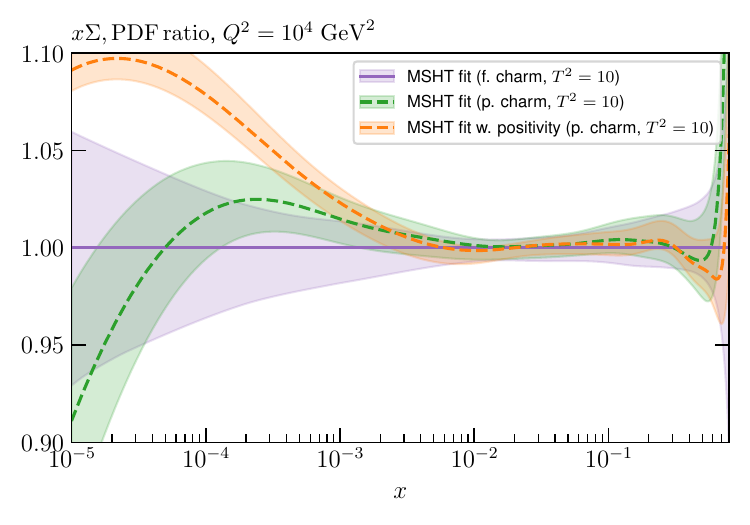}
\includegraphics[scale=0.55]{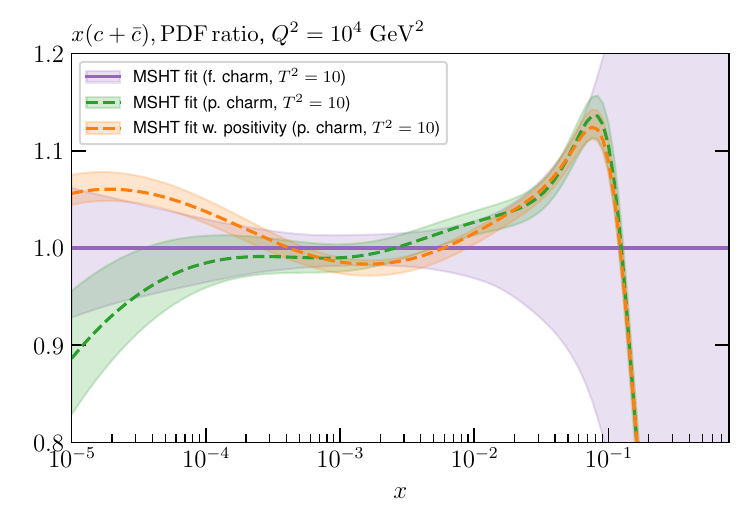}
\caption{\sf A comparison of the PDFs  at $Q^2=10^4 \, {\rm GeV^2}$  that result from  MSHT fits  to a  global PDF fit to the NNPDF4.0 dataset/theory with perturbative and fitted charm. In the perturbative charm case, results both with and without positivity imposed are shown, whereas for the fitted charm result only the case with positivity imposed are shown.}
\label{fig:fit_pch_vs_fch}
\end{center}
\end{figure}

In this appendix we compare directly the MSHT fits with perturbative charm (both with and without positivity imposed) and with fitted charm, as shown in Fig.~\ref{fig:fit_pch_vs_fch}.

\subsection{NNPDF Fits Without Positivity Imposed}\label{app:NNPDFnopos}

\begin{figure}[H]
\begin{center}
\includegraphics[scale=0.52]{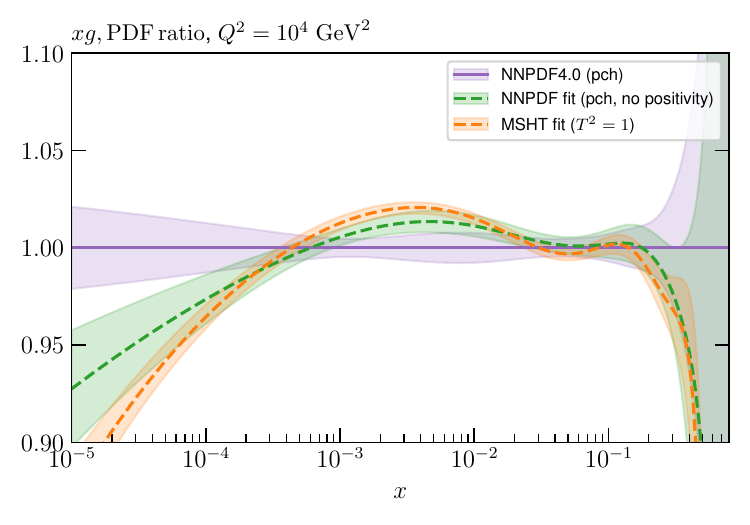}
\includegraphics[scale=0.52]{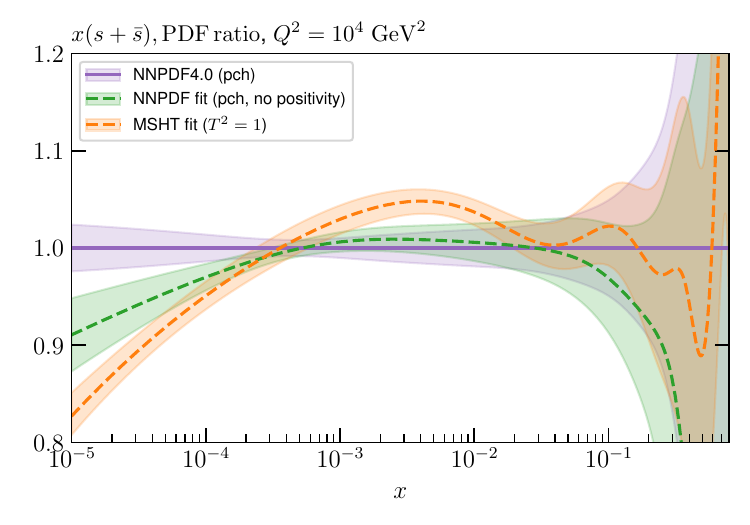}
\includegraphics[scale=0.52]{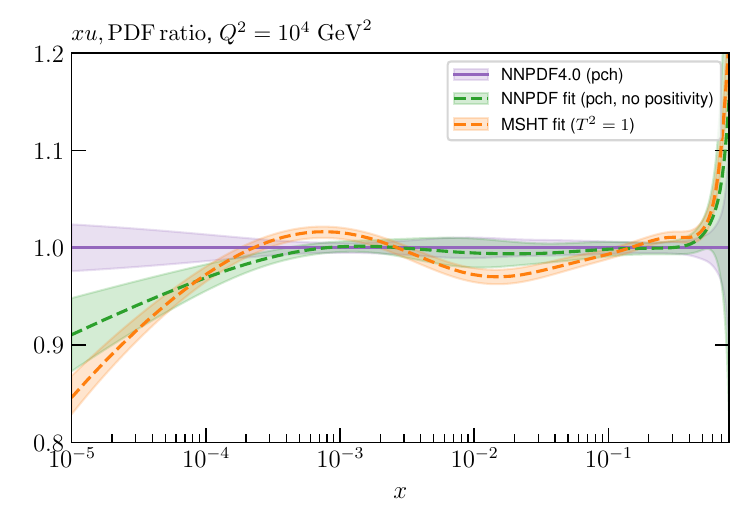}
\includegraphics[scale=0.52]{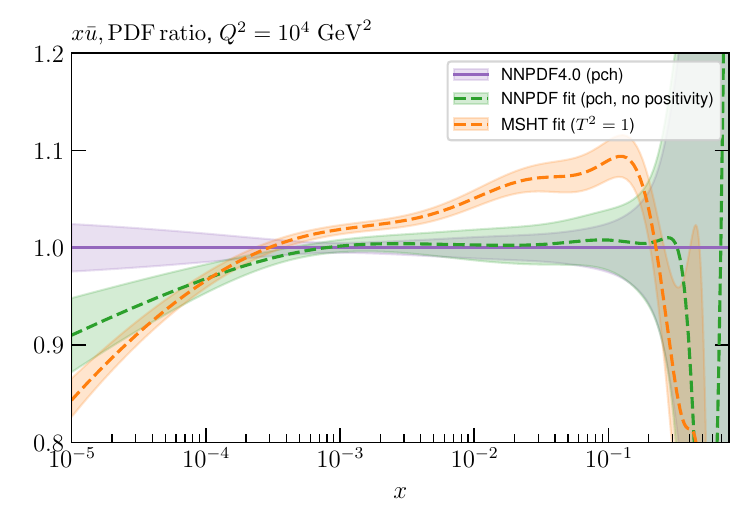}
\includegraphics[scale=0.52]{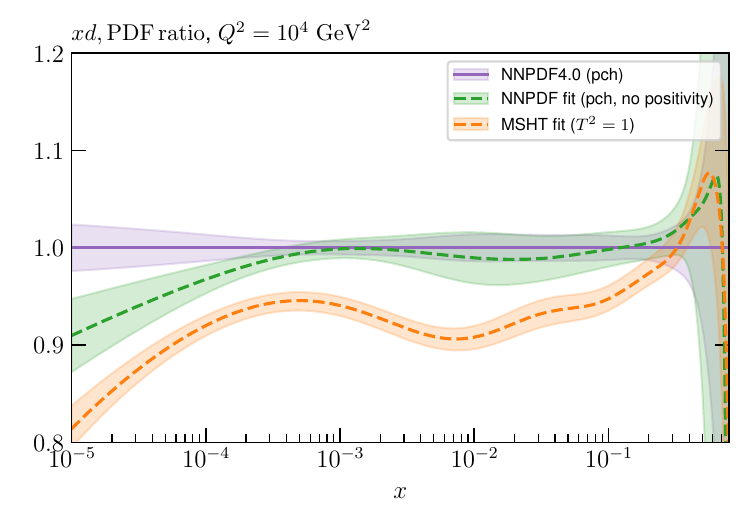}
\includegraphics[scale=0.52]{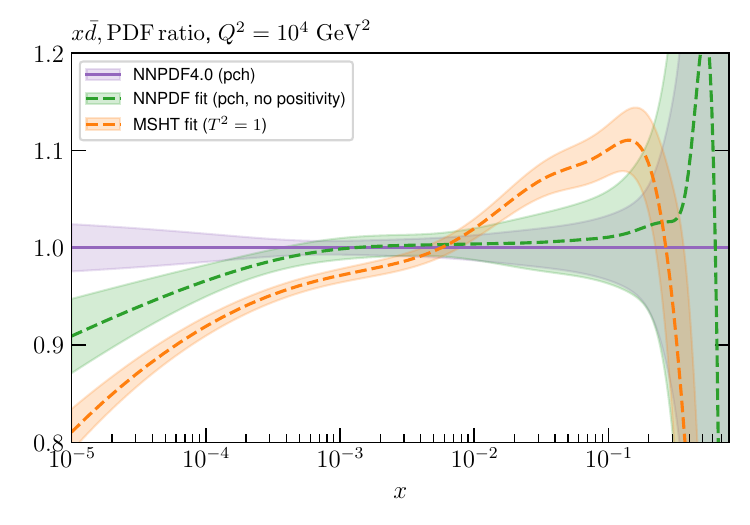}
\includegraphics[scale=0.52]{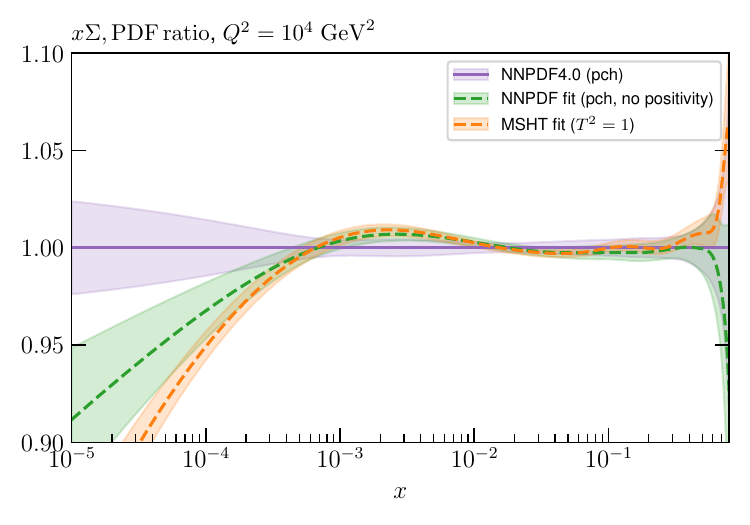}
\includegraphics[scale=0.52]{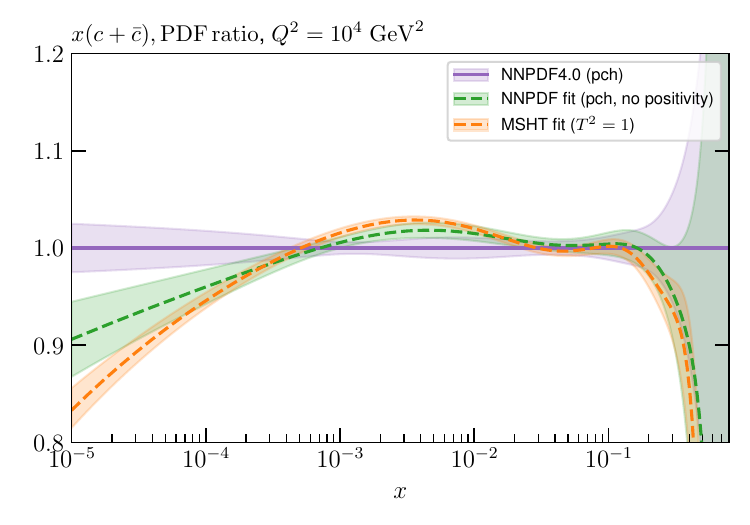}
\caption{\sf A selection of PDFs at $Q^2=10^4 \, {\rm GeV^2}$ due to the baseline NNPDF4.0 (perturbative charm) fit, a fit within the same NNPDF4.0 framework but without positivity imposed, as well as the MSHT fit   to the NNPDF4.0 dataset/theory with $T^2=1$ (and no positivity imposed).}
\label{fig:fit_NNpch_wpos}
\end{center}
\end{figure}

\begin{figure}[H]
\begin{center}
\includegraphics[scale=0.52]{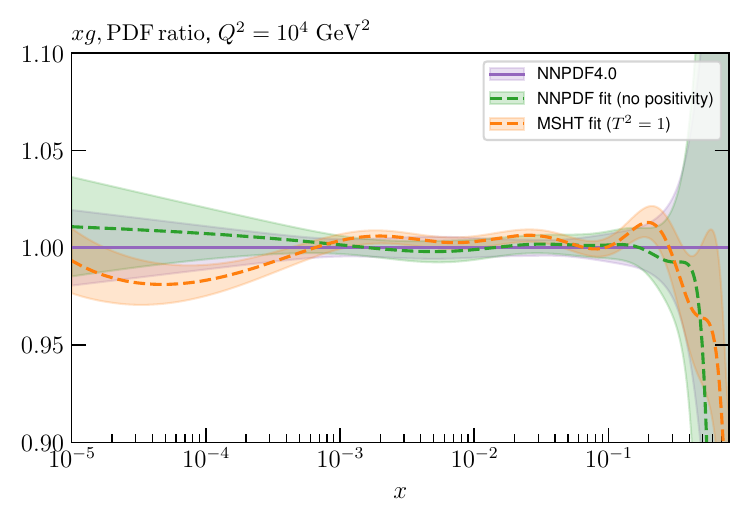}
\includegraphics[scale=0.52]{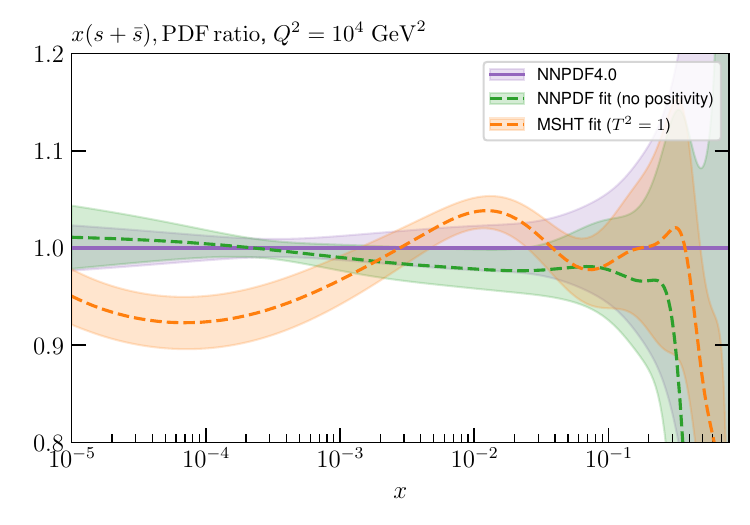}
\includegraphics[scale=0.52]{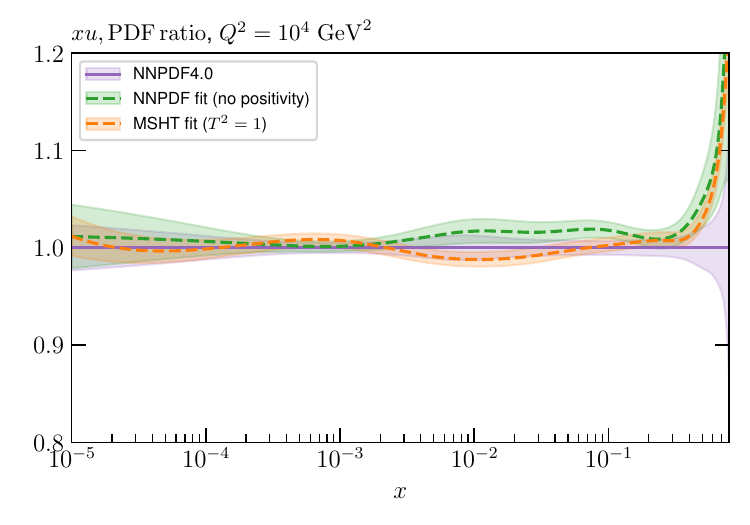}
\includegraphics[scale=0.52]{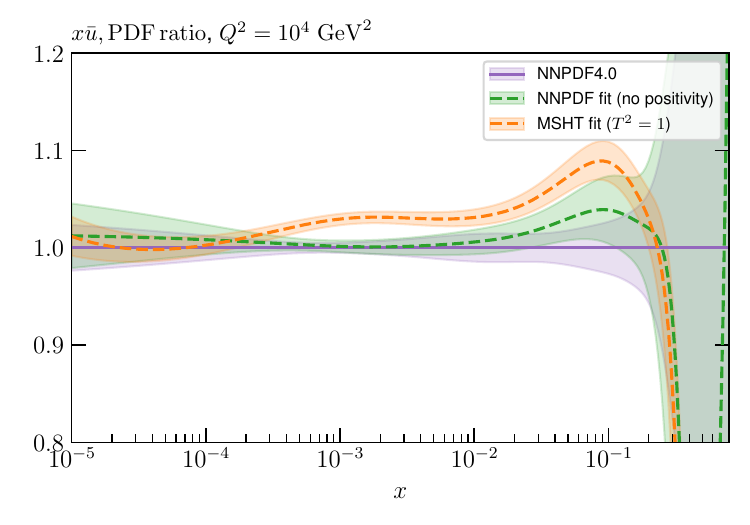}
\includegraphics[scale=0.52]{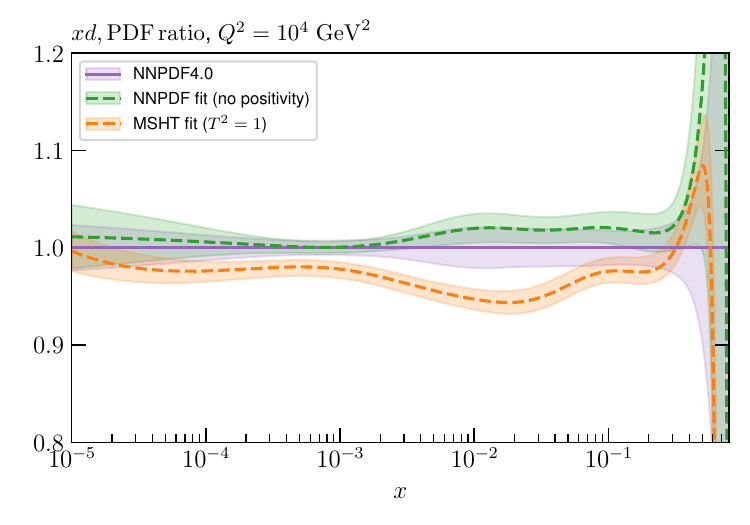}
\includegraphics[scale=0.52]{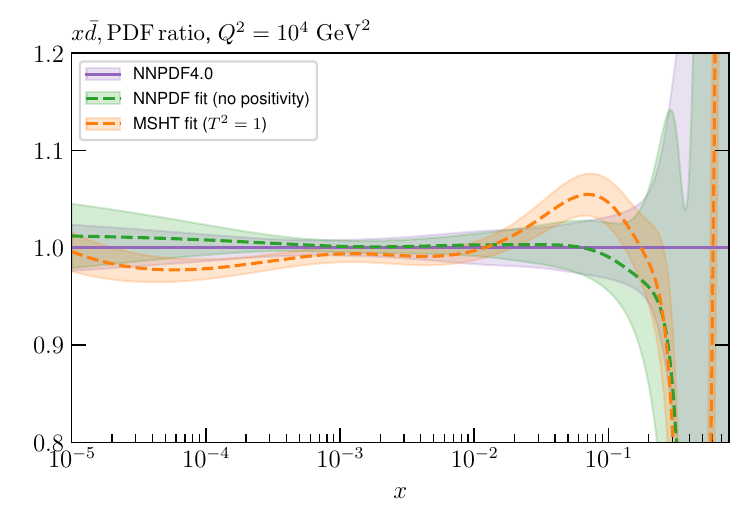}
\includegraphics[scale=0.52]{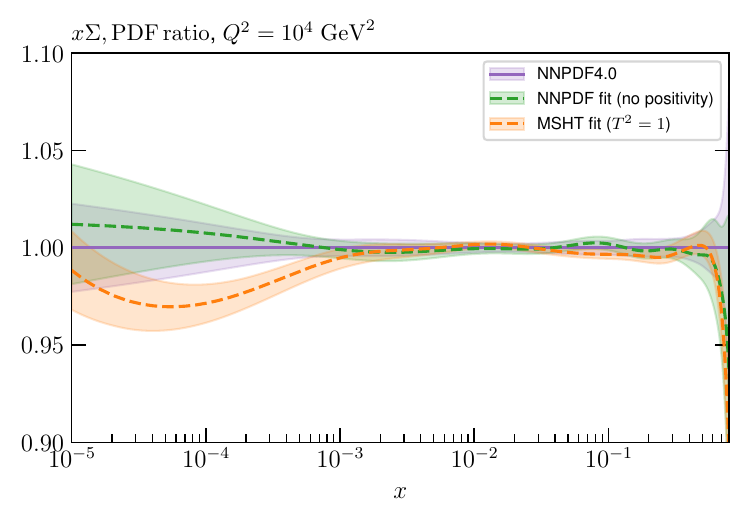}
\includegraphics[scale=0.52]{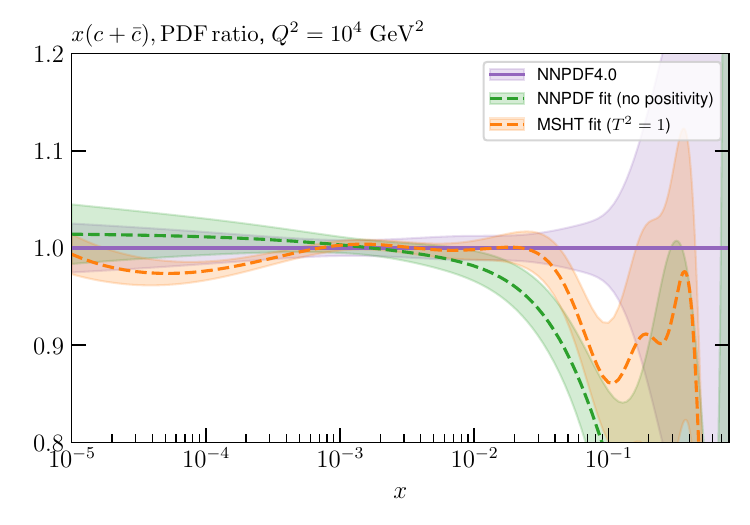}
\caption{\sf As in Fig.~\ref{fig:fit_NNpch_wpos} but for the fitted charm case.}
\label{fig:fit_NNfch_wpos}
\end{center}
\end{figure}

In this appendix we compare in Figs.~\ref{fig:fit_NNpch_wpos} and~\ref{fig:fit_NNfch_wpos} the baseline NNPDF4.0 fits with perturbative and fitted charm, respectively, to the case where a fit within the same NNPDF4.0 framework is performed, but positivity is not imposed. The MSHT fit, with $T^2=1$ and no positivity imposed, is shown for guidance.

\bibliography{references}{}
\bibliographystyle{h-physrev}

\end{document}